\documentclass[12pt]{article}

\usepackage{cite}
\usepackage{epsfig}
\usepackage{graphicx}
\usepackage{amsmath}
\usepackage{amssymb}
\usepackage{ulem}
\usepackage{mdwlist} 
\usepackage{color}  
\usepackage[OT2, T1]{fontenc}
\usepackage[russian,english]{babel}

\usepackage{a41}
\usepackage{color}
\usepackage[rflt]{floatflt}
\usepackage{float}
\usepackage{slashed}

\newcommand{\AXF}{{\rm Li}_5\left(\frac{1}{2}\right)}
\newcommand{\AXS}{{\rm Li}_7\left(\frac{1}{2}\right)}
\newcommand{\AXN}{{\rm Li}_9\left(\frac{1}{2}\right)}

\usepackage{array}

\usepackage[english]{babel}

\usepackage{url}

\usepackage{amsmath, amsthm, amssymb}
\newtheorem{thm}{Theorem}[section]

\def\asr{\left( \frac{\alpha_s}{4 \pi} \right)}
\def\asb{\left( \frac{\bar{\alpha}_s}{4 \pi} \right)}
\def\b0{\beta_0}

\newtheorem{definition}[thm]{Definition}

\newtheorem{remark}[thm]{Remark}

\newcommand{\lsim}{\raisebox{-0.07cm}
{$\, \stackrel{<}{{\scriptstyle\sim}}\, $}}

\renewcommand{\P}{\mathbb P}

\newcommand{\Li}{{\rm Li}}

\newcommand{\Dx}{D_x}
\newcommand{\KK}{\mathbb{K}}
\newcommand{\QQ}{\mathbb{Q}}
\newcommand{\NNN}{\mathbb{N}}
\newcommand{\ZZ}{\mathbb{Z}}

\newcommand{\ep}{\varepsilon}

\def\asr{\left( \frac{\alpha_s}{4 \pi} \right)}
\def\asb{\left( \frac{\bar{\alpha}_s}{4 \pi} \right)}
\def\b0{\beta_0}

\usepackage{rotating}

\usepackage{graphicx}

\newcounter{mmacnt}
\def\restartmma{\setcounter{mmacnt}{0}}
\restartmma \catcode`|=\active
\def|#1|{\mathrm{#1}}
\catcode`|=12
\newenvironment{mma}{
 \par\smallskip
 \catcode`|=\active
 \parskip=0pt\parindent=0pt 
 \small
 \def\In##1\\{%
\def\linebreak{\hfill\break\null\qquad}%
\refstepcounter{mmacnt}
\hangindent=2.5em\hangafter=0
\leavevmode
\llap{\tiny\sffamily n[\arabic{mmacnt}]:=\kern.5em}%
\mathversion{bold}\footnotesize$\displaystyle##1$\normalsize
\mathversion{normal}\par
 }%
 \def\Print##1\\{%
\def\linebreak{\hfill\break}%
\hangindent=2.5em\hangafter=0
\leavevmode ##1\par}%
 \def\Out##1\\{%
\def\linebreak{$\hfill\break\null\hfill$}%
\kern\abovedisplayskip\par
\hangindent=2.5em\hangafter=0
\leavevmode
\llap{\tiny\sffamily Out[\arabic{mmacnt}]=\kern.5em}
\footnotesize$\displaystyle##1$\normalsize\hfill\null\par
\kern\belowdisplayskip
 }%
 \def\Warning##1##2\\{%
\def\linebreak{\hfill\break}%
\hangindent=2.5em\hangafter=0
\leavevmode
{\scriptsize##1 : ##2}\par}%
}{%
 \par\smallskip
}

\usepackage{color}

\newenvironment{fshaded}{%
\MakeFramed {\FrameRestore}
}%
{\endMakeFramed}

\catcode`,\active

\catcode`\,12

\allowdisplaybreaks[4]

\begin{document}
\setlength{\baselineskip}{0.515cm}
\sloppy
\thispagestyle{empty}
\begin{flushleft}
DESY 19--095
\\
DO--TH 19/08\\
SAGEX-19-10\\ 
TIF-UNIMI-2019-13\\
July 2019\\
\end{flushleft}

\mbox{}
\vspace*{\fill}
\begin{center}

{\LARGE\bf  }

\vspace*{2mm}
{\LARGE\bf  The Heavy Fermion Contributions to the} 

\vspace*{3mm}
{\LARGE\bf Massive Three Loop Form Factors}

\vspace{3cm}
\large 
J.~Bl\"umlein$^a$,
P.~Marquard$^a$, 
N.~Rana$^{a,b}$, and
C.~Schneider$^c$

\vspace{1.cm}
\normalsize
{\it  $^a$ Deutsches Elektronen--Synchrotron, DESY,}\\
{\it  Platanenallee 6, D-15738 Zeuthen, Germany}

\vspace*{3mm}
{\it  $^b$ INFN, Sezione di Milano,}\\ 
{\it Via Celoria 16, I-20133 Milano, Italy}

\vspace*{3mm}
{\it $^c$~Research Institute for Symbolic Computation (RISC),\\
  Johannes Kepler University, Altenbergerstra{\ss}e 69,
  A--4040, Linz, Austria}\\


\end{center}
\normalsize
\vspace{\fill}
\begin{abstract}
\noindent
We compute the $n_h$ terms to the massive three loop vector-, axialvector-, scalar- and pseudoscalar form factors 
in a direct analytic calculation using the method of large moments. This method has the advantage, that the master 
integrals have to be dealt with only in their moment representation, allowing to also consider quantities which obey 
differential equations, which are not first order factorizable (elliptic and higher), already at this level. To obtain 
all the associated recursions, up to 8000 moments had to be calculated. A new technique has been applied to 
solve the associated differential equation systems. Here the decoupling is performed such, that only minimal depth 
$\ep$--expansions had to be performed for non--first-order factorizing systems, minimizing the calculation of 
initial 
values. The pole terms in the dimensional parameter $\ep$ can be completely predicted using renormalization 
group methods, as confirmed by the present results. A series of contributions at $O(\ep^0)$ have first order factorizable 
representations. For a smaller number of color--zeta projections this is not the case. All first order factorizing 
terms can be represented by harmonic polylogarithms. We also obtain analytic results for the non--first-order 
factorizing
terms by Taylor series in a variable $x$, for which we have calculated at least 2000 expansion coefficients, in an 
approximation. Based on this representation the form factors can be given in the Euclidean region and in the region 
$q^2 \approx 0$. Numerical results are presented.
\end{abstract}

\vspace*{\fill}
\noindent
\numberwithin{equation}{section}
\newpage 
\section{Introduction}
\label{sec:1}

\vspace*{1mm}
\noindent
The knowledge of the massive three--loop form factor is essential ingredient to the calculation for a series of 
massive
processes at $e^+e^-$ and hadron colliders, determined by  vector, axialvector, scalar and pseudoscalar currents. It 
has been calculated to two--loop order in Refs.~\cite{Bernreuther:2004ih,Bernreuther:2004th,Bernreuther:2005rw,
Bernreuther:2005gw,Gluza:2009yy,Ablinger:2017hst}. At three--loop order the color planar contributions have 
been 
computed in Refs.~\cite{Henn:2016tyf,Henn:2016kjz,Ablinger:2018yae,Lee:2018nxa,Lee:2018rgs,Ablinger:2018zwz} and its 
asymptotic behaviour has been studied in \cite{Ahmed:2017gyt,Blumlein:2018tmz}, including partial results at 
four--loop order.

In the present paper, we compute the $n_h$ contributions of the massive three--loop form factor for vector, 
axialvector, 
scalar and pseudoscalar currents. As the basic computational method we use the method of arbitrarily large moments  
\cite{Blumlein:2017dxp}. Here the differential equations given by the integration by parts (IBP) 
relations~\cite{Lagrange:IBP,Gauss:IBP, Green:IBP,Ostrogradski:IBP,Chetyrkin:1981qh,Laporta:2001dd,REDUZE,CRUSHER} are 
transformed into recursions, through which a large number of moments for the master integrals and the form factors are 
generated using the package {\tt SolveCoupledSystems} \cite{Blumlein:2017dxp}.
These moments are sequences in 
$\mathbb{Q}$ parameterized by multiple zeta values (MZVs) \cite{Blumlein:2009cf} and color factors. Using the method 
of guessing \cite{GSAGE}\footnote{For an early application of this method in perturbative calculations in Quantum 
Field Theory, cf.~\cite{Blumlein:2009tj}.} we determine the associated difference equations, which are finally solved 
using {\tt Sigma} \cite{Schneider:2007a,Schneider:2013a}. In the expansion of the form factors to master 
integrals usually higher order terms 
in $\ep = 2-D/2$ are 
contributing. These are containing, however, elliptic and more involved contributions. Although being present, these 
terms cannot be distinguished from the simpler contributions considering moments since they appear only 
encoded as rational 
numbers. The advantage of the present method is that it always allows to obtain difference equations for all MZV and 
color projections. In the case of the pole terms and a large number of contributions at $O(\ep^0)$ the corresponding
difference equations are first order factorizable and can therefore be solved by {\tt Sigma}. In case of the remaining 
terms we are able to factorize the first order factors. The remainder terms need other methods to be solved. 
The first order factorizable contributions are given by iterative integrals, cf.~\cite{Blumlein:2018cms}. In the 
present case these iterative integrals are harmonic polylogarithms (HPLs) \cite{Remiddi:1999ew}. 

The paper is organized as follows. After some basic definitions given in Section~\ref{sec:2} main steps of the 
calculation are described in Section~\ref{sec:3}. In Section~\ref{sec:4} the universal infrared structure of the
form factors is presented which is later compared with the unrenormalized three--loop form factors providing
a check of the calculation. In Section~\ref{sec:5} we describe a new decoupling strategy, which allows to work
with a minimal--depth expansion concerning the initial values. A brief summary on the recurrences, which are not 
factorizing to first order, is given in Section~\ref{sec:6}. In Section~\ref{sec:7} we present the analytic results 
for the $n_h$--contributions to the different form factors and also give numerical illustrations. Section~\ref{sec:8}
contains the conclusions. In the appendix we present a series of deeper $\ep$--expansions for some integrals 
defining
the initial conditions.
\section{The Form Factors}
\label{sec:2}

\vspace*{1mm}
\noindent
The basic structure of the massive form factors has been described in Ref.~\cite{Ablinger:2017hst} before.
We consider vector, axialvector, scalar and pseudoscalar currents coupling to a heavy quark pair of mass $m$
\begin{eqnarray}
\overline{u}_c(q_1) X_{cd} v_d(q_2),
\end{eqnarray}
with $q = q_1 + q_2$. The main variable considered is $x$ given by
\begin{eqnarray}
\frac{q^2}{m^2} = z = - \frac{(1-x)^2}{x}.
\end{eqnarray}
We work in $D = 4 - 2\ep$ dimensions.
For the treatment of $\gamma_5$ in the axialvector and pseudoscalar case we consider here only the non-singlet 
contributions, where $\gamma_5$ can be treated anticommuting.

We consider the decay amplitude ($\Gamma^{\mu}$) of the $Z$-boson
into a pair of heavy quarks.
The general structure of $\Gamma^{\mu}$ consists of six form factors,
two of which are CP odd. As we consider only higher order QCD effects
and Standard Model (SM) neutral current interactions to lowest order, the CP invariance holds.
This implies that $\Gamma^{\mu}$ has four form factors
$F_{V,i} (s), F_{A,i} (s) ~ i = 1,2$ comprising the following general form
\begin{align}
 \Gamma_{cd}^{\mu} &= \Gamma_{V,cd}^{\mu} + \Gamma_{A,cd}^{\mu}
 \nonumber\\
 &= -i \delta_{cd}\Big[ v_Q \Big( \gamma^{\mu}~F_{V,1} + \frac{i}{2 m} \sigma^{\mu\nu} q_{\nu} ~F_{V,2} \Big)
+ a_Q \Big( \gamma^{\mu} \gamma_5~F_{A,1}
         + \frac{1}{2 m} q^{\mu} \gamma_5 ~ F_{A,2}  \Big) \Big]
\end{align} 
where $\sigma^{\mu\nu} = \frac{i}{2} [\gamma^{\mu},\gamma^{\nu}]$, $q=q_1+q_2$, and $v_Q$ and $a_Q$ are the
SM vector and axial-vector coupling constants as defined by
\begin{equation}
 v_Q = \frac{e}{\sin \theta_w \cos \theta_w} \Big( \frac{T_3^Q}{2} - \sin^2 \theta_w Q_Q \Big) \,,
 \qquad
 a_Q = - \frac{e}{\sin \theta_w \cos \theta_w} \frac{T_3^Q}{2} \,.
\end{equation}
$e$ is the charge of positron, $\theta_w$ is the weak mixing angle, $T_3^Q$ is the third component
of the weak isospin, and $Q_Q$ is the charge of the heavy quark.

In case of the vector and axialvector form factors it is convenient 
to use their decomposition into two parts, respectively, which are labeled by the functions
\begin{align}
\label{eq:G1}
g_{V,1}^1 &= \frac{x}{4 (1-\ep) (1+x)^2} & g_{V,1}^2 &= \frac{(3-2 \ep) x^2}{(1-\ep) (1+x)^4}\\
g_{V,2}^1 &= \frac{x^2}{(1-\ep) (1-x^2)^2}
& g_{V,2}^2 &= \frac{2 x^2 [-1+\ep (1-x)^2+(4-x) x]}{(1-\ep) (1-x)^2 (1+x)^4}\\
g_{A,1}^1 &= \frac{x}{4 (1 - \ep) (1 + x)^2} & g_{A,1}^2 &= \frac{x^2}{(1 - \ep) (1 - x^2)^2}\\
\label{eq:G4}
g_{A,2}^1 &= \frac{x^2}{(1-\ep) (1-x^2)^2}   & g_{A,2}^2 &= \frac{2 x^2 [1-\ep (1+x)^2+x (4+x)]}{(1-\ep) 
(1-x)^4 (1+x)^2}.
\end{align}
We will use this decomposition throughout the present calculation. 

Furthermore, we consider a general neutral particle $h$ that couples to
heavy quarks through the following Yukawa interaction
\begin{equation}
 {\cal L}_{int} = - \frac{m}{v} \Big[ s_Q \bar{Q}Q + i p_Q \bar{Q} \gamma_5 Q  \Big] h,
\end{equation}
where $m$ denotes the heavy quark mass, $v = (\sqrt{2} G_F)^{-1/2}$ is the SM Higgs vacuum expectation value, with
$G_F$ being the Fermi constant, $s_Q$ and $p_Q$ are the scalar/pseudo-scalar coupling, respectively,
and $Q$ and $h$ are the heavy quark and scalar and pseudo-scalar field, respectively.
The decay amplitude of $h \rightarrow \bar{Q} + Q$, $X_{cd} \equiv \Gamma_{cd}$, consists of two form factors 
with the 
following
general structure
\begin{align}
 \Gamma_{cd} &= \Gamma_{S,cd} +  \Gamma_{P,cd}
 \nonumber\\
 &= - \frac{m}{v} \delta_{ij} ~ \Big[ s_Q \, F_{S} + i p_Q \gamma_5 \, F_{P} \Big] \,,
\end{align}
where $F_S$ and $F_P$ denote the renormalized scalar and pseudo-scalar form factors, respectively.
The form factors obey the expansion
\begin{align}
F_{i,l}(x,a_s) = \delta_{l,1} + \sum_{k=1}^\infty a_s^k F_i^{(k)}(x),
\end{align}
with $i = V,A,S,P$ and $l = 1,2$ for $i = A,V$ and $a_s = \alpha_s/(4\pi)$ denotes the strong coupling 
constant. 

\section{The Calculation}
\label{sec:3}

\vspace*{1mm}
\noindent
The diagrams of the $n_h$--contributions of the different massive three loop form factors are generated using {\tt QGRAF} 
\cite{Nogueira:1991ex} and the color structures are evaluated using {\tt Color} \cite{vanRitbergen:1998pn}. 
Furthermore, we 
used {\tt q2e/exp} \cite{Harlander:1997zb,Seidensticker:1999bb} and perform the Dirac-algebra using {\tt Form} 
\cite{Vermaseren:2000nd,Tentyukov:2007mu}. In the present paper we perform the calculation of the form factor for QCD, 
setting $C_A = N_c = 3, C_F = (N_c^2-1)/(2N_c) = 4/3$ and $T_F =1/2$, to reduce the complexity of the problem. The 
only free parameters are the numbers of equal mass heavy flavors $n_h$ and massless flavors $n_l$, which partly occur 
together with zeta-values. In the expansions, in which we mainly work, this decomposition is unique. By 
transforming 
to $x$--space additional $\zeta$--terms may occur e.g. due to regularizations. The IBP reduction is performed using the 
package {\tt Crusher} \cite{CRUSHER} and systems of linear ordinary differential equations are obtained for the master 
integrals. To the $n_h$ case 14 families with a total of 103 master integrals contribute. 
We map to the new variable $y$ 
\begin{eqnarray}
x = 1 - y
\end{eqnarray}
in which the master integrals obey the Taylor expansions
\begin{eqnarray}
M_k(\ep,x) = \sum_{l=0}^\infty \tilde{m}_{k,l}(\ep) y^l.
\end{eqnarray}
The systems of differential equations are thus mapped into associated systems of difference equations. To work with 
moments for the master integrals has the advantage that also general non--first order factorizing cases can be 
treated. 
The difference equations are now solved consecutively starting from the highest pole in $\ep$ 
working through to 
the required power in the dimensional parameter $\ep$. The way of decoupling is here very 
important, cf. \cite{Schneider:2019} and Section~\ref{sec:5} for 
details, since the required depth in expanding in $\ep$ for the associated initial values, may easily go beyond the level,
which is currently known. We have performed those extensions in a series of cases, see Appendix~\ref{sec:A}. In 
general those 
extensions are costly and time consuming. In the present case they could be avoided and the initial values are 
either known 
from the literature \cite{Melnikov:2000zc} or having been calculated for other previous applications, 
cf.~\cite{Marquard:2006qi,Marquard:2007uj,Marquard:2015qpa,Kurz:2016bau}, 
turned out to be sufficient. We could work with a minimal number of additional terms in the $\ep$-expansion, here in the 
non--first order factorizing case. 

For each color--zeta projection of the given massive form factor one obtains a series of rational numbers 
$m_{k,l,n}$, where $-n$ labels the power in $\ep$. We seek now a minimal recurrence determined by the set 
\begin{eqnarray}
\{m_{k,l,n}|l = 0..N_{\rm max}\}.
\end{eqnarray}

For each family we obtain a system of differential 
equations, the largest of which has a coefficient matrix of $7 \times 7$. 
The form factor can then be rewritten as
\begin{eqnarray}
\label{eq:F3exp}
F^{(3)}(x) = \sum_{k=3}^0 \frac{1}{\ep^k} F_{-k}^{(3)}(y),~~~~~~~F_{-k}^{(3)} (y) = \sum_{l=0}^\infty a_{-k,l} y^l.
\end{eqnarray}
Here the expansion coefficients $a_{-k,l}$ obey difference equations in $l$ for $k \in \{3,2,1,0\}$, 
which are parameterized
by polynomials of color factors and multiple zeta values (MZVs) \cite{Blumlein:2009cf} over $\mathbb Q$, 
irrespective of the fact 
whether in the general $l$ representation elliptic or higher structures contribute or not, see also 
\cite{Bierenbaum:2009mv,Ablinger:2017ptf}. 

Also the master integrals are rewritten in moment--form (\ref{eq:F3exp}). Their recurrences are finally used to calculate a large 
set of moments, assembled to $a_{k,l}$, using the method of large moments of Ref.~\cite{Blumlein:2017dxp}, which are projected 
to the different contributing monomials in MZVs and $n_l, n_h$. The number of moments, now given by sequences of rational numbers, 
needs to be large enough to allow the determination of their recurrence by the method of guessing \cite{GSAGE}, implemented in 
{\tt Sage} \cite{SAGE}, which has been successfully applied in different calculations in 
Refs.~\cite{Blumlein:2009tj,Blumlein:2017dxp} before.

In the case of the pole terms of the three--loop form factors all the corresponding recurrences have 
to be solvable in difference
rings 
\cite{Karr:1981,Bron:00,Schneider:01,Schneider:04a,Schneider:05a,Schneider:05b,Schneider:07d,Schneider:10b,Schneider:10c,
Schneider:15a,Schneider:08c,Schneider:08d,Schneider:08e,Schneider:2007a,Schneider:2013a},
since they are expected to factorize at first order. The solution can therefore be found using the package {\tt 
Sigma} \cite{Schneider:2007a,Schneider:2013a}. This may be as well the case for some of the MZV-factors contributing to the constant 
term, as in the 
case for the massive operator matrix element $A_{Qg}^{(3)}$, cf. Ref.~\cite{Blumlein:2017dxp}.

As result one obtains representations of $a_{k,l}$ in terms of harmonic sums \cite{Vermaseren:1998uu,Blumlein:1998if}
and generalized harmonic sums \cite{Moch:2001zr,Ablinger:2013cf} in $l$. The latter ones occur because of the 
necessary 
transformation $x \rightarrow (1-y)$ to also deal with the logarithmic contributions $\propto \ln^m(x)$. 
Cyclotomic or finite binomial sums
\cite{Ablinger:2011te,Ablinger:2014bra} do not contribute.

The infinite sums appearing in (\ref{eq:F3exp}) can now be performed using a series of procedures of the package
{\tt HarmonicSums} \cite{Vermaseren:1998uu,Blumlein:1998if,
Ablinger:2014rba, Ablinger:2010kw, Ablinger:2013hcp, Ablinger:2011te, Ablinger:2013cf, Ablinger:2014bra,Ablinger:2017Mellin}
and one obtains harmonic polylogarithms (HPLs) \cite{Remiddi:1999ew} or Kummer--Poincar\'e iterated integrals 
\cite{KUMMER,POINCARE,Moch:2001zr,Ablinger:2013cf} in the variable $y$. The transformation $y = 1 - x$ then yields representations 
in terms of HPLs in $x$, which are further reduced to suitable bases, to reduce the numbers of contributing functions as much as 
possible.

We remark, that the different master integrals for the present representation partly need deep expansion in the dimensional variable 
$\varepsilon$. If we needed to write them in real terms, also elliptic and even higher integral representations would be needed
in explicit form. It is an advantage of the present method that these contributions do earliest show up in the 
factorization of some of the recursions to be solved for the physical quantity under consideration, but cancel 
otherwise.  The automated
solution of differential equations over general bases, presented in Ref.~\cite{Ablinger:2018zwz}, working in 
the case of first order 
factorization, can therefore not be applied here.

In the following we discuss a sample calculation to illustrate the general method. For all computation we used 
the {\tt qftquad}--cluster equipped with Xeon Gold 6128 and Xeon 6C E5-2643v4 processors. 
For this we choose all contributions to the
massive three--loop vector form factors $\propto n_l$. They have been calculated up to $O(\varepsilon^0)$ by using different 
methods in Refs.~\cite{Ablinger:2018zwz,Lee:2018nxa}. By setting $N_c = 3$,~60 different recurrences have to be found and solved.
A linear combination of these quantities yields the vector current form factors $F_{V,1}$ and $F_{V,2}$. The reduction to master 
integrals led  to maximally  $5 \times 5$ systems, which were decoupled by using the Gauss approach 
implemented in the package {\tt Oresys} \cite{ORESYS} with a decoupling time of 3.1 min.
28 integrals contribute. The largest depth of the initial values to be provided were 10 
moments with an $\ep$-expansion up to $\ep^{3}$.

The most demanding terms were $g_1 n_l$ and $g_2 n_l$. All other contributions are significantly simpler. For $g_1 n_l$ 3000 
generated moments were not enough to determine the associated recurrence. Therefore we generated 4000 moments, which took
$12.88$ days instead of $9.96$ days. The guessing time amounted to 16.1 min and 23.2 min and led to recurrences of 
degree and order ${\sf (d = 474, o = 22), (d = 537, o = 35)}$ for the two largest cases, respectively.
Their solution using {\tt Sigma} and representation in terms of HPLs in the variable $x$ took $3.28$ days.
In the representation in the variable $y$ 13 harmonic sums and 185 generalized harmonic sums, 
transcendent to each other, contributed. In $x$--space we obtain representations in terms of 55 HPLs. The results agree with 
those given in Refs.~\cite{Ablinger:2018zwz,Lee:2018nxa}.

The needed computational times of the present example let it appear feasible to compute the pole terms and some of the terms of 
$\ep^0$ of the massive three--loop form factor. Here a wider range of initial values, to high order in $\varepsilon$, is needed.
We therefore had to extend the results given in \cite{Melnikov:2000zc} significantly.
\section{The universal infrared structure of QCD amplitudes}
\label{sec:4}

\vspace*{1mm}
\noindent
Scattering amplitudes in perturbative QCD contain infrared singularities arising from soft gluon contributions and 
collinear parton divergences. In this respect much work has been performed for  massless scattering amplitudes 
\cite{COLLINS,Catani:1998bh,Sterman:2002qn,Becher:2009cu,Gardi:2009qi,Ravindran:2004mb}. Especially in the case of 
massless 
QCD amplitudes with two partons, i.e. the form factors, the infrared (IR) structure becomes interesting due to its 
prominent form in terms of anomalous dimensions. The interplay of the collinear and soft anomalous dimensions to 
shape the singular structure of the massless form factors was noticed in \cite{Ravindran:2004mb} at two--loop order 
and later established at three--loop order in \cite{Moch:2005tm}.

In the case of massive form factors, the finding of a Sudakov type integro-differential equation was a challenge as 
the massive form factors do not exponentiate. However, in the asymptotic limit i.e in the limit where the quark 
mass
is small compared to the center of mass energy, the massless QCD corrections to the massive form factor do 
exponentiate.
In \cite{Mitov:2006xs}, the first step was taken by obtaining the singular behavior of massive QCD amplitudes 
in the asymptotic limit. Meanwhile, a factorization theorem was also proposed in \cite{Penin:2005eh, Becher:2007cu}, 
also in the asymptotic limit.
Recently, following the method proposed for massless form factors in \cite{Ravindran:2005vv, Ravindran:2006cg}, 
a rigorous study has been performed in \cite{Blumlein:2018tmz} in the asymptotic limit
to obtain all the poles and also all logarithmic contributions to finite pieces of the three loop heavy quark form factors
for vector, axial-vector, scalar and pseudo-scalar currents. In this scenario, one can relate the massless form factors 
to the massive ones and hence use the massless results \cite{Ahmed:2014cla, Ahmed:2014cha} to obtain these predictions.

A general IR structure is needed for the exact computation, which
was obtained in \cite{Becher:2009kw}, following a soft--collinear effective theory
(SCET) approach at two-loop. However, the argument can be extended to three--loop 
appropriately. The IR singularities of the massive form factors can be factorized as
a multiplicative renormalization factor, whose structure is
constrained by the renormalization group equation (RGE), as follows,

\begin{equation}
 F_{I} (\alpha_s, x, \ep)  = Z (\alpha_s, x, \ep, \mu) F_{I}^{\mathrm{fin}} ( \alpha_s, x, \ep, \mu)\, ,
\end{equation}
where $F_{I}^{\mathrm{fin}}$ is finite as $\ep \rightarrow 0$. 
We note that $Z$ does not carry any process dependent information $(I)$.
Here $\mu$ is the scale introduced corresponding to this particular factorization.
Now one can write down the renormalization group equation (RGE) for $Z$ which is characterized
by the massive cusp anomalous dimension, $\Gamma$. However, before we proceed, 
we note that $\Gamma$ does not contain any contributions from internal heavy quark loops, i.e. like the QCD $\beta$ function,
or light quark mass anomalous dimension, the massive cusp anomalous dimension also has been computed considering 
the massless QCD corrections with $n_l$ light quark flavors. 
On the other hand, the form factors are defined for ($n_l+1$) flavors. 
Hence, ${\Gamma}$ cannot describe the singularities arising in case of a massive quark loop contributions 
to the heavy quark form factors. To overcome the hurdle, the immediate solution is to use the 
decoupling relations 
\cite{Weinberg:1980wa,Ovrut:1980dg,Wetzel:1981qg,Bernreuther:1981sg,Bernreuther:1983zp,Chetyrkin:1997un}.
To obtain these decoupling relations, one constructs an effective theory with $n_l$ light quark flavors 
and then demands consistency with the full theory of $n_l+n_h$ flavors
by relating the couplings and light quark masses in the two cases. 
For our case, we need the decoupling relation for the strong coupling constant i.e.
the relation between $\bar\alpha_s$ and $\alpha_s$, where $\bar\alpha_s$ is defined 
for an effective theory with $n_l$ light quark only and $\alpha_s$ is defined for the full 
theory with $n_l+n_h$ quark flavors.

Keeping this in mind, we now write down the RGE for $\bar Z$, the equivalent of $Z$ in the effective theory 
with $n_l$ light quark,
which reads
\begin{equation} \label{eq:rgeZ}
 \frac{d}{d \ln \mu} \ln \bar Z( \alpha_s, x, \ep, \mu)  = - \Gamma ( \alpha_s, x, \mu) \,.
\end{equation}
$\Gamma$ is by now available up to the three--loop level
\cite{Korchemsky:1987wg,Korchemsky:1991zp,Grozin:2014hna,Grozin:2015kna}.
Both $\bar Z$ and $\Gamma$ can be expanded in a perturbative series in
$\alpha_s$ as follows
\begin{equation}
 \bar Z = \sum_{n=0}^{\infty} \asr^n \bar Z^{(n)} \,, \qquad
 \Gamma = \sum_{n=0}^{\infty} \asr^{n+1} \Gamma_{n} \,.
\end{equation}
Next, we solve the RGE, Eq.~(\ref{eq:rgeZ}), in massless QCD,
i.e. considering only $n_l$ light quarks. 
\begin{align} \label{eq:solnZ}
 \bar{Z} &= 1 + \asb \Bigg[ \frac{\Gamma_0}{2 \ep} \Bigg] 
   + \asb^2 \Bigg[ \frac{1}{\ep^2} \Big( \frac{\Gamma_0^2}{8} - \frac{\bar{\beta}_0 \Gamma_0}{4} \Big) + \frac{\Gamma_1}{4 \ep} \Bigg] 
   \nonumber\\
  &+ \asb^3 \bigg[ \frac{1}{\ep^3} \left( \frac{\Gamma_0^3}{48} - \frac{\bar{\beta}_0 \Gamma_0^2}{8} + \frac{\bar{\beta}_0^2 \Gamma_0}{6} \right)
                 + \frac{1}{\ep^2} \left( \frac{\Gamma_0 \Gamma_1}{8} - \frac{\bar{\beta}_1 \Gamma_0+\bar{\beta}_0 \Gamma_1}{6} \right) 
                 + \frac{1}{\ep} \left( \frac{\Gamma_2}{6} \right) \bigg]
   + {\cal O} (\bar{\alpha}_s^4) \,.
\end{align}
$\bar\beta$ is the QCD $\beta$ function for $n_l$ light quark flavors.
Now to obtain $Z$ from $\bar Z$. We use the following decoupling relation obtained using 
the background field method \cite{Chetyrkin:1997un,Grozin:2007fh,Grozin:2011nk}
to obtain the relation between $\bar \alpha_s$ and $\alpha_s$
\begin{equation}
\bar{\alpha}_s = \alpha_s \Big[ 1 - \asr \Big( \frac{2}{3} \zeta_2 n_h T_F \ep + {\cal O}(\ep^2) \Big) 
                 + \asr^2 \Big( \frac{32}{9} C_A T_F n_h - 15 C_F T_F n_h  + {\cal O}(\ep) \Big) + {\cal 
O}(\alpha_s^3)  \Big] 
\end{equation}
and obtain
\begin{align} \label{eq:solnzZ}
 Z &= \bar{Z} \big|_{\bar{\alpha}_s \rightarrow \alpha_s} 
\nonumber\\ 
 &= 1 + \asr \Bigg[ \frac{\Gamma_0}{2 \ep} \Bigg] 
   + \asr^2 \Bigg[ \frac{1}{\ep^2} \Big( \frac{\Gamma_0^2}{8} - \frac{\bar{\beta}_0 \Gamma_0}{4} \Big) + 
\frac{\Gamma_1}{4 \ep} \Bigg] 
   \nonumber\\
  &+ \asr^3 \bigg[ \frac{1}{\ep^3} \left( \frac{\Gamma_0^3}{48} - \frac{\bar{\beta}_0 \Gamma_0^2}{8} + 
\frac{\bar{\beta}_0^2 \Gamma_0}{6} \right)
                 + \frac{1}{\ep^2} \left( \frac{\Gamma_0 \Gamma_1}{8} - \frac{\bar{\beta}_1 \Gamma_0 + 
\bar{\beta}_0 \Gamma_1}{6} \right) 
\nonumber\\&                 
                 + \frac{1}{\ep} \left( \frac{\Gamma_2}{6} + 2 \Big( \frac{\Gamma_0^2}{8} - \frac{\bar{\beta}_0 
\Gamma_0}{4} \Big) \Big( -\frac{2}{3} \zeta_2 n_h T_F
\Big)
                 + \frac{\Gamma_0}{2} \Big( \frac{32}{9} C_A T_F n_h - 15 C_F T_F n_h \Big)  \right) \bigg]
   + {\cal O} (\bar{\alpha}_s^4) \,.
\end{align}
The above representation leads to the prediction of the pole terms for all the massive form factors to three--loop 
order, which has been an open problem in Ref.~\cite{Mitov:2006xs}.

\section{Refined versions of the large moment method}
\label{sec:5}

\vspace*{1mm}
\noindent
We will now describe a general toolbox that enables one to calculate large numbers of moments in the integer 
variable, say 
$n=0,1,2,\dots,\mu$, for a finite number of Feynman integrals $F_i(n,\ep)$ with $1\leq i\leq\lambda$. Here the moments 
$F_i(n,\ep)$ depend also on the dimensional parameter $\ep$ and the corresponding $\ep$-expansion 
\begin{equation}\label{Equ:FiEpMoments}
F_i(n,\ep)=\sum_{k=l}^{r_i}F_{i,k}(n)\,\ep^k+O(\ep^{r_{i}+1})
\end{equation}
is calculated  for each moment $n=0,\dots,\mu$ up to the order $r_i$.
Standard procedures like {\tt Mincer} \cite{Larin:1991fz} or {\tt MATAD} \cite{Steinhauser:2000ry} allow 
the calculation of a 
comparable small number of moments, e.g., $\mu \lsim 50$. The idea to calculate expansions 
from differential equations has already frequently been used, see 
e.g.~\cite{Boughezal:2006uu,Maier:2007yn,Maier:2011jd,Maier:2015jxa,Maierhofer:2012vv,Lee:2017qql,Kudashkin:2017skd}.
Only recently, we obtained a new method in~\cite{Blumlein:2017dxp} 
that can compute thousands of such moments.
In general, this method assumes that their (formal) power series representations
\begin{equation}\label{Equ:FiEpxMoments}
f_i(x,\ep)=\sum_{n=0}^{\infty}F_i(n,\ep) x^n
\end{equation}
are a solution of a given coupled system
\begin{eqnarray}
\label{eq:DEQEp}
\hspace*{2cm}\Dx 
\left(
\begin{matrix}
f_1(x,\ep)\\ \vdots \\ f_{\lambda}(x,\ep)\end{matrix}\right) 
= A(x,\ep)
\left(\begin{matrix} f_1(x,\ep)\\ \vdots \\ f_{\lambda}(x,\ep)\end{matrix}\right) 
+ \left(\begin{matrix} g_1(x,\ep)\\ \vdots \\ g_{\lambda}(x,\ep)\end{matrix}\right),
\end{eqnarray}
with $A(x,\ep)$ being an invertible $\lambda\times\lambda$ matrix with entries from the polynomial 
ring\footnote{We suppose that $\KK$ is a computable field containing the rational numbers $\QQ$ as a sub--field.} $\KK[x,\ep]$ 
and where the inhomogeneous parts $g_i(x,\ep)$ are given in terms of linear combinations of simpler master integrals. Here we 
assume that their moments $G_{i,k}(n)$, $n=1,\dots,\mu$, with
\begin{equation}\label{Equ:GiDef}
\begin{split}
g_i(x,\ep)&=\sum_{n=0}^{\infty}G_i(x,\ep)x^n,\\
G_i(n,\ep)&=\sum_{k=l}^{r'_i}G_{i,k}(n)\,ep^k+O(\ep^{r'_{i}+1})
\end{split}
\end{equation}
can be determined by 
\begin{itemize}
\item other coupled systems to which the large moment method under consideration is applied recursively;
\item symbolic summation or integration methods~\cite{Schneider:2007a,Schneider:2013a,Ablinger:2015tua} that yield 
representations in terms of indefinite nested sums or integrals from which one can produce a large number of moments;
\item by standard procedures like {\tt Mincer} \cite{Larin:1991fz} or {\tt MATAD} \cite{Steinhauser:2000ry} 
if only a small 
number of moments contributes.
\end{itemize}
Summarizing, it is assumed that already $\mu+1$ moments for the inhomogeneous parts $g_i(x,\ep)$ in~\eqref{eq:DEQEp} are 
computed. Then given such an input, we propose the following strategy to compute the first $\mu+1$ moments for 
$f_1(x,\ep),\dots,f_{\lambda}(x,\ep)$.

\medskip

\noindent\textbf{Strategy 1:} 
\begin{enumerate}
\item Uncouple the system~\eqref{eq:DEQEp}. Experiments showed that the implementation of Gauss's elimination method in 
\texttt{OreSys}~\cite{ORESYS} is an excellent choice. In general one obtains $k$ linear differential equations of the form
\begin{equation}\label{Equ:UncoupledDEQ}
\begin{split}
\sum_{i=0}^{o_1}b_{1,i}(x,\ep)\Dx^{i} f_1(x,\ep)&=\sum_{i,j}d_{1,i,j}(x,\ep)\Dx^ig_j(x,\ep)\\
\vdots&\\
\sum_{i=0}^{o_k}b_{k,i}(x,\ep)\Dx^{i} f_{k}(x,\ep)&=\sum_{i,j}d_{k,i,j}(x,\ep)\Dx^ig_j(x,\ep)+\sum_{j=1}^{k-1}
\sum_{i}e_{k,i,j}(x,\ep)\Dx^if_j(x,\ep)\\
\vdots&\\
\sum_{i=0}^{o_{\lambda}}b_{\lambda,i}(x,\ep)\Dx^{i} f_{\lambda}(x,\ep)&=\sum_{i,j}d_{\lambda,i,j}(x,\ep)
\Dx^ig_j(x,\ep)+\sum_{j=1}^{\lambda-1}\sum_{i}e_{\lambda,i,j}(x,\ep)\Dx^if_j(x,\ep)\\
\end{split}
\end{equation}
for explicitly given polynomials $b_{k,i}(x,\ep)\in\KK[x,\ep]$ and rational functions $d_{k,i,j}(x,\ep)$, 
$e_{k,i,j}(x,\ep)$
$\in \KK(x,\ep)$. Here the $k$th equation, $1\leq k\leq\lambda$, is considered as a linear differential 
equation of order 
$o_k$ in $f_k(x,\ep)$ where the right hand side is given in terms of the inhomogeneous parts $g_j(x,\ep)$, the functions 
$f_1(x,\ep),\dots,f_{k-1}(x,\ep)$, that will be treated already within our iterative method, and their derivatives. 
In general it can happen that the orders $o_i$ might be larger than $\lambda$. However, in all our calculations we ended 
up at surprisingly nice orders; see also Remark~\ref{Remark:Improvements} below.

\item We suppose that the greatest common divisor of the coefficients $b_{1,0}(x,\ep),\dots,b_{1,o_1}(x,\ep)$ in the 
first equation~\eqref{Equ:UncoupledDEQ} is $1$, i.e., there is no common polynomial factor that depends on $x$ or $\ep$. 
This implies that there is at least one $i$ such that $b_{1,i}(x,0)\neq0$. In addition, let $p_1(x)\in\KK[x]$ be the 
greatest common divisor of $b_{1,0}(x,0),\dots,b_{1,o_1}(x,0)$, i.e., $p_1(x)$ contains all common polynomial factors in 
$x$ of the $b_{1,i}(x,0)$. 
Dividing the first equation of~\eqref{Equ:UncoupledDEQ} by $p_1(x)$ yields
\begin{equation}\label{Equ:NormalizedEq}
\sum_{i=0}^{o_1}\frac{b_{1,i}(x,\ep)}{p_1(x)}\Dx^{i} f_1(x,\ep)=\sum_{i,j}\frac{d_{1,i,j}(x,\ep)}{p_1(x)}\Dx^ig_j(x,\ep).
\end{equation}
Plugging~\eqref{Equ:GiDef} into the right-hand side of~\eqref{Equ:NormalizedEq} (with $r'_1$ sufficiently large) and performing 
its $\ep$-expansion up to $r_i$ and its $x$-expansion up to $\mu$ produces\footnote{Efficient (and 
parallelized) methods have 
been implemented in the package \texttt{SolveCoupledSystem.m} in order to calculate the moments $\tilde{G}_k(n)$ efficiently 
from the given moments $G_{i,k}(n)$ in~\eqref{Equ:GiDef}.}
\begin{equation}\label{Equ:RHSExpanded}
\sum_{i=0}^{o_1}\frac{b_{1,i}(x,\ep)}{p_1(x)}\Dx^{i} f_1(x,\ep)=\sum_{k=l}^{r_1}\sum_{n=0}^{\mu}
\tilde{G}_k(n)\ep^kx^n+O(\ep^{r_1+1}x^{\mu+1}).
\end{equation}
Consequently, by coefficient comparison in~\eqref{Equ:RHSExpanded} w.r.t.\ $\ep^l$ we obtain the linear differential equation
$$\sum_{i=0}^{o_1}\tilde{b}_i(x)\Dx^i\sum_{n=0}^{\infty}F_{1,l}(n)x^n=\sum_{n=0}^{\infty}\tilde{G}_{l}(n)\,x^n$$
with 
$$\tilde{b}_i(x)=\frac{b_i(x,0)}{p_i(x)}\in\KK[x],$$
not all $0$. Finally, by coefficient comparison w.r.t.\ $x^n$ in the last equation we get a linear recurrence relation
\begin{equation}\label{Equ:Rec1}
\sum_{i=0}^{\nu_1}\beta_i(n)F_{1,l}(n+i)=\tilde{G}_{l+l_1}(n)
\end{equation}
of order $\nu_1$ for some polynomials $\beta_i(n)\in\KK[n]$ with $\beta_{\nu_1}(n)\neq0$ and an integer $l_1\in\ZZ$. In 
particular, the following bound on the order of the recurrence holds: 
\begin{equation}\label{Equ:RecOrderBound}
\nu_1\leq o_1+\max_i\deg_x(\tilde{b}_{i}(x))= o_1+\max_i\deg_x(b_{1,i}(x,0))-\deg_x(p_1(x));
\end{equation}
in the generic case we have equality -- in any case the above bound is a good indication which order we can expect.
Using this recurrence plus the first $\nu_1$ initial values of $F_{1,l}$, one can now calculate\footnote{Special care has 
to be taken if $\beta_{n_1}(n)$ has integer zeroes: this might require extra initial values. Furthermore, it might happen 
that $|l_1|$ extra moments for $G_{1,l}(n)$ are needed.} in linear time the moments for $F_{1,l}(n)$ for $n=0,\dots,\mu$.

\item Next, we plug in these moments $F_{1,l}(n)$ with $n=0,\dots,\mu$ into~\eqref{Equ:RHSExpanded} and update its right-hand 
side. In this way, $l$ is replaced by $l+1$ and $F_{1,l+1}$ takes over the role of $F_{1,l}$. Now we repeat this method for 
$k=l+1,\dots,r_1$ to get the moments of the remaining $\ep$-contributions; we remark that the coefficients on the left-hand 
side of the recurrence~\eqref{Equ:Rec1} remain unchanged - only the inhomogeneous part on the right-hand side has to be updated.

\item Finally, we repeat the steps (2)--(3) for the second equation ($k=2$) in~\eqref{Equ:UncoupledDEQ} using the moments 
of $F_{1,k}(n)$. Together with sufficiently many initial values, say $\nu_2$, we can calculate the moments $F_2(n,\ep)$ 
for $n=0\dots,\mu$. Similarly, we calculate iteratively the moments for $F_k(n,\ep)$ with $k=3,\dots,\lambda$.  
\end{enumerate}

\noindent Note that the orders $\nu_i$ (for $i=1$ see~\eqref{Equ:Rec1}) are usually small (in many cases $\leq 5$ and in 
harder cases $\sim 25$), but $\mu$ can be arbitrarily large (e.g., $\mu=2000$ or $\mu=8000$).

\begin{remark}\label{Remark:Improvements}
\normalfont
In the original version of the large moment method~\cite{Blumlein:2017dxp} we followed a slightly different approach. (1) 
For the uncoupling task we used Z{\"u}rcher's algorithm~\cite{Zuercher:94} from \texttt{OreSys}~\cite{ORESYS}. This method 
provides a system~\eqref{eq:DEQEp} where usually only the first equation is of higher-order (namely of order $\lambda$). 
All other equations are of order 0, i.e., $o_2=\dots=o_{\lambda}=0$. In general, such an uncoupled system leads to
a simpler method to calculate the desired moments. On the other side, in all our calculations it turned out that the 
Gaussian elimination method delivered much smaller orders than $\lambda$. As a consequence, switching to the Gaussian 
elimination tactic, the recurrence orders can be reduced considerably (compare the order bound~\eqref{Equ:RecOrderBound}). 
In addition, using the Gaussian uncoupling method, the coefficients $b_{1,i}(x,\ep)$ are less complicated and factors of 
the form $\frac1{\ep^{q}}$ arise with smaller $q\in\NNN$. As a consequence, using the Gaussian uncoupling strategy we 
could decrease notably the necessary orders $r_i$ of the $\ep$-expansions.\\
(2) Furthermore, our original method from~\cite{Blumlein:2017dxp} differs as follows: the first differential equation 
in~\eqref{Equ:UncoupledDEQ} (and usually the only differential equation) is directly transformed to a linear recurrence 
in $n$ whose coefficients also depend on $\ep$. This allows one to utilize a rather efficient 
machinery~\cite{BKSF:12,Blumlein:2017dxp} to compute the moments $F_{1,k}(n)$. However, in order to compute such a 
recurrence as a preprocessing step, operations have to be carried out in $\KK(x,\ep)$ which are rather costly. In 
addition, this approach misses 
the opportunity to cancel the polynomial $p_1(x)$ (which in applications is usually large) and thus to reduce the 
recurrence order substantially (compare again the order bound~\eqref{Equ:RecOrderBound}).\\
Summarizing, with our improved Strategy~1, the recurrence orders could be reduced significantly and the required 
$\ep$-orders $r_i$ can be kept rather small. 
\end{remark}

\noindent Strategy~1 has been implemented in the package \texttt{SolveCoupledSystem.m} and works for the calculations 
of the $n_h$ contributions in many cases highly efficient.

\medskip

\noindent\textit{Example.} For a system~\eqref{eq:DEQEp} with $\lambda=4$ we needed the moments~\eqref{Equ:FiEpMoments} 
for $n=0,\dots,\mu$ up to the orders $r_1=r_3=r_4=4$ and $r_2=3$. The uncoupled system has the form~\eqref{Equ:UncoupledDEQ} 
with $o_1=4$, $o_2=2$, $o_3=0$ and $o_4=0$. Using this output, we obtain a recurrence of the form~\eqref{Equ:Rec1} for 
$F_{1,l}(n)$ of order $\nu_1=4$. We note that this small recurrence order was possible by sneaking in the polynomial 
$p_1(x)\in\QQ[x]$ of degree $13$ within the linear differential equation~\eqref{Equ:RHSExpanded}; setting $p_1(x)=1$ would have 
delivered a recurrence of order $4+13=17$. 
Similarly, we obtain recurrences for $F_{2,l}(n),F_{3,l}(n),F_{4,l}(n)$ of orders $2,0,0$, respectively (again 
we reduced the 
recurrence order from 9 to 2 for $F_2(n,\ep)$ by factoring out a polynomial $p_2(x)$ of degree $7$). Finally, with the 
corresponding initial values, we calculated the moments~\eqref{Equ:FiEpMoments} for $n=0,\dots,\mu$. E.g., for $\mu=4000$ we 
needed $14000$ seconds ($257723$ CPU seconds\footnote{Adding up the calculation time of the 15 CPUs that we used in our 
parallelized implementation.}), for $\mu=6000$ we needed $25598$ seconds ($257723$ CPU seconds) and for $\mu=8000$ we 
needed $62063$ seconds ($678321$ CPU seconds) in order to get the moments of the rational contribution (ignoring the moments 
that depend on $\zeta_2,\zeta_3$ etc.).

\medskip

In general, Strategy~1 turned out to be optimal for systems with the dimension $\lambda\leq4$. However, if $\lambda\geq5$, 
the uncoupling step (in particular, using Gaussian elimination or Z{\"u}rcher's algorithm) failed by space-time resources
or produced a not digestible output: the degrees of the polynomials $b_{i,j}(x,\ep)$ in~\eqref{Equ:UncoupledDEQ} were very 
high yielding linear recurrences with orders close to 1000. 
\medskip

For these more complicated systems $\lambda\geq5$ we developed another variant of our large moment method, that is 
also implemented in our package \texttt{SolveCoupledSystem.m}.

\medskip

\noindent \textbf{Strategy 2.} Here we assume that not only the matrix $A(x,\ep)$ in~\eqref{eq:DEQEp} but also $A(x,0)$ 
is invertible. Consider the Laurent series expansions
\begin{align*}
f_i(x,\ep)&=\sum_{k=l}^{\infty}f_{i,k}(x)\ep^k,\\
g_i(x,\ep)&=\sum_{k=l}^{\infty}g_{i,k}(x)\ep^k
\end{align*}
for $1\leq i\leq\lambda$ with $l\in\ZZ$. Then by coefficient comparison w.r.t.\ $\ep^l$ in~\eqref{eq:DEQEp} we get the system
\begin{eqnarray}
\label{eq:DEQEp0}
\hspace*{2cm}\Dx 
\left(
\begin{matrix}
f_{1,l}(x)\\ \vdots \\ f_{\lambda,l}(x)\end{matrix}\right) 
= A(x,0)
\left(\begin{matrix} f_{1,l}(x)\\ \vdots \\ f_{\lambda,1}(x)\end{matrix}\right) 
+ \left(\begin{matrix} g_{1,l}(x)\\ \vdots \\ g_{\lambda,l}(x)\end{matrix}\right)
\end{eqnarray}
which is free of $\ep$; see also~\cite{Ablinger:2018zwz}.
Now we carry out steps\footnote{Since $A(x,0)$ is invertible, the available uncoupling methods from \texttt{OreSys}, 
in particular Gauss' method, are applicable.} (1)--(2) of Strategy~1 to this simpler system.  This leads to two central 
improvements: In step (1) the uncoupling method only deals with univariate rational functions in $\KK(x)$ and not in the 
very expensive multivariate case $\KK(x,\ep)$. Furthermore, the rational functions in~\eqref{Equ:UncoupledDEQ} (now 
free of $\ep$!) are much smaller and usually have lower degree. As a consequence, this leads in step~(2) to linear 
recurrences with rather small orders. Furthermore, no factors of the form $\frac1{\ep^q}$ can occur, i.e., the 
$\ep$--order 
cannot be increased by carrying out steps (1) and (2). After calculating the moments $F_{i,l}(n)$ for $n=0,\dots,\mu$ 
and $1\leq i\leq\lambda$, one plugs them into~\eqref{eq:DEQEp} and obtains a new system (similar to~\eqref{eq:DEQEp0}) 
where $f_{i,l+1}(x)$ takes over the role $f_{i,l}(x)$ and the right hand side is adapted accordingly by taking into 
account the computed moments of $F_{i,l}(n)$. Now we repeat this calculation iteratively looping through $k=l+1,\dots,\rho$ 
with
\begin{equation}\label{Equ:UpperLoopBound}
\rho=\max(r_1,\dots,r_{\lambda})
\end{equation}
in order to get the moments for $F_{i,k}(n)$, $n=0,\dots,\mu$.

\medskip

\noindent\textit{Example.} For a system~\eqref{eq:DEQEp} with $\lambda=6$ we needed the moments~\eqref{Equ:FiEpMoments} 
for $n=0,\dots,\mu$ up to the orders $r_i=4$ for $1\leq i\leq 6$. The uncoupled system of~\eqref{eq:DEQEp0} (free of $\ep$) 
has the form~\eqref{Equ:UncoupledDEQ} with $o_1=o_2=4$, $o_3=2$ and $o_4=o_5=o_6=0$. Using this information, we obtain 
recurrences for $F_{i,l}(n)$ of orders $\nu_1 = 28$, $\nu_2=11$, $\nu_3=10$ and $\nu_4=\nu_5=\nu_6=0$ with $l=-3$. Finally, with 
the corresponding initial values, we can calculate the moments $F_{i,l}(n)$  for $1\leq i\leq 6$. Plugging them 
into~\eqref{eq:DEQEp} one obtains  a new system of the form~\eqref{eq:DEQEp} and repeats this process for $k=l+1,\dots,4$.
E.g., for $\mu=4000$ we needed $115209$ seconds (637678 CPU seconds), for $\mu=6000$ we 
needed $262156$ seconds ($1.764\cdot10^6$ CPU seconds), and for 
$\mu=8000$ we needed ${3.724 \cdot 10^6}$ seconds (${5.161}\cdot 10^6$ CPU seconds) in 
order to get the moments of the rational contribution (ignoring the moments that depend on 
$\zeta_2,\zeta_3$ etc.).

\begin{remark}
\normalfont
For all systems~\eqref{eq:DEQEp} with $\lambda\geq 5$ the following modification was sufficient to obtain a matrix $A(x,\ep)$ 
such that $A(x,0)$ was invertible. We simply carried out a substitution $f_i(x,\ep)\mapsto 
\frac1{\ep^{\lambda_i}}\bar{f}_i(x,\ep)$ 
for some suitable $\lambda_i\in\NNN$ with $\bar{f}_i(x,\ep):=x^{\lambda_i}f_i(x,\ep)$, cleared the arising denominators and 
cancelled common factors. As a side-effect, this modification increased the necessary $\ep$-orders $r_i$ for $\bar{f}_i(x,\ep)$ 
to obtain in the end the required $\ep$-order for $f_i(x,\ep)$. However, this phenomenon arises only in this initialization 
phase when the matrix $A$ is usually simple. In our cases the $\lambda_i$ turned out to be rather small ($\lambda_i\leq2$).
\end{remark}

If Strategy~1 was applicable (which was the case for $\lambda\leq 4$), it was superior to Strategy~2: 
One has to loop up simultaneously for all $f_1(x,\ep),\dots,f_{\lambda}(x,\ep)$ to $\ep^{\rho}$ with~\eqref{Equ:UpperLoopBound}, 
while in Strategy~1 one computes the $\ep$-expansions for smaller recurrences taking care individually to which order $r_i$ the 
$f_i(x,\ep)$ have to be expanded.  The latter approach with individual $r_i$ reduces substantially the calculation cost. 
Furthermore, taking $\rho$ with~\eqref{Equ:UpperLoopBound} (instead of considering the individual orders $r_i$) often implied 
that the right hand sides $g_i(x,\ep)$ in~\eqref{eq:DEQEp}, and thus the simpler master integrals arising in the $g_i(x,\ep)$, 
have to be calculated to a higher $\ep$-expansion. As it turns out, this enlargement of the required $\ep$-orders even raised 
to a higher power when one applies the large moment method iteratively to the recursively defined systems coming from IBP 
relations.

\medskip

In order to tackle the master integrals coming from the $n_h$-contributions, we applied the two strategies (Strategy~1 if the 
dimension of the system is $\leq 4$ and Strategy~2 if the dimension is $\geq5$) to 41 systems, more precisely, we tackled $16$ 
systems with $\lambda=1$, $15$ with $\lambda=2$, $2$ with $\lambda=3$, $3$ with $\lambda=4$, $3$ with $\lambda=5$, $1$ with 
$\lambda=6$, and $1$ with $\lambda=7$. Here the systems depend recursively on each other: the inhomogeneous parts $g_i(x,\ep)$ 
in~\eqref{eq:DEQEp} depend on master integrals that are solutions of simpler systems. 
The calculation time of the moments up to the required $\ep$-order for these $92$ master integrals can be summarized as follows. 
For $\mu=2000$ moments we needed in total $438432$ second ($2.630\cdot 10^6$ CPU seconds). From these $2000$ moments we produced 
the corresponding number of moments of the color-factors and succeeded in guessing recurrences for almost all cases: only the 
recurrences for the $\zeta_3$-contributions and the constant free contributions could not be guessed with the given number of 
moments. Thus we restarted our large moment method to produce $\mu=4000$ moments in $720874$ seconds ($6.751\cdot 10^6$ CPU 
seconds) ignoring all constants that have been tackled already. This time we succeeded in computing the recurrences for the 
$\zeta_3$-contribution but not yet for the constant-free terms. Therefore we restarted our method for $\mu=6000$ moments 
ignoring also the $\zeta_3$ contributions and obtained them in $1.735 \cdot 10^6$
seconds (${2.091}\cdot10^6$ CPU seconds). However, we failed to derive further recurrences. Finally, we 
produced 8000 moments in ${3.724}\cdot10^6$ seconds (${5.161}\cdot10^6$ CPU seconds) and succeeded in 
guessing the remaining recurrences. 
\section{Non-First Order Factorizing Contributions}
\label{sec:6}

\vspace*{1mm}
\noindent
Using the method of arbitrarily high moments \cite{Blumlein:2017dxp} and guessing \cite{GSAGE}, the recurrences for all pole 
terms and of a series of color-zeta contributions at the $O(\ep^0)$ can be obtained using 2000 moments, while 
for 
other 
color-zeta projections the corresponding recurrences can be obtained from 4000  moments. For the purely rational 
contributions we tried the guessing method first with 6000  moments, which were not sufficient. As the next step, we 
generated 8000 moments by which we obtained the recurrences for the purely rational terms.
\begin{table}[H]\centering
\begin{tabular}{|c|r|r|r|r|}
\hline
\multicolumn{1}{|c|}{     }     &
\multicolumn{1}{|c|}{color}     &
\multicolumn{1}{|c|}{degree}    &
\multicolumn{1}{|c|}{order}     &
\multicolumn{1}{|c|}{remaining} \\
\multicolumn{1}{|c|}{     }     &
\multicolumn{1}{|c|}{     }     &
\multicolumn{1}{|c|}{     }     &
\multicolumn{1}{|c|}{     }     &
\multicolumn{1}{|c|}{order}     \\
\hline
$F_{V}$   & $g_1 n_h        $ &1288 & 54 & 15 \\
          & $g_1 n_h \zeta_3$ & 409 & 29 & 10 \\
          & $g_1 n_h \zeta_2$ & 295 & 24 &  6 \\
          & $g_2 n_h        $ &1324 & 55 & 15 \\
          & $g_2 n_h \zeta_3$ & 430 & 30 & 10 \\
          & $g_2 n_h \zeta_2$ & 273 & 23 &  6 \\
\hline
$F_{A}$   & $g_1 n_h        $ &1314 & 54 & 15 \\
          & $g_1 n_h \zeta_3$ & 419 & 29 & 10 \\
          & $g_1 n_h \zeta_2$ & 280 & 23 &  6 \\
          & $g_2 n_h        $ &1130 & 52 & 15 \\
          & $g_2 n_h \zeta_3$ & 352 & 28 & 10 \\
          & $g_2 n_h \zeta_2$ & 232 & 23 &  6 \\
\hline
$F_{S}$   & $n_h$             &1114 & 50 & 15 \\
          & $n_h \zeta_3$     & 350 & 27 & 10 \\
          & $n_h \zeta_2$     & 230 & 22 &  6 \\
\hline
$F_{P}$   & $n_h$             &1130 & 52 & 15 \\
          & $n_h \zeta_3$     & 352 & 28 & 10 \\
          & $n_h \zeta_2$     & 232 & 23 &  6 \\
\hline
\end{tabular}
\caption[]{\sf \small Structure of the recurrences for the non--first order contributions
\label{TAB1}}
\end{table}

\noindent
While for many of these projections the corresponding recurrences are first order factorizable and can thus 
be solved 
using the difference ring methods encoded in the package {\tt Sigma}, for a smaller number non--first order 
factorizing terms contribute. In Table~\ref{TAB1} we characterize those recurrences. Separating the first order 
factorizing parts, we find  remaining non--first order factorizing contributions of order $o = 6, 10$ 
and 15. This is uniformly the case for all currents. The remaining recurrences have to be studied with 
other techniques. One may translate these remaining equations into systems of ordinary 
differential equations again. In other cases it has been observed \cite{Ablinger:2017bjx,Ablinger:2017ptf} that systems 
of this kind may decompose into series of smaller systems. This has to be investigated in further studies.

We also analyzed the leading color case for the scalar current. Here 10 color-$\zeta$ structures contribute, out of 
which 8 have a representation, which results from a difference equation which factorizes at first order. The 
largest recurrences could be found using 6000 moments. In Table~\ref{TAB2} we summarize the characteristics for the 
recurrences for the constant term, which contain non--first order factorizing parts of order {\sf o = 5} and 
{\sf o = 
4}, respectively. The solution for the $\zeta_2$--term at leading color, unlike in the full color case, can be 
represented in terms of nested sums.
\begin{table}[H]\centering
\begin{tabular}{|c|r|r|r|r|}
\hline
\multicolumn{1}{|c|}{     }     &
\multicolumn{1}{|c|}{color}     &
\multicolumn{1}{|c|}{degree}    &
\multicolumn{1}{|c|}{order}     &
\multicolumn{1}{|c|}{remaining} \\
\multicolumn{1}{|c|}{     }     &
\multicolumn{1}{|c|}{     }     &
\multicolumn{1}{|c|}{     }     &
\multicolumn{1}{|c|}{     }     &
\multicolumn{1}{|c|}{order}     \\
\hline
$F_{S}$   & $N_c^2 n_h        $ & 901 & 46 & 5 \\
          & $N_c^2 n_h \zeta_3$ & 257 & 23 & 4 \\
\hline
\end{tabular}
\caption[]{\sf \small  Structure of the recurrences for the non-first order contributions in leading color 
approximation
for the scalar three loop massive form factor
\label{TAB2}}
\end{table}
The corresponding recurrence is of degree {\sf d = 150} and order {\sf o = 17} and one obtains 
\begin{eqnarray}
F_S^{n_h N_c^2 \zeta_2} &=& 
-\frac{N_C^2 n_h \zeta_2}{2(1+x)^2} \Biggl\{
        \frac{8 R_{28}}{9 x (1+x)^4} \ln(2)
        +\frac{R_{27}}{54 (1-x)^2  (1+x)^4}
        +\Biggl[
                -\frac{8 R_{23}}{3 (1-x)  (1+x)^3} \ln(2)
\nonumber\\ &&
                -\frac{R_{29}}{18 (1-x)^3  (1+x)^4}
                +\Biggl(
                        \frac{16 R_2}{(1-x^2) } \ln(2)
                        -\frac{R_{24}}{3 (1-x)  (1+x)^3}
                \Biggr) H_{-1}
\nonumber\\ &&
                +\frac{8 R_8}{3 (1-x^2) } H_{-1}^2
                -\frac{4 R_{17}}{3 (1-x)^2  (1+x)} \zeta_2
                +\frac{48 x^2}{(1-x^2)} \zeta_3
        \Biggr] H_0
        +\Bigl(
                \frac{24 x^2}{(1-x^2)} \zeta_2
\nonumber\\ &&                
+                \frac{R_{26}}{36 (1-x)^2  (1+x)^3}
+\frac{8 R_{15}}{(1-x)^2  (1+x)} \ln(2)
                -\frac{8 R_{14}}{3 (1-x)^2  (1+x)} H_{-1}
        \Biggr) H_0^2
\nonumber\\ &&        
+\frac{4 R_{16}}{9 (1-x)^2  (1+x)} H_0^3
        -\frac{2 x^2}{(1-x^2)} H_0^4
        +\Biggl[
                \Biggl(
                        -\frac{80 (1+x) \big(
                                1+x^2\big)}{(1-x) } \ln(2)
\nonumber\\ &&                        
-\frac{8 R_{22}}{9 (1-x)  (1+x)^3}
                        -\frac{16 R_5}{3 (1-x^2) } H_{-1}
                \Biggr) H_0
                +\frac{8 \big(
                        5-13 x+54 x^2-x^3-x^4\big)}{3 (1-x^2) } H_0^2
\nonumber\\ &&             
   -\frac{16 R_6}{3 (1-x^2) } \zeta_2
        \Biggr] H_1
        -\frac{32 (1+x) \big(
                1-3 x+x^2\big)}{3 (1-x) } H_0 H_1^2
        +\Biggl(
                -\frac{8 R_{28}
                }{9 x (1+x)^4}
\nonumber\\ &&                
-
                \frac{16 R_7}{3 (1-x^2) } \zeta_2
        \Biggr) H_{-1}
        +\Biggl(
                 \frac{80 (1+x) \big(
                        1+x^2\big)}{(1-x) } \ln(2)
                +\frac{8 R_{22}}{9 (1-x)  (1+x)^3}
\nonumber\\ &&                
-\frac{8 R_{19}}{3 (1-x)^2  (1+x)} H_0
                -\frac{32 x^2 H_0^2}{(1-x^2)}
                +\frac{64 (1+x) \big(
                        1-3 x+x^2\big)}{3 (1-x) } H_1
                +\frac{128 R_1}{3 (1-x^2) } 
\nonumber\\ && \times
H_{-1}
        \Biggr) H_{0,1}
        +\Biggl(
                \frac{8  R_{11}}{3 (1-x)^2 } H_0
                -\frac{16 R_2}{(1-x^2) } \ln(2)
                -\frac{R_{21}}{3 (1-x)  (1+x)^3}
                +\frac{64 x^2}{(1-x^2)} H_0^2
\nonumber\\ &&                
+\frac{32 R_4}{3 (1-x^2) } H_1
                -\frac{32 R_9}{3 (1-x^2) } H_{-1}
        \Biggr) H_{0,-1}
        -\frac{16 R_2}{(1-x^2) } H_0 H_{-1,1}
\nonumber\\ &&     
   +\Biggl(
                \frac{16 x R_{13}}{3 (1-x)^2 (1+x)}
                +\frac{192 x^2}{(1-x^2)} H_0
        \Biggr) H_{0,0,1}
        -\Biggl(
                \frac{16 R_{20}}{3 (1-x)^2  (1+x)}
                +\frac{288 x^2}{(1-x^2)} H_0
        \Biggr) 
\nonumber\\ && \times
H_{0,0,-1}
        - \frac{64 (1+x) \big(
                1-3 x+x^2\big)}{3 (1-x) } H_{0,1,1}
        -\frac{16 R_{10}}{3 (1-x^2)} H_{0,1,-1}
        -\frac{32 R_4}{3 (1-x^2) } H_{0,-1,1}
\nonumber\\ &&        
+\frac{16 R_{12}}{3 (1-x^2) } H_{0,-1,-1}
        -\frac{384 x^2 }{(1-x^2)} H_{0,0,0,1}
        +\frac{480 x^2 }{(1-x^2)} H_{0,0,0,-1}
        +\Biggl(
                 \frac{16 R_3}{(1-x^2)} \ln(2)
\nonumber\\ &&                
-\frac{R_{25}}{18 (1-x)  (1+x)^3}
        \Biggr) \zeta_2
        -\frac{72 x^2}{5 (1-x^2)} \zeta_2^2
        +\frac{2 R_{18}}{3 (1-x)^2  (1+x)} \zeta_3
\Biggr\},
\end{eqnarray}
with the polynomials
\begin{eqnarray}
R_1 &=& 7 x^4+3 x^3+28 x^2+3 x+7,
\\
R_2 &=& 11 x^4+4 x^3+34 x^2+4 x+11,
\\
R_3 &=& 17 x^4-2 x^3+58 x^2-2 x+17,
\\
R_4 &=& 19 x^4+21 x^3+76 x^2+21 x+19,
\\
R_5 &=& 23 x^4+12 x^3+122 x^2+12 x+23,
\\
R_6 &=& 23 x^4+17 x^3+60 x^2+17 x+23,
\\
R_7 &=& 24 x^4+5 x^3+130 x^2+5 x+24,
\\
R_8 &=& 31 x^4+26 x^3+86 x^2+26 x+31,
\\
R_9 &=&32 x^4+19 x^3+94 x^2+19 x+32,
\\
R_{10} &=& 71 x^4+54 x^3+254 x^2+54 x+71,
\\
R_{11} &=& 83 x^4-96 x^3+214 x^2-96 x+83,
\\
R_{12} &=& 97 x^4+50 x^3+290 x^2+50 x+97,
\\
R_{13} &=& 100 x^4-19 x^3+197 x^2-105 x+11,
\\
R_{14} &=& 3 x^5+23 x^4-22 x^3+148 x^2+7 x+57,
\\
R_{15} &=& 15 x^5-3 x^4+14 x^3-16 x^2-x-1,
\\
R_{16} &=& 19 x^5-5 x^4+212 x^3-112 x^2+43 x+19,
\\
R_{17} &=& 31 x^5+39 x^4+258 x^3-70 x^2+91 x+27,
\\
R_{18} &=& 89 x^5-289 x^4+22 x^3-134 x^2-55 x-1,
\\
R_{19} &=& 101 x^5-19 x^4+142 x^3-38 x^2-7 x+5,
\\
R_{20} &=& 125 x^5-45 x^4+182 x^3-78 x^2-23 x+23,
\\
R_{21} &=& -617 x^6-690 x^5+2033 x^4+3572 x^3+2033 x^2-690 x-617,
\\
R_{22} &=& 45 x^6+398 x^5-481 x^4-2148 x^3-481 x^2+398 x+45,
\\
R_{23} &=& 48 x^6+131 x^5-28 x^4+18 x^3-28 x^2+131 x+48,
\\
R_{24} &=& 233 x^6-358 x^5-1809 x^4-3716 x^3-1809 x^2-358 x+233,
\\
R_{25} &=& 2571 x^6+8438 x^5-13795 x^4-45084 x^3-13795 x^2+8438 x+2571,
\\
R_{26} &=& 917 x^7+12853 x^6-3767 x^5-24311 x^4-11289 x^3+9223 x^2-325 x-709,
\\
R_{27} &=& -100331 x^8-173294 x^7-40776 x^6+237806 x^5+226918 x^4+237806 x^3
\nonumber\\ &&
-40776 x^2
-173294 x-100331,
\\
R_{28} &=& 124 x^8-5 x^7-609 x^6-2635 x^5-3950 x^4-2635 x^3-609 x^2-5 x+124,
\\
R_{29} &=& 1984 x^{10}-5471 x^9-11702 x^8-28981 x^7+9401 x^6+43521 x^5+12817 x^4
\nonumber\\ &&
-24727 x^3-13829 x^2-8150 x+561.
\end{eqnarray}

\noindent
Here the functions $H_{\vec{a}}(x)$ denote the harmonic polylogarithms \cite{Remiddi:1999ew}, which are defined 
by
\begin{equation}
H_{b,\vec{a}}(x) = \int_0^x dy f_b(y) H_{\vec{a}}(y),~~~H_\emptyset = 1,~~a_i,b \in \{0,1,-1\}
\end{equation}
with
\begin{equation}
f_0(x) = \frac{1}{x},~~~~~ 
f_1(x) = \frac{1}{1-x},~~~~~ 
f_{-1}(x) = \frac{1}{1+x}.
\end{equation}
In the case of the color--$\zeta$ terms  the recurrences of which do not factorize to first order we can use the 
finite analytic representation given by the number of analytic moments $N_{\rm max} = 2000$ for numerical results 
and graphical illustrations given below, which already gives a good representation. In some cases they have a 
logarithmic divergence near $x = 0$, however.
\section{The Results}
\label{sec:7}

\vspace*{1mm}
\noindent
In the following we present the results for the contribution $\propto n_h^{k}, k = 1,2,$ of the massive three--loop form 
factors in the vector-, axialvector-, scalar- and pseudoscalar case. The pole terms are given in terms of HPLs of the variable 
$x$. For the constant contributions, for a series of color--zeta terms also first order factorizing solutions are obtained,
which are expressed using HPLs. Other contributions contain non--first order factorizing parts. For these we 
have derived
the corresponding recurrences of their Taylor coefficients also using the method of arbitrarily high 
moments \cite{Blumlein:2017dxp}. 
In the present paper we present these contributions in terms of analytic series expansions, for which we derived
at least 2000 terms and in some cases 4000 or 8000 terms, cf.~Section~\ref{sec:6}.

The renormalized form factors, which are lengthy expressions, except the functions $F_{C,i}^{(0)}, i = 1...3, C = 
V_1, V_2, A_1, A_2, P, S$ defined below, are given in an attachment to this paper, where we also present the 
other results in computer-readable form.
\subsection{The Vector Form Factors}
\label{sec:51}

\vspace*{1mm}
\noindent
The vector form factor $F_{V,1}$ is given by
\begin{eqnarray}
F_{V,1} &=& \frac{1}{\ep^3} \Biggl\{
n_h^2 \Biggl[
        \frac{16}{27}
        -\frac{32 \big(
                1+x^2\big) H_0}{27 (1-x^2)}
\Biggr]
-n_h \Biggl[
        \frac{2 \big(
                589+602 x+589 x^2\big)}{27 (1+x)^2}
        - n_l \Biggl(
                \frac{16}{27}
                -\frac{32 \big(
                        1+x^2\big) H_0}{27 (1-x^2)}
        \Biggr)
\nonumber\\ &&
        -\frac{16
\big(
                41+190 x+154 x^2+190 x^3+41 x^4\big) H_0}{27 (1-x) (1+x)^3} 
        +\frac{128 \big(
                1+x^2\big)^2}{27 (1-x^2)^2 } H_0^2
\Biggr]
\Biggr\}
\nonumber\\ &&
+\frac{1}{\ep^2} \Biggl\{
n_h^2 \Biggl[
        \frac{16}{27}
        -\Biggl(
                \frac{16 \big(
                        3+2 x+3 x^2\big)}{27 (1-x^2)}
                -\frac{64 \big(
                        1+x^2\big)}{27 (1-x^2)} H_{-1}
        \Biggr) H_0
        -\frac{16 \big(
                1+x^2\big)}{27 (1-x^2)} H_0^2
\nonumber\\ &&
        -\frac{64 \big(
                1+x^2\big)}{27 (1-x^2)} H_{0,-1}
        +\frac{32 \big(
                1+x^2\big)}{27 (1-x^2)} \zeta_2
\Biggr]
+ n_h \Biggl[
        -\frac{4 \big(
                263-66 x+263 x^2\big)}{27 (1+x)^2}
        + n_l \Biggl(
                \frac{128}{81}
\nonumber\\ &&                
+\Biggl(
                        -\frac{64 \big(
                                7+3 x+7 x^2\big)}{81 (1-x^2)}
                        +\frac{128 \big(
                                1+x^2\big)}{27 (1-x^2)} H_{-1}
                \Biggr) H_0
                -\frac{32 \big(
                        1+x^2\big)}{27 (1-x^2)} H_0^2
                -\frac{128 \big(
                        1+x^2\big)}{27 (1-x^2)} H_{0,-1}
\nonumber\\ &&
                +\frac{64 \big(
                        1+x^2\big)}{27 (1-x^2)} \zeta_2
        \Biggr]
        +\Biggl(
                \frac{8\big(
                        371+1622 x+1334 x^2+1622 x^3+371 x^4\big)
}{27 (1-x) (1+x)^3} 
                +\frac{64 \big(
                        1+x^2\big)}{3 (1-x^2)} H_1
\nonumber\\ &&
                -\frac{128
\big(
                        23+73 x+64 x^2+73 x^3+23 x^4\big)} {27 (1-x) (1+x)^3} 
H_{-1}
                +\frac{64 \big(
                        1+x^2\big)^2}{3 (1-x^2)^2} 
H_{0,1}
                -\frac{1088 \big(
                        1+x^2\big)^2}{27 (1-x^2)^2}
\nonumber\\ && \times
                 H_{0,-1}
        \Biggr) H_0
        +\Biggl(
                \frac{16
\big(
                        13+51 x-46 x^2-10 x^3-167 x^4-97 x^5\big)}{27 (1-x)^2 (1+x)^3} 
                +\frac{512 \big(
                        1+x^2\big)^2}{27 (1-x)^2 (1+x)^2} 
\nonumber\\ && \times
H_{-1}
        \big) H_0^2
        +\frac{64 \big(
                -2+x^2
        \big)
\big(1+x^2\big)}{27 (1-x)^2 (1+x)^2} H_0^3
        -\frac{64 \big(
                1+x^2\big)}{3 (1-x^2)} H_{0,1}
        +\frac{128}{27 (1-x) (1+x)^3} 
\nonumber\\ && \times
\big(
                23+73 x+64 x^2+73 x^3+23 x^4\big) H_{0,-1}
        -\frac{128 \big(
                1+x^2\big)^2}{3 (1-x)^2 (1+x)^2} H_{0,0,1}
        +\frac{128 \big(
                1+x^2\big)^2}{3 (1-x)^2 (1+x)^2} 
\nonumber\\ && \times
H_{0,0,-1}
        +\big(
                -\frac{64}{27 (1-x) (1+x)^3} \big(
                        23+73 x+46 x^2+37 x^3+5 x^4\big)
\nonumber\\ &&
                +\frac{32 \big(
                        1+x^2
                \big)
\big(-1+35 x^2\big)}{27 (1-x)^2 (1+x)^2} H_0
        \big) \zeta_2
        +\frac{32 \big(
                1+x^2\big)^2}{3 (1-x)^2 (1+x)^2} \zeta_3
\big)
\Biggr\}
\nonumber\\ &&
+ \frac{1}{\ep} \Biggl\{
n_h^2 \Biggl[
        -\frac{16 H_0 P^{(1)}_4}{243 (1-x) (1+x)^3}
        +\frac{8 H_0^2 P^{(1)}_9}{81 (1-x) (1+x)^4}
        -\frac{16 H_0^3 P^{(1)}_{13}}{27 (1-x) (1+x)^5}
\nonumber\\ &&
        -\frac{64 \big(
                13-268 x+13 x^2\big)}{243 (1+x)^2}
        -\frac{64 \big(
                1+x^2\big)}{27 (1-x^2)} H_{-1}^2 H_0
        +\Biggl(
                \frac{32 \big(
                        3+2 x+3 x^2\big)}{27 (1-x^2)} H_0
                +\frac{32 \big(
                        1+x^2\big)}{27 (1-x^2)} H_0^2
\nonumber\\ &&
                +\frac{128 \big(
                        1+x^2\big)}{27 (1-x^2)} H_{0,-1}
        \Biggr) H_{-1}
        -\frac{32 \big(
                3+2 x+3 x^2\big)}{27 (1-x^2)} H_{0,-1}
        -\frac{64 \big(
                1+x^2\big)}{27 (1-x^2)} H_{0,0,-1}
\nonumber\\ &&
        -\frac{128 \big(
                1+x^2\big)}{27 (1-x^2)} H_{0,-1,-1}
        +\Biggl(
                \frac{8 P^{(1)}_{11}}{27 (1-x) (1+x)^4}
                -\frac{16 H_0 P^{(1)}_{16}}{27 (1-x) (1+x)^5}
                -\frac{64 \big(
                        1+x^2\big)}{27 (1-x^2)} H_{-1}
        \Biggr) \zeta_2
\nonumber\\ &&
        +\frac{64 \big(
                1+x^2\big)}{27 (1-x^2)} \zeta_3
\Biggr]
+n_h \Biggl[
        -\frac{64 H_{0,1} P^{(1)}_1}{27 (1-x) (1+x)^3}
        +\frac{128 H_{0,-1,-1} P^{(1)}_3}{27 (1-x) (1+x)^3}
        -\frac{4 P^{(1)}_6}{243 (1+x)^4}
\nonumber\\ &&
        +\frac{128 H_{0,0,1} P^{(1)}_{18}}{27 (1-x)^2 (1+x)^4}
        -
        \frac{64 H_{0,0,-1} P^{(1)}_{21}}{27 (1-x)^2 (1+x)^4}
        +\frac{128 H_{0,0,0,-1} P^{(1)}_{27}}{9 (1-x)^2 (1+x)^5}
        -\frac{128 H_{0,0,0,1} P^{(1)}_{30}}{27 (1-x)^2 (1+x)^5}
\nonumber\\ &&
        -\frac{16 \zeta_2^2 P^{(1)}_{37}}{135 (1-x)^2 (1+x)^5}
        -\frac{8 H_0^4 P^{(1)}_{39}}{81 (1-x)^2 (1+x)^6}
        +n_l \Biggl(
                -\frac{32 \big(
                        1-194 x+x^2\big)}{81 (1+x)^2}
\nonumber\\ &&
                -\frac{32 H_0 P^{(1)}_2}{81 (1-x) (1+x)^3}
                -\frac{16 H_0^2 P^{(1)}_7}{81 (1-x) (1+x)^4}
                -\frac{32 H_0^3 P^{(1)}_{14}}{27 (1-x) (1+x)^5}
                -\frac{256 \big(
                        1+x^2\big)}{27 (1-x^2)} H_{-1}^2 H_0
\nonumber\\ &&
                +\Biggl(
                        \frac{256 \big(
                                7+3 x+7 x^2\big)}{81 (1-x^2)} H_0
                        +\frac{128 \big(
                                1+x^2\big)}{27 (1-x^2)} H_0^2
                        +\frac{512 \big(
                                1+x^2\big)}{27 (1-x^2)} H_{0,-1}
                \Biggr) H_{-1}
\nonumber\\ &&
                -\frac{256 \big(
                        7+3 x+7 x^2\big)}{81 (1-x^2) } H_{0,-1}
                -\frac{256 \big(
                        1+x^2\big)}{27 (1-x^2)} H_{0,0,-1}
                +\frac{512 \big(
                        1+x^2\big)}{27 (1-x^2)} H_{0,-1,-1}
\nonumber\\ &&
                +\Biggl(
                        \frac{8 P^{(1)}_{12}}{81 (1-x) (1+x)^4}
                        -\frac{16 H_0 P^{(1)}_{17}}{27 (1-x) (1+x)^5}
                        -\frac{256 \big(
                                1+x^2\big)}{27 (1-x^2)}
                         H_{-1}
                \Biggr) \zeta_2
                +
                \frac{256 \big(
                        1+x^2\big)}{27 (1-x^2)} \zeta_3
        \Biggr)
\nonumber\\ &&
        +\Biggl(
                \frac{4 P^{(1)}_{23}}{243 (1-x) (1+x)^5}
                -\frac{128 H_{0,1} P^{(1)}_{15}}{27 (1-x)^2 (1+x)^4}
                -\frac{64 H_{0,-1} P^{(1)}_{19}}{27 (1-x)^2 (1+x)^4}
\nonumber\\ &&
                -\frac{512 H_{0,0,1} P^{(1)}_{24}}{27 (1-x)^2 (1+x)^5}
\nonumber\\ &&
                +\frac{128 H_{0,0,-1} P^{(1)}_{29}}{27 (1-x)^2 (1+x)^5}
                +\Biggl(
                        \frac{64 P^{(1)}_1}{27 (1-x) (1+x)^3}
                        +\frac{256 \big(
                                1+x^2\big)^2}{27 (1-x^2)^2} [H_{0,1} - H_{0,-1}]
                \Biggr) 
\nonumber\\ &&
\times H_1
+                \frac{64 \big(
                        1+x^2\big)}{3 (1-x^2) } H_1^2
                +\frac{640 \big(
                        1+x^2\big)^2}{27 (1-x^2)^2} H_{0,1,1}
                -\frac{896 \big(
                        1+x^2\big)^2}{27 (1-x^2)^2} H_{0,1,-1}
\nonumber\\ &&
                -\frac{896 \big(
                        1+x^2\big)^2}{27 (1-x^2)^2} H_{0,-1,1}
                +\frac{128 \big(
                        1+x^2\big)^2}{27 (1-x^2)^2} H_{0,-1,-1}
        \Biggr) H_0
        +\Biggl(
                \frac{64 H_1 P^{(1)}_8}{27 (1-x) (1+x)^4}
\nonumber\\ &&
                +\frac{64 H_{0,1} P^{(1)}_{31}}{27 (1-x)^2 (1+x)^5}
                -\frac{64 H_{0,-1} P^{(1)}_{33}}{27 (1-x)^2 (1+x)^5}
                -\frac{4 P^{(1)}_{41}}{243 (1-x)^2 (1+x)^6}
        \Biggr) H_0^2
\nonumber\\ &&
        +\Biggl(
                -\frac{64 H_1 P^{(1)}_{28}
                }{81 (1-x)^2 (1+x)^5}
                -
                \frac{16 P^{(1)}_{35}}{81 (1-x)^2 (1+x)^5}
        \Biggr) H_0^3
        +\Biggl(
                -\frac{128 \big(
                        1+x^2\big)}{3 (1-x^2)} H_{0,1}
\nonumber\\ &&
                +\frac{128 \big(
                        1+x^2\big)}{(1-x^2)} H_{0,-1}
                -\frac{512 \big(
                        1+x^2\big)^2}{27 (1-x^2)^2} H_{0,0,1}
                +\frac{512 \big(
                        1+x^2\big)^2}{27 (1-x^2)^2} H_{0,0,-1}
        \Biggr) H_1
\nonumber\\ &&
        +\Biggl(
                -\frac{128 H_{0,-1} P^{(1)}_3}{27 (1-x) (1+x)^3}
                +\frac{32 H_0^2 P^{(1)}_{22}}{27 (1-x)^2 (1+x)^4}
                -\Biggl(
                        \frac{16 P^{(1)}_5}{27 (1-x) (1+x)^3}
                        +\frac{128 \big(
                                1+x^2\big)}{(1-x^2)} 
\nonumber\\ &&
\times H_1
                        +\frac{256 \big(
                                1+x^2\big)^2}{3 (1-x^2)^2} H_{0,1}
                        -\frac{4352 \big(
                                1+x^2\big)^2}{27 (1-x^2)^2} H_{0,-1}
                \Biggr) H_0
                +\frac{64 \big(
                        1+x^2
                \big)
\big(-1+5 x^2\big)}{27 (1-x^2)^2} H_0^3
\nonumber\\ &&
                +\frac{128 \big(
                        1+x^2\big)}{(1-x^2)} H_{0,1}
                +\frac{512 \big(
                        1+x^2\big)^2}{3 (1-x^2)^2} H_{0,0,1}
                -\frac{512 \big(
                        1+x^2\big)^2}{3 (1-x^2)^2} H_{0,0,-1}
        \Biggr) H_{-1}
\nonumber\\ &&
        +\Biggl(
                \frac{64 H_0 P^{(1)}_3}{27 (1-x) (1+x)^3}
                -\frac{1024 \big(
                        1+x^2\big)^2}{27 (1-x^2)^2} H_0^2
        \Biggr) H_{-1}^2
        -
        \frac{448 \big(
                1+x^2\big)^2}{27 (1-x^2)^2} H_{0,1}^2
\nonumber\\ &&
        +\Biggl(
                \frac{16 P^{(1)}_5}{27 (1-x) (1+x)^3}
                +\frac{3200 \big(
                        1+x^2\big)^2}{27 (1-x^2)^2} H_{0,1}
        \Biggr) H_{0,-1}
        -\frac{2240 \big(
                1+x^2\big)^2}{27 (1-x^2)^2} H_{0,-1}^2
\nonumber\\ &&
        +\frac{128 \big(
                1+x^2\big)}{3 (1-x^2)} H_{0,1,1}
        -\frac{128 \big(
                1+x^2\big)}{(1-x^2)} H_{0,1,-1}
        -\frac{128 \big(
                1+x^2\big)}{(1-x^2)} H_{0,-1,1}
        +\frac{512 \big(
                1+x^2\big)^2}{27 (1-x^2)^2} 
\nonumber\\ &&
\times H_{0,0,1,1}
        -\frac{512 \big(
                1+x^2\big)^2}{3 (1-x^2)^2} H_{0,0,1,-1}
        -\frac{512 \big(
                1+x^2\big)^2}{3 (1-x^2)^2} H_{0,0,-1,1}
        +\frac{512 \big(
                1+x^2\big)^2}{3 (1-x^2)^2} H_{0,0,-1,-1}
\nonumber\\ &&
        -\frac{2048 \big(
                1+x^2\big)^2}{27 (1-x^2)^2} H_{0,-1,0,1}
        +\Biggl(
                -\frac{128 \ln(2) \big(
                        1+x^2\big)}{3 (1+x)^2}
                -\frac{64 H_{0,-1} P^{(1)}_{25}}{27 (1-x)^2 (1+x)^5}
\nonumber\\ &&
                +\frac{64 H_{0,1} P^{(1)}_{32}}{27 (1-x)^2 (1+x)^5}
                -\frac{P^{(1)}_{38}}{27 (1-x) (1+x)^6}
                -\frac{32 H_0^2 P^{(1)}_{40}}{27 (1-x)^2 (1+x)^6}
\nonumber\\ &&
                -\Biggl(
                        \frac{256 H_1 P^{(1)}_{26}
                        }{27 (1-x)^2 (1+x)^5}
                        -\frac{8 P^{(1)}_{36}}{27 (1-x)^2 (1+x)^5}
                \Biggr) H_0
                -\frac{64 \big(
                        1+x^2\big)}{3 (1-x) (1+x)} H_1
\nonumber\\ &&
                +\Biggl(
                        \frac{64 P^{(1)}_{10}}{27 (1-x) (1+x)^4}
                        -\frac{256 \big(
                                1+x^2
                        \big)
\big(13+4 x^2\big)}{27 (1-x)^2 (1+x)^2} H_0
                \big) H_{-1}
        \Biggr) \zeta_2
        +\Biggl(
                \frac{32 P^{(1)}_{20}}{27 (1-x)^2 (1+x)^4}
\nonumber\\ &&
                +\frac{32 H_0 P^{(1)}_{34}}{27 (1-x)^2 (1+x)^5}
                +\frac{128 \big(
                        1+x^2\big)^2}{27 (1-x^2)^2} H_1
                -\frac{128 \big(
                        1+x^2\big)^2}{3 (1-x^2)^2} H_{-1}
        \Biggr) \zeta_3
\Biggr]
\Biggr\} + F_{V,1}^{(0)},
\end{eqnarray}
and the polynomials 
\begin{eqnarray}
P^{(1)}_1 &=&37 x^4+104 x^3+86 x^2+104 x+37
\\
P^{(1)}_2 &=& 103 x^4+98 x^3+702 x^2+98 x+103
\\
P^{(1)}_3 &=& 115 x^4+284 x^3+266 x^2+284 x+115
\\
P^{(1)}_4 &=& 337 x^4+20 x^3+3638 x^2+20 x+337
\\
P^{(1)}_5 &=& 913 x^4+1462 x^3+6506 x^2+1462 x+913
\\
P^{(1)}_6 &=& 14921 x^4+130220 x^3+280134 x^2+130220 x+14921
\\
P^{(1)}_7 &=& 9 x^5+117 x^4-60 x^3+356 x^2+75 x+47
\\
P^{(1)}_8 &=& 14 x^5-89 x^4+761 x^3-617 x^2+197 x+22
\\
P^{(1)}_9 &=& 29 x^5-75 x^4+362 x^3-470 x^2+9 x-47
\\
P^{(1)}_{10} &=& 55 x^5+264 x^4+277 x^3+391 x^2+210 x+67
\\
P^{(1)}_{11} &=& 87 x^5+x^4+1502 x^3-1430 x^2+43 x-75
\\
P^{(1)}_{12} &=& 331 x^5+249 x^4+4942 x^3-3758 x^2+519 x-107
\\
P^{(1)}_{13} &=& x^6+4 x^5+3 x^4+48 x^3+3 x^2+4 x+1
\\
P^{(1)}_{14} &=& x^6+4 x^5+5 x^4+28 x^3+5 x^2+4 x+1
\\
P^{(1)}_{15} &=& 2 x^6+123 x^5-870 x^4+1234 x^3-870 x^2+123 x+2
\\
P^{(1)}_{16} &=& 5 x^6+20 x^5+11 x^4+280 x^3+11 x^2+20 x+5
\\
P^{(1)}_{17} &=& 7 x^6+28 x^5+25 x^4+296 x^3+25 x^2+28 x+7
\\
P^{(1)}_{18} &=& 18 x^6+143 x^5-890 x^4+1090 x^3-926 x^2+71 x-18
\\
P^{(1)}_{19} &=& 43 x^6-685 x^5+5403 x^4-8498 x^3+5403 x^2-685 x+43
\\
P^{(1)}_{20} &=& 137 x^6+1490 x^5-6779 x^4+12512 x^3-7257 x^2+606 x-197
\\
P^{(1)}_{21} &=& 142 x^6+1359 x^5-6912 x^4+11114 x^3-7268 x^2+683 x-142
\\
P^{(1)}_{22} &=& 228 x^6-11 x^5+3894 x^4-5882 x^3+3538 x^2-687 x-56
\\
P^{(1)}_{23} &=& 46991 x^6+217814 x^5+668705 x^4+1193908 x^3+668705 x^2+217814 x
\nonumber\\ &&
+46991
\\
P^{(1)}_{24} &=& 5 x^7-37 x^6+421 x^5-1247 x^4+1268 x^3-406 x^2+46 x-2
\\
P^{(1)}_{25} &=& 7 x^7+15 x^6+17 x^5-239 x^4+x^3-187 x^2-117 x-41
\\
P^{(1)}_{26} &=& 13 x^7-14 x^6+449 x^5-1326 x^4+1319 x^3-454 x^2+11 x-14
\\
P^{(1)}_{27} &=& 15 x^7+201 x^6-1135 x^5+3933 x^4-3709 x^3+1295 x^2-105 x+17
\\
P^{(1)}_{28} &=& 25 x^7-31 x^6+893 x^5-2659 x^4+2631 x^3-913 x^2+19 x-29
\\
P^{(1)}_{29} &=& 30 x^7-171 x^6+2127 x^5-6392 x^4+6448 x^3-2087 x^2+195 x-22
\\
P^{(1)}_{30} &=& 41 x^7+382 x^6-1814 x^5+6397 x^4-5949 x^3+2134 x^2-190 x+23
\\
P^{(1)}_{31} &=& 42 x^7-31 x^6+1377 x^5-3742 x^4+3966 x^3-1217 x^2+127 x-10
\\
P^{(1)}_{32} &=& 43 x^7-83 x^6+1751 x^5-5367 x^4+5213 x^3-1861 x^2+17 x-65
\\
P^{(1)}_{33} &=& 50 x^7+45 x^6+1011 x^5-2442 x^4+2750 x^3-791 x^2+87 x-6
\\
P^{(1)}_{34} &=& 71 x^7-207 x^6+3511 x^5-10527 x^4+10653 x^3-3421 x^2+261 x-53
\\
P^{(1)}_{35} &=& 229 x^7-245 x^6+6189 x^5-4655 x^4-731 x^3+399 x^2-167 x+5
\\
P^{(1)}_{36} &=& 281 x^7+4495 x^6-12289 x^5+14449 x^4-953 x^3+2801 x^2-671 x+79
\\
P^{(1)}_{37} &=& 809 x^7+655 x^6+16429 x^5-41801 x^4+43971 x^3-14879 x^2+275 x-499
\\
P^{(1)}_{38} &=& 18177 x^7+16453 x^6+426009 x^5+662997 x^4-333541 x^3-273065 x^2
\nonumber\\ &&
+23243 x-8305
\\
P^{(1)}_{39} &=& 8 x^8+31 x^7+42 x^6+693 x^5+212 x^4+795 x^3+214 x^2+145 x+36
\\
P^{(1)}_{40} &=& 71 x^8+232 x^7+832 x^6-74 x^5+498 x^4+1490 x^3-256 x^2+80 x+7
\\
P^{(1)}_{41} &=& 13421 x^8+46436 x^7+167926 x^6+195276 x^5+214672 x^4-46932 x^3-22982 x^2
\nonumber\\ &&
-25564 x-5677,
\end{eqnarray}

\noindent
and 
\begin{eqnarray}
F_{V,2} &=& \frac{x}{\ep^3} \Biggl\{
n_h 128 \Biggl[
        -\frac{1}{3 (1+x)^2}
        +\frac{\big(
                1+x^2\big)}{3 (-1+x) (1+x)^3} H_0
\Biggr] \Biggr\}
\nonumber\\ &&
+ \frac{x}{\ep^2} \Biggl\{
- n_h^2 \frac{64  H_0}{27 (1-x^2)}
+ n_h \Biggl[
        \frac{64 }{27 (1-x) (1+x)^3} \left(
                -73 (1-x^2)
                +18 (1+x^2) \zeta_2
        \right)
\nonumber\\ &&
        -n_l \frac{128 H_0}{27 (1-x^2)}
        -\left(
                \frac{128  \big(
                        18-73 x+18 x^2\big)}{27 (1-x) (1+x)^3}
                -\frac{256  \big(
                        1+x^2\big)}{3 (1-x) (1+x)^3} H_{-1}
        \right) H_0
\nonumber\\  &&
        +\frac{64 (-17+x)  \big(
                1+x^2\big)}{27 (1-x)^2 (1+x)^3} H_0^2
        -\frac{256  \big(
                1+x^2\big)}{3 (1-x) (1+x)^3} H_{0,-1}
\Biggr]
\Biggr\}
\nonumber\\ &&
+ \frac{x}{\ep} \Biggl\{
n_h^2 \Biggl[
        -
        \frac{2176 }{27 (1+x)^2}
        +\Biggl(
                -\frac{64  \big(
                        37-298 x+37 x^2\big)}{81 (1-x) (1+x)^3}
                +\frac{512 x^2 \zeta_2}{3 (1-x) (1+x)^5}
        \Biggr) H_0
\nonumber\\ &&
        +\frac{128  H_{-1} H_0}{27 (1-x^2)}
        +\frac{32  \big(
                -3+39 x-45 x^2+x^3\big)}{27 (1-x) (1+x)^4} H_0^2
        +\frac{256 x^2 H_0^3}{9 (1-x) (1+x)^5}
        -\frac{128  H_{0,-1}}{27 (1-x^2)}
\nonumber\\ &&
        -\frac{64  \big(
                -7-213 x+207 x^2+5 x^3\big)}{27 (1-x) (1+x)^4} \zeta_2
\Biggr]
+n_h \Biggl[
        -\frac{512  H_{0,0,1} P^{(2)}_4}{27 (1-x)^2 (1+x)^4}
        -\frac{256  \zeta_3 P^{(2)}_5}{27 (1-x)^2 (1+x)^4}
\nonumber\\ &&
        +\frac{128  H_{0,0,-1} P^{(2)}_9}{27 (1-x)^2 (1+x)^4}
        -\frac{64  \zeta_2 P^{(2)}_{18}}{27 (1-x)^2 (1+x)^6}
        +\frac{160  \big(
                635+1654 x+635 x^2\big)}{81 (1+x)^4}
\nonumber\\ &&
        + n_l \Biggl(
                -\frac{2176 }{27 (1+x)^2}
                +\Biggl(
                        -\frac{128  \big(
                                37-112 x+37 x^2\big)}{81 (1-x) (1+x)^3}
                        +\frac{512 x^2 \zeta_2}{3 (1-x) (1+x)^5}
                \Biggr) H_0
                +\frac{512  H_{-1} H_0}{27 (1-x^2)}
\nonumber\\ &&
                -\frac{64  \big(
                        3-15 x+27 x^2+x^3\big)}{27 (1-x) (1+x)^4} H_0^2
                +\frac{256 x^2 H_0^3}{9 (1-x) (1+x)^5}
                -
                \frac{512  H_{0,-1}}{27 (1-x^2)}
\nonumber\\ &&
                -\frac{128  \big(
                        -5-111 x+99 x^2+x^3\big)}{27 (1-x) (1+x)^4} \zeta_2
        \Biggr)
        +\Biggl(
                -\frac{128  H_{0,-1} P^{(2)}_6}{9 (1-x)^2 (1+x)^4}
                -\frac{16  P^{(2)}_{12}}{81 (1-x) (1+x)^5}
\nonumber\\ &&
                +\frac{64  \zeta_2 P^{(2)}_{16}}{27 (1-x)^3 (1+x)^5}
                +\Biggl(
                        \frac{256  \big(
                                5+22 x+5 x^2\big)}{27 (1-x) (1+x)^3}
                        -\frac{5120 x \big(
                                7-16 x+7 x^2\big)}{9 (1-x) (1+x)^5} \zeta_2
                \Biggr) H_1
\nonumber\\ &&
                +\frac{512  \big(
                        6-55 x+6 x^2\big)}{9 (1+x)^4} H_{0,1}
                -\frac{26624 x \big(
                        8-17 x+8 x^2\big)}{27 (1-x) (1+x)^5} H_{0,0,1}
\nonumber\\ &&
                +\frac{512 x \big(
                        521-1142 x+521 x^2\big)}{27 (1-x) (1+x)^5} H_{0,0,-1}
                +\frac{512 x \big(
                        209-479 x+209 x^2\big)}{27 (1-x) (1+x)^5} \zeta_3
        \Biggr) H_0
\nonumber\\ &&
        -\frac{256  \big(
                1+x^2\big)}{3 (1-x) (1+x)^3} H_{-1}^2 H_0
        +\Biggl(
                -\frac{256  H_1 P^{(2)}_3}{27 (1-x)^2 (1+x)^4}
                -\frac{64  P^{(2)}_{17}}{81 (1-x)^2 (1+x)^6}
\nonumber\\ &&
                +\frac{256 x \big(
                        313-688 x+313 x^2\big)}{27 (1-x) (1+x)^5} H_{0,1}
                -\frac{256 x P^{(2)}_8}{27 (1-x)^3 (1+x)^5} H_{0,-1}
                -\frac{128 x P^{(2)}_{14}}{27 (1-x)^3 (1+x)^6}
                 \zeta_2
        \Biggr) 
\nonumber\\ && \times
H_0^2
        +\Biggl(
                \frac{64 P^{(2)}_{15}}{81 (1-x)^3 (1+x)^5}
                -\frac{2560 x \big(
                        7-16 x+7 x^2\big)}{27 (1-x) (1+x)^5} H_1
        \Biggr) H_0^3
        -\frac{32 x P^{(2)}_{13}}{81 (1-x)^3 (1+x)^6} H_0^4
\nonumber\\ &&
        +\Biggl(
                \frac{128  \zeta_2 P^{(2)}_2}{9 (1-x)^2 (1+x)^4}
                +\frac{64  H_0^2 P^{(2)}_7}{27 (1-x)^2 (1+x)^4}
                -\frac{128  \big(
                        143-318 x+143 x^2\big)}{27 (1-x) (1+x)^3} H_0
\nonumber\\ &&
                +\frac{512  \big(
                        1+x^2\big)}{3 (1-x) (1+x)^3} H_{0,-1}
        \Biggr) H_{-1}
        +\Biggl(
               - \frac{256  \big(
                        5+22 x+5 x^2\big)}{27 (1-x) (1+x)^3}
                +\frac{5120 x \big(
                        7-16 x+7 x^2\big)}{9 (1-x) (1+x)^5} \zeta_2
        \Biggr) 
\nonumber\\ && \times
H_{0,1}
        +\Biggl(
                \frac{128  \big(
                        143-318 x+143 x^2\big)}{27 (1-x) (1+x)^3}
                -\frac{512 x P^{(2)}_{1}}{9 (1-x)^3 (1+x)^5} \zeta_2
        \Biggr) H_{0,-1}
\nonumber\\ &&
        -\frac{512  \big(
                1+x^2\big)}{3 (1-x) (1+x)^3} H_{0,-1,-1}
        +\frac{512 x \big(
                173-356 x+173 x^2\big)}{9 (1-x) (1+x)^5} H_{0,0,0,1}
\nonumber\\ &&
        -\frac{512 x P^{(2)}_{10}}{9 (1-x)^3 (1+x)^5} H_{0,0,0,-1}
        -\frac{128 x P^{(2)}_{11}}{45 (1-x)^3 (1+x)^5} \zeta_2^2
\Biggr]
\Biggr\} + F_{V,2}^{(0)},
\end{eqnarray}
with 
\begin{eqnarray}
P^{(2)}_1&=&x^4-9 x^3+10 x^2-9 x+1
\\
P^{(2)}_2 &=&11 x^4-12 x^3+26 x^2-12 x-1
\\
P^{(2)}_3 &=&14 x^4-215 x^3+370 x^2-215 x+14
\\
P^{(2)}_4 &=&22 x^4-187 x^3+362 x^2-187 x+22
\\
P^{(2)}_5 &=&58 x^4-937 x^3+1616 x^2-937 x+40
\\
P^{(2)}_6 &=&65 x^4-876 x^3+1578 x^2-876 x+65
\\
P^{(2)}_7 &=&125 x^4-1772 x^3+3262 x^2-1772 x+161
\\
P^{(2)}_8&=&209 x^4-887 x^3+1354 x^2-887 x+209
\\
P^{(2)}_9 &=&265 x^4-3484 x^3+6206 x^2-3484 x+229
\\
P^{(2)}_{10}&=&312 x^4-1297 x^3+1972 x^2-1297 x+312
\\
P^{(2)}_{11}&=&590 x^4-2447 x^3+3760 x^2-2447 x+590
\\
P^{(2)}_{12}&=&609 x^4+21136 x^3+56414 x^2+21136 x+609
\\
P^{(2)}_{13}&=&x^5+148 x^4-143 x^3+145 x^2-140 x+1
\\
P^{(2)}_{14}&=&104 x^5-263 x^4+173 x^3+317 x^2-407 x+104
\\
P^{(2)}_{15}&=&86 x^6-1756 x^5+2685 x^4-1008 x^3-470 x^2+76 x-45
\\
P^{(2)}_{16}&=&157 x^6-2540 x^5+3393 x^4-1728 x^3+299 x^2-532 x+87
\\
P^{(2)}_{17}&=&671 x^6+284 x^5-322 x^4-2888 x^3+863 x^2+188 x+212
\\
P^{(2)}_{18}&=&24 \ln(2) (x-1)^2 (x+1)^4-61 x^6+7514 x^5+2979 x^4-17684 x^3
\nonumber\\ &&
+1601 x^2+7194 x+185.
\end{eqnarray}

The constant parts $F_{V,1(2)}^{(0)}$ are given by the genuine terms and  more simple terms, 
$F_{V,1(2)}^{(0),r}$, resulting
from the pole-terms in the $\ep$-expansion. The latter terms are not displayed explicitly, because of space reasons.
In an attachment we will present the complete renormalized form factors. Because of contributions due to non 
first order factorizing terms (elliptic and higher) in some color--zeta combinations higher functions contribute,
for which we have calculated analytically at least 2000 moments, and in some cases 4000, 6000 and 8000 to determine 
their recurrence relations. They are used to define the corresponding numeric representations.
The corresponding recurrences, which do not factorize in first order completely will be studied 
elsewhere. 
\begin{eqnarray}
F_{V,1}^{(0)} &=& n_h^2 \Biggl\{
        -\frac{64 H_{0,-1,-1} P^{(3)}_3}{27 (1-x) (1+x)^3}
        +\frac{256  P^{(3)}_5}{27 (1+x)^4} \ln(2) \zeta_2
        -\frac{128 H_{0,0,1} P^{(3)}_8}{81 (1+x)^4}
        -\frac{64 H_{0,-1} P^{(3)}_{12}}{243 (1-x) (1+x)^3}
\nonumber\\ &&
        +\frac{8 P^{(3)}_{14}}{729 (1+x)^4}
        +\frac{32 H_{0,0,-1} P^{(3)}_{17}}{81 (1-x) (1+x)^4}
        +\frac{64 \zeta_2^2 P^{(3)}_{22}}{9 (1-x) (1+x)^5}
        -\frac{64 H_{0,0,0,-1} P^{(3)}_{25}}{27 (1-x) (1+x)^5}
\nonumber\\ &&
        -\frac{4 H_0^4 P^{(3)}_{31}}{27 (1-x) (1+x)^5}
        +\frac{8 H_0^3 P^{(3)}_{52}}{243 (1-x) (1+x)^7}
        +\Biggl(
                \frac{128 H_{0,1} P^{(3)}_8}{81 (1+x)^4}
                -\frac{16 P^{(3)}_{46}}{729 (1-x) (1+x)^5}
\nonumber\\ &&
                +\frac{512\big(
                        1+6 x+x^2\big) P^{(3)}_1}{27 (1-x) (1+x)^5} H_{0,0,1} - \frac{128(1+x^2)}{27(1-x^2)} 
H_{0,0,-1}
        \Biggr) H_0
        +\frac{128 \big(
                1+x^2\big)}{81 (1-x^2)} H_{-1}^3 H_0
\nonumber\\ &&
        +\Biggl(
                -\frac{64 H_1 P^{(3)}_8}{81 (1+x)^4}
                +\frac{32 P^{(3)}_{47}}{243 (1-x) (1+x)^6}
                -\frac{128 \big(
                        1+6 x+x^2\big) P^{(3)}_1}{27 (1-x) (1+x)^5} H_{0,1}
\nonumber\\ &&             
   +\frac{64 \big(
                        1+x^2\big)}{27 (1-x^2)} H_{0,-1}
        \Biggr) H_0^2
        +\Biggl(
                \frac{64 H_{0,-1} P^{(3)}_3}{27 (1-x) (1+x)^3}
                +\frac{64 H_0 P^{(3)}_{12}
                }{243 (1-x) (1+x)^3}
\nonumber\\ &&               
  -
                \frac{16 H_0^2 P^{(3)}_{17}}{81 (1-x) (1+x)^4}
                +\frac{32 H_0^3 P^{(3)}_{25}}{81 (1-x) (1+x)^5}
                +\frac{128 \big(
                        1+x^2\big)}{27 (1-x^2)} [H_{0,0,-1} + 2  H_{0,-1,-1}]
        \Biggr) 
\nonumber\\ &&  \times
H_{-1}
        +\Biggl(
                -\frac{32 H_0 P^{(3)}_3}{27 (1-x) (1+x)^3}
                -\frac{32 \big(
                        1+x^2\big)}{27 (1-x^2)} H_0^2
                -\frac{128 \big(
                        1+x^2\big)}{27 (1-x^2)} H_{0,-1}
        \Biggr) H_{-1}^2
\nonumber\\ &&         
-\frac{256 \big(
                1+6 x+x^2\big) P^{(3)}_1}{9 (1-x) (1+x)^5} H_{0,0,0,1}
        -\frac{128 \big(
                1+x^2\big)}{27 (1-x^2)} H_{0,0,-1,-1}
        -\frac{256 \big(
                1+x^2\big)}{27 (1-x^2)} H_{0,-1,-1,-1}
\nonumber\\ &&
        +\Bigg(
                -\frac{8 H_0^2 P^{(3)}_{26}}{3 (1-x) (1+x)^5}
                +\frac{32 H_{0,-1} P^{(3)}_{32}}{9 (1-x) (1+x)^5}
                +\frac{32 P^{(3)}_{50}}{1215 (1-x) (1+x)^6}
                +\frac{8 H_0 P^{(3)}_{53}}{81 (1-x) (1+x)^7}
\nonumber\\ &&              
  -\Biggl(
                        \frac{32 P^{(3)}_{18}}{27 (1-x) (1+x)^4}
                        -\frac{32 H_0 P^{(3)}_{24}}{27 (1-x) (1+x)^5}
                \Biggr) H_{-1}
                +\frac{64 \big(
                        1+x^2\big)}{27 (1-x^2)} H_{-1}^2
        \Biggr) \zeta_2
\nonumber\\ &&
        +\Biggl(
                \frac{16 P^{(3)}_{20}}{81 (1-x) (1+x)^4}
                +\frac{32 H_0 P^{(3)}_{34}
                }{27 (1-x) (1+x)^5}
                -
                \frac{128 \big(
                        1+x^2\big)}{27 (1-x^2)} H_{-1}
        \Biggr) \zeta_3
\Biggr\}
\nonumber\\ &&
+ n_h \Biggl\{ 
       \Biggl(
                \frac{128 P^{(3)}_9}{27 (1+x)^4}
                +\big(
                        -\frac{2048 P^{(3)}_{30}}{27 (1-x) (1+x)^5}
                        -\frac{1024 x^2 \big(
                                5-2 x+5 x^2\big)}{(1-x) (1+x)^5} H_1
                \big) H_0
\nonumber\\ &&              
  -\frac{256 x^2 \big(
                        5-2 x+5 x^2\big)}{(1-x) (1+x)^5} H_0^2
                +\frac{1024 x^2 \big(
                        5-2 x+5 x^2\big)}{(1-x) (1+x)^5} H_{0,1}
        \Biggr)
\Li_4\left(\frac{1}{2}\right)
\nonumber\\ && 
        + \Biggl(
                \frac{16 P^{(3)}_9}{81 (1+x)^4}
                +\Biggl(
                        -\frac{256 P^{(3)}_{30}}{81 (1-x) (1+x)^5}
                        -\frac{128 x^2 \big(
                                5-2 x+5 x^2\big)}{3 (1-x) (1+x)^5} H_1
                \Biggr) H_0
\nonumber\\ &&
                -\frac{32 x^2 \big(
                        5-2 x+5 x^2\big)}{3 (1-x) (1+x)^5} H_0^2
                +\frac{128 x^2 \big(
                        5-2 x+5 x^2\big)}{3 (1-x) (1+x)^5} H_{0,1}
                -\frac{128 x^2 \big(
                        5-2 x+5 x^2\big)}{3 (1-x) (1+x)^5} \zeta_2
        \Biggr) 
\nonumber\\ &&
\times \ln^4(2)
        +  \Biggl[
                \frac{64 \zeta_2^2 P^{(3)}_{41}}{9 (1-x) (1+x)^5}
                +\Biggl(
                        -\frac{128 H_{0,1} P^{(3)}_{36}}{9 (1-x) (1+x)^5}
                        -\frac{128 H_{0,-1} P^{(3)}_{40}}{9 (1-x) (1+x)^5}
\nonumber\\ &&              
          +\frac{64 H_0^2 P^{(3)}_{51}}{9 (1-x)^2 (1+x)^6}
                        +\frac{16 P^{(3)}_{54}}{81 x (1+x)^6}
                        +\Biggl(
                                \frac{128 H_1 P^{(3)}_{36}
                                }{9 (1-x) (1+x)^5}
                  -
                                \frac{32 P^{(3)}_{43}}{27 (1-x) (1+x)^5}
                        \Biggr) H_0
\nonumber\\ &&              
                        -\frac{64 x^2 \big(
                                7-22 x+7 x^2\big)}{(1-x) (1+x)^5} H_0^3
                        +\Biggl(
                                -\frac{128 P^{(3)}_4}{3 (1+x)^4}
                                +\frac{128 H_0 P^{(3)}_{39}}{9 (1-x) (1+x)^5}
                        \Biggr) H_{-1}
  \Biggr) \zeta_2
        \Biggr] \ln(2)
\nonumber\\ && 
        + \Biggl[
                \Biggl(
                        \frac{32 P^{(3)}_{11}}{27 (1+x)^4}
                        +\Biggl(
                                \frac{128 P^{(3)}_{35}}{27 (1-x) (1+x)^5}
                                +\frac{256 x^2 \big(
                                        5-2 x+5 x^2\big)}{(1-x) (1+x)^5} H_1
                        \Biggr) H_0 
\nonumber\\ &&                
         +\frac{64 x^2 \big(
                                5-2 x+5 x^2\big)}{(1-x) (1+x)^5} H_0^2
                        -\frac{256 x^2 \big(
                                5-2 x+5 x^2\big)}{(1-x) (1+x)^5} H_{0,1}
                \Biggr) \zeta_2
                +\frac{256 x^2 \big(
                        5-2 x+5 x^2\big)}{(1-x) (1+x)^5} 
\nonumber\\ &&  \times
\zeta_2^2
        \Biggr] 
\ln^2(2)
        +n_l \Biggl[
                \frac{256 P^{(3)}_5}{27 (1+x)^4} \ln(2) \zeta_2
                -\frac{512 H_{0,-1,-1} P^{(3)}_6}{81 (1-x) (1+x)^3}
                -\frac{128 H_{0,0,1} P^{(3)}_8}{81 (1+x)^4}
\nonumber\\ &&                
 -\frac{128 H_{0,-1} P^{(3)}_{10}}{81 (1-x) (1+x)^3}
                +\frac{64 P^{(3)}_{13}}{729 (1+x)^4}
                -\frac{128 H_{0,0,-1} P^{(3)}_{16}}{81 (1-x) (1+x)^4}
                -\frac{128 H_{0,0,0,-1} P^{(3)}_{28}}{9 (1-x) (1+x)^5}
\nonumber\\ &&
                -\frac{8 H_0^4 P^{(3)}_{33}}{27 (1-x) (1+x)^5}
                +\frac{64 \zeta_2^2 P^{(3)}_{37}
                }{135 (1-x) (1+x)^5}
                +
                \frac{16 H_0^3 P^{(3)}_{38}}{81 (1-x) (1+x)^5}
                +\Biggl(
                        \frac{128 H_{0,1} P^{(3)}_8}{81 (1+x)^4}
\nonumber\\ &&               
          -\frac{64 P^{(3)}_{45}}{729 (1-x) (1+x)^5}
                        +\frac{\big(
                                1+6 x+x^2\big) P^{(3)}_1}{(1-x) (1+x)^5} \left[\frac{512}{27} H_{0,0,1}
                        -\frac{128}{9} H_{0,0,-1} \right]
                \Biggr) H_0
\nonumber\\ &&                
 +\frac{1024 \big(
                        1+x^2\big)}{81 (1-x^2)} H_{-1}^3 H_0
                -\Biggl(
                        \frac{64 H_1 P^{(3)}_8}{81 (1+x)^4}
                        +\frac{32 P^{(3)}_{48}}{81 (1-x) (1+x)^6}
                        +\frac{128 \big(
                                1+6 x+x^2\big) P^{(3)}_1}{27 (1-x) (1+x)^5} 
\nonumber\\ &&                
\times H_{0,1}
         -\frac{64 \big(
                                1+6 x+x^2\big) P^{(3)}_1}{9 (1-x) (1+x)^5} H_{0,-1}
                \Biggr) H_0^2
                +\Biggl(
                        \frac{512 H_{0,-1} P^{(3)}_6}{81 (1-x) (1+x)^3}
                        +\frac{128 H_0 P^{(3)}_{10}}{81 (1-x) (1+x)^3}
\nonumber\\ &&                
                        +\frac{64 H_0^2 P^{(3)}_{16}}{81 (1-x) (1+x)^4}
         +\frac{64 H_0^3 P^{(3)}_{28}}{27 (1-x) (1+x)^5}
                        +\frac{1024 \big(
                                1+x^2\big)}{27 (1-x^2)} H_{0,0,-1}
                        +\frac{2048 \big(
                                1+x^2\big)}{27 (1-x^2)} 
\nonumber\\ && \times 
H_{0,-1,-1}
                \Biggr) H_{-1}
                -\Bigg(
                        \frac{256 H_0 P^{(3)}_6}{81 (1-x) (1+x)^3}
                        +\frac{256 \big(
                                1+x^2\big)}{27 (1-x^2)} H_0^2
                        +\frac{1024 \big(
                                1+x^2\big)}
                        {27 (1-x^2)} H_{0,-1}
                \Biggr) 
\nonumber\\ && \times 
H_{-1}^2
                -
                \frac{256 \big(
                        1+6 x+x^2\big) P^{(3)}_1}{9 (1-x) (1+x)^5} H_{0,0,0,1}
                -\frac{1024 \big(
                        1+x^2\big)}{27 (1-x^2)} H_{0,0,-1,-1}
                -\frac{2048 \big(
                        1+x^2\big)}{27 (1-x^2)} 
\nonumber\\ && \times 
H_{0,-1,-1,-1}
                +\Biggl(
                         \frac{64 H_{0,-1} P^{(3)}_{23}}{9 (1-x) (1+x)^5}
                        -\frac{16 H_0^2 P^{(3)}_{27}}{3 (1-x) (1+x)^5}
                        +\frac{16 H_0 P^{(3)}_{42}}{81 (1-x) (1+x)^5}
\nonumber\\ &&                       
  +\frac{8 P^{(3)}_{49}}{81 (1-x) (1+x)^6}
                        +\big(
                                -\frac{128 P^{(3)}_{19}}{81 (1-x) (1+x)^4}
                                +\frac{64 H_0 P^{(3)}_{29}}{27 (1-x) (1+x)^5}
                        \big) H_{-1}
\nonumber\\ && 
                        +\frac{512 \big(
                                1+x^2\big)}{27 (1-x^2)} 
H_{-1}^2
 \Biggr) \zeta_2
+\Biggl(
                         \frac{16 P^{(3)}_{21}}{81 (1-x) (1+x)^4}
                        +\frac{32 \big(
                                1+x^2\big)}{3 (1-x^2)} H_0
                        -\frac{1024 \big(
                                1+x^2\big)}{27 (1-x^2)} 
\nonumber\\ && 
\times
H_{-1}
                \Biggr) \zeta_3
        \Biggr]
        +\Biggl(
                -\frac{1024  x^2 \big(
                        5-2 x+5 x^2\big)}{(1-x) (1+x)^5} \Li_4\left(\frac{1}{2}\right)
                +\Biggl(
                        -\frac{32 x^2 \big(
                                5-32 x+5 x^2\big)}{(1-x) (1+x)^5} H_0
\nonumber\\  && 
                        -\frac{8 P^{(3)}_7}{(1+x)^4}
                \Biggr) \zeta_3
        \Biggr) \zeta_2
        +\Biggl(
                -\frac{8 P^{(3)}_{15}}{135 (1+x)^4}
                +\Biggl(
                        \frac{64 x^2 \big(
                                43+98 x+43 x^2\big)}
                        {5 (1-x) (1+x)^5} H_1
\nonumber\\ && 
                        -\frac{32 P^{(3)}_{44}}{135 (1-x) (1+x)^5}
                \Biggr) H_0
                +\frac{16 x^2 \big(
                        43+98 x+43 x^2\big)}{5 (1-x) (1+x)^5} H_0^2
                -\frac{64 x^2 \big(
                        43+98 x+43 x^2\big)}{5 (1-x) (1+x)^5} 
\nonumber\\ &&       
\times H_{0,1}
        \Biggr) \zeta_2^2
 +\frac{64 x^2 \big(
                43+98 x+43 x^2\big)}{5 (1-x) (1+x)^5} \zeta_2^3
        +\Biggl(
                \frac{80 x^2 \big(
                        1-16 x+x^2\big)}{(1-x) (1+x)^5} H_0
\nonumber\\ &&                
+ \frac{20 P^{(3)}_2}{(1+x)^4}
        \Biggr) \zeta_5
\Biggr\}
+ n_h F_{V,1,1}^{(0)}(x)
+ n_h \zeta_2 F_{V,1,2}^{(0)}(x)
+ n_h \zeta_3 F_{V,1,3}^{(0)}(x) 
+ F_{V,1}^{(0),r}(x),
\end{eqnarray}
with the polynomials
\begin{align}
P^{(3)}_1 &=x^4-2 x^3+12 x^2-2 x+1,
\\
P^{(3)}_2 &=x^4+8 x^3+50 x^2+8 x+1,
\\
P^{(3)}_3 &=x^4+12 x^3+6 x^2+12 x+1,
\\
P^{(3)}_4 &=2 x^4-7 x^3+12 x^2-7 x+2,
\\
P^{(3)}_5 &=11 x^4+14 x^3+238 x^2+14 x+11,
\\
P^{(3)}_6 &=11 x^4+40 x^3+34 x^2+40 x+11,
\\
P^{(3)}_7 &=13 x^4+56 x^3+170 x^2+56 x+13,
\\
P^{(3)}_8 &=19 x^4-2 x^3+206 x^2-2 x+19,
\\
P^{(3)}_9 &=53 x^4+184 x^3+502 x^2+184 x+53,
\\
P^{(3)}_{10} &=89 x^4+108 x^3+710 x^2+108 x+89,
\\
P^{(3)}_{11} &=91 x^4-94 x^3-70 x^2-94 x+91,
\\
P^{(3)}_{12} &=155 x^4+10 x^3+1846 x^2+10 x+155,
\\
P^{(3)}_{13} &=377 x^4+3452 x^3+6006 x^2+3452 x+377,
\\
P^{(3)}_{14} &=1367 x^4+26372 x^3+46362 x^2+26372 x+1367,
\\
P^{(3)}_{15} &=5363 x^4+15190 x^3+15322 x^2+15190 x+5363,
\\
P^{(3)}_{16} &=3 x^5+123 x^4-60 x^3+356 x^2+81 x+41,
\\
P^{(3)}_{17} &=35 x^5-81 x^4+362 x^3-470 x^2+3 x-41,
\\
P^{(3)}_{18} &=39 x^5-29 x^4+434 x^3-398 x^2+55 x-37,
\\
P^{(3)}_{19} &=79 x^5+39 x^4+772 x^3-476 x^2+165 x-35,
\\
P^{(3)}_{20} &=253 x^5+435 x^4+7450 x^3-7234 x^2-279 x-241,
\\
P^{(3)}_{21} &=463 x^5+837 x^4+7414 x^3-5046 x^2+795 x-111,
\\
P^{(3)}_{22} &=x^6+4 x^5+78 x^3+4 x+1,
\\
P^{(3)}_{23} &=x^6+4 x^5-21 x^4+288 x^3-21 x^2+4 x+1,
\\
P^{(3)}_{24} &=x^6+4 x^5-17 x^4+248 x^3-17 x^2+4 x+1,
\\
P^{(3)}_{25} &=x^6+4 x^5-5 x^4+128 x^3-5 x^2+4 x+1,
\\
P^{(3)}_{26} &=x^6+4 x^5-x^4+88 x^3-x^2+4 x+1,
\\
P^{(3)}_{27} &=x^6+4 x^5+3 x^4+48 x^3+3 x^2+4 x+1,
\\
P^{(3)}_{28} &=x^6+4 x^5+9 x^4-12 x^3+9 x^2+4 x+1,
\\
P^{(3)}_{29} &=x^6+4 x^5+19 x^4-112 x^3+19 x^2+4 x+1,
\\
P^{(3)}_{30} &=x^6+10 x^5+36 x^4+84 x^3+36 x^2+10 x+1,
\\
P^{(3)}_{31} &=3 x^6+12 x^5+x^4+224 x^3+x^2+12 x+3,
\\
P^{(3)}_{32} &=3 x^6+12 x^5+5 x^4+184 x^3+5 x^2+12 x+3,
\\
P^{(3)}_{33} &=3 x^6+12 x^5+11 x^4+124 x^3+11 x^2+12 x+3,
\\
P^{(3)}_{34} &=11 x^6+44 x^5+5 x^4+808 x^3+5 x^2+44 x+11,
\\
P^{(3)}_{35} &=13 x^6+67 x^5+243 x^4+228 x^3+243 x^2+67 x+13,
\\
P^{(3)}_{36} &=21 x^6-76 x^5-193 x^4-36 x^3-193 x^2-76 x+21,
\\
P^{(3)}_{37} &=31 x^6+124 x^5+304 x^4-622 x^3+304 x^2+124 x+31,
\\
P^{(3)}_{38} &=35 x^6-118 x^5+325 x^4-404 x^3-423 x^2-110 x-41,
\\
P^{(3)}_{39} &=99 x^6+70 x^5+61 x^4+24 x^3+61 x^2+70 x+99,
\\
P^{(3)}_{40} &=105 x^6+88 x^5+127 x^4-48 x^3+127 x^2+88 x+105,
\\
P^{(3)}_{41} &=147 x^6-64 x^5-259 x^4-120 x^3-259 x^2-64 x+147,
\\
P^{(3)}_{42} &=189 x^6-390 x^5+1827 x^4-1684 x^3-1165 x^2-358 x-115,
\\
P^{(3)}_{43} &=381 x^6+316 x^5-2779 x^4+13392 x^3-3139 x^2+568 x+309,
\\
P^{(3)}_{44} &=806 x^6+2501 x^5+5574 x^4+9924 x^3+5574 x^2+2501 x+806,
\\
P^{(3)}_{45} &=1471 x^6+5920 x^5+19873 x^4+30272 x^3+19873 x^2+5920 x+1471,
\\
P^{(3)}_{46} &=2945 x^6+9044 x^5+56687 x^4+91384 x^3+56687 x^2+9044 x+2945,
\\
P^{(3)}_{47} &=19 x^7-43 x^6+476 x^5-2038 x^4-3695 x^3-2817 x^2-432 x-174,
\\
P^{(3)}_{48} &=62 x^7+269 x^6+965 x^5+2426 x^4+2876 x^3+1637 x^2+481 x+116,
\\
P^{(3)}_{49} &=1969 x^7+7833 x^6+31477 x^5+37933 x^4+4483 x^3-10661 x^2-1833 x-545,
\\
P^{(3)}_{50} &=4452 x^7+17001 x^6+81141 x^5+78032 x^4-20702 x^3-57731 x^2-12251 x-2902,
\\
P^{(3)}_{51} &=70 x^8+173 x^7-342 x^6+245 x^5+860 x^4+241 x^3-458 x^2+45 x-2,
\\
P^{(3)}_{52} &=181 x^8+240 x^7+1444 x^6+1952 x^5-3522 x^4-4000 x^3-1788 x^2-336 x-123,
\\
P^{(3)}_{53} &=213 x^8+240 x^7+1452 x^6+1264 x^5-6642 x^4-4688 x^3-1780 x^2-336 x-91,
\\
P^{(3)}_{54} &=2768 x^8-60837 x^7-222044 x^6-979155 x^5-1790040 x^4-979155 x^3-222044 x^2
\nonumber\\ &
-60837 x+2768.
\end{align}

\noindent
In the variable $y = 1-x$ the first expansion coefficients of the functions $F_{V,1,i}^{(0)}(x), i = 1..3$ are 
given by
\begin{eqnarray}
F_{V,1,1}^{(0)}(x) &=& 
-\frac{2222242}{243} + \frac{1047067 y^2}{729} + \frac{1047067 y^3}{729} +
  \frac{3436873681 y^4}{3499200} 
\nonumber\\ &&
+ \frac{923912881 y^5}{1749600} 
+ O(y^6) 
\\
F_{V,1,2}^{(0)}(x) &=& 
\frac{2390434}{243} - \frac{8763197 y^2}{4860} - \frac{8763197 y^3}{4860} -
  \frac{4103868673 y^4}{3402000} 
\nonumber\\ &&
- \frac{345583241 y^5}{567000} + O(y^6)
\\
F_{V,1,3}^{(0)}(x) &=& 
\frac{311488}{81} - \frac{259276 y^2}{243} - \frac{259276 y^3}{243} -
  \frac{571282067 y^4}{777600} - \frac{156440467 y^5}{388800} 
+ O(y^6)
\nonumber\\
\end{eqnarray}
and
\begin{eqnarray}
F_{V,2}^{(0)} &=&  x \Biggl\{n_h^2 \Biggl\{
        -
        \frac{128 \big(
                451+830 x+451 x^2\big)}{243 (1+x)^4}
        +\Biggl(
                \frac{128 P^{(4)}_{11}}{243 (x-1) (1+x)^5}
                -\frac{512 \big(
                        7-79 x+7 x^2\big)}{81 (x-1) (1+x)^3} H_{-1}
\nonumber\\ &&
                -\frac{128 (x-1) H_{-1}^2}{27 (1+x)^3}
        \Biggr) H_0
        +\Biggl(
                -\frac{128 P^{(4)}_{14}}{81 (x-1) (1+x)^6}
                +\frac{64 \big(
                        1-40 x+3 x^2\big)}{27 (1+x)^4} H_{-1}
        \Biggr) H_0^2
\nonumber\\ &&              
        +\Biggl(
                -\frac{32 P^{(4)}_{18}}{81 (x-1) (1+x)^7}
  +\frac{512 x^2 H_{-1}}{9 (x-1) (1+x)^5}
        \Biggr) H_0^3
        -\frac{320 x^2 H_0^4}{9 (x-1) (1+x)^5}
\nonumber\\ &&
        -\frac{256 \big(
                1-20 x+x^2\big) H_0^2 H_1}{27 (1+x)^4}
        +\Biggl(
                \frac{512 \big(
                        1-20 x+x^2\big) H_0}{27 (1+x)^4}
\nonumber\\ &&              
                -\frac{1024 x^2 H_0^2}{3 (x-1) (1+x)^5}
        \Biggr) H_{0,1}
        +\Biggl(
                \frac{512 \big(
                        7-79 x+7 x^2\big)}{81 (x-1) (1+x)^3}
  +\frac{256 (x-1) H_{-1}}{27 (1+x)^3}
        \Biggr) H_{0,-1}
\nonumber\\ &&     
        +\Biggl(
                -\frac{512 \big(
                        1-20 x+x^2\big)}{27 (1+x)^4}
                +\frac{4096 x^2 H_0}{3 (x-1) (1+x)^5}
        \Biggr) H_{0,0,1}
   -\frac{128 \big(
                1-40 x+3 x^2\big)}{27 (1+x)^4} H_{0,0,-1}
\nonumber\\ &&      
        -\frac{256 (x-1) H_{0,-1,-1}}{27 (1+x)^3}
        -\frac{2048 x^2 H_{0,0,0,1}
        }{(x-1) (1+x)^5}
  -
        \frac{1024 x^2 H_{0,0,0,-1}}{3 (x-1) (1+x)^5}
\nonumber\\ &&              
        +\Biggl(
                -\frac{1024 \big(
                        1+22 x+x^2\big)}{9 (1+x)^4} \ln(2)
                -\frac{32 P^{(4)}_{16}}{405 (x-1) (1+x)^6}
  +\Biggl(
                        -\frac{32 P^{(4)}_{19}}{27 (x-1) (1+x)^7}
\nonumber\\ &&
                        +\frac{1024 x^2 H_{-1}}{3 (x-1) (1+x)^5}
                \Biggr) H_0
                -\frac{256 x^2 H_0^2}{(x-1) (1+x)^5}
+\frac{128 \big(
                        7-120 x+5 x^2\big)}{27 (1+x)^4} H_{-1}
\nonumber\\ &&
                +\frac{2048 x^2 H_{0,-1}}{3 (x-1) (1+x)^5}
        \Biggr) \zeta_2
        +\frac{1792 x^2 \zeta_2^2}{3 (x-1) (1+x)^5}
        +\big(
                \frac{128 \big(
                        27+364 x+29 x^2\big)}{27 (1+x)^4}
\nonumber\\ &&           
     +\frac{1024 x^2 H_0}{(x-1) (1+x)^5}
        \Biggr) \zeta_3
\Biggr\}
\nonumber\\ &&        
+ n_h \Biggl\{
        -\frac{160 \zeta_5 P^{(4)}_1}{(x-1)^2 (1+x)^4}
+\ln^4(2) \Biggl(
                \frac{128 x H_0 P^{(4)}_7}{27 (x-1)^3 (1+x)^5}
                -\frac{128 P^{(4)}_8}{81 (x-1)^2 (1+x)^4}
\nonumber\\ &&              
                -\frac{128 x \big(
                        1-x+x^2\big)}{3 (x-1) (1+x)^5} H_0^2
  -\frac{512 x \big(
                        1-x+x^2\big)}{3 (x-1) (1+x)^5} H_0 H_1
                +\frac{512 x \big(
                        1-x+x^2\big)}{3 (x-1) (1+x)^5} H_{0,1}
        \Biggr)
\nonumber\\ &&               
        +\Li_4\left(\frac{1}{2}\right) \Biggl(
                \frac{1024 x H_0 P^{(4)}_7}{9 (x-1)^3 (1+x)^5}
 -\frac{1024 P^{(4)}_8}{27 (x-1)^2 (1+x)^4}
                -
                \frac{1024 x \big(
                        1-x+x^2\big)}{(x-1) (1+x)^5} H_0^2
\nonumber\\ &&              
                -\frac{4096 x \big(
                        1-x+x^2\big)}{(x-1) (1+x)^5} H_0 H_1
  +\frac{4096 x \big(
                        1-x+x^2\big)}{(x-1) (1+x)^5} H_{0,1}
        \Biggr)
        + n_l \Biggl[
                \Biggl(
                        \frac{256 P^{(4)}_9}{81 (x-1) (1+x)^5}
\nonumber\\ &&
          -\frac{512 \big(
                                23-130 x+23 x^2\big)}{81 (x-1) (1+x)^3} H_{-1}
                        +\frac{2048 x H_{-1}^2}{27 (x-1) (1+x)^3}
                \Biggr) H_0
                -\frac{1024 \big(
                        14+27 x+14 x^2\big)}{81 (1+x)^4}
\nonumber\\ &&              
                +\Biggl(
                        \frac{128 P^{(4)}_{15}}{81 (x-1) (1+x)^6}
          +\frac{256 \big(
                                -1+17 x-25 x^2+x^3\big)}{27 (x-1) (1+x)^4} H_{-1}
                \Biggr) H_0^2
\nonumber\\ &&              
                +\Biggl(
                        -\frac{64 P^{(4)}_4}{27 (x-1) (1+x)^5}
                        -\frac{512 x^2 H_{-1}}{9 (x-1) (1+x)^5}
                \Biggr) H_0^3
  -\frac{320 x^2 H_0^4}{9 (x-1) (1+x)^5}
\nonumber\\ &&              
                -\frac{256 \big(
                        1-20 x+x^2\big) H_0^2 H_1}{27 (1+x)^4}
                +\Biggl(
                        \frac{512 \big(
                                1-20 x+x^2\big) H_0}{27 (1+x)^4}
          -\frac{1024 x^2 H_0^2}{3 (x-1) (1+x)^5}
                \Biggr) H_{0,1}
\nonumber\\ &&              
                +\Biggl(
                        \frac{512 \big(
                                23-130 x+23 x^2\big)}{81 (x-1) (1+x)^3}
                        +\frac{512 x^2 H_0^2}{(x-1) (1+x)^5}
                        -\frac{4096 x H_{-1}}{27 (x-1) (1+x)^3}
                \Biggr) H_{0,-1}
\nonumber\\ &&              
  +\Biggl(
                        -
                        \frac{512 \big(
                                1-20 x+x^2\big)}{27 (1+x)^4}
                        +\frac{4096 x^2 H_0}{3 (x-1) (1+x)^5}
                \Biggr) H_{0,0,1}
                -\Biggl(
	          \frac{1024 x^2 H_0}{(x-1) (1+x)^5}
\nonumber\\ &&              
+\frac{512 \big(
                                -1+17 x-25 x^2+x^3\big)}{27 (x-1) (1+x)^4}
                \Biggr) H_{0,0,-1}
                +\frac{4096 x H_{0,-1,-1}}{27 (x-1) (1+x)^3}
                -\frac{2048 x^2 H_{0,0,0,1}}{(x-1) (1+x)^5}
\nonumber\\ &&              
  +\frac{1024 x^2 H_{0,0,0,-1}}{3 (x-1) (1+x)^5}
                +\Biggl(
                        -\frac{1024 \big(
                                1+22 x+x^2\big)}{9 (1+x)^4} \ln(2)
                        -\frac{32 P^{(4)}_{17}}{81 (x-1) (1+x)^6}
\nonumber\\ &&              
          +\Biggl(
                                -\frac{128 P^{(4)}_6}{27 (x-1) (1+x)^5}
                                -\frac{1024 x^2 H_{-1}}{3 (x-1) (1+x)^5}
                        \Biggr) H_0
                        -\frac{256 x^2 H_0^2}{(x-1) (1+x)^5}
\nonumber\\ &&              
          +\frac{512 \big(
                                -3+67 x-59 x^2+3 x^3\big)}{27 (x-1) (1+x)^4} H_{-1}
                        +\frac{7168 x^2 H_{0,-1}}{3 (x-1) (1+x)^5}
                \Biggr) \zeta_2
                -\frac{7424 x^2 \zeta_2^2}{15 (x-1) (1+x)^5}
\nonumber\\ &&              
  +\frac{256 \big(
                        -7-155 x+139 x^2+7 x^3\big)}{27 (x-1) (1+x)^4} \zeta_3
        \Biggr]
        +\Biggl(
                -\frac{4096 x \big(
                        1-x+x^2\big)}{(x-1) (1+x)^5} \Li_4\left(\frac{1}{2}\right)
\nonumber\\ &&              
                -\frac{512 x \big(
                        1-x+x^2\big)}{3 (x-1) (1+x)^5} \ln^4(2)
  +\ln(2)
                 \Biggl(
                        -
                        \frac{256 H_{-1} P^{(4)}_5}{3 (x-1)^2 (1+x)^4}
                        +\frac{1024 H_{0,-1} P^{(4)}_{20}}{9 (x-1)^3 (1+x)^5}
\nonumber\\ &&              
                        -\frac{512 H_0 H_1 P^{(4)}_{23}}{9 (x-1)^3 (1+x)^5}
          +\frac{512 H_{0,1} P^{(4)}_{23}}{9 (x-1)^3 (1+x)^5}
                        +\frac{256 x H_0^2 P^{(4)}_{24}}{9 (x-1)^3 (1+x)^6}
\nonumber\\ &&              
                        +\frac{64 P^{(4)}_{26}}{81 (x-1)^2 x (1+x)^6}
          +\Biggl(
                                -\frac{1024 H_{-1} P^{(4)}_{22}}{9 (x-1)^3 (1+x)^5}
                                +\frac{256 P^{(4)}_{25}}{27 (x-1)^3 (1+x)^5}
                        \Biggr) H_0
\nonumber\\ &&
                        -\frac{256 x \big(
                                1-7 x+x^2\big)}{(x-1) (1+x)^5} H_0^3
                \Biggr)
                +\ln^2(2) \Biggl(
                        -\frac{256 x H_0 P^{(4)}_3}{9 (x-1)^3 (1+x)^5}
                        +\frac{128 P^{(4)}_{10}}{27 (x-1)^2 (1+x)^4}
\nonumber\\ &&                        
                        +\frac{256 x \big(
                                1-x+x^2\big)}{(x-1) (1+x)^5} H_0^2
+\frac{1024 x \big(
                                1-x+x^2\big)}{(x-1) (1+x)^5} H_0 H_1
                        -\frac{1024 x \big(
                                1-x+x^2\big)}{(x-1) (1+x)^5} H_{0,1}
                \Biggr)
\nonumber\\ &&              
                +\Biggl(
                        \frac{64 P^{(4)}_2}{(x-1)^2 (1+x)^4}
          +\frac{768 x^2 \big(
                                2-3 x+2 x^2\big)}{(x-1)^3 (1+x)^5} H_0
                \Biggr) \zeta_3
        \Biggr) \zeta_2
        +\Biggl(
                -\frac{64 x H_0 P^{(4)}_{12}}{45 (x-1)^3 (1+x)^5}
\nonumber\\ &&              
                -\frac{32 P^{(4)}_{13}}{135 (x-1)^2 (1+x)^4}
  -\frac{1024 P^{(4)}_{21}}{9 (x-1)^3 (1+x)^5} \ln(2)
                +\frac{1024 x \big(
                        1-x+x^2\big)}{(x-1) (1+x)^5} \ln^2(2)
\nonumber\\ &&              
  +\frac{64 x \big(
                        11+25 x+11 x^2\big)}{5 (x-1) (1+x)^5} H_0^2
  +\frac{256 x \big(
                        11+25 x+11 x^2\big)}{5 (x-1) (1+x)^5} H_0 H_1
\nonumber\\ &&    
                -\frac{256 x \big(
                        11+25 x+11 x^2\big)}{5 (x-1) (1+x)^5} H_{0,1}
        \Biggr) \zeta_2^2
        +\frac{256 x \big(
                11+25 x+11 x^2\big)}{5 (x-1) (1+x)^5} \zeta_2^3
\nonumber\\ &&
    -\frac{640 (-2+x) x^2 (-1+2 x)}{(x-1)^3 (1+x)^5} H_0 \zeta_5
\Biggr\}\Biggr\}
\nonumber\\
&& + n_h F_{V,2,1}^{(0)}(x) + n_h F_{V,2,2}^{(0)}(x) \zeta_2 + n_h F_{V,2,3}^{(0)}(x) \zeta_3 + 
F_{V,2}^{(0),r}(x),
\end{eqnarray}
with the polynomials
\begin{align}
  P^{(4)}_1&=x^4+5 x^3-10 x^2+5 x+1,\\
  P^{(4)}_2&=x^4+9 x^3-26 x^2+9 x+1,\\P^{(4)}_3&=x^4+82 x^3+62 x^2+82 x+1,\\P^{(4)}_4&=3 x^4-64 x^3-6 
x^2+16 x-1,\\P^{(4)}_5&=5 x^4-12 x^3+26 x^2-12 x+5,\\P^{(4)}_6&=7 x^4-147 x^3-8 x^2+13 x-1,\\P^{(4)}_7&=7 x^4+28 x^3+122 x^2+28 
x+7,\\P^{(4)}_8&=17 x^4+102 x^3+50 x^2+102 x+17,\\P^{(4)}_9&=67 x^4-10 x^3-138 x^2-10 x+67,\\P^{(4)}_{10}&=115 x^4+96 x^3+262 x^2+96 
x+115,\\P^{(4)}_{11}&=187 x^4-1300 x^3-2590 x^2-1300 x+187,\\P^{(4)}_{12}&=671 x^4+1042 x^3-702 x^2+1042 x+671,\\P^{(4)}_{13}&=1145 
x^4-960 x^3-8542 x^2-960 x+1145,\\P^{(4)}_{14}&=2 x^5-61 x^4+137 x^3+281 x^2+177 x-16,\\P^{(4)}_{15}&=14 x^5-35 x^4-271 
x^3-325 x^2-87 x+32,\\P^{(4)}_{16}&=77 x^5-19649 x^4-22606 x^3+5886 x^2+15009 x+483,\\P^{(4)}_{17}&=361 x^5-3677 x^4-5558 
x^3+790 x^2+2701 x+7,\\P^{(4)}_{18}&=13 x^6-174 x^5-381 x^4+60 x^3+243 x^2+114 x-3,\\P^{(4)}_{19}&=17 x^6-206 x^5-449 x^4+124 
x^3+175 x^2+82 x+1,\\P^{(4)}_{20}&=27 x^6-31 x^5+21 x^4-10 x^3+21 x^2-31 x+27,\\P^{(4)}_{21}&=27 x^6-28 x^5-24 x^4+56 x^3-24 
x^2-28 x+27,\\P^{(4)}_{22}&=27 x^6-28 x^5-6 x^4+20 x^3-6 x^2-28 x+27,\\P^{(4)}_{23}&=27 x^6-25 x^5-69 x^4+122 x^3-69 x^2-25 
x+27,\\P^{(4)}_{24}&=27 x^6+136 x^5-245 x^4-221 x^3+202 x^2+223 x-134,\\P^{(4)}_{25}&=165 x^6-101 x^5-2215 x^4+5984 x^3-2404 
x^2+7 x+120,\\P^{(4)}_{26}&=1280 x^8-7735 x^7+155816 x^6+33391 x^5-401120 x^4+33391 x^3+155816 x^2
\nonumber\\ &
-7735 x+1280,
\end{align}

\noindent
and
\begin{eqnarray}
F_{V,2,1}^{(0)}(x) 
&=&  \frac{1083812}{243} - \frac{32229191 y^2}{18225} - \frac{32229191 y^3}{18225} -
  \frac{488527686149 y^4}{428652000} 
\nonumber\\ &&
- \frac{109512399989 y^5}{214326000} + O(y^6)
\\
F_{V,2,2}^{(0)}(x) 
&=& -\frac{1126457}{243} + \frac{150475907 y^2}{72900} + \frac{150475907 y^3}{72900} +
  \frac{462427840529 y^4}{357210000} 
\nonumber\\ &&
+ \frac{93761868379 y^5}{178605000} + O(y^6)
\\
F_{V,2,3}^{(0)}(x) 
&=& -\frac{334736}{81} + \frac{8944046 y^2}{6075} + \frac{8944046 y^3}{6075} +
  \frac{92909575471 y^4}{95256000} 
\nonumber\\ &&
+ \frac{22788254831 y^5}{47628000} + O(y^6).
\end{eqnarray}
\subsection{The Scalar Form Factor}
\label{sec:52}

\vspace*{1mm}
\noindent
The scalar form factor is given by
\begin{eqnarray}
F_S  &=& 
- \frac{1}{\ep^3}
\frac{1}{2 (1+x)^2} \Biggl\{
        n_h^2 \Biggl[
                -\frac{64}{27} (1+x)^2
                +\frac{64 (1+x) \big(
                        1+x^2\big)}{27 (1-x)} H_0
        \Biggr]
        + n_h \Biggl[
                \frac{4}{27} \big(
                        997+1418 x
\nonumber\\ && 
+997 x^2\big)
                -\frac{32 H_0 P^{(5)}_{8}}{27 (1-x^2)}
                - n_l \Biggl[
                        \frac{32}{9} (1+x)^2
                        -\frac{64 (1+x) \big(
                                1+x^2\big)}{27 (1-x)} H_0
                \Biggr]
                +\frac{256 \big(
                        1+x^2\big)^2}{27 (1-x)^2} H_0^2
        \Biggr]
\Biggr\}
\nonumber\\ && 
- \frac{1}{\ep^2} 
\frac{1}{2 (1+x)^2} \Biggl\{
        n_h^2 \Biggl[
                -\frac{832}{81} (1+x)^2
                -\frac{256 x (1+x) H_0}{27 (1-x)}
                -\frac{128 (1+x) \big(
                        1+x^2\big)}{27 (1-x)} H_{-1} H_0
\nonumber\\ && 
                +\frac{32 (1+x) \big(
                        1+x^2\big)}{27 (1-x)} H_0^2
                +\frac{128 (1+x) \big(
                        1+x^2\big)}{27 (1-x)} H_{0,-1}
                -\frac{64 (1+x) \big(
                        1+x^2\big)}{27 (1-x)} \zeta_2
        \Biggr]
\nonumber\\ && 
        + n_h \Biggl[
                \frac{16}{27} \big(
                        897+1786 x+897 x^2\big)
                + n_l \Biggl[
                        -\frac{64}{3} (1+x)^2
                        +\frac{64 (1+x) \big(
                                5-24 x+5 x^2\big)}{81 (1-x)} H_0
\nonumber\\ &&                
         -\frac{256 (1+x) \big(
                                1+x^2\big)}{27 (1-x)} H_{-1} H_0
                        +\frac{64 (1+x) \big(
                                1+x^2\big)}{27 (1-x)} H_0^2
                        +\frac{256 (1+x) \big(
                                1+x^2\big)}
                        {27 (1-x)} H_{0,-1}
\nonumber\\ && 
                        -
                        \frac{128 (1+x) \big(
                                1+x^2\big)}{27 (1-x)} \zeta_2
                \Biggr]
                +\Biggl(
                        \frac{128 H_{-1} P^{(5)}_{7}}{27 (1-x^2)}
                        -\frac{16 P^{(5)}_{13}}{27 (1-x^2)}
                \Biggr) H_0
                +\Biggl(
                        \frac{64 P^{(5)}_{26}}{27 (1-x)^2 (1+x)}
\nonumber\\ &&                
         -\frac{1024 \big(
                                1+x^2\big)^2}{27 (1-x)^2} H_{-1}
                \Biggr) H_0^2
                -\frac{128 \big(
                        -2+x^2
                \big)
\big(1+x^2\big)}{27 (1-x)^2} H_0^3
                -\frac{128 (1+x) \big(
                        1+x^2\big)}{3 (1-x)} H_0 H_1
\nonumber\\ &&                
 +\Biggl(
                        \frac{128 (1+x) \big(
                                1+x^2\big)}{3 (1-x)}
                        -\frac{128 \big(
                                1+x^2\big)^2}{3 (1-x)^2} H_0
                \Biggr) H_{0,1}
                -\Biggl(
                        \frac{128 P^{(5)}_{7}}{27 (1-x^2)}
                        - \frac{2176 \big(
                                1+x^2\big)^2}{27 (1-x)^2} H_0
                \Biggr) 
\nonumber\\ &&                
\times H_{0,-1}
                +\frac{256 \big(
                        1+x^2\big)^2}{3 (1-x)^2} H_{0,0,1}
                -\frac{256 \big(
                        1+x^2\big)^2}{3 (1-x)^2} H_{0,0,-1}
                +\Biggl(
                        \frac{64 P^{(5)}_{5}}{27 (1-x^2)}
\nonumber\\ && 
                        -\frac{64 \big(
                                1+x^2
                        \big)
\big(-1+35 x^2\big)}{27 (1-x)^2} H_0
                \Biggr) \zeta_2
                -\frac{64 \big(
                        1+x^2\big)^2}{3 (1-x)^2} \zeta_3
        \Biggr]
\Biggr\}
\nonumber\\ && 
-\frac{1}{\ep} \frac{1}{2 (1+x)^2} \Biggl\{
        n_h^2 \Biggl[
                -\frac{32}{81} \big(
                        107+294 x+107 x^2\big)
                +\Biggl(
                        \frac{64 P^{(5)}_{10}
                        }{243 (1-x^2)}
                        +
                        \frac{512 x (1+x) H_{-1}}{27 (1-x)}
\nonumber\\ && 
                        +\frac{128 (1+x) \big(
                                1+x^2\big)}{27 (1-x)} H_{-1}^2
                \Biggr) H_0
                -\Biggl(
                        \frac{64 P^{(5)}_{17}}{81 (1-x) (1+x)^2}
                        +\frac{64 (1+x) \big(
                                1+x^2\big)}{27 (1-x)} H_{-1}
                \Biggr) H_0^2
\nonumber\\ && 
                +\frac{32 (1+x) \big(
                        1+x^2\big)}{27 (1-x)} H_0^3
                -\Biggl(
                        \frac{512 x (1+x)}{27 (1-x)}
                        +\frac{256 (1+x) \big(
                                1+x^2\big)}{27 (1-x)} H_{-1}
                \Biggr) H_{0,-1}
\nonumber\\ &&                
 +\frac{128 (1+x) \big(
                        1+x^2\big)}{27 (1-x)} H_{0,0,-1}
                +\frac{256 (1+x) \big(
                        1+x^2\big)}{27 (1-x)} H_{0,-1,-1}
                +\Biggl(
                        -\frac{32 P^{(5)}_{25}}{27 (1-x) (1+x)^2}
\nonumber\\ && 
                        +\frac{160 (1+x) \big(
                                1+x^2\big)}{27 (1-x)} H_0
                        +\frac{128 (1+x) \big(
                                1+x^2\big)}{27 (1-x)} H_{-1}
                \Biggr) \zeta_2
                -\frac{128 (1+x) \big(
                        1+x^2\big)}{27 (1-x)} \zeta_3
        \Biggr]
\nonumber\\ && 
        + n_h \Biggl[
                \frac{64 P^{(5)}_{15}}{243 (1+x)^2}
                + n_l \Biggl[
                        -\frac{640}{243} \big(
                                37+86 x+37 x^2\big)
                        +\Biggl(
                                \frac{512 (1+x) \big(
                                        1+x^2\big)}{27 (1-x)} H_{-1}^2
\nonumber\\ && 
                                +\frac{128 P^{(5)}_{4}}{81 (1-x^2) }
                                -\frac{256 (1+x) \big(
                                        5-24 x+5 x^2\big)}{81 (1-x)} H_{-1}
                        \Biggr) H_0
                        +\Biggl(
                                -\frac{128 P^{(5)}_{2}
                                }{81 (1-x) (1+x)^2}
\nonumber\\ && 
 -
                                \frac{256 (1+x) \big(
                                        1+x^2\big)}{27 (1-x)} H_{-1}
                        \Biggr) H_0^2
                        +\frac{64 (1+x) \big(
                                1+x^2\big)}{27 (1-x)} H_0^3
                        +\Biggl(
                                \frac{256 (1+x) \big(
                                        5-24 x+5 x^2\big)}{81 (1-x)}
\nonumber\\ && 
                                -\frac{1024 (1+x) \big(
                                        1+x^2\big)}{27 (1-x)} H_{-1}
                        \Biggr) H_{0,-1}
                        +\frac{512 (1+x) \big(
                                1+x^2\big)}{27 (1-x)} [H_{0,0,-1} + 2 H_{0,-1,-1}]
\nonumber\\ &&               
                        +\Biggl(
                                -\frac{16 P^{(5)}_{32}}{81 (1-x) (1+x)^2}
                                +\frac{224 (1+x) \big(
                                        1+x^2\big)}{27 (1-x)} H_0
                                +\frac{512 (1+x) \big(
                                        1+x^2\big)}{27 (1-x)} H_{-1}
                        \Biggr) \zeta_2
\nonumber\\ && 
                        -\frac{512 (1+x) \big(
                                1+x^2\big)}{27 (1-x)} \zeta_3
                \Biggr]
                +\Biggl(
                        -\frac{256 H_{-1}^2 P^{(5)}_{9}}{27 (1-x^2)}
                        +\frac{32 H_{-1} P^{(5)}_{14}}{27 (1-x^2)}
                        -\frac{32 P^{(5)}_{37}}{243 (1-x) (1+x)^3}
                \Biggr) H_0
\nonumber\\ && 
                +\Biggl(
                        \frac{8 P^{(5)}_{41}}{243 (1-x)^2 (1+x)^4}
                        -\frac{32 P^{(5)}_{34}}{27 (1-x)^2 (1+x)} H_{-1}
                        +\frac{2048 \big(
                                1+x^2\big)^2}{27 (1-x)^2} H_{-1}^2
                \Biggr) H_0^2
\nonumber\\ && 
                +\Biggl(
                        \frac{32 P^{(5)}_{39}}{81 (1-x)^2 (1+x)^3}
                        -\frac{128 \big(
                                1+x^2
                        \big)
\big(-1+5 x^2\big)}{27 (1-x)^2} H_{-1}
                \Biggr) H_0^3
                +\frac{16 P^{(5)}_{19}
                }{81 (1-x)^2 (1+x)} H_0^4
\nonumber\\ && 
                +\Biggl(
                        \frac{256 H_0^2 P^{(5)}_{1}}{27 (1-x)^2}
                        -\Biggl(
                                \frac{256 (1+x) \big(
                                        7-32 x+7 x^2\big)}{27 (1-x)}
                                -\frac{256 (1+x) \big(
                                        1+x^2\big)}{1-x} H_{-1}
                        \Biggr) H_0
\nonumber\\ && 
                        +\frac{128 P^{(5)}_{24}}{81 (1-x)^2 (1+x)} H_0^3
                \Biggr) H_1
                -\frac{128 (1+x) \big(
                        1+x^2\big)}{3 (1-x)} H_0 H_1^2
                +\Biggl(
                        \frac{256 (1+x) \big(
                                7-32 x+7 x^2\big)}{27 (1-x)}
\nonumber\\ &&                
         +\Biggl(
                                \frac{512 P^{(5)}_{3}}{27 (1-x)^2}
                                +\frac{512 \big(
                                        1+x^2\big)^2}{3 (1-x)^2} H_{-1}
                        \Biggr) H_0 
                        -\frac{256 P^{(5)}_{23} H_0^2}{27 (1-x)^2 (1+x)}
                        +\Biggl(
                                \frac{256 (1+x) \big(
                                        1+x^2\big)}{3 (1-x)}
\nonumber\\ && 
                                -\frac{512 \big(
                                        1+x^2\big)^2}{27 (1-x)^2} H_0
                        \Biggr) H_1
                        -\frac{256 (1+x) \big(
                                1+x^2\big)}{1-x} H_{-1}
                        -\frac{6400 \big(
                                1+x^2\big)^2}{27 (1-x)^2} H_{0,-1}
                \big) H_{0,1}
\nonumber\\ &&                
 +\frac{896 \big(
                        1+x^2\big)^2}{27 (1-x)^2} H_{0,1}^2
                +\Biggl(
                        -\frac{512 H_{-1} P^{(5)}_{9}}{27 (-1+x) (1+x)}
                        +\frac{32 P^{(5)}_{14}}{27 (-1+x) (1+x)}
                        -\Biggl(
                                \frac{64 P^{(5)}_{12}}{27 (1-x)^2}
\nonumber\\ &&                               
  +\frac{8704 \big(
                                        1+x^2\big)^2}{27 (1-x)^2} H_{-1}
                        \Biggr) H_0
                        +\frac{128 P^{(5)}_{29}}{27 (1-x)^2 (1+x)} H_0^2
                        +\Biggl(
                                -\frac{256 (1+x) \big(
                                        1+x^2\big)}
                                {1-x}
\nonumber\\ &&                
                 +
                                \frac{512 \big(
                                        1+x^2\big)^2}{27 (1-x)^2} H_0
                        \Biggr) H_1
                \Biggr) H_{0,-1}
                +\frac{4480 \big(
                        1+x^2\big)^2}{27 (1-x)^2} H_{0,-1}^2
                +\Biggl(
                        -\frac{512 P^{(5)}_{6}}{27 (1-x)^2}
\nonumber\\ &&                 
        +\frac{1024 P^{(5)}_{16}}{27 (1-x)^2 (1+x)} H_0
                        +\frac{1024 \big(
                                1+x^2\big)^2}{27 (1-x)^2} H_1
                        -\frac{1024 \big(
                                1+x^2\big)^2}{3 (1-x)^2} H_{-1}
                \Biggr) H_{0,0,1}
\nonumber\\ && 
                +\Biggl(
                        \frac{64 P^{(5)}_{35}}{27 (1-x)^2 (1+x)}
                        -\frac{512 P^{(5)}_{21}}{27 (1-x)^2 (1+x)} H_0
                        -\frac{1024 \big(
                                1+x^2\big)^2}{27 (1-x)^2} H_1
                        +\frac{1024 \big(
                                1+x^2\big)^2}{3 (1-x)^2} 
\nonumber\\ &&  \times
H_{-1}
                \Biggr) H_{0,0,-1}
                +\Biggl(
                        -\frac{256 (1+x) \big(
                                1+x^2\big)}{3 (1-x)}
                        -\frac{1280 \big(
                                1+x^2\big)^2}{27 (1-x)^2} H_0
                \Biggr) H_{0,1,1}
\nonumber\\ && 
                +\Biggl(
                        \frac{256 (1+x) \big(
                                1+x^2\big)}{1-x}
                        +\frac{1792 \big(
                                1+x^2\big)^2}{27 (1-x)^2} H_0
                \Biggr) [H_{0,1,-1} + H_{0,-1,1}]
                +\Biggl(
                        \frac{512 P^{(5)}_{9}}{27 (-1+x) (1+x)}
\nonumber\\ &&    
                        -\frac{256 \big(
                                1+x^2\big)^2}{27 (1-x)^2} H_0
                \Biggr) H_{0,-1,-1}
                +\frac{256 P^{(5)}_{27}
                }{27 (1-x)^2 (1+x)} H_{0,0,0,1}
                -
                \frac{256 P^{(5)}_{22}}{9 (1-x)^2 (1+x)} H_{0,0,0,-1}
\nonumber\\ &&            
                +\frac{1024 \big(
                        1+x^2\big)^2}{3 (1-x)^2} [-\frac{1}{9} H_{0,0,1,1} + H_{0,0,1,-1}+H_{0,0,-1,1}-H_{0,0,-1,-1}
                + \frac{4}{9} H_{0,-1,0,1}]
\nonumber\\ &&                
 +\Biggl(
                        \frac{256}{3} \ln(2) \big(
                                1+x^2\big)
                        -\frac{64 H_{-1} P^{(5)}_{11}}{27 (1-x^2) }
                        +\frac{2 P^{(5)}_{40}}{27 (1-x) (1+x)^4}
                        +\Biggl(
                                -\frac{16 P^{(5)}_{38}}{27 (1-x)^2 (1+x)^3}
\nonumber\\ && 
+\frac{512 \big(
                                        1+x^2
                                \big)
\big(13+4 x^2\big)}{27 (1-x)^2} H_{-1}
                        \Biggr) H_ 0
                        +\frac{64 P^{(5)}_{31}}{27 (1-x)^2 (1+x)} H_0^2
                        +\Biggl(
                                \frac{128 (1+x) \big(
                                        1+x^2\big)}{3 (1-x)}
\nonumber\\ && 
                                +\frac{512 P^{(5)}_{20}}{27 (1-x)^2 (1+x)} H_0
                        \Biggr) H_1
                        -\frac{128 P^{(5)}_{28}}{27 (1-x)^2 (1+x)} H_{0,1}
                        +\frac{128 P^{(5)}_{18}}{27 (1-x)^2 (1+x)} H_{0,-1}
                \Biggr) \zeta_2
\nonumber\\ && 
                +\frac{32 P^{(5)}_{36}}{135 (1-x)^2 (1+x)} \zeta_2^2
                -\Biggl(
                        \frac{64 (P^{(5)}_{33}+P^{(5)}_{30} H_0)}{27 (1-x)^2 (1+x)}
                        +\frac{256 \big(
                                1+x^2\big)^2}
                        {27 (1-x)^2} H_1
                        -\frac{256 \big(
                                1+x^2\big)^2}{3 (1-x)^2} 
\nonumber\\ && \times
H_{-1}
                \Biggr) \zeta_3
        \Biggr]
\Biggr\} + F_S^{(0)},
\end{eqnarray}
with the polynomials
\begin{align}
P^{(5)}_{1}&=x^4-32 x^3+48 x^2-32 x-17,
\\
P^{(5)}_{2}&=3 x^4+42 x^3+10 x^2+6 x-5,
\\
P^{(5)}_{3}&=13 x^4+47 x^3-56 x^2+47 x+13,
\\
P^{(5)}_{4}&=17 x^4-40 x^3-34 x^2-40 x+17,
\\
P^{(5)}_{5}&=19 x^4+92 x^3+110 x^2+164 x+55,
\\
P^{(5)}_{6}&=27 x^4+62 x^3-64 x^2+62 x+9,
\\
P^{(5)}_{7}&=55 x^4+164 x^3+146 x^2+164 x+55,
\\
P^{(5)}_{8}&=59 x^4+226 x^3+190 x^2+226 x+59,
\\
P^{(5)}_{9}&=62 x^4+151 x^3+142 x^2+151 x+62,
\\
P^{(5)}_{10}&=65 x^4-68 x^3+214 x^2-68 x+65,
\\
P^{(5)}_{11}&=137 x^4+436 x^3+352 x^2+340 x+143,
\\
P^{(5)}_{12}&=163 x^4+1050 x^3-1146 x^2+1050 x+163,
\\
P^{(5)}_{13}&=255 x^4-214 x^3-1530 x^2-214 x+255,
\\
P^{(5)}_{14}&=499 x^4-730 x^3-3050 x^2-730 x+499,
\\
P^{(5)}_{15}&=11134 x^4+45469 x^3+67950 x^2+45469 x+11134,
\\
P^{(5)}_{16}&=5 x^5-11 x^4+89 x^3-83 x^2+14 x-2,
\\
P^{(5)}_{17}&=5 x^5+3 x^4+50 x^3-14 x^2+9 x-5,
\\
P^{(5)}_{18}&=7 x^5+7 x^4-52 x^3-16 x^2-41 x-41,
\\
P^{(5)}_{19}&=8 x^5+8 x^4+35 x^3+53 x^2+36 x+36,
\\
P^{(5)}_{20}&=13 x^5-3 x^4+79 x^3-81 x^2+2 x-14,
\\
P^{(5)}_{21}&=15 x^5-25 x^4+216 x^3-208 x^2+29 x-11,
\\
P^{(5)}_{22}&=15 x^5+63 x^4-225 x^3+289 x^2-31 x+17,
\\
P^{(5)}_{23}&=21 x^5-3 x^4+142 x^3-110 x^2+19 x-5,
\\
P^{(5)}_{24}&=25 x^5-7 x^4+156 x^3-164 x^2+3 x-29,
\\
P^{(5)}_{25}&=33 x^5+19 x^4+130 x^3-178 x^2-35 x-33,
\\
P^{(5)}_{26}&=41 x^5+36 x^4-59 x^3-41 x^2-82 x-23,
\\
P^{(5)}_{27}&=41 x^5+121 x^4-372 x^3+500 x^2-57 x+23,
\\
P^{(5)}_{28}&=43 x^5-21 x^4+298 x^3-342 x^2-x-65,
\\
P^{(5)}_{29}&=50 x^5+18 x^4+211 x^3-123 x^2+26 x-6,
\\
P^{(5)}_{30}&=71 x^5-57 x^4+682 x^3-646 x^2+75 x-53,
\\
P^{(5)}_{31}&=71 x^5+55 x^4+157 x^3-x^2+23 x+7,
\\
P^{(5)}_{32}&=133 x^5-225 x^4+298 x^3-1130 x^2+81 x-53,
\\
P^{(5)}_{33}&=221 x^5+799 x^4-194 x^3-122 x^2+213 x-149,
\\
P^{(5)}_{34}&=273 x^5-425 x^4-196 x^3-124 x^2-1245 x-331,
\\
P^{(5)}_{35}&=599 x^5+2001 x^4-388 x^3-316 x^2+1181 x-5,
\\
P^{(5)}_{36}&=809 x^5+265 x^4+3072 x^3-2452 x^2+45 x-499,
\\
P^{(5)}_{37}&=5411 x^6+5015 x^5-11707 x^4-28382 x^3-11707 x^2+5015 x+5411,
\\
P^{(5)}_{38}&=155 x^7+1277 x^6+1073 x^5-2993 x^4-1563 x^3-1661 x^2-433 x+49,
\\
P^{(5)}_{39}&=181 x^7+191 x^6-288 x^5-210 x^4-1103 x^3-1097 x^2-646 x-100,
\\
P^{(5)}_{40}&=13209 x^7+22973 x^6+44113 x^5-24083 x^4-106013 x^3-82753 x^2-11469 x-7017,
\\
P^{(5)}_{41}&=5069 x^8-892 x^7-75926 x^6-69132 x^5+50716 x^4+90780 x^3+17638 x^2-5140 x-4921.
\end{align}

The constant part, $F_S^{(0)}$, reads
\begin{eqnarray}
F_S^{(0)} &=& -\frac{1}{2(1+x)^2} \Biggl\{
n_h^2 \Biggl[
        \frac{256 P^{(6)}_2  H_0^2 H_1}{81 (1+x)^2} 
        -\frac{16 P^{(6)}_{14}}{243 (1+x)^2}
        +\Biggl(
                \frac{32 P^{(6)}_{26}}{729 (1-x) (1+x)^3}
                -\frac{128 P^{(6)}_{12} H_{-1}}{243 (1-x^2)} 
\nonumber\\ && 
                -\frac{512 x (1+x)}{27 (1-x)} H_{-1}^2
                -\frac{256 (1+x) \big(
                        1+x^2\big)}{81 (1-x)} H_{-1}^3
        \Biggr) H_0
        +\Biggl(
                \frac{64 P^{(6)}_{27}}{243 (1-x) (1+x)^4}
\nonumber\\ && 
                +\frac{128 P^{(6)}_{16}}{81 (1-x) (1+x)^2} H_{-1}
                +\frac{64 (1+x) \big(
                        1+x^2\big)}{27 (1-x)} H_{-1}^2
        \Biggr) H_0^2
        +\Biggl(
                -\frac{64 P^{(6)}_{30}}{243 (1-x) (1+x)^5}
\nonumber\\ &&                
 -\frac{64 (1+x) \big(
                        1+x^2\big)}{81 (1-x)} H_{-1}
        \Biggr) H_0^3
        +\frac{8 (1+x) \big(
                1+x^2\big)}{9 (1-x)} H_0^4
        +\Biggl(
                +\frac{256 (1+x) \big(
                        1+x^2\big)}{27 (1-x)} H_0^2
\nonumber\\ && 
                -\frac{512 H_0 P^{(6)}_2}{81 (1+x)^2}
        \Biggr) H_{0,1}
        +\Biggl(
                \frac{128 P^{(6)}_{12}}{243 (1-x) (1+x)}
                -\frac{128 (1+x) \big(
                        1+x^2\big)}{27 (1-x)} H_0^2
                +\frac{1024 x (1+x) H_{-1}}{27 (1-x)}
\nonumber\\ &&                
 +\frac{256 (1+x) \big(
                        1+x^2\big)}{27 (1-x)} H_{-1}^2
        \Biggr) H_{0,-1}
        +\Biggl(
                \frac{512 P^{(6)}_2}{81 (1+x)^2}
                -\frac{1024 (1+x) \big(
                        1+x^2\big)}{27 (1-x)}
                 H_0
        \Biggr) H_{0,0,1}
\nonumber\\ && 
        +\Biggl(
                -
                \frac{256 P^{(6)}_{16}}{81 (1-x) (1+x)^2}
                +\frac{256 (1+x) \big(
                        1+x^2\big)}{27 (1-x)} H_0
                -\frac{256 (1+x) \big(
                        1+x^2\big)}{27 (1-x)} H_{-1}
        \Biggr) H_{0,0,-1}
\nonumber\\ && 
        +\Biggl(
                -\frac{1024 x (1+x)}{27 (1-x)}
                -\frac{512 (1+x) \big(
                        1+x^2\big)}{27 (1-x)} H_{-1}
        \Biggr) H_{0,-1,-1}
        +\frac{512 (1+x) \big(
                1+x^2\big)}{9 (1-x)} H_{0,0,0,1}
\nonumber\\ &&         
+\frac{128 (1+x) \big(
                1+x^2\big)}{27 (1-x)} H_{0,0,0,-1}
        +\frac{256 (1+x) \big(
                1+x^2\big)}{27 (1-x)} H_{0,0,-1,-1}
        +\frac{512 (1+x) \big(
                1+x^2\big)}{27 (1-x)} 
\nonumber\\ &&  \times
H_{0,-1,-1,-1}
        +\Biggl(
                -\frac{512 P^{(6)}_5}{27 (1+x)^2} \ln(2)
                -\frac{32 P^{(6)}_{28}}{1215 (1-x) (1+x)^4}
                +\Biggl(
                        -\frac{128 P^{(6)}_{29}}{81 (1-x) (1+x)^5}
\nonumber\\ &&                
         -\frac{64 (1+x) \big(
                                1+x^2\big)}{27 (1-x)} H_{-1}
                \Biggr) H_0
                +\frac{16 (1+x) \big(
                        1+x^2\big)}{3 (1-x)} H_0^2
 +\frac{256 P^{(6)}_{15}}{27 (1-x) (1+x)^2} H_{-1}
\nonumber\\ && 
                -\frac{128 (1+x) \big(
                        1+x^2\big)}{27 (1-x)} H_{-1}^2
                -\frac{64 (1+x) \big(
                        1+x^2\big)}{3 (1-x)} H_{0,-1}
        \Biggr) \zeta_2
        -\frac{128 (1+x) \big(
                1+x^2\big)}{9 (1-x)} \zeta_2^2
\nonumber\\ && 
        +\Biggl(
                -\frac{256 P^{(6)}_{20}}{81 (1-x) (1+x)^2}
                -
                \frac{704 (1+x) \big(
                        1+x^2\big)}{27 (1-x)} H_0
                +\frac{256 (1+x) \big(
                        1+x^2\big)}{27 (1-x)} H_{-1}
        \Biggr) \zeta_3
\Biggr]
\nonumber\\ && 
+ n_h \Biggl[
        \ln^4(2) \Biggl(
                -\frac{32}{81} \big(
                        35+106 x+35 x^2\big)
                +\frac{64 P^{(6)}_4}{81 (1-x^2)} H_0
        \Biggr)
        +\Li_4\left(\frac{1}{2}\right) \Biggl(
                -\frac{256}{27} \big(
                        35
+106 x
\nonumber\\ && 
+35 x^2\big)
                +\frac{512 P^{(6)}_4}{27 (1-x) (1+x)} H_0
        \Biggr)
        + n_l \Biggl[
                -\frac{1664}{243} \big(
                        77+142 x+77 x^2\big)
                +\frac{256 H_0^2 H_1 P^{(6)}_2}{81 (1+x)^2}
\nonumber\\ &&                
 +\Biggl(
                        \frac{2048 P^{(6)}_6}{729 (1-x) (1+x)}
                        -\frac{512 P^{(6)}_7}{81 (1-x) (1+x)} H_{-1}
                        +\frac{512 (1+x) \big(
                                5-24 x+5 x^2\big)}{81 (1-x)} H_{-1}^2
\nonumber\\ &&                
         -\frac{2048 (1+x) \big(
                                1+x^2\big)}{81 (1-x)} H_{-1}^3
                \Biggr) H_0
                +\Biggl(
                        \frac{128 P^{(6)}_{19}}{81 (1-x) (1+x)^2}
                        +\frac{512 P^{(6)}_1}{81 (1-x) (1+x)^2} H_{-1}
\nonumber\\ &&                
         +\frac{512 (1+x) \big(
                                1+x^2\big)}{27 (1-x)} H_{-1}^2
                \Biggr) H_0^2
                +\Biggl(
                        -\frac{128 P^{(6)}_{17}}{81 (1-x) (1+x)^2}
                        -\frac{128 (1+x) \big(
                                1+x^2\big)}{27 (1-x)} H_{-1}
                \Biggr) 
\nonumber\\ &&  \times
H_0^3
                +\frac{16 (1+x) \big(
                        1+x^2\big)}{9 (1-x)} H_0^4
                +\Biggl(
                        -\frac{512 H_0 P^{(6)}_2}
                        {81 (1+x)^2}
                        +
                        \frac{256 (1+x) \big(
                                1+x^2\big)}{27 (1-x)} H_0^2
                \Biggr) H_{0,1}
\nonumber\\ &&                
 +\Biggl(
                        \frac{512 P^{(6)}_7}{81 (1-x) (1+x)}
                        -\frac{128 (1+x) \big(
                                1+x^2\big)}{9 (1-x)} H_0^2
                        -\frac{1024 (1+x) \big(
                                5-24 x+5 x^2\big)}{81 (1-x)} H_{-1}
\nonumber\\ &&                
         +\frac{2048 (1+x) \big(
                                1+x^2\big)}{27 (1-x)} H_{-1}^2
                \Biggr) H_{0,-1}
                +\Biggl(
                        \frac{512 P^{(6)}_2}{81 (1+x)^2}
                        -\frac{1024 (1+x) \big(
                                1+x^2\big)}{27 (1-x)} H_0
                \Biggr) H_{0,0,1}
\nonumber\\ &&                
 +\Biggl(
                        -\frac{1024 P^{(6)}_1}{81 (1-x) (1+x)^2}
                        +\frac{256 (1+x) \big(
                                1+x^2\big)}{9 (1-x)} H_0
                        -\frac{2048 (1+x) \big(
                                1+x^2\big)}{27 (1-x)} H_{-1}
                \Biggr) H_{0,0,-1}
\nonumber\\ &&                
 +\Biggl(
                        \frac{1024 (1+x) \big(
                                5-24 x+5 x^2\big)}{81 (1-x)}
                        -\frac{4096 (1+x) \big(
                                1+x^2\big)}{27 (1-x)} H_{-1}
                \Biggr) H_{0,-1,-1}
\nonumber\\ && 
                +\frac{512 (1+x) \big(
                        1+x^2\big)}{9 (1-x)} H_{0,0,0,1}
                +\frac{256 (1+x) \big(
                        1+x^2\big)}{9 (1-x)} H_{0,0,0,-1}
                +\frac{2048 (1+x) \big(
                        1+x^2\big)}{27 (1-x)} 
\nonumber\\ &&  \times
H_{0,0,-1,-1}
                +\frac{4096 (1+x) \big(
                        1+x^2\big)}{27 (1-x)} H_{0,-1,-1,-1}
                +\Biggl(
                        -\frac{512 P^{(6)}_5}{27 (1+x)^2} \ln(2)
\nonumber\\ &&                
         -\frac{32 P^{(6)}_{24}}{81 (1-x) (1+x)^2}
                        -\Biggl(
                                \frac{32 P^{(6)}_{22}}{81 (1-x) (1+x)^2}
                                +\frac{128 (1+x) \big(
                                        1+x^2\big)}{27 (1-x)} H_{-1}
                        \Biggr) H_0
\nonumber\\ && 
                        +\frac{32 (1+x) \big(
                                1+x^2\big)}{3 (1-x)} H_0^2
                        +\frac{1024 P^{(6)}_{18}}{81 (1-x) (1+x)^2} H_{-1}
                        -\frac{1024 (1+x) \big(
                                1+x^2\big)}{27 (1-x)} H_{-1}^2
\nonumber\\ && 
                        -\frac{128 (1+x) \big(
                                1+x^2\big)}{9 (1-x)} H_{0,-1}
                \Biggr) \zeta_2
                -\frac{3968 (1+x) \big(
                        1+x^2\big)}{135 (1-x)} \zeta_2^2
                +\Biggl(
                        -\frac{32 P^{(6)}_{23}}{81 (1-x) (1+x)^2}
\nonumber\\ &&                
         -\frac{64 (1+x) \big(
                                1+x^2\big)}{3 (1-x)} H_0
                        +\frac{2048 (1+x) \big(
                                1+x^2\big)}{27 (1-x)} H_{-1}
                \Biggr) \zeta_3
        \Biggr]
        +\Biggl(
                \ln^2(2) \Biggl(
                        -\frac{128}{27} \big(
                                41-125 x
\nonumber\\ && 
+41 x^2\big)
                        -\frac{128 P^{(6)}_8}{27 (1-x) (1+x)} H_0
                \Biggr)
                +\ln(2) \Biggl(
                        -\frac{128 P^{(6)}_{31}}{81 x (1+x)^4}
                        +\frac{64 P^{(6)}_{25}}{27 (1-x) (1+x)^3} H_0
\nonumber\\ &&                 
        -\frac{128 P^{(6)}_{21}}{9 (1-x)^2 (1+x)} H_0^2
                        -\frac{256 P^{(6)}_3}{3 (1-x) (1+x)} H_0 H_1
         +\Biggl(
                                \frac{128}{3} \big(
                                        1-16 x+x^2\big)
\nonumber\\ && 
                                -\frac{256 P^{(6)}_9}{3 (1-x) (1+x)} H_0
                        \Biggr) H_{-1}
                        +\frac{256 P^{(6)}_3}{3 (1-x) (1+x)} H_{0,1}
                        +\frac{256 P^{(6)}_{10}}{3 (1-x) (1+x)}
                         H_{0,-1}
                \Biggr)
\nonumber\\ &&                
 +\Biggl(
                        16 \big(
                                13+20 x+13 x^2\big)
                        +\frac{192 x^2}{(1-x) (1+x)} H_0
                \Biggr) \zeta_3
        \Biggr) \zeta_2
        +
                \frac{32}{135} \big(
                        2146+3383 x+2146 x^2\big)
\nonumber\\ &&          
       -\frac{128 P^{(6)}_{11}}{3 (1-x) (1+x)} \ln(2)
                +\frac{32 P^{(6)}_{13}}{135 (1-x) (1+x)} H_0
        \Biggr)  \zeta_2^2
        -40 \Biggl(
                1+4 x+x^2\Biggr) \zeta_5
\nonumber\\ &&
        +\frac{160 x^2}{(1-x) (1+x)} H_0 \zeta_5
\Biggr)
\Biggr\} 
+ n_h F_{S,1}^{(0)}(x) 
+ n_h \zeta_2 F_{S,2}^{(0)}(x) 
+ n_h \zeta_3 F_{S,3}^{(0)}(x), 
\end{eqnarray}
with
\begin{align}
P^{(6)}_1    &=3 x^4+42 x^3+10 x^2+6 x-5,
\\
P^{(6)}_2    &=5 x^4+2 x^3+34 x^2+2 x+5,
\\
P^{(6)}_3    &=7 x^4-13 x^3+34 x^2-13 x+7,
\\
P^{(6)}_4    &=8 x^4-11 x^3-2 x^2-11 x+8,
\\
P^{(6)}_5    &=11 x^4+26 x^3+70 x^2+26 x+11,
\\
P^{(6)}_6    &=16 x^4-193 x^3-445 x^2-193 x+16,
\\
P^{(6)}_7    &=17 x^4-40 x^3-34 x^2-40 x+17,
\\
P^{(6)}_8    &=26 x^4+25 x^3-20 x^2+25 x+26,
\\
P^{(6)}_9    &=33 x^4+12 x^3+100 x^2+12 x+33,
\\
P^{(6)}_{10} &=35 x^4+16 x^3+98 x^2+16 x+35,
\\
P^{(6)}_{11} &=49 x^4-10 x^3+166 x^2-10 x+49,
\\
P^{(6)}_{12} &=65 x^4-68 x^3+214 x^2-68 x+65,
\\
P^{(6)}_{13} &=1612 x^4+3737 x^3+6068 x^2+3737 x+1612,
\\
P^{(6)}_{14} &=3859 x^4+15236 x^3+22882 x^2+15236 x+3859,
\\
P^{(6)}_{15} &=5 x^5-5 x^4+26 x^3-38 x^2+x-5,
\\
P^{(6)}_{16} &=5 x^5+3 x^4+50 x^3-14 x^2+9 x-5,
\\
P^{(6)}_{17} &=5 x^5+3 x^4+86 x^3+22 x^2+9 x-5,
\\
P^{(6)}_{18} &=10 x^5-9 x^4+22 x^3-74 x^2-5,
\\
P^{(6)}_{19} &=16 x^5+13 x^4-25 x^3-123 x^2-59 x+18,
\\
P^{(6)}_{20} &=35 x^5+54 x^4+107 x^3-143 x^2-66 x-35,
\\
P^{(6)}_{21} &=70 x^5-11 x^4+71 x^3-70 x^2-2 x-2,
\\
P^{(6)}_{22} &=105 x^5+51 x^4+1548 x^3+524 x^2+147 x-55,
\\
P^{(6)}_{23} &=325 x^5+87 x^4+274 x^3-1938 x^2-375 x-165,
\\
P^{(6)}_{24} &=391 x^5-979 x^4-1930 x^3+746 x^2+611 x-119,
\\
P^{(6)}_{25} &=261 x^6+790 x^5-1127 x^4+1600 x^3-569 x^2+1042 x+243,
\\
P^{(6)}_{26} &=599 x^6-4456 x^5-20887 x^4-30512 x^3-20887 x^2-4456 x+599,
\\
P^{(6)}_{27} &=11 x^7+202 x^6+445 x^5+304 x^4+131 x^3-240 x^2-75 x+54,
\\
P^{(6)}_{28} &=7239 x^7+16287 x^6+11607 x^5+7399 x^4+1301 x^3-7507 x^2-13747 x-5939,
\\
P^{(6)}_{29} &=15 x^8+57 x^7+264 x^6+779 x^5+1062 x^4+539 x^3+112 x^2+9 x-5,
\\
P^{(6)}_{30} &=25 x^8+84 x^7+358 x^6+944 x^5+1122 x^4+464 x^3+54 x^2-12 x-15,
\\
P^{(6)}_{31} &=536 x^8-15135 x^7-63923 x^6-147609 x^5-194394 x^4-147609 x^3-63923 x^2
\nonumber\\ &
-15135 x+536.
\end{align}

\noindent
The first expansion coefficients of $F_{S,i}^{(0)}, i = 1..3$ are given by
\begin{align}
F_{S,1}^{(0)}(x) & =  -\frac{96756433 y^5}{218700}-\frac{316061833 y^4}{437400}-\frac{731018 
y^3}{729}-\frac{731018
   y^2}{729}-\frac{874750}{243}
+ O(y^6)
\\
F_{S,2}^{(0)}(x) & = \frac{3932123 y^5}{18225}+\frac{16041283 y^4}{36450}+\frac{2421832 
y^3}{3645}+\frac{2421832
   y^2}{3645}+\frac{343864}{81}
+ O(y^6)
\\
F_{S,3}^{(0)}(x) & = -\frac{7752703 y^5}{48600}-\frac{21262303 y^4}{97200}-\frac{22516 y^3}{81}-\frac{22516
   y^2}{81}+\frac{62968}{27}
+ O(y^6).
\end{align}

\subsection{The Pseudoscalar Form Factor}
\label{sec:53}

\vspace*{1mm}
\noindent
The unrenormalized pseudoscalar form factor reads\\
\begin{eqnarray}
F_P &=& \frac{1}{\ep^3}\frac{1}{2 (1-x)^2} \Biggl\{
        n_h^2 \Biggl[
                \frac{64}{27} (1-x)^2
                -\frac{64 (1-x) \big(
                        1+x^2\big) H_0}{27 (1+x)}
        \Biggr]
        + n_h \Biggl[
                -\frac{3988}{27} (1-x)^2
\nonumber\\ &&
                +n_l \Biggl(
                        \frac{32}{9} (1-x)^2
                        -\frac{64 (1-x) \big(
                                1+x^2\big) H_0}{27 (1+x)}
                \Biggr)
                +\frac{32 (1-x) \big(
                        59+36 x+59 x^2\big)}{27 (1+x)} H_0
\nonumber\\ &&               
 -\frac{256 \big(
                        1+x^2\big)^2 H_0^2}{27 (1+x)^2}
        \Biggr]
\Biggr\}
+ \frac{1}{\ep^2} \frac{1}{2 (1-x)^2} \Biggl\{
        n_h^2 \Biggl[
                \frac{832}{81} (1-x)^2
                +\frac{128 (1-x) \big(
                        1+x^2\big) H_{-1} H_0}{27 (1+x)}
\nonumber\\ &&
                -\frac{32 (1-x) \big(
                        1+x^2\big) H_0^2}{27 (1+x)}
                -\frac{128 (1-x) \big(
                        1+x^2\big) H_{0,-1}}{27 (1+x)}
                +\frac{64 (1-x) \big(
                        1+x^2\big) \zeta_2}{27 (1+x)}
        \Biggr]
\nonumber\\ &&
        + n_h \Biggl[
                -\frac{16 (1-x)^2 \big(
                        299+694 x+299 x^2\big)}{9 (1+x)^2}
                + n_l \Biggl[
                        \frac{64}{3} (1-x)^2
                        -\frac{320 (1-x) \big(
                                1+x^2\big) H_0}{81 (1+x)}
\nonumber\\ &&
                        +\frac{256 (1-x) \big(
                                1+x^2\big) H_{-1} H_0}{27 (1+x)}
                        -\frac{64 (1-x) \big(
                                1+x^2\big) H_0^2}{27 (1+x)}
                        -\frac{256 (1-x) \big(
                                1+x^2\big) H_{0,-1}
                        }{27 (1+x)}
\nonumber\\ &&
                        +
                        \frac{128 (1-x) \big(
                                1+x^2\big) \zeta_2}{27 (1+x)}
                \Biggr]
                +\Biggl(
                        \frac{16 (1-x) P^{(7)}_{12}}{27 (1+x)^3}
                        -\frac{128 (1-x) \big(
                                55+18 x+55 x^2\big)}{27 (1+x)} H_{-1}
                \Biggr) H_0
\nonumber\\ &&
                +\Biggl(
                        \frac{64 (1-x) \big(
                                23+9 x+41 x^2\big)}{27 (1+x)}
                        +\frac{1024 \big(
                                1+x^2\big)^2 H_{-1}}{27 (1+x)^2}
                \Biggr) H_0^2
                -\frac{128 \big(
                        2-x^2
                \big)
\big(1+x^2\big)}{27 (1+x)^2} H_0^3
\nonumber\\ &&
                +\frac{128 (1-x) \big(
                        1+x^2\big) H_0 H_1}{3 (1+x)}
                -\Biggl(
                        \frac{128 (1-x) \big(
                                1+x^2\big)}{3 (1+x)}
                        +\frac{128 \big(
                                1+x^2\big)^2 H_0}{3 (1+x)^2}
                \Biggr) H_{0,1}
\nonumber\\ &&
                +\Biggl(
                        \frac{128 (1-x) \big(
                                55+18 x+55 x^2\big)}{27 (1+x)}
                        -\frac{2176 \big(
                                1+x^2\big)^2 H_0}{27 (1+x)^2}
                \Biggr) H_{0,-1} 
                -\frac{256 \big(
                        1+x^2\big)^2 H_{0,0,1}}{3 (1+x)^2}
\nonumber\\ && 
                +\frac{256 \big(
                        1+x^2\big)^2 H_{0,0,-1}}{3 (1+x)^2}
                -\Biggl(
                        \frac{64 (1-x) \big(
                                55+18 x+19 x^2\big)}{27 (1+x)}
                        +\frac{64 \big(
                                1+x^2
                        \big)
\big(1-35 x^2\big)}{27 (1+x)^2} H_0
                \Biggr) \zeta_2
\nonumber\\ &&
                +\frac{64 \big(
                        1+x^2\big)^2 \zeta_3}{3 (1+x)^2}
        \big)
\Biggr\}
+\frac{1}{\ep} \frac{1}{2 (1-x)^2} \Biggl\{
        n_h^2 \Biggl[
                \frac{32 (1-x)^2 \big(
                        107+198 x+107 x^2\big)}{81 (1+x)^2}
\nonumber\\ &&                       
 -\Biggl(
                        \frac{64 (1-x) P^{(7)}_{8}}{243 (1+x)^3}
 +\frac{128 (1-x) \big(
                                1+x^2\big) H_{-1}^2}{27 (1+x)}
                \Biggr) H_0
                +\Biggl(
                        \frac{64 (1-x) \big(
                                1+x^2\big) H_{-1}}{27 (1+x)}
\nonumber\\ &&
                        -\frac{64 (1-x)^2 \big(
                                1+x^2
                        \big)\big(5+14 x+5 x^2\big)}{81 (1+x)^4}
                \Biggr) H_0^2
                -\frac{32 (1-x) \big(
                        1+x^2\big) H_0^3}{27 (1+x)}
\nonumber\\ && 
                +\frac{256 (1-x) \big(
                        1+x^2\big) H_{-1} H_{0,-1}}{27 (1+x)}
                -\frac{128 (1-x) \big(
                        1+x^2\big) H_{0,0,-1}}{27 (1+x)}
                -\frac{256 (1-x) \big(
                        1+x^2\big)}{27 (1+x)}
\nonumber\\ &&   \times H_{0,-1,-1}        
 -\Biggl(
                        \frac{32 (1-x)^2 P^{(7)}_{6}}{27 (1+x)^4}
                        +\frac{160 (1-x) \big(
                                1+x^2\big) H_0}{27 (1+x)}
                        +\frac{128 (1-x) \big(
                                1+x^2\big) H_{-1}}{27 (1+x)}
                \Biggr) \zeta_2
\nonumber\\ && 
 + \frac{128 (1-x) \big(
                        1+x^2\big) \zeta_3}{27 (1+x)}
        \Biggr]
        + n_h \Biggl[
                -\frac{128 (1-x)^2 P^{(7)}_{16}}{243 (1+x)^4}
\nonumber\\ &&                
 + n_l \Biggl[
                        \frac{128 (1-x)^2 \big(
                                185+358 x+185 x^2\big)}{243 (1+x)^2}
                        -\Biggl(
                                \frac{128 (1-x) P^{(7)}_{3}}{81 (1+x)^3}
                                -\frac{1280 (1-x) \big(
                                        1+x^2\big) H_{-1}}{81 (1+x)}
\nonumber\\ && 
                                +
                                \frac{512 (1-x) \big(
                                        1+x^2\big) H_{-1}^2}{27 (1+x)}
                        \Biggr) H_0
                        +\Biggl( -
                                \frac{128 (1-x) \big(
                                        1+x^2
                                \big)
\big(5+12 x+3 x^2\big)}{81 (1+x)^4}
\nonumber\\ && 
                                +\frac{256 (1-x) \big(
                                        1+x^2\big) H_{-1}}{27 (1+x)}
                        \Biggr) H_0^2
                        -\frac{64 (1-x) \big(
                                1+x^2\big) H_0^3}{27 (1+x)}
                        -\Biggl(
                               \frac{1280 (1-x) \big(
                                        1+x^2\big)}{81 (1+x)}
\nonumber\\ &&                                 
-\frac{1024 (1-x) \big(
                                        1+x^2\big) H_{-1}}{27 (1+x)}
                        \Biggr) H_{0,-1}
                        -\frac{512 (1-x) \big(
                                1+x^2\big) H_{0,0,-1}}{27 (1+x)}
\nonumber\\ &&                
         -\frac{1024 (1-x) \big(
                                1+x^2\big) H_{0,-1,-1}}{27 (1+x)}
                        +\Biggl(
                                \frac{16 (1-x) P^{(7)}_{27}}{81 (1+x)^4}
                                -\frac{224 (1-x) \big(
                                        1+x^2\big) H_0}{27 (1+x)}
\nonumber\\ &&                
                 -\frac{512 (1-x) \big(
                                        1+x^2\big) H_{-1}}{27 (1+x)}
                        \Biggr) \zeta_2
                        +\frac{512 (1-x) \big(
                                1+x^2\big)}{27 (1+x)} \zeta_3
                \Biggr)
                -\Biggl(
                        \frac{32 (1-x) H_{-1} P^{(7)}_{14}}{27 (1+x)^3}
\nonumber\\ &&                         
-\frac{32 (1-x) P^{(7)}_{29}}{243 (1+x)^5}
                        -\frac{256 (1-x) \big(
                                62+9 x+62 x^2\big)}{27 (1+x)} H_{-1}^2
                \Biggr) H_0
                +\Biggl(
                        \frac{32 H_{-1} P^{(7)}_{13}}{27 (1+x)^2}
                        -\frac{8 P^{(7)}_{32}}{243 (1+x)^6}
\nonumber\\ &&                
         -\frac{2048 \big(
                                1+x^2\big)^2 H_{-1}^2}{27 (1+x)^2}
                \Biggr) H_0^2
                +\Biggl(
                        \frac{32 P^{(7)}_{31}}{81 (1-x) (1+x)^5}
                        +\frac{128 \big(
                                1+x^2
                        \big)
\big(-1+5 x^2\big)}{27 (1+x)^2} H_{-1}
                \Biggr) H_0^3
\nonumber\\ && 
                +\frac{16 P^{(7)}_{19}}{81 (1-x) (1+x)^2} H_0^4
                +\Biggl(
                        -\frac{128 H_0^3 P^{(7)}_{4}}{81 (1+x)^2}
                        +\Biggl(
                                \frac{1792 (1-x) \big(
                                        1+x^2\big)}{27 (1+x)}
\nonumber\\ &&                
                 -\frac{256 (1-x) \big(
                                        1+x^2\big) H_{-1}}{1+x}
                        \Biggr) H_0
                        -\frac{256 \big(
                                -17-6 x-6 x^3+x^4\big)}{27 (1+x)^2} H_0^2
                \Biggr) H_1
\nonumber\\ &&                
 +\frac{128 (1-x) \big(
                        1+x^2\big) H_0 H_1^2}{3 (1+x)}
                +\Biggl(
                        -\frac{1792 (1-x) \big(
                                1+x^2\big)}{27 (1+x)}
                        -\Biggl(
                                \frac{512 P^{(7)}_{2}}{27 (1+x)^2}
\nonumber\\ &&               
                 +\frac{512 \big(
                                        1+x^2\big)^2 H_{-1}}{3 (1+x)^2}
                        \Biggr) H_0
                        -\frac{256 P^{(7)}_{22}}{27 (1-x) (1+x)^2} H_0^2
                        -\Biggl(
                                \frac{256 (1-x) \big(
                                        1+x^2\big)}{3 (1+x)}
\nonumber\\ &&                                 
-\frac{512 \big(
                                        1+x^2\big)^2 H_0}{27 (1+x)^2}
                        \Biggr) H_1
                        +\frac{256 (1-x) \big(
                                1+x^2\big) H_{-1}}{1+x}
                        +\frac{6400 \big(
                                1+x^2\big)^2 H_{0,-1}}{27 (1+x)^2}
                \Biggr) H_{0,1}
\nonumber\\ &&                
 -\frac{896 \big(
                        1+x^2\big)^2 H_{0,1}^2}{27 (1+x)^2}
                +\Biggl(
                        \frac{32 (1-x) P^{(7)}_{14}}{27 (1+x)^3}
                        +\Biggl(
                                \frac{64 P^{(7)}_{10}
                                }{27 (1+x)^2}
                                +
                                \frac{8704 \big(
                                        1+x^2\big)^2 H_{-1}}{27 (1+x)^2}
                        \Biggr) H_0
\nonumber\\ &&                
         +\frac{128 P^{(7)}_{24}}{27 (1-x) (1+x)^2} H_0^2
                        +\Biggl(
                                \frac{256 (1-x) \big(
                                        1+x^2\big)}{1+x}
                 -\frac{512 \big(
                                        1+x^2\big)^2 H_0}{27 (1+x)^2}
                        \Biggr) H_1
\nonumber\\ &&                
                        -\frac{512 (1-x) \big(
                                62+9 x+62 x^2\big)}{27 (1+x)} H_{-1}
                \Biggr) H_{0,-1} 
                -\frac{4480 \big(
                        1+x^2\big)^2 H_{0,-1}^2}{27 (1+x)^2}
                +\Biggl(
                        \frac{512 P^{(7)}_{5}}{27 (1+x)^2}
\nonumber\\ && 
                        +\frac{1024 P^{(7)}_{17}}{27 (1-x) (1+x)^2} H_0
                        -\frac{1024 \big(
                                1+x^2\big)^2 H_1}{27 (1+x)^2}
                        +\frac{1024 \big(
                                1+x^2\big)^2 H_{-1}}{3 (1+x)^2}
                \Biggr) H_{0,0,1}
\nonumber\\ &&                
 -\Biggl(
                        \frac{64 P^{(7)}_{15}}{27 (1+x)^2}
                        +\frac{512 P^{(7)}_{20}}{27 (1-x) (1+x)^2} H_0
                        -\frac{1024 \big(
                                1+x^2\big)^2 H_1}{27 (1+x)^2}
                        +\frac{1024 \big(
                                1+x^2\big)^2 H_{-1}}{3 (1+x)^2}
                \Biggr) 
\nonumber\\ &&  \times
H_{0,0,-1}
                +\Biggl(
                        \frac{256 (1-x) \big(
                                1+x^2\big)}{3 (1+x)}
                        +\frac{1280 \big(
                                1+x^2\big)^2 H_0}{27 (1+x)^2}
                \Biggr) H_{0,1,1}
                -\Biggl(
                        \frac{256 (1-x) \big(
                                1+x^2\big)}{1+x}
\nonumber\\ &&                         
+\frac{1792 \big(
                                1+x^2\big)^2 H_0}{27 (1+x)^2}
                \Biggr) H_{0,1,-1}
                +\Biggl(
                        -\frac{256 (1-x) \big(
                                1+x^2\big)}
                        {1+x}
                        -
                        \frac{1792 \big(
                                1+x^2\big)^2 H_0}{27 (1+x)^2}
                \Biggr) H_{0,-1,1}
\nonumber\\ && 
                +\Biggl(
                        \frac{512 (1-x) \big(
                                62+9 x+62 x^2\big)}{27 (1+x)}
                        +\frac{256 \big(
                                1+x^2\big)^2 H_0}{27 (1+x)^2}
                \Biggl) H_{0,-1,-1} +
                \frac{256 P^{(7)}_{23} H_{0,0,0,1}}{27 (1-x) (1+x)^2} 
\nonumber\\ &&                
 -\frac{256 P^{(7)}_{21}}{9 (1-x) (1+x)^2} H_{0,0,0,-1}
                +\frac{1024 \big(
                        1+x^2\big)^2 H_{0,0,1,1}}{27 (1+x)^2}
                -\frac{1024 \big(
                        1+x^2\big)^2 H_{0,0,1,-1}}{3 (1+x)^2}
\nonumber\\ &&                
 -\frac{1024 \big(
                        1+x^2\big)^2 H_{0,0,-1,1}}{3 (1+x)^2}
                +\frac{1024 \big(
                        1+x^2\big)^2 H_{0,0,-1,-1}}{3 (1+x)^2}
                -\frac{4096 \big(
                        1+x^2\big)^2 H_{0,-1,0,1}}{27 (1+x)^2}
\nonumber\\ &&                
 +\Biggl(
                        \frac{128 H_{0,1} P^{(7)}_{7}}{27 (1+x)^2}
                        -\frac{64 H_{-1} P^{(7)}_{9}}{27 (1+x)^2}
                        +\frac{2 P^{(7)}_{33}}{27 (1+x)^6}
                        -\frac{256 (1-x)^2 \big(
                                3+4 x+3 x^2\big)}{9 (1+x)^2} \ln(2)
\nonumber\\ && 
                        +\Biggl(
                                -\frac{16 P^{(7)}_{30}}{27 (1-x) (1+x)^5}
                                -\frac{512 \big(
                                        1+x^2
                                \big)
\big(13+4 x^2\big)}{27 (1+x)^2} H_{-1}
                        \Biggr) H_0
                        +\frac{64 P^{(7)}_{26}}{27 (1-x) (1+x)^2} H_0^2
\nonumber\\ && 
                        +\Biggl(
                                -\frac{128 (1-x) \big(
                                        1+x^2\big)}{3 (1+x)}
                                -\frac{512 H_0 P^{(7)}_{1}
                                }{27 (1+x)^2}
                        \Bigg) H_1
                        +\frac{128 P^{(7)}_{18}}{27 (1-x) (1+x)^2} H_{0,-1}
                \Biggr) \zeta_2
\nonumber\\ && 
                +\frac{32 P^{(7)}_{28}}{135 (1-x) (1+x)^2} \zeta_2^2
                +\Biggl(
                        \frac{64 P^{(7)}_{11}}{27 (1+x)^2}
                        -\frac{64 P^{(7)}_{25}}{27 (1-x) (1+x)^2} H_0
                        +\frac{256 \big(
                                1+x^2\big)^2 H_1}{27 (1+x)^2}
\nonumber\\ &&               
         -\frac{256 \big(
                                1+x^2\big)^2 H_{-1}}{3 (1+x)^2}
                \Biggr) \zeta_3
        \Biggr]
\Biggr\} + F_P^{(0)},
\end{eqnarray}
with
\begin{align}
P^{(7)}_{1}&=13 x^4-16 x^3-x^2+16 x-14,
\\
P^{(7)}_{2}&=13 x^4+5 x^3-8 x^2+5 x+13,
\\
P^{(7)}_{3}&=17 x^4+48 x^3+46 x^2+48 x+17,
\\
P^{(7)}_{4}&=25 x^4-32 x^3-4 x^2+32 x-29,
\\
P^{(7)}_{5}&=27 x^4+4 x^3-16 x^2+4 x+9,
\\
P^{(7)}_{6}&=33 x^4+108 x^3+118 x^2+108 x+33,
\\
P^{(7)}_{7}&=43 x^4-64 x^3-22 x^2+64 x-65,
\\
P^{(7)}_{8}&=65 x^4+196 x^3+166 x^2+196 x+65,
\\
P^{(7)}_{9}&=137 x^4+54 x^3-54 x^2-18 x-143,
\\
P^{(7)}_{10}&=163 x^4+174 x^3+54 x^2+174 x+163,
\\
P^{(7)}_{11}&=221 x^4+166 x^3+164 x^2+94 x-149,
\\
P^{(7)}_{12}&=255 x^4+370 x^3+806 x^2+370 x+255,
\\
P^{(7)}_{13}&=273 x^4-94 x^3-170 x^2-166 x-331,
\\
P^{(7)}_{14}&=499 x^4+998 x^3+1574 x^2+998 x+499,
\\
P^{(7)}_{15}&=599 x^4+254 x^3-62 x^2+182 x-5,
\\
P^{(7)}_{16}&=5567 x^4+21986 x^3+33054 x^2+21986 x+5567,
\\
P^{(7)}_{17}&=5 x^5-21 x^4+21 x^3+15 x^2-18 x+2,
\\
P^{(7)}_{18}&=7 x^5-7 x^4-28 x^3+40 x^2-41 x+41,
\\
P^{(7)}_{19}&=8 x^5-8 x^4+39 x^3-49 x^2+36 x-36,
\\
P^{(7)}_{20}&=15 x^5-55 x^4+48 x^3+40 x^2-51 x+11,
\\
P^{(7)}_{21}&=15 x^5+33 x^4-21 x^3-85 x^2+65 x-17,
\\
P^{(7)}_{22}&=21 x^5-45 x^4+42 x^3+10 x^2-29 x+5,
\\
P^{(7)}_{23}&=41 x^5+39 x^4-28 x^3-156 x^2+103 x-23,
\\
P^{(7)}_{24}&=50 x^5-82 x^4+79 x^3-9 x^2-38 x+6,
\\
P^{(7)}_{25}&=71 x^5-199 x^4+154 x^3+118 x^2-181 x+53,
\\
P^{(7)}_{26}&=71 x^5-87 x^4+89 x^3-67 x^2-9 x-7,
\\
P^{(7)}_{27}&=133 x^5+255 x^4+10 x^3+310 x^2-15 x-53,
\\
P^{(7)}_{28}&=809 x^5-1353 x^4+776 x^3+156 x^2-1043 x+499,
\\
P^{(7)}_{29}&=5411 x^6+25439 x^5+53201 x^4+69802 x^3+53201 x^2+25439 x+5411,
\\
P^{(7)}_{30}&=155 x^8+794 x^7+852 x^6-778 x^5-2634 x^4-346 x^3+396 x^2+74 x-49,
\\
P^{(7)}_{31}&=181 x^8+266 x^7-359 x^6-726 x^5+123 x^4+278 x^3+275 x^2+246 x+100,
\\
P^{(7)}_{32}&=5069 x^8+25148 x^7+56746 x^6+50652 x^5+21436 x^4-10044 x^3-13562 x^2
\nonumber\\ &
-14812 x-4921,
\\
P^{(7)}_{33}&=13209 x^8+44196 x^7+34324 x^6+5148 x^5-48650 x^4-38052 x^3-12044 x^2
\nonumber\\ &
+19428 x+7017,
\end{align}

\noindent
and
\begin{eqnarray}
F_P^{(0)} &=&  
\frac{1}{4(1-x)^2} \Biggl\{
n_h^2 \Biggl[
        \frac{32 (1-x)^2 P^{(8)}_{14}}{243 (1+x)^4}
        +(1-x) \Biggl(
                \frac{256 H_{-1} P^{(8)}_9}{243 (1+x)^3}
                +\frac{512  \big(
                        1+x^2\big) H_{-1}^3}{81 (1+x)}
\nonumber\\ &&  
                -\frac{64 P^{(8)}_{21}}{729 (1+x)^5}
        \Biggr) H_0
        +\Biggl(
                -\frac{128 (1-x) P^{(8)}_{22}}{243 (1+x)^6}
                +\frac{256 (1-x)^2 \big(
                        1+x^2
                \big)
\big(5+14 x+5 x^2\big)}{81 (1+x)^4} H_{-1}
\nonumber\\ &&                 
-\frac{128 (1-x) \big(
                        1+x^2\big) H_{-1}^2}{27 (1+x)}
        \Biggr) H_0^2
        +\Biggl(
                \frac{128 (1-x) \big(
                        1+x^2\big) P^{(8)}_{20}}{243 (1+x)^7}
                +\frac{128 (1-x) \big(
                        1+x^2\big) H_{-1}}{81 (1+x)}
        \Biggr) 
\nonumber\\ &&  \times
H_0^3
        -\frac{16 (1-x) \big(
                1+x^2\big) H_0^4}{9 (1+x)}
        -\frac{512 (1-x)^2 \big(
                1+x^2
        \big)
\big(5+14 x+5 x^2\big)}{81 (1+x)^4} H_0^2 H_1
\nonumber\\ && 
        +\Biggl(
                \frac{1024 (1-x)^2 \big(
                        1+x^2
                \big)
\big(5+14 x+5 x^2\big)}{81 (1+x)^4} H_0
                -\frac{512 (1-x) \big(
                        1+x^2\big) H_0^2}{27 (1+x)}
        \Biggr) H_{0,1}
\nonumber\\ && 
        -\Biggl(
                \frac{256 (1-x) P^{(8)}_9}{243 (1+x)^3}
                -\frac{256 (1-x) \big(
                        1+x^2\big) H_0^2}{27 (1+x)}
                +\frac{512 (1-x) \big(
                        1+x^2\big) H_{-1}^2}{27 (1+x)}
        \Biggr) H_{0,-1}
\nonumber\\ && 
        -\Biggl(
                \frac{1024 (1-x)^2 \big(
                        1+x^2
                \big)
\big(5+14 x+5 x^2\big)}{81 (1+x)^4}
                -\frac{2048 (1-x) \big(
                        1+x^2\big) H_0}{27 (1+x)}
        \Biggr) H_{0,0,1}
\nonumber\\ && 
        +\Biggl(
                -\frac{512 (1-x)^2 \big(
                        1+x^2
                \big)
\big(5+14 x+5 x^2\big)}{81 (1+x)^4}
                -\frac{512 (1-x) \big(
                        1+x^2\big) H_0}{27 (1+x)}
\nonumber\\ &&  
                +\frac{512 (1-x) \big(
                        1+x^2\big) H_{-1}}{27 (1+x)}
        \Biggr) H_{0,0,-1}
        +\frac{1024 (1-x) \big(
                1+x^2\big)}{27(1+x)} [H_{-1} H_{0,-1,-1} 
- 3 H_{0,0,0,1} 
\nonumber\\ && 
- \frac{1}{4} (H_{0,0,0,-1} + 2 H_{0,0,-1,-1}) -  H_{0,-1,-1,-1}]
        +\Biggl(
                \frac{1024 (1-x)^2 P^{(8)}_3}{27 (1+x)^4} \ln(2)
                +\frac{64 (1-x) P^{(8)}_{23}}{1215 (1+x)^6}
\nonumber\\ && 
                +\Biggl(
                        \frac{256 (1-x) \big(
                                1+x^2\big) P^{(8)}_{19}}{81 (1+x)^7}
                        +\frac{128 (1-x) \big(
                                1+x^2\big) H_{-1}}{27 (1+x)}
                \Biggr) H_0
                -\frac{32 (1-x) \big(
                        1+x^2\big) H_0^2}{3 (1+x)}
\nonumber\\ &&               
 +\frac{512 (1-x)^2 \big(
                        1+x^2
                \big)
\big(5+14 x+5 x^2\big)}{27 (1+x)^4}
                 H_{-1}
                +
                \frac{256 (1-x) \big(
                        1+x^2\big) H_{-1}^2}{27 (1+x)}
\nonumber\\ && 
                +\frac{128 (1-x) \big(
                        1+x^2\big) H_{0,-1}}{3 (1+x)}
        \Biggr) \zeta_2
        +\frac{256 (1-x) \big(
                1+x^2\big)}{9 (1+x)} \zeta_2^2
        +\Biggl(
                \frac{1408 (1-x) \big(
                        1+x^2\big) H_0}{27 (1+x)}
\nonumber\\ &&              
  -\frac{512 (1-x) \big(
                        1+x^2\big) H_{-1}}{27 (1+x)}
                -\frac{2560 (1-x)^2 \big(
                        1+x+x^2
                \big)
\big(7+18 x+7 x^2\big)}{81 (1+x)^4}
        \Biggr) \zeta_3
\Biggr]
\nonumber\\ && 
+n_h \Biggl[
        \ln^4(2) \Biggl(
                \frac{64}{81} \big(
                        35-22 x+35 x^2\big)
                -\frac{128 P^{(8)}_2}{81 (1-x) (1+x)} H_0
        \Biggr)
        +\Li_4\left(\frac{1}{2}\right) 
\nonumber\\ && \times
\Biggl(
                \frac{512}{27} \big(
                        35
-22 x
+35 x^2\big)
                -\frac{1024 P^{(8)}_2}{27 (1-x) (1+x)} H_0
        \Biggr)
\nonumber\\ && 
+ n_l \Biggl[-\frac{512 (1-x) (1+x^2)}{9(1+x)} H_0
               + \frac{256 (1-x)^2 \big(
                        1001+1966 x+1001 x^2\big)}{243 (1+x)^2}
\nonumber\\ &&                
 +\Biggl(
                        \frac{1024 (1-x) H_{-1} P^{(8)}_4}{81 (1+x)^3}
                        -\frac{1024 (1-x) P^{(8)}_8}{729 (1+x)^3}
                        -\frac{5120 (1-x) \big(
                                1+x^2\big) H_{-1}^2}{81 (1+x)}
\nonumber\\ &&                
         +\frac{4096 (1-x) \big(
                                1+x^2\big) H_{-1}^3}{81 (1+x)}
                \Biggr) H_0
                -\Biggl(
                        -\frac{1024 (1-x) \big(
                                1+x^2
                        \big)
\big(5+12 x+3 x^2\big)}
                        {81 (1+x)^4} H_{-1}
\nonumber\\ &&                 
        +\frac{256 (1-x) P^{(8)}_{15}}{81 (1+x)^4}
                        +
                        \frac{1024 (1-x) \big(
                                1+x^2\big) H_{-1}^2}{27 (1+x)}
                \Biggr) H_0^2
                +\Biggl(
                        +\frac{256 (1-x) \big(
                                1+x^2\big) H_{-1}}{27 (1+x)}
\nonumber\\ &&                
         -\frac{256 (1-x)^2 \big(
                                1+x^2
                        \big)
\big(5+14 x+5 x^2\big)}{81 (1+x)^4}
                \Biggr) H_0^3
                -\frac{32 (1-x) \big(
                        1+x^2\big) H_0^4}{9 (1+x)}
\nonumber\\ && 
 -\frac{512 (1-x)^2 \big(
                        1+x^2
                \big)
\big(5+14 x+5 x^2\big)}{81 (1+x)^4} H_0^2 H_1
                +\Biggl(
                        \frac{1024 (1-x)^2 \big(
                                1+x^2
                        \big)
\big(5+14 x+5 x^2\big)}{81 (1+x)^4} 
\nonumber\\ &&                 
\times 
H_0
        +\frac{512 (-1+x) \big(
                                1+x^2\big) H_0^2}{27 (1+x)}
                \Biggr) H_{0,1}
                +\Biggl(
                        -\frac{1024 (1-x) P^{(8)}_4}{81 (1+x)^3}
                        +\frac{256 (1-x) \big(
                                1+x^2\big) H_0^2}{9 (1+x)}
\nonumber\\ &&                       
 +\frac{10240 (1-x) \big(
                                1+x^2\big) H_{-1}}{81 (1+x)}
                        -\frac{4096 (1-x) \big(
                                1+x^2\big) H_{-1}^2}{27 (1+x)}
                \Biggr) H_{0,-1}
 +\Biggl(
                        \frac{2048 (1-x) \big(
                                1+x^2\big)}{27 (1+x)}
\nonumber\\ && \times H_0
                        -\frac{1024 (1-x)^2 \big(
                                1+x^2
                        \big)
\big(5+14 x+5 x^2\big)}{81 (1+x)^4}
                \Biggr) H_{0,0,1}
                        +\frac{512 (1-x) \big(
                                1+x^2\big) H_0}
                        {9 (1+x)}
\nonumber\\ && 
                -\Biggl(
                        \frac{2048 (1-x) \big(
                                1+x^2
                        \big)
\big(5+12 x+3 x^2\big)}{81 (1+x)^4}
                        -
                        \frac{4096 (1-x) \big(
                                1+x^2\big) H_{-1}}{27 (1+x)}
                \Biggr) H_{0,0,-1}
\nonumber\\ && 
                +\Biggl(
                        -\frac{10240 (1-x) \big(
                                1+x^2\big)}{81 (1+x)}
                        +\frac{8192 (1-x) \big(
                                1+x^2\big) H_{-1}}{27 (1+x)}
                \Biggr) H_{0,-1,-1}
\nonumber\\ &&                
 -\frac{512 (1-x) \big(
                        1+x^2\big)}{9 (1+x)} \Biggl[2 H_{0,0,0,1} + H_{0,0,0,-1} + \frac{8}{3} H_{0,0,-1,-1} + \frac{16}{3} 
H_{0,-1,-1,-1}\Biggr]
\nonumber\\ && 
                +\Biggl(
                        \frac{1024 (1-x)^2 P^{(8)}_3}{27 (1+x)^4} \ln(2)
                        +\frac{64 (1-x) P^{(8)}_{18}}{81 (1+x)^4}
                        +\Biggl(
                                +\frac{256 (1-x) \big(
                                        1+x^2\big) H_{-1}}{27 (1+x)}
\nonumber\\ &&                                 
+\frac{64 (1-x) \big(
                                        1+x^2
                                \big)
\big(-55-69 x+219 x^2+105 x^3\big)}{81 (1+x)^4}
                        \Biggr) H_0
                        -\frac{64 (1-x) \big(
                                1+x^2\big) H_0^2}{3 (1+x)}
\nonumber\\ && 
                        -\frac{2048 (1-x) \big(
                                1+x^2
                        \big)
\big(-5-6 x+21 x^2+10 x^3\big)}{81 (1+x)^4} H_{-1}
                        +\frac{2048 (1-x) \big(
                                1+x^2\big) H_{-1}^2}{27 (1+x)}
\nonumber\\ && 
        +\frac{256 (1-x) \big(
                                1+x^2\big) H_{0,-1}}{9 (1+x)}
                \Biggr) \zeta_2
                +\frac{7936 (1-x) \big(
                        1+x^2\big)}
                {135 (1+x)} \zeta_2^2
                +\Biggl(
                        \frac{64 (1-x) P^{(8)}_{17}}{81 (1+x)^4}
\nonumber\\ &&                       
 +\frac{128 (1-x) \big(
                                1+x^2\big) H_0}{3 (1+x)}
                        -\frac{4096 (1-x) \big(
                                1+x^2\big) H_{-1}}{27 (1+x)}
                \Biggr) \zeta_3
        \Biggr)
        +\Biggl(
                \ln^2(2) \Biggl(
                        \frac{256 P^{(8)}_7}{27 (1+x)^2}
\nonumber\\ &&                
         +\frac{256 P^{(8)}_6}{27 (1-x) (1+x)} H_0
                \Biggl)
                +\ln(2) \Biggl(
                        \frac{256 P^{(8)}_{25}}{81 x (1+x)^6}
                        -\frac{128 P^{(8)}_{24}}{27 (1-x) (1+x)^5} H_0
\nonumber\\ &&                
         -\frac{256 P^{(8)}_{16}}{9 (1-x) (1+x)^2} H_0^2
                        +\frac{512 P^{(8)}_5}{9 (1-x) (1+x)} H_0 H_1
                        -\Biggl(
                                \frac{256 P^{(8)}_1}{3 (1+x)^2}
                                -\frac{512 P^{(8)}_{10}}{9 (1-x) (1+x)} H_0
                        \Biggr) 
\nonumber\\ &&  \times
H_{-1}
                        -\frac{512 P^{(8)}_5}{9 (1-x) (1+x)} H_{0,1}
                        -\frac{512 P^{(8)}_{11}}{9 (1-x) (1+x)} H_{0,-1}
                \Biggr)
                +\Biggl(
                        -32 \big(
                                13-12 x+13 x^2\big)
\nonumber\\ &&                 
        -\frac{896 x^2}{(1-x) (1+x)} H_0
                \Biggr) \zeta_3
        \Biggr) \zeta_2
        -\Biggl(
                \frac{64}{135} \big(
                        2146-3893 x+2146 x^2\big)
                -\frac{256 P^{(8)}_{12}}{9 (1-x) (1+x)} \ln(2)
\nonumber\\ && 
                +\frac{64 P^{(8)}_{13}}{135 (1-x) (1+x)} H_0
        \Biggl) \zeta_2^2
        +80 \big(
                1+4 x+x^2\big) \zeta_5
        +\frac{960 x^2}{(1-x) (1+x)} H_0 \zeta_5
\Biggr]
\nonumber\\ &&
- \frac{512 (1-x) (1+x^2)}{9(1+x)} H_0 H_{0,0,-1}
\Biggr\}  
+ n_h F_{P,1}^{(0)}(x)
+ n_h \zeta_2 F_{P,2}^{(0)}(x)
+ n_h \zeta_3 F_{P,3}^{(0)}(x),
\nonumber\\ 
\end{eqnarray}
with
\begin{align}
P^{(8)}_1    &=x^4-6 x^3+18 x^2-6 x+1,
\\
P^{(8)}_2    &=8 x^4-43 x^3+22 x^2-43 x+8,
\\
P^{(8)}_3    &=11 x^4+38 x^3+46 x^2+38 x+11,
\\
P^{(8)}_4    &=17 x^4+48 x^3+46 x^2+48 x+17,
\\
P^{(8)}_5    &=21 x^4-69 x^3+94 x^2-69 x+21,
\\
P^{(8)}_6    &=26 x^4-79 x^3+76 x^2-79 x+26,
\\
P^{(8)}_7    &=41 x^4-87 x^3+32 x^2-87 x+41,
\\
P^{(8)}_8    &=64 x^4+263 x^3+218 x^2+263 x+64,
\\
P^{(8)}_9    &=65 x^4+196 x^3+166 x^2+196 x+65,
\\
P^{(8)}_{10} &=99 x^4-252 x^3+308 x^2-252 x+99,
\\
P^{(8)}_{11} &=105 x^4-264 x^3+326 x^2-264 x+105,
\\
P^{(8)}_{12} &=147 x^4-402 x^3+514 x^2-402 x+147,
\\
P^{(8)}_{13} &=1612 x^4-2711 x^3+2996 x^2-2711 x+1612,
\\
P^{(8)}_{14} &=3859 x^4+15092 x^3+22402 x^2+15092 x+3859,
\\
P^{(8)}_{15} &=16 x^5+57 x^4+107 x^3+81 x^2+73 x+18,
\\
P^{(8)}_{16} &=70 x^5-97 x^4+13 x^3+12 x^2-2 x+2,
\\
P^{(8)}_{17} &=325 x^5+807 x^4+418 x^3+222 x^2-327 x-165,
\\
P^{(8)}_{18} &=391 x^5+1357 x^4+854 x^3+650 x^2-317 x-119,
\\
P^{(8)}_{19} &=15 x^6+78 x^5+159 x^4+152 x^3+27 x^2-18 x-5,
\\
P^{(8)}_{20} &=25 x^6+126 x^5+225 x^4+152 x^3-39 x^2-66 x-15,
\\
P^{(8)}_{21} &=599 x^6+3824 x^5+7193 x^4+7360 x^3+7193 x^2+3824 x+599,
\\
P^{(8)}_{22} &=11 x^7+70 x^6+361 x^5+712 x^4+635 x^3+588 x^2+321 x+54,
\\
P^{(8)}_{23} &=7239 x^7+36399 x^6+60903 x^5+38023 x^4-11083 x^3-41923 x^2-28579 x-5939,
\\
P^{(8)}_{24} &=261 x^8+1096 x^7-702 x^6-3888 x^5-1232 x^4-4392 x^3-810 x^2+1168 x+243,
\\
P^{(8)}_{25} &=848 x^{10}-14847 x^9-55535 x^8-37906 x^7+58575 x^6+123650 x^5+58575 x^4
\nonumber\\ &
-37906 x^3-55535 x^2-14847 x+848.
\end{align}

\noindent
The first expansion coefficients of  $F_{P,i}^{(0)}, i = 1...3$, are given by
\begin{align}
F_{P,1}^{(0)}(x) &= \frac{19068183229 y^5}{85730400}+\frac{22625094013 y^4}{171460800}+\frac{756146 
y^3}{18225}+\frac{756146
   y^2}{18225}-\frac{5529994}{729} 
+ O(y^6)
\\
F_{P,2}^{(0)}(x) &= -\frac{524338481 y^5}{1488375}-\frac{804767441 y^4}{2976750}-\frac{381536 
y^3}{2025}-\frac{381536
   y^2}{2025}+\frac{4990072}{729}
+ O(y^6)
\\
F_{P,3}^{(0)}(x) &= -\frac{4050340711 y^5}{19051200}-\frac{622908463 y^4}{4233600}-\frac{496121 
y^3}{6075}-\frac{496121
   y^2}{6075}+\frac{426952}{243}
+ O(y^6).
\end{align}
\subsection{The Axialvector Form Factors}
\label{sec:54}

\vspace*{1mm}
\noindent
The axialvector form factors are given by
\begin{eqnarray}
F_{A,1} &=&   
\frac{1}{\ep^3} \Biggl\{
n_h^2 \Biggl[
        \frac{16}{27}
        -\frac{32 \big(
                1+x^2\big)}{27 (1-x) (1+x)} H_0
\Biggr]
+ n_h \Biggl[
        -\frac{2 \big(
                589+602 x+589 x^2\big)}{27 (1+x)^2}
        + n_l \Biggl[
                \frac{16}{27}
\nonumber\\ && 
                -\frac{32 \big(
                        1+x^2\big)}{27 (1-x) (1+x)} H_0
        \Biggr]
        +\frac{16 P^{(9)}_8}{27 (1-x) (1+x)^3} H_0
        -\frac{128 \big(
                1+x^2\big)^2}{27 (1-x^2)^2} H_0^2
\Biggr]\Biggr\}
\nonumber\\ && 
+
\frac{1}{\ep^2} \Biggl\{
n_h^2 \Biggl[
        \frac{16(1+x^2)}{27 (1-x^2)}
        +\Biggl(
                -\frac{16 \big(
                        3-2 x+3 x^2\big)}{27 (1-x^2)}
                +\frac{64 \big(
                        1+x^2\big) H_{-1}}{27 (1-x^2)}
        \Biggr) H_0
        -\frac{16 \big(
                1+x^2\big) H_0^2}{27 (1-x^2)}
\nonumber\\ && 
        -\frac{64 \big(
                1+x^2\big) H_{0,-1}}{27 (1-x^2)}
        + 
                \frac{32(1+x^2)}{27 (1-x^2)} \zeta_2
\Biggr]
+ n_h \Biggl[ \frac{1}{9 (1-x)^2(1+x)^3} \Biggl(-\frac{1052}{3} - \frac{988}{3} x 
\nonumber\\ && 
+ 680 x^2 + 680 x^3 - \frac{988}{3} x^4 - 
\frac{1052}{3} x^5\Biggr)  
        +n_l 
         \Biggl[
                \frac{128 (1+x^2) }{81 (1-x^2)}
                +\Biggl(
                        -\frac{64 \big(
                                7-3 x+7 x^2\big)}{81 (1-x^2)}
\nonumber\\ && 
                        +\frac{128 \big(
                                1+x^2\big) H_{-1}}{27 (1-x^2)}
                \Biggr) H_0
                -\frac{32 \big(
                        1+x^2\big) H_0^2}{27 (1-x^2)}
                -\frac{128 \big(
                        1+x^2\big) H_{0,-1}}{27 (1-x^2)}
                +\frac{64(1+x^2)}{27 (1-x^2)}\zeta_2
        \Biggr]
\nonumber\\ && 
        +\frac{2968}{27(1-x)^2(1+x)^3}(1+2x-x^2+x^3-2x^4-x^5) H_0
        -\frac{128 P^{(9)}_{7}}{27 (1-x) (1+x)^3} H_{-1} H_0
\nonumber\\ && 
        +\Biggl(
                -\frac{16 P^{(9)}_{22}}{27 (1-x)^2 (1+x)^3}
                +\frac{512 \big(
                        1+x^2\big)^2}{27 (1-x)^2 (1+x)^2} H_{-1}
        \Biggr) H_0^2
        +\frac{64 \big(
                -2+x^2
        \big)
\big(1+x^2\big)}{27 (1-x^2)^2} H_0^3
\nonumber\\ && 
        +\frac{64 \big(
                1+x^2\big) H_0 H_1}{3 (1-x^2)}
        -\frac{64 \big(
                1+x^2\big) H_{0,1}}{3 (1-x^2)}
        +\frac{64 \big(
                1+x^2\big)^2}{3 (1-x^2)^2}
         H_0 H_{0,1}
        +
        \frac{128 P^{(9)}_{7}}{27 (1-x) (1+x)^3} H_{0,-1}
\nonumber\\ && 
        -\frac{1088 \big(
                1+x^2\big)^2}{27 (1-x^2)^2} H_0 H_{0,-1}
        -\frac{128 \big(
                1+x^2\big)^2}{3 (1-x^2)^2} H_{0,0,1}
        +\frac{128 \big(
                1+x^2\big)^2}{3 (1-x^2)^2} H_{0,0,-1}
\nonumber\\ && 
+\frac{1}{(1-x)^2(1+x)^3}
\Biggl(-\frac{1472}{27}
       -\frac{3200}{27} x 
       + 64 x^2 
       +\frac{64}{3} x^3
       +\frac{2048}{27} x^4
       +\frac{320}{27} x^5\Biggr) 
\nonumber\\ && 
-\Biggl[\frac{32(1+x+x^2+x^3) H_0}{27 (1-x)^2(1+x)^3} 
-\frac{1120}{27 (1-x)^2(1+x)^3} x^2 (1+x+x^2+x^3) H_0\Biggr]\zeta_2
+\frac{32}{3(1-x^2)}
\nonumber\\ && \times 
\frac{1}{(1+x)^3}
(1+x+2x^2+2x^3+x^4+x^5) \zeta_3
\Biggr\}
+\frac{1}{\ep} \Biggl\{
n_h^2 \Biggl[
        -\frac{64 \big(
                13-34 x+13 x^2\big)}
        {243 (1+x)^2}
\nonumber\\ && 
        -
        \frac{16 P^{(9)}_{12}}{243 (1-x) (1+x)^3} H_0
        +\frac{32 \big(
                3-2 x+3 x^2\big)}{27 (1-x^2)} H_{-1} H_0
        -\frac{64 \big(
                1+x^2\big) H_{-1}^2 H_0}{27 (1-x^2)}
\nonumber\\ && 
        +\frac{8 P^{(9)}_{18}}{81 (1-x) (1+x)^4} H_0^2
        +\frac{32 \big(
                1+x^2\big) H_{-1} H_0^2}{27 (1-x^2)}
        -\frac{16 P^{(9)}_{1}}{27 (1-x) (1+x)^3} H_0^3
\nonumber\\ &&
        -\frac{32 \big(
                3-2 x+3 x^2\big)}{27 (1-x^2)} H_{0,-1}
        +\frac{128 \big(
                1+x^2\big) H_{-1} H_{0,-1}}{27 (1-x^2)}
        -\frac{64 \big(
                1+x^2\big) H_{0,0,-1}}{27 (1-x^2)}
\nonumber\\ && 
        -\frac{128 \big(
                1+x^2\big) H_{0,-1,-1}}{27 (1-x^2)}
        +\Biggl(
                \frac{8 P^{(9)}_{21}}{27 (1-x) (1+x)^4}
                -\frac{16 P^{(9)}_{3}}{27 (1-x) (1+x)^3} H_0
\nonumber\\ &&              
   -\frac{64 \big(
                        1+x^2\big) H_{-1}}{27 (1-x^2) }
        \Biggr) \zeta_2
        +\frac{64 \big(
                1+x^2\big)}{27 (1-x^2)} \zeta_3
\Biggr]
+n_h \Biggl[
        -\frac{128 H_0 H_{0,1} P^{(9)}_{2}}{27 (1-x^2)^2}
        -\frac{64 H_0^2 H_1 P^{(9)}_{5}}{27 (1-x^2)^2} 
\nonumber\\ && 
        +\frac{128 H_{0,0,1} P^{(9)}_{6}}{27 (1-x^2)^2}
        -\frac{64 H_0 H_{0,-1} P^{(9)}_{9}}{27 (1-x^2)^2}
        -
        \frac{4 P^{(9)}_{14}}{243 (1+x)^4}
        -\frac{64 H_{0,0,-1} P^{(9)}_{24}}{27 (1-x)^2 (1+x)^3}
\nonumber\\ && 
        +\frac{32 H_{-1} H_0^2 P^{(9)}_{25}}{27 (1-x)^2 (1+x)^3}
        +\frac{512 H_0 H_{0,0,1} P^{(9)}_{27}}{27 (1-x^2)^3}
        -\frac{128 H_{0,0,0,-1} P^{(9)}_{29}}{9 (1-x^2)^3}
        -\frac{128 H_0 H_{0,0,-1} P^{(9)}_{30}}{27 (1-x^2)^3}
\nonumber\\ && 
        +\frac{128 H_{0,0,0,1} P^{(9)}_{31}}{27 (1-x^2)^3}
        -\frac{64 H_0^2 H_{0,1} P^{(9)}_{32}}{27 (1-x^2)^3}
        +\frac{64 H_0^2 H_{0,-1} P^{(9)}_{33}}{27 (1-x^2)^3}
        +\frac{16 P^{(9)}_{168}}{135 (1-x^2)^3} \zeta_2^2
\nonumber\\ && 
        +\frac{8 H_0^4 P^{(9)}_{37}}{81 (1-x)^3 (1+x)^4}
        -\frac{4 H_0^2 P^{(9)}_{41}}{243 (1-x)^2 (1+x)^6}
        +n_l \Biggl[
                -\frac{32 \big(
                        1-38 x+x^2\big)}{81 (1+x)^2}
\nonumber\\ &&                
 -\frac{32 P^{(9)}_{10}}{81 (1-x) (1+x)^3} H_0
                +\frac{256 \big(
                        7-3 x+7 x^2\big)}{81 (1-x^2)} H_{-1} H_0
                -\frac{256 \big(
                        1+x^2\big) H_{-1}^2 H_0}{27 (1-x^2)}
\nonumber\\ &&               
 -\frac{16 P^{(9)}_{15}}{81 (1-x) (1+x)^4} H_0^2
                +\frac{128 \big(
                        1+x^2\big) H_{-1} H_0^2}{27 (1-x^2)}
                -\frac{32 \big(
                        1+2 x+2 x^3+x^4\big)}
                {27 (1-x) (1+x)^3} H_0^3
\nonumber\\ &&                
 -
                \frac{256 \big(
                        7-3 x+7 x^2\big)}{81 (1-x^2)} H_{0,-1}
                +\frac{512 \big(
                        1+x^2\big) H_{-1} H_{0,-1}}{27 (1-x^2)}
                -\frac{256 \big(
                        1+x^2\big) H_{0,0,-1}}{27 (1-x^2)}
\nonumber\\ &&                
 -\frac{512 \big(
                        1+x^2\big) H_{0,-1,-1}}{27 (1-x^2)}
                +\Biggl(
                        \frac{8 P^{(9)}_{26}}{81 (1-x) (1+x)^4}
                        -\frac{16 P^{(9)}_{4}}{27 (1-x) (1+x)^3} H_0
\nonumber\\ &&                
         -\frac{256 \big(
                                1+x^2\big) H_{-1}}{27 (1-x^2)}
                \Biggr) \zeta_2
                +\frac{256 \big(
                        1+x^2\big) \zeta_3}{27 (1-x^2)}
        \Biggr]
        +\frac{4 P^{(9)}_{36}}{243 (1-x) (1+x)^5} H_0
\nonumber\\ &&      
   -\frac{16 P^{(9)}_{13}}{27 (1-x) (1+x)^3} H_{-1} H_0
        +\Biggl(
                -\frac{64 H_1 P^{(9)}_{17}}{81 (1-x)^2 (1+x)^3}
                +\frac{16 P^{(9)}_{39}}{81 (1-x)^3 (1+x)^5}
        \Biggr) H_0^3
\nonumber\\ && 
        +\frac{64 \big(
                1+x^2
        \big)
\big(-1+5 x^2\big)}{27 (1-x^2)^2} H_{-1} H_0^3
        +\frac{64 \big(
                37-14 x+37 x^2\big)}{27 (1-x^2)} H_0 H_1
\nonumber\\ && 
        -\frac{128 \big(
                1+x^2\big) H_{-1} H_0 H_1}{(1-x^2)}
        +\frac{64 \big(
                1+x^2\big) H_0 H_1^2}{3 (1-x^2)}
        +\Biggl(
                \frac{64 P^{(9)}_{11}}{27 (1-x) (1+x)^3} H_0
\nonumber\\ &&                
 -\frac{1024 \big(
                        1+x^2\big)^2}
                {27 (1-x^2)^2} H_0^2
        \Biggr) H_{-1}^2
        -
        \frac{64 \big(
                37-14 x+37 x^2\big)}{27 (1-x^2)} H_{0,1}
        -\frac{256 \big(
                1+x^2\big)^2}{3 (1-x^2)^2} H_{-1} H_0 H_{0,1}
\nonumber\\ &&      
   -\frac{128 \big(
                1+x^2\big) H_1 H_{0,1}}{3 (1-x^2)}
        +\frac{256 \big(
                1+x^2\big)^2}{27 (1-x^2)^2} H_0 H_1 H_{0,1}
        +\frac{128 \big(
                1+x^2\big) H_{-1} H_{0,1}}{(1-x^2)}
\nonumber\\ && 
        +\frac{3200 \big(
                1+x^2\big)^2}{27 (1-x^2)^2} H_{0,-1} H_{0,1}
        -\frac{448 \big(
                1+x^2\big)^2}{27 (1-x^2)^2} H_{0,1}^2
        +\frac{16 P^{(9)}_{13}}{27 (1-x) (1+x)^3} H_{0,-1}
\nonumber\\ && 
        +\frac{4352 \big(
                1+x^2\big)^2}{27 (1-x^2)^2} H_{-1} H_0 H_{0,-1}
        +\frac{128 \big(
                1+x^2\big) H_1 H_{0,-1}}{1-x^2}
        -\frac{256 \big(
                1+x^2\big)^2}{27 (1-x^2)^2} H_0 H_1 H_{0,-1}
\nonumber\\ && 
        -\frac{128 P^{(9)}_{11}}{27 (1-x) (1+x)^3} H_{-1} H_{0,-1}
        -\frac{2240 \big(
                1+x^2\big)^2}{27 (1-x^2)^2} H_{0,-1}^2
        -\frac{512 \big(
                1+x^2\big)^2}{27 (1-x^2)^2} H_1 H_{0,0,1}
\nonumber\\ && 
        +\frac{512 \big(
                1+x^2\big)^2}{3 (1-x^2)^2} H_{-1} H_{0,0,1}
        +
        \frac{512 \big(
                1+x^2\big)^2}{27 (1-x^2)^2} H_1 H_{0,0,-1}
        -\frac{512 \big(
                1+x^2\big)^2}{3 (1-x^2)^2} H_{-1} H_{0,0,-1}
\nonumber\\ && 
        +\frac{128 \big(
                1+x^2\big) H_{0,1,1}}{3 (1-x^2)}
        +\frac{640 \big(
                1+x^2\big)^2}{27 (1-x^2)^2} H_0 H_{0,1,1}
        -\frac{128 \big(
                1+x^2\big) H_{0,1,-1}}{(1-x^2)}
\nonumber\\ &&    
     -\frac{896 \big(
                1+x^2\big)^2}{27 (1-x^2)^2} H_0 H_{0,1,-1}
        -\frac{128 \big(
                1+x^2\big) H_{0,-1,1}}{1-x^2}
        -\frac{896 \big(
                1+x^2\big)^2}{27 (1-x^2)^2} H_0 H_{0,-1,1}
\nonumber\\ && 
        +\frac{128 P^{(9)}_{11}}{27 (1-x) (1+x)^3} H_{0,-1,-1}
        +\frac{128 \big(
                1+x^2\big)^2}{27 (1-x^2)^2} H_0 H_{0,-1,-1}
        +\frac{512 \big(
                1+x^2\big)^2}{27 (1-x^2)^2} H_{0,0,1,1}
\nonumber\\ && 
        -\frac{512 \big(
                1+x^2\big)^2}{3 (1-x^2)^2} H_{0,0,1,-1}
        -\frac{512 \big(
                1+x^2\big)^2}{3 (1-x^2)^2} H_{0,0,-1,1}
        +\frac{512 \big(
                1+x^2\big)^2}{3 (1-x^2)^2} H_{0,0,-1,-1}
\nonumber\\ && 
        -\frac{2048 \big(
                1+x^2\big)^2}{27 (1-x^2)^2} H_{0,-1,0,1}
        +\Biggl[
                -\frac{256 H_0 H_1 P^{(9)}_{16}
                }{27 (1-x)^2 (1+x)^3}
                +\frac{64 H_{0,1} P^{(9)}_{19}}{27 (1-x)^2 (1+x)^3}
\nonumber\\ &&                
 -\frac{64 H_{-1} P^{(9)}_{20}}{27 (1-x)^2 (1+x)^3}
                +\frac{64 H_{0,-1} P^{(9)}_{28}}{27 (1-x^2)^3}
                +\frac{32 H_0^2 P^{(9)}_{38}}{27 (1-x)^3 (1+x)^4}
                -\frac{8 H_0 P^{(9)}_{40}}{27 (1-x)^3 (1+x)^5}
\nonumber\\ &&                
 +\frac{P^{(9)}_{42}}{27 (1-x)^2 (1+x)^6}
                -\frac{256 \big(
                        1+x^2
                \big)
\big(13+4 x^2\big)}{27 (1-x^2)^2} H_{-1} H_0
                -\frac{64 \big(
                        1+x^2\big) H_1}{3 (1-x^2)}
        \Biggr] \zeta_2
\nonumber\\ && 
        +\Biggl[
                \frac{32 P^{(9)}_{23}}{27 (1-x)^2 (1+x)^3}
                -\frac{32 H_0 P^{(9)}_{34}}{27 (1-x^2)^3}
                +\frac{128 \big(
                        1+x^2\big)^2}{27 (1-x^2)^2} H_1
\nonumber\\ &&
                -\frac{128 \big(
                        1+x^2\big)^2}{3 (1-x^2)^2} H_{-1}
        \Biggr] \zeta_3
\Biggr\}
+ F_{A,1}^{(0)},
\end{eqnarray}
with
\begin{align}
P^{(9)}_{1} &= x^4+2 x^3-2 x^2+2 x+1,
\\
P^{(9)}_{2} &= 2 x^4+125 x^3-346 x^2+125 x+2,
\\
P^{(9)}_{3} &=5 x^4+10 x^3-14 x^2+10 x+5,
\\
P^{(9)}_{4} &=7 x^4+14 x^3-10 x^2+14 x+7,
\\
P^{(9)}_{5} &=14 x^4-121 x^3+310 x^2-121 x-22,
\\
P^{(9)}_{6} &=18 x^4+129 x^3-382 x^2+129 x-18,
\\
P^{(9)}_{7} &=23 x^4+73 x^3+64 x^2+73 x+23,
\\
P^{(9)}_{8} &=41 x^4+190 x^3+154 x^2+190 x+41,
\\
P^{(9)}_{9} &=43 x^4-777 x^3+1970 x^2-777 x+43,
\\
P^{(9)}_{10} &=103 x^4+70 x^3+94 x^2+70 x+103,
\\
P^{(9)}_{11} &=115 x^4+284 x^3+266 x^2+284 x+115,
\\
P^{(9)}_{12} &=337 x^4-64 x^3+158 x^2-64 x+337,
\\
P^{(9)}_{13} &=913 x^4+1014 x^3-390 x^2+1014 x+913,
\\
P^{(9)}_{14} &=14921 x^4+59084 x^3+82566 x^2+59084 x+14921,
\\
P^{(9)}_{15} &=9 x^5+105 x^4+48 x^3+104 x^2+39 x+47,
\\
P^{(9)}_{16} &=13 x^5-38 x^4+226 x^3-228 x^2+37 x-14,
\\
P^{(9)}_{17} &=25 x^5-77 x^4+450 x^3-458 x^2+73 x-29,
\\
P^{(9)}_{18} &=29 x^5-87 x^4+38 x^3-74 x^2+45 x-47,
\\
P^{(9)}_{19} &=43 x^5-161 x^4+886 x^3-930 x^2+139 x-65,
\\
P^{(9)}_{20} &=55 x^5+136 x^4-27 x^3+9 x^2-94 x-67,
\\
P^{(9)}_{21} &=87 x^5-151 x^4+86 x^3-62 x^2+179 x-75,
\\
P^{(9)}_{22} &=97 x^5+135 x^4-22 x^3+14 x^2-83 x-13,
\\
P^{(9)}_{23} &=137 x^5+1241 x^4-1412 x^3-1340 x^2+691 x-197,
\\
P^{(9)}_{24} &=142 x^5+1207 x^4-1655 x^3-1619 x^2+815 x-142,
\\
P^{(9)}_{25} &=228 x^5-261 x^4+731 x^3+767 x^2-653 x-56,
\\
P^{(9)}_{26} &=331 x^5-279 x^4+478 x^3+130 x^2+855 x-107,
\\
P^{(9)}_{27} &=5 x^6-52 x^5+277 x^4-458 x^3+274 x^2-52 x+2,
\\
P^{(9)}_{28} &=7 x^6+6 x^5-65 x^4+24 x^3-31 x^2+6 x+41,
\\
P^{(9)}_{29} &=15 x^6+156 x^5-823 x^4+1378 x^3-855 x^2+156 x-17,
\\
P^{(9)}_{30} &=30 x^6-259 x^5+1378 x^4-2298 x^3+1370 x^2-259 x+22,
\\
P^{(9)}_{31} &=41 x^6+261 x^5-1367 x^4+2282 x^3-1431 x^2+261 x-23,
\\
P^{(9)}_{32} &=42 x^6-155 x^5+828 x^4-1374 x^3+796 x^2-155 x+10,
\\
P^{(9)}_{33} &=50 x^6-103 x^5+554 x^4-920 x^3+510 x^2-103 x+6,
\\
P^{(9)}_{34} &=71 x^6-412 x^5+2199 x^4-3696 x^3+2181 x^2-412 x+53,
\\
P^{(9)}_{35} &=809 x^6-1764 x^5+8941 x^4-15196 x^3+8631 x^2-1764 x+499,
\\
P^{(9)}_{36} &=46991 x^6+160202 x^5+253361 x^4+257260 x^3+253361 x^2+160202 x+46991,
\\
P^{(9)}_{37} &=8 x^7+9 x^6-21 x^5+102 x^4-38 x^3+31 x^2-35 x-36,
\\
P^{(9)}_{38} &=71 x^7+19 x^6+211 x^5-135 x^4-261 x^3+181 x^2-59 x-7,
\\
P^{(9)}_{39} &=229 x^8-318 x^7+98 x^6+1828 x^5-1004 x^4-1266 x^3-54 x^2+300 x-5,
\\
P^{(9)}_{40} &=281 x^8+3062 x^7+48 x^6-6750 x^5+2006 x^4+1130 x^3+688 x^2+382 x-79,
\\
P^{(9)}_{41} &=13421 x^8+46988 x^7+29734 x^6+3876 x^5+15184 x^4+26916 x^3-11702 x^2
\nonumber\\ &
-13780 x-5677,
\\
P^{(9)}_{42} &=-1152 \ln(2) (1-x)^2 (x+1)^4 \left(x^2+1\right)+18177 x^8-28796 x^7-86188 x^6+6780 x^5
\nonumber\\ &
+175430 x^4
+42236 x^3
           -91148 x^2-57084 x+8305,
\end{align}
\noindent 
and

\begin{eqnarray}
F_{A,1}^{(0)} &=& 
n_h^2 \Biggl\{
        \frac{64 H_{0,0,0,-1} P^{(10)}_2}{27 (x-1) (1+x)^3}
        -\frac{64 \zeta_2^2 P^{(10)}_3}{9 (x-1) (1+x)^3}
        +\frac{256 H_{0,0,0,1} P^{(10)}_4}{9 (x-1) (1+x)^3}
\nonumber\\ && 
       +\frac{4 H_0^4 P^{(10)}_9}{27 (x-1) (1+x)^3}
        -\frac{64 H_0^2 H_1 P^{(10)}_{14}}{81 (1+x)^4}
        +\frac{8 P^{(10)}_{24}}{729 (1+x)^4}
\nonumber\\ && 
        +\Biggl(
                -\frac{320 H_{-1} P^{(10)}_{15}}{243 (x-1) (1+x)^3}
                +\frac{16 P^{(10)}_{40}}{729 (x-1) (1+x)^5}
                +\frac{32 (x-1) H_{-1}^2}{27 (1+x)}
\nonumber\\ && 
                -\frac{128 \big(
                        1+x^2\big)}{81 (x-1) (1+x)} H_{-1}^3
        \Biggr) H_0
        +\Biggl(
                \frac{16 H_{-1} P^{(10)}_{17}}{81 (1+x)^4}
                -\frac{32 P^{(10)}_{41}}{243 (x-1) (1+x)^6}
\nonumber\\ && 
                +\frac{32 \big(
                        1+x^2\big)}{27 (x-1) (1+x)} H_{-1}^2
        \Biggr) H_0^2
        +\Biggl(
                -\frac{32 H_{-1} P^{(10)}_2}{81 (x-1) (1+x)^3}
                -\frac{8 P^{(10)}_{44}}{243 (x-1) (1+x)^7}
        \Biggr) H_0^3
\nonumber\\ && 
        +\Biggl(
                \frac{128 H_0^2 P^{(10)}_4}{27 (x-1) (1+x)^3}
                +\frac{128 H_0 P^{(10)}_{14}}{81 (1+x)^4}
        \Biggr) H_{0,1}
        +\Biggl(
                \frac{320 P^{(10)}_{15}}{243 (x-1) (1+x)^3}
\nonumber\\ && 
                -\frac{64 \big(
                        1+x^2\big)}{27 (x-1) (1+x)} H_0^2
                -\frac{64 (x-1) H_{-1}}{27 (1+x)}
                +\frac{128 \big(
                        1+x^2\big)}{27 (x-1) (1+x)} H_{-1}^2
        \Biggr) H_{0,-1}
\nonumber\\ && 
        +\Biggl(
                -
                \frac{512 H_0 P^{(10)}_4}{27 (x-1) (1+x)^3}
                -\frac{128 P^{(10)}_{14}}{81 (1+x)^4}
        \Biggr) H_{0,0,1}
        +\Biggl(
                -\frac{32 P^{(10)}_{17}}{81 (1+x)^4}
\nonumber\\ && 
                +\frac{128 \big(
                        1+x^2\big)}{27 (x-1) (1+x)} H_0
                -\frac{128 \big(
                        1+x^2\big)}{27 (x-1) (1+x)} H_{-1}
        \Biggr) H_{0,0,-1}
        +\Biggl(
                \frac{64 (x-1)}{27 (1+x)}
\nonumber\\ && 
                -\frac{256 \big(
                        1+x^2\big)}{27 (x-1) (1+x)} H_{-1}
        \Biggr) H_{0,-1,-1}
        +\frac{128 \big(
                1+x^2\big)}{27 (x-1) (1+x)} H_{0,0,-1,-1}
\nonumber\\ && 
        +\frac{256 \big(
                1+x^2\big)}{27 (x-1) (1+x)} H_{0,-1,-1,-1}
        +\frac{256 (x-1)^2 \big(
                11+12 x+11 x^2\big)}{27 (1+x)^4} \ln(2) \zeta_2
\nonumber\\ && 
        +\Biggl(
                -\frac{32 H_{0,-1} P^{(10)}_{10}}{9 (x-1) (1+x)^3}
                +\frac{32 H_{-1} P^{(10)}_{18}}{27 (1+x)^4}
                -\frac{64 P^{(10)}_{43}}{1215 (x-1) (1+x)^6}
\nonumber\\ && 
                +\Biggl(
                        -\frac{8 P^{(10)}_{45}}{81 (x-1) (1+x)^7}
                        -\frac{32 \big(
                                1-4 x+x^2
                        \big)
\big(1+6 x+x^2\big)}{27 (x-1) (1+x)^3} H_{-1}
                \Biggr) H_0
\nonumber\\ &&          
      +\frac{8 (x-1) \big(
                        1+4 x+x^2\big)}{3 (1+x)^3} H_0^2
                -\frac{64 \big(
                        1+x^2\big)}{27 (x-1) (1+x)} H_{-1}^2
        \Biggr) \zeta_2
        +\Biggl(
                -\frac{32 H_0 P^{(10)}_{12}}{27 (x-1) (1+x)^3}
\nonumber\\ && 
                -\frac{16 P^{(10)}_{23}
                }{81 (1+x)^4}
                +
                \frac{128 \big(
                        1+x^2\big)}{27 (x-1) (1+x)} H_{-1}
        \Biggr) \zeta_3
\Biggr\}
+ n_h \Biggl\{
        \Li_4\left(\frac{1}{2}\right) \Biggl(
                \frac{128 P^{(10)}_{19}}{27 (x-1)^2 (1+x)^2}
\nonumber\\ && 
  +\frac{1024 H_0 P^{(10)}_{34}}{27 (x-1)^3 (1+x)^3}
                +\frac{256 x^2 H_0^2}{(x-1) (1+x)^3}
                +\frac{1024 x^2 H_0 H_1}{(x-1) (1+x)^3}
                -\frac{1024 x^2 H_{0,1}}{(x-1) (1+x)^3}
        \Biggr)
\nonumber\\ && 
        +\ln^4(2) \Biggl(
                \frac{16 P^{(10)}_{19}}{81 (x-1)^2 (1+x)^2}
                +\frac{128 H_0 P^{(10)}_{34}}{81 (x-1)^3 (1+x)^3}
                +\frac{32 x^2 H_0^2}{3 (x-1) (1+x)^3}
\nonumber\\ && 
                +\frac{128 x^2 H_0 H_1}{3 (x-1) (1+x)^3}
                -\frac{128 x^2 H_{0,1}}{3 (x-1) (1+x)^3}
                +\frac{128 x^2 \zeta_2}{3 (x-1) (1+x)^3}
        \Biggr)
\nonumber\\ && 
        +\ln(2) \Biggl(
                -\frac{64 \zeta_2^2 P^{(10)}_{38}}{9 (x-1)^3 (1+x)^3}
                +\Biggl(
                        -\frac{128 H_{-1} P^{(10)}_8}{3 (x-1)^2 (1+x)^2}
                        -\frac{128 H_0 H_1 P^{(10)}_{36}}{9 (x-1)^3 (1+x)^3}
\nonumber\\ && 
                        +\frac{128 H_{0,1} P^{(10)}_{36}}{9 (x-1)^3 (1+x)^3}
                        +\frac{64 H_0^2 P^{(10)}_{42}}{9 (x-1)^3 (1+x)^4}
                        +\frac{16 P^{(10)}_{47}}{81 (x-1)^2 x (1+x)^6}
\nonumber\\ && 
                        +\big(
                                -\frac{128 H_{-1} P^{(10)}_{37}}{9 (x-1)^3 (1+x)^3}
                                +\frac{32 P^{(10)}_{46}}{27 (x-1)^3 (1+x)^5}
                        \Biggl) H_0
                        +
                        \frac{192 x^2 H_0^3}{(x-1) (1+x)^3}
\nonumber\\ && 
                        +\frac{128 \big(
                                1+x^2\big) P^{(10)}_{22}}{9 (x-1)^3 (1+x)^3} H_{0,-1}
                \Biggr) \zeta_2
        \Biggr)
        +\ln^2(2) \Biggl(
                \Biggl(
                        \frac{32 P^{(10)}_{21}}{27 (x-1)^2 (1+x)^2}
                        -\frac{128 H_0 P^{(10)}_{35}}{27 (x-1)^3 (1+x)^3}
\nonumber\\ &&
                        -\frac{64 x^2 H_0^2}{(x-1) (1+x)^3}
                        -\frac{256 x^2 H_0 H_1}{(x-1) (1+x)^3}
                        +\frac{256 x^2 H_{0,1}}{(x-1) (1+x)^3}
                \Biggr) \zeta_2
                -\frac{256 x^2 \zeta_2^2}{(x-1) (1+x)^3}
        \Biggr)
\nonumber\\ &&
        + n_l \Biggl[
                \frac{256 H_{0,0,0,1} P^{(10)}_4}{9 (x-1) (1+x)^3}
                +\frac{128 H_{0,0,0,-1} P^{(10)}_6}{9 (x-1) (1+x)^3}
                +\frac{8 H_0^4 P^{(10)}_{11}}{27 (x-1) (1+x)^3}
\nonumber\\ &&
                -\frac{64 H_0^2 H_1 P^{(10)}_{14}}{81 (1+x)^4}
                -\frac{64 \zeta_2^2 P^{(10)}_{16}}{135 (x-1) (1+x)^3}
                +\frac{64 \big(
                        377+106 x+377 x^2\big)}{729 (1+x)^2}
\nonumber\\ &&
                +\Biggl(
                        -\frac{128 H_{-1} P^{(10)}_{20}}{81 (x-1) (1+x)^3}
                        +\frac{64 P^{(10)}_{25}}{729 (x-1) (1+x)^3}
                        +\frac{256 \big(
                                11-6 x+11 x^2\big)}{81 (x-1) (1+x)} H_{-1}^2
\nonumber\\ &&
                        -\frac{1024 \big(
                                1+x^2\big)}{81 (x-1) (1+x)} H_{-1}^3
                \Biggr) H_0
                +\Biggl(
                        -\frac{64 H_{-1} P^{(10)}_{27}}{81 (x-1) (1+x)^4}
                        +
                        \frac{32 P^{(10)}_{29}}{81 (x-1) (1+x)^4}
\nonumber\\ &&
                        +\frac{256 \big(
                                1+x^2\big)}{27 (x-1) (1+x)} H_{-1}^2
                \Biggr) H_0^2
                +\Biggl(
                        -\frac{64 H_{-1} P^{(10)}_6}{27 (x-1) (1+x)^3}
                        -\frac{16 P^{(10)}_{28}}{81 (x-1) (1+x)^4}
                \Biggr) H_0^3
\nonumber\\ &&
                +\Biggl(
                        \frac{128 H_0^2 P^{(10)}_4}{27 (x-1) (1+x)^3}
                        +\frac{128 H_0 P^{(10)}_{14}}{81 (1+x)^4}
                \Biggr) H_{0,1}
                +\Biggl(
                        -\frac{64 H_0^2 P^{(10)}_4}{9 (x-1) (1+x)^3}
\nonumber\\ &&
                        +\frac{128 P^{(10)}_{20}}{81 (x-1) (1+x)^3}
                        -\frac{512 \big(
                                11-6 x+11 x^2\big)}{81 (x-1) (1+x)} H_{-1}
                        +\frac{1024 \big(
                                1+x^2\big)}{27 (x-1) (1+x)} H_{-1}^2
                \Biggr) H_{0,-1}
\nonumber\\ &&
                +\Biggl(
                        -\frac{512 H_0 P^{(10)}_4}{27 (x-1) (1+x)^3}
                        -\frac{128 P^{(10)}_{14}}{81 (1+x)^4}
                \Biggr) H_{0,0,1}
                +\Biggl(
                        \frac{128 H_0 P^{(10)}_4}{9 (x-1) (1+x)^3}
\nonumber\\ &&
                        +\frac{128 P^{(10)}_{27}}{81 (x-1) (1+x)^4}
                        -\frac{1024 \big(
                                1+x^2\big)}{27 (x-1) (1+x)} H_{-1}
                \Biggr) H_{0,0,-1}
\nonumber\\ &&
                +\Biggl(
                        \frac{512 \big(
                                11-6 x+11 x^2\big)}{81 (x-1) (1+x)}
                        -\frac{2048 \big(
                                1+x^2\big)}{27 (x-1) (1+x)} H_{-1}
                \Biggr) H_{0,-1,-1}
\nonumber\\ &&
                +\frac{1024 \big(
                        1+x^2\big)}{27 (x-1) (1+x)} H_{0,0,-1,-1}
                +\frac{2048 \big(
                        1+x^2\big)}
                {27 (x-1) (1+x)} H_{0,-1,-1,-1}
\nonumber\\ &&               
 +
                \frac{256 (x-1)^2 \big(
                        11+12 x+11 x^2\big)}{27 (1+x)^4} \ln(2) \zeta_2
                +\Biggl(
                        -\frac{64 H_{0,-1} P^{(10)}_1}{9 (x-1) (1+x)^3}
\nonumber\\ &&
                        +\frac{16 H_0^2 P^{(10)}_5}{3 (x-1) (1+x)^3}
                        +\frac{128 H_{-1} P^{(10)}_{30}}{81 (x-1) (1+x)^4}
                        -\frac{8 P^{(10)}_{33}}{81 (x-1) (1+x)^4}
\nonumber\\ &&
                        +\big(
                                -\frac{64 H_{-1} P^{(10)}_7}{27 (x-1) (1+x)^3}
                                -\frac{16 P^{(10)}_{31}}{81 (x-1) (1+x)^4}
                        \big) H_0
                        -\frac{512 \big(
                                1+x^2\big)}{27 (x-1) (1+x)} H_{-1}^2
                \Biggr) \zeta_2
\nonumber\\ &&
                +\Biggl(
                        -\frac{16 P^{(10)}_{32}}{81 (x-1) (1+x)^4}
                        -\frac{32 \big(
                                1+x^2\big)}{3 (x-1) (1+x)} H_0
                        +\frac{1024 \big(
                                1+x^2\big)}{27 (x-1) (1+x)} H_{-1}
                \Biggr) \zeta_3
        \Biggr]
\nonumber\\ &&
        +\Biggl(
                \frac{1024 x^2}{(x-1) (1+x)^3} \Li_4\left(\frac{1}{2}\right)
                +\Biggl(
                        -\frac{8 P^{(10)}_{13}}{(x-1)^2 (1+x)^2}
                        +\frac{32 x^2 \big(
                                5-24 x+5 x^2\big)}{(x-1)^3 (1+x)^3} H_0
                \Biggl) \zeta_3
        \Biggl) \zeta_2
\nonumber\\ &&
        +\Biggl(
                -\frac{8 P^{(10)}_{26}}{135 (x-1)^2 (1+x)^2}
                +\frac{32 H_0 P^{(10)}_{39}}{135 (x-1)^3 (1+x)^3}
                +\frac{16 x^2 H_0^2}{5 (x-1) (1+x)^3}
\nonumber\\ &&
                +\frac{64 x^2 H_0 H_1}{5 (x-1) (1+x)^3}
                -\frac{64 x^2 H_{0,1}}{5 (x-1) (1+x)^3}
        \Biggr) \zeta_2^2
        +\frac{64 x^2 \zeta_2^3}
        {5 (x-1) (1+x)^3}
\nonumber\\ &&
        +\Biggl(
                \frac{20 \big(
                        1+x^2
                \big)
\big(1-8 x+x^2\big)}{(x-1)^2 (1+x)^2}
                -\frac{80 x^2 \big(
                        1-8 x+x^2\big)}{(x-1)^3 (1+x)^3} H_0
        \Biggr) \zeta_5
\Biggr\} 
\nonumber\\ &&
+ F_{A,1,1}^{(0)} + F_{A,1,2}^{(0)} \zeta_2 + F_{A,1,3}^{(0)}  \zeta_3 + F_{A,1}^{(0),r},
\end{eqnarray}
with the polynomials
\begin{align}
P^{(10)}_1&=x^4+2 x^3-26 x^2+2 x+1,\\P^{(10)}_2&=x^4+2 x^3-10 x^2+2 x+1,\\P^{(10)}_3&=x^4+2 x^3-5 x^2+2 x+1,\\P^{(10)}_4&=x^4+2 x^3-4 x^2+2 
x+1,\\P^{(10)}_5&=x^4+2 x^3-2 x^2+2 x+1,\\P^{(10)}_6&=x^4+2 x^3+4 x^2+2 x+1,\\P^{(10)}_7&=x^4+2 x^3+14 x^2+2 x+1,\\P^{(10)}_8&=2 x^4-9 x^3+12 x^2-9 
x+2,\\P^{(10)}_9&=3 x^4+6 x^3-14 x^2+6 x+3,\\P^{(10)}_{10}&=3 x^4+6 x^3-10 x^2+6 x+3,\\P^{(10)}_{11}&=3 x^4+6 x^3-4 x^2+6 x+3,\\P^{(10)}_{12}&=11 
x^4+22 x^3-50 x^2+22 x+11,\\P^{(10)}_{13}&=13 x^4-24 x^3-6 x^2-24 x+13,\\P^{(10)}_{14}&=19 x^4-14 x^3+14 x^2-14 x+19,\\P^{(10)}_{15}&=31 
x^4-10 x^3+14 x^2-10 x+31,\\P^{(10)}_{16}&=31 x^4+62 x^3+149 x^2+62 x+31,\\P^{(10)}_{17}&=35 x^4-34 x^3+28 x^2-22 x+41,\\P^{(10)}_{18}&=39 
x^4-26 x^3+28 x^2-30 x+37,\\P^{(10)}_{19}&=53 x^4-188 x^3+174 x^2-188 x+53,\\P^{(10)}_{20}&=89 x^4+48 x^3+78 x^2+48 x+89,\\P^{(10)}_{21}&=91 
x^4-190 x^3+258 x^2-190 x+91,\\P^{(10)}_{22}&=105 x^4-208 x^3+198 x^2-208 x+105,\\P^{(10)}_{23}&=253 x^4-296 x^3-310 x^2-320 
x+241,\\P^{(10)}_{24}&=1367 x^4+3428 x^3+4506 x^2+3428 x+1367,\\P^{(10)}_{25}&=1471 x^4+2042 x^3-10 x^2+2042 x+1471,\\P^{(10)}_{26}&=5363 
x^4-4754 x^3-2814 x^2-4754 x+5363,\\P^{(10)}_{27}&=3 x^5+87 x^4+24 x^3+80 x^2+21 x+41,\\P^{(10)}_{28}&=35 x^5-141 x^4+34 x^3-78 
x^2-9 x-41,\\P^{(10)}_{29}&=62 x^5+125 x^4+205 x^3+47 x^2+149 x+116,\\P^{(10)}_{30}&=79 x^5-45 x^4+136 x^3-32 x^2+153 
x-35,\\P^{(10)}_{31}&=189 x^5-579 x^4+222 x^3-226 x^2-51 x-115,\\P^{(10)}_{32}&=463 x^5-219 x^4+598 x^3+234 x^2+1083 
x-111,\\P^{(10)}_{33}&=1969 x^5+1147 x^4-1734 x^3+3750 x^2+1045 x-545,\\P^{(10)}_{34}&=2 x^6+15 x^5-37 x^4+64 x^3-37 x^2+15 
x+2,\\P^{(10)}_{35}&=13 x^6+39 x^5-119 x^4+164 x^3-119 x^2+39 x+13,\\P^{(10)}_{36}&=21 x^6-110 x^5+195 x^4-208 x^3+195 x^2-110 
x+21,\\P^{(10)}_{37}&=99 x^6-214 x^5+333 x^4-440 x^3+333 x^2-214 x+99,\\P^{(10)}_{38}&=147 x^6-428 x^5+693 x^4-832 x^3+693 x^2-428 
x+147,\\P^{(10)}_{39}&=806 x^6-951 x^5+1172 x^4-2852 x^3+1172 x^2-951 x+806,\\P^{(10)}_{40}&=2945 x^6+6836 x^5+287 x^4-6056 x^3+287 
x^2+6836 x+2945,\\P^{(10)}_{41}&=19 x^7-133 x^6-466 x^5-238 x^4+73 x^3+81 x^2-282 x-174,\\P^{(10)}_{42}&=70 x^7+15 x^6-69 x^5-94 
x^4+6 x^3+127 x^2-53 x+2,\\P^{(10)}_{43}&=2226 x^7+4218 x^6-1317 x^5-2729 x^4+3554 x^3+3242 x^2-2143 x-1451,\\P^{(10)}_{44}&=181 
x^8+228 x^7-212 x^6+44 x^5+942 x^4+428 x^3+12 x^2-156 x-123,\\P^{(10)}_{45}&=213 x^8+276 x^7-300 x^6+268 x^5+1854 x^4+652 
x^3-76 x^2-108 x-91,\\P^{(10)}_{46}&=381 x^8+214 x^7-7742 x^6+5550 x^5+7394 x^4+5154 x^3-7526 x^2+394 x+309,\\P^{(10)}_{47}&=2768 
x^{10}-63429 x^9-44774 x^8+54896 x^7+46326 x^6-43414 x^5+46326 x^4
\nonumber\\ &
+54896 x^3-44774 x^2-63429 x+2768.
\end{align}

\noindent
The first expansion coefficients of $F_{A,1,i}^{(0)}, i = 1...3$ are given by
\begin{eqnarray}
F_{A,1,1}^{(0)}(x) &=& 
-\frac{1105690}{729} - \frac{19976198 y^2}{18225} - \frac{19976198 y^3}{18225} -
  \frac{647051207603 y^4}{857304000} 
\nonumber\\ &&
- \frac{177211030643 y^5}{428652000} 
+O(y^6)
\\
F_{A,1,2}^{(0)}(x) &=& 
\frac{1979131}{729} + \frac{17033692 y^2}{18225} + \frac{17033692 y^3}{18225} +
  \frac{27237088943 y^4}{44651250} 
\nonumber\\ &&
+ \frac{6370816243 y^5}{22325625}    
+O(y^6)
\\
F_{A,1,3}^{(0)}(x) &=& 
\frac{24544}{243} + \frac{1061573 y^2}{6075} + \frac{1061573 y^3}{6075} +
  \frac{255928217 y^4}{2352000} 
\nonumber\\ &&
+ \frac{4084720937 y^5}{95256000} 
+O(y^6).
\end{eqnarray}

The form factor $F_{A,2}$ is given by
\begin{eqnarray}
F_{A,2} &=&
x \Biggl\{
-\frac{128 n_h}{\ep^3} \Biggl\{
         \frac{1}{3 (1+x)^2}
        +\frac{\big(
                1+x^2\big)}{3 (1-x) (1+x)^3} H_0
\Biggr\}
+\frac{1}{\ep^2} \Biggl\{
n_h^2 \Biggl[
        -\frac{128}{27 (1-x)^2}
\nonumber\\ && 
 -\frac{64 \big(
                3-2 x+3 x^2\big)}{27 (1-x)^3 (1+x)} H_0
\Biggr]
+ n_h \Biggl[
        \frac{4672 }{27 (1-x)^2 (1+x)^3}
        +\frac{23488 x(1+x)}{27 (1-x)^2 (1+x)^3}
\nonumber\\ &&
        +\frac{4672 x^3}{27 (1-x)^2 (1+x)^3}
        -\frac{64 H_0^2 P^{(11)}_{24}}{27 (1-x)^4 (1+x)^3}
        - n_l \Biggl[
                \frac{256}{27 (1-x)^2}
                +\frac{128  \big(
                        3-2 x+3 x^2\big)}{27 (1-x)^3 (1+x)} H_0
        \Biggr] 
\nonumber\\ &&
        +\Biggl(
                \frac{128  P^{(11)}_{4}}{27 (1-x)^3 (1+x)^3}
                +\frac{256  \big(
                        1+x^2\big)}{3 (1-x) (1+x)^3} H_{-1}
        \Biggr) H_0
        -\frac{256  \big(
                1+x^2\big)}{3 (1-x) (1+x)^3} H_{0,-1}
\nonumber\\ &&
        +
                \frac{128 (1+x^2)}{3 (1-x) (1+x)^3} \zeta_2
\Biggr\}
+\frac{1}{\ep}
\Biggl\{
n_h^2 \Biggl[
        -\frac{64 H_0 P^{(11)}_{7}}{81 (1-x^2)^3}
        +\frac{32 H_0^2 P^{(11)}_{20}
        }{27 (1-x)^3 (1+x)^4}
        +
\nonumber\\ && 
        \frac{256  \big(
                1+26 x+x^2\big)}{81 (1-x)^2 (1+x)^2}
        +\frac{128  \big(
                3-2 x+3 x^2\big)}{27 (1-x)^3 (1+x)} H_{-1} H_0
        +\frac{256 x^2 H_0^3}{27 (1-x^2)^3}
        -\frac{128 \big(
                3-2 x+3 x^2\big)}{27 (1-x)^3 (1+x)} 
\nonumber\\ && \times 
H_{0,-1}
        +\Biggl(
                \frac{64  P^{(11)}_{26}}{27 (1-x)^3 (1+x)^4}
                +\frac{512 x^2 H_0}{9 (1-x^2)^3}
        \Biggr) \zeta_2
\Biggr]
+ n_h \Biggl[
        -\frac{256  H_0^2 H_1 P^{(11)}_{2}}{27 (1-x)^4 (1+x)^2}
\nonumber\\ && 
        +\frac{512  H_0 H_{0,1} P^{(11)}_{3}}{27 (1-x)^4 (1+x)^2}
        -\frac{512  H_{0,0,1} P^{(11)}_{6}}{27 (1-x)^4 (1+x)^2}
        +\frac{128  P^{(11)}_{13}}{81 (10x)^2 (1+x)^4}
        +\frac{128  H_{-1} H_0 P^{(11)}_{16}}{27 (1-x)^3 (1+x)^3}
\nonumber\\ && 
        +\frac{128  H_{0,-1} P^{(11)}_{16}}{27 (1-x^2)^3}
        -\frac{128  H_0 H_{0,-1} P^{(11)}_{17}}{27 (1-x)^4 (1+x)^2}
        +\frac{64  H_{-1} H_0^2 P^{(11)}_{28}}{27 (1-x)^4 (1+x)^3}
        +\frac{128  H_{0,0,-1} P^{(11)}_{29}}{27 (1-x)^4 (1+x)^3}
\nonumber\\ && 
        +\frac{16  H_0 P^{(11)}_{30}}{81 (1-x)^3 (1+x)^5}
        -\frac{64  H_0^2 P^{(11)}_{33}}{81 (1-x)^4 (1+x)^6}
        + n_l \Biggl[
                -\frac{128  H_0 P^{(11)}_{8}
                }{81 (1-x^2)^3}
                -
                \frac{64  H_0^2 P^{(11)}_{22}}{27 (1-x)^3 (1+x)^4}
\nonumber\\ && 
                -\frac{128  \big(
                        23-2 x+23 x^2\big)}{81 (1-x^2)^2}
                +\frac{512  \big(
                        3-2 x+3 x^2\big)}{27 (1-x)^3 (1+x)} H_{-1} H_0
                +\frac{256 x^2 H_0^3}{27 (1-x^2)^3}
\nonumber\\ &&          
  -\frac{512  \big(
                        3-2 x+3 x^2\big)}{27 (1-x)^3 (1+x)} H_{0,-1}
                +\Biggl(
                        \frac{128  P^{(11)}_{25}}{27 (1-x)^3 (1+x)^4}
                        +\frac{512 x^2 H_0}{9 (1-x^2)^3}
                \Biggr) \zeta_2
        \Biggr]
\nonumber\\ && 
        -\frac{256  \big(
                1+x^2\big)}{3 (1-x) (1+x)^3} H_{-1}^2 H_0
        +\Biggl(
                +\frac{64  P^{(11)}_{31}}{81 (1-x^2)^5}
                -\frac{512 x \big(
                        35-208 x+35 x^2\big)}{81 (1-x^2)^3} H_1
        \Biggr) H_0^3
\nonumber\\ && 
        +\frac{32 x P^{(11)}_{19}}{81 (1-x)^5 (1+x)^4} H_0^4
        +\frac{256  \big(
                23-14 x+23 x^2\big)}{27 (1-x)^3 (1+x)} H_0 H_1
        -\frac{256  \big(
                23-14 x+23 x^2\big)}{27 (1-x)^3 (1+x)} H_{0,1}
\nonumber\\ && 
        +\frac{256 x P^{(11)}_{11}}{27 (1-x)^5 (1+x)^3} H_0^2 H_{0,1}
        -\frac{256 x P^{(11)}_{10}}{27 (1-x)^5 (1+x)^3} H_0^2 H_{0,-1}
        +\frac{512  \big(
                1+x^2\big)}{3 (1-x) (1+x)^3} H_{-1} 
\nonumber\\ &&  \times
H_{0,-1}
        -\frac{2048 x P^{(11)}_{5}
        }{27 (1-x)^5 (1+x)^3} H_0 H_{0,0,1}
        +\frac{512 x P^{(11)}_{14}}{27 (1-x)^5 (1+x)^3} H_0 H_{0,0,-1}
\nonumber\\ && 
        -\frac{512  \big(
                1+x^2\big)}{3 (1-x) (1+x)^3} H_{0,-1,-1}
        +\frac{512 x P^{(11)}_{15}}{27 (1-x)^5 (1+x)^3} H_{0,0,0,1}
        -\frac{512 x^2 P^{(11)}_{12}}{9 (1-x)^5 (1+x)^3} H_{0,0,0,-1}
\nonumber\\ && 
        +\Biggl(
                \frac{128  H_{-1} P^{(11)}_{21}}{3 (1-x)^4 (1+x)^3}
                -\frac{64  H_0 P^{(11)}_{32}}{27 (1-x)^5 (1+x)^5}
                -\frac{64  P^{(11)}_{34}}{27 (1-x)^4 (1+x)^6}
  -\frac{128 x P^{(11)}_{23} H_0^2}{9 (1-x)^5 (1+x)^4} 
\nonumber\\ &&               
                -\frac{1024 x \big(
                        35-208 x+35 x^2\big)}{27 (1-x^2)^3} H_0 H_1
                +\frac{1024 x \big(
                        35-208 x+35 x^2\big)}{27 (1-x^2)^3} H_{0,1}
\nonumber\\ && 
                +\frac{512 x P^{(11)}_{1}}{9 (1-x)^5 (1+x)^3} H_{0,-1}
        \Biggr) \zeta_2
        -\frac{128 x P^{(11)}_{18}}{135 (1-x)^5 (1+x)^3} \zeta_2^2
        +\Biggl(
                -\frac{256  P^{(11)}_{27}}{27 (1-x)^4 (1+x)^3}
\nonumber\\ &&
                +\frac{512 x P^{(11)}_{9}}{27 (1-x)^5 (1+x)^3} H_0
        \Biggr) \zeta_3
\Biggr]
\Biggr\} + F_{A,2}^{(0)}
\end{eqnarray}
with the polynomials
\begin{align}
P^{(11)}_{1} &=x^4-5 x^3+2 x^2-5 x+1,
\\
P^{(11)}_{2} &=12 x^4-109 x^3+310 x^2-109 x+12,
\\
P^{(11)}_{3} &=24 x^4-115 x^3+330 x^2-115 x+24,
\\
P^{(11)}_{4} &=29 x^4+167 x^3-16 x^2+167 x+29,
\\
P^{(11)}_{5} &=36 x^4-277 x^3+494 x^2-277 x+36,
\\
P^{(11)}_{6} &=36 x^4-121 x^3+350 x^2-121 x+36,
\\
P^{(11)}_{7} &=69 x^4-152 x^3-58 x^2-152 x+69,
\\
P^{(11)}_{8} &=69 x^4-26 x^3+2 x^2-26 x+69,
\\
P^{(11)}_{9} &=71 x^4-561 x^3+992 x^2-561 x+71,
\\
P^{(11)}_{10} &=71 x^4-557 x^3+990 x^2-557 x+71,
\\
P^{(11)}_{11} &=107 x^4-834 x^3+1478 x^2-834 x+107,
\\
P^{(11)}_{12} &=108 x^4-835 x^3+1484 x^2-835 x+108,
\\
P^{(11)}_{13} &=158 x^4+469 x^3+1006 x^2+469 x+158,
\\
P^{(11)}_{14} &=179 x^4-1392 x^3+2474 x^2-1392 x+179,
\\
P^{(11)}_{15} &=181 x^4-1378 x^3+2466 x^2-1378 x+181,
\\
P^{(11)}_{16} &=207 x^4+8 x^3-982 x^2+8 x+207,
\\
P^{(11)}_{17} &=249 x^4-1380 x^3+3994 x^2-1380 x+249,
\\
P^{(11)}_{18} &=610 x^4-4759 x^3+8064 x^2-4759 x+610,
\\
P^{(11)}_{19} &=x^5-52 x^4+81 x^3-15 x^2+44 x+1,
\\
P^{(11)}_{20} &=3 x^5-41 x^4+18 x^3-30 x^2+27 x-9,
\\
P^{(11)}_{21} &=3 x^5-9 x^4+6 x^3-10 x^2+3 x-1,
\\
P^{(11)}_{22} &=3 x^5+31 x^4+24 x^2-3 x+9,
\\
P^{(11)}_{23} &=12 x^5-75 x^4+53 x^3+69 x^2-91 x+12,
\\
P^{(11)}_{24} &=15 x^5+35 x^4-4 x^3+68 x^2-19 x+33,
\\
P^{(11)}_{25} &=21 x^5-31 x^4+48 x^3-24 x^2+59 x-9,
\\
P^{(11)}_{26} &=33 x^5-83 x^4+78 x^3-66 x^2+97 x-27,
\\
P^{(11)}_{27} &=48 x^5-421 x^4+859 x^3+787 x^2-367 x+30,
\\
P^{(11)}_{28} &=183 x^5-701 x^4+1726 x^3+1870 x^2-809 x+219,
\\
P^{(11)}_{29} &=315 x^5-1561 x^4+3502 x^3+3358 x^2-1453 x+279,
\\
P^{(11)}_{30} &=8449 x^6+19482 x^5+13519 x^4-7316 x^3+13519 x^2+19482 x+8449,
\\
P^{(11)}_{31} &=48 x^8-584 x^7+457 x^6+2554 x^5-1127 x^4-1544 x^3-329 x^2+54 x-105,
\\
P^{(11)}_{32} &=63 x^8+1134 x^7-402 x^6-2986 x^5+2320 x^4+738 x^3+146 x^2+154 x-15,
\\
P^{(11)}_{33} &=696 x^8+1820 x^7-2251 x^6-3898 x^5-521 x^4+3080 x^3+155 x^2+86 x-63,
\\
P^{(11)}_{34} &=24 \ln(2) \left(x^2-1\right)^4-843 x^8+2432 x^7+5402 x^6+2028 x^5-8680 x^4-2888 x^3
\nonumber\\ &
+2814 x^2+2652 
x-613
\end{align}

\noindent
and
\begin{eqnarray}
F_{A,2}^{(0)} &=& x \Biggl\{ 
n_h^2 \Biggl\{
        -
        \frac{256 H_0^2 H_1 P^{(12)}_2}{27 (x-1)^2 (1+x)^4}
        -\frac{128 H_{0,0,-1} P^{(12)}_3}{27 (x-1)^2 (1+x)^4}
        +\frac{512 P^{(12)}_5}{243 (x-1)^2 (1+x)^4}
\nonumber\\ &&
        +\Biggl(
                -\frac{512 H_{-1} P^{(12)}_6}{81 (x-1)^3 (1+x)^3}
                +\frac{128 P^{(12)}_{28}}{243 (x-1)^3 (1+x)^5}
                +\frac{128 H_{-1}^2}{27 (x-1) (1+x)}
        \Biggr) H_0
\nonumber\\ &&
        +\Biggl(
                \frac{64 H_{-1} P^{(12)}_3}{27 (x-1)^2 (1+x)^4}
                -\frac{128 P^{(12)}_{29}}{81 (x-1)^3 (1+x)^6}
        \Biggr) H_0^2
        +\Biggl(
                -\frac{32 P^{(12)}_{31}}{81 (x-1)^3 (1+x)^7}
\nonumber\\ &&
                +\frac{512 x^2 H_{-1}}{27 (x-1)^3 (1+x)^3}
        \Biggr) H_0^3
        -\frac{320 x^2 H_0^4}{27 (x-1)^3 (1+x)^3}
        +\Biggl(
                \frac{512 H_0 P^{(12)}_2}{27 (x-1)^2 (1+x)^4}
\nonumber\\ &&
                -\frac{1024 x^2 H_0^2}{9 (x-1)^3 (1+x)^3}
        \Biggr) H_{0,1}
        +\Biggl(
                \frac{512 P^{(12)}_6}{81 (x-1)^3 (1+x)^3}
                -\frac{256 H_{-1}}{27 (x-1) (1+x)}
        \Biggr) H_{0,-1}
\nonumber\\ &&
        +\Biggl(
                -\frac{512 P^{(12)}_2}{27 (x-1)^2 (1+x)^4}
                +\frac{4096 x^2 H_0}{9 (x-1)^3 (1+x)^3}
        \Biggr) H_{0,0,1}
        +\frac{256 H_{0,-1,-1}}{27 (x-1) (1+x)}
\nonumber\\ &&
        -\frac{2048 x^2 H_{0,0,0,1}}{3 (x-1)^3 (1+x)^3}
        -\frac{1024 x^2 H_{0,0,0,-1}}{9 (x-1)^3 (1+x)^3}
        +\Biggl(
                \frac{128 H_{-1} P^{(12)}_8}{27 (x-1)^2 (1+x)^4}
\nonumber\\ &&
                -\frac{32 P^{(12)}_{30}}{405 (x-1)^3 (1+x)^6}
                +\frac{1024 \big(
                        3+2 x+3 x^2\big)}{9 (1+x)^4} \ln(2)
                +\Biggl(
                        -\frac{32 P^{(12)}_{32}}{27 (x-1)^3 (1+x)^7}
\nonumber\\ &&
                        +\frac{1024 x^2 H_{-1}}{9 (x-1)^3 (1+x)^3}
                \Biggr) H_0
                -\frac{256 x^2 H_0^2}{3 (x-1)^3 (1+x)^3}
                +\frac{2048 x^2 H_{0,-1}}{9 (x-1)^3 (1+x)^3}
        \Biggr) \zeta_2
\nonumber\\ &&
        +\frac{1792 x^2 \zeta_2^2}{9 (x-1)^3 (1+x)^3}
        +\Biggl(
                -\frac{128 P^{(12)}_{11}}{27 (x-1)^2 (1+x)^4}
                +\frac{1024 x^2 H_0}{3 (x-1)^3 (1+x)^3}
        \Biggr) \zeta_3
\Biggr\}
\nonumber\\ &&
+n_h \Biggl\{
        \frac{160 \zeta_5 P^{(12)}_1}{(x-1)^4 (1+x)^2}
        +\Li_4\left(\frac{1}{2}\right) \Biggl(
                \frac{1024 P^{(12)}_7}{27 (x-1)^4 (1+x)^2}
                +\frac{1024 x H_0 P^{(12)}_{17}}{27 (x-1)^5 (1+x)^3}
\nonumber\\ &&
                +\frac{1024 x^2 H_0^2}{(x-1)^3 (1+x)^3}
                +\frac{4096 x^2 H_0 H_1}{(x-1)^3 (1+x)^3}
                -\frac{4096 x^2 H_{0,1}}{(x-1)^3 (1+x)^3}
        \Biggr)
\nonumber\\ &&
        +\ln^4(2) \Biggl(
                \frac{128 P^{(12)}_7}{81 (x-1)^4 (1+x)^2}
                +\frac{128 x H_0 P^{(12)}_{17}}{81 (x-1)^5 (1+x)^3}
                +\frac{128 x^2 H_0^2}{3 (x-1)^3 (1+x)^3}
\nonumber\\ &&
                +\frac{512 x^2 H_0 H_1}{3 (x-1)^3 (1+x)^3}
                -
                \frac{512 x^2 H_{0,1}}{3 (x-1)^3 (1+x)^3}
        \Biggr)
        + n_l \Biggl[
                -\frac{256 H_0^2 H_1 P^{(12)}_2}{27 (x-1)^2 (1+x)^4}
\nonumber\\ &&
                -\frac{256 \zeta_3 P^{(12)}_{27}}{27 (x-1)^3 (1+x)^4}
                -\frac{512 \big(
                        11+32 x+11 x^2\big)}{27 (x-1)^2 (1+x)^2}
                +\Biggl(
                        \frac{256 P^{(12)}_{19}}{81 (x-1)^3 (1+x)^3}
\nonumber\\ &&
                        -\frac{512 \big(
                                55+62 x+55 x^2\big)}{81 (x-1) (1+x)^3} H_{-1}
                        +\frac{2048 \big(
                                1-x+x^2\big)}{27 (x-1)^3 (1+x)} H_{-1}^2
                \Biggr) H_0
                +\Biggl(
                        \frac{128 P^{(12)}_{10}}{81 (x-1)^2 (1+x)^4}
\nonumber\\ &&
                        -\frac{256 H_{-1} P^{(12)}_{22}}{27 (x-1)^3 (1+x)^4}
                \Biggr) H_0^2
                +\Biggl(
                        -\frac{64 P^{(12)}_{25}}{81 (x-1)^3 (1+x)^4}
                        -\frac{512 x^2 H_{-1}}{27 (x-1)^3 (1+x)^3}
                \Biggr) H_0^3
\nonumber\\ &&
                -\frac{320 x^2 H_0^4}{27 (x-1)^3 (1+x)^3}
                +\Biggl(
                        \frac{512 H_0 P^{(12)}_2}{27 (x-1)^2 (1+x)^4}
                        -\frac{1024 x^2 H_0^2}{9 (x-1)^3 (1+x)^3}
                \Biggr) H_{0,1}
\nonumber\\ &&
                +\Biggl(
                        \frac{512 \big(
                                55+62 x+55 x^2\big)}{81 (x-1) (1+x)^3}
                        +\frac{512 x^2 H_0^2}{3 (x-1)^3 (1+x)^3}
                        -\frac{4096 \big(
                                1-x+x^2\big)}{27 (x-1)^3 (1+x)} H_{-1}
                \Biggr) H_{0,-1}
\nonumber\\ &&
                +\Biggl(
                        -\frac{512 P^{(12)}_2}{27 (x-1)^2 (1+x)^4}
                        +
                        \frac{4096 x^2 H_0}{9 (x-1)^3 (1+x)^3}
                \Biggr) H_{0,0,1}
                +\Biggl(
                        \frac{512 P^{(12)}_{22}}{27 (x-1)^3 (1+x)^4}
\nonumber\\ &&
                        -\frac{1024 x^2 H_0}{3 (x-1)^3 (1+x)^3}
                \Biggr) H_{0,0,-1}
                +\frac{4096 \big(
                        1-x+x^2\big)}{27 (x-1)^3 (1+x)} H_{0,-1,-1}
                -\frac{2048 x^2 H_{0,0,0,1}}{3 (x-1)^3 (1+x)^3}
\nonumber\\ &&
                +\frac{1024 x^2 H_{0,0,0,-1}}{9 (x-1)^3 (1+x)^3}
                +\Biggl(
                        -\frac{32 P^{(12)}_{20}}{81 (x-1)^2 (1+x)^4}
                        +\frac{512 H_{-1} P^{(12)}_{23}}{27 (x-1)^3 (1+x)^4}
\nonumber\\ &&
                        +\frac{1024 \big(
                                3+2 x+3 x^2\big)}{9 (1+x)^4} \ln(2)
                        +\Biggl(
                                -\frac{128 P^{(12)}_{24}}{27 (x-1)^3 (1+x)^4}
                                -\frac{1024 x^2 H_{-1}}{9 (x-1)^3 (1+x)^3}
                        \Biggr) H_0
\nonumber\\ &&
                        -\frac{256 x^2 H_0^2}{3 (x-1)^3 (1+x)^3}
                        +\frac{7168 x^2 H_{0,-1}}{9 (x-1)^3 (1+x)^3}
                \Biggr) \zeta_2
                -\frac{7424 x^2 \zeta_2^2}{45 (x-1)^3 (1+x)^3}
        \Biggr]
\nonumber\\ &&
        +\Biggl(
                \frac{4096 x^2}{(x-1)^3 (1+x)^3} \Li_4\left(\frac{1}{2}\right)
                +\frac{512 x^2}{3 (x-1)^3 (1+x)^3} \ln^4(2)
                +\ln(2) 
\nonumber\\ &&
\times \Biggl(
                        -\frac{1024 x H_{0,-1} P^{(12)}_{13}}{9 (x-1)^5 (1+x)^3}
                        +\frac{512 x H_0 H_1 P^{(12)}_{16}}{9 (x-1)^5 (1+x)^3}
                        -\frac{512 x H_{0,1}
                         P^{(12)}_{16}}{9 (x-1)^5 (1+x)^3}
\nonumber\\ &&
                        -\frac{64 P^{(12)}_{34}}{27 (x-1)^4 x (1+x)^6}
                        +\Biggl(
                                \frac{1024 x H_{-1} P^{(12)}_{14}}{9 (x-1)^5 (1+x)^3}
                                +\frac{256 P^{(12)}_{33}}{27 (x-1)^5 (1+x)^5}
                        \Biggr) H_0
\nonumber\\ &&
                        -
                        \frac{256 x H_0^2 P^{(12)}_{26}}{9 (x-1)^5 (1+x)^4}
                        +\frac{768 x^2 H_0^3}{(x-1)^3 (1+x)^3}
                        -\frac{256 \big(
                                1+x^2
                        \big)
\big(1-4 x+x^2\big)}{(x-1)^4 (1+x)^2} H_{-1}
                \Biggr)
\nonumber\\ &&
                +\ln^2(2) \Biggl(
                        \frac{128 P^{(12)}_9}{27 (x-1)^4 (1+x)^2}
                        -\frac{256 x H_0 P^{(12)}_{18}}{27 (x-1)^5 (1+x)^3}
                        -\frac{256 x^2 H_0^2}{(x-1)^3 (1+x)^3}
\nonumber\\ &&
                        -\frac{1024 x^2 H_0 H_1}{(x-1)^3 (1+x)^3}
                        +\frac{1024 x^2 H_{0,1}}{(x-1)^3 (1+x)^3}
                \Biggr)
                +\Biggl(
                        -\frac{64 P^{(12)}_4}{(x-1)^4 (1+x)^2}
\nonumber\\ &&
                        -\frac{256 x^2 \big(
                                1+19 x+x^2\big)}{(x-1)^5 (1+x)^3} H_0
                \Biggr) \zeta_3
        \Biggr) \zeta_2
        +\Biggl(
                \frac{32 P^{(12)}_{12}}{135 (x-1)^4 (1+x)^2}
	                -\frac{64 x H_0 P^{(12)}_{21}}{135 (x-1)^5 (1+x)^3}
\nonumber\\ &&
                +\frac{1024 x P^{(12)}_{15}}{9 (x-1)^5 (1+x)^3} \ln(2)
                -\frac{1024 x^2}{(x-1)^3 (1+x)^3} \ln^2(2)
                +\frac{64 x^2 H_0^2}{5 (x-1)^3 (1+x)^3}
\nonumber\\ &&
                +\frac{256 x^2 H_0 H_1}{5 (x-1)^3 (1+x)^3}
                -\frac{256 x^2 H_{0,1}
                }{5 (x-1)^3 (1+x)^3}
        \Biggr) \zeta_2^2
        +\frac{256 x^2 \zeta_2^3}{5 (x-1)^3 (1+x)^3}
\nonumber\\ &&
        +\frac{640 x^2 \big(
                1+7 x+x^2\big)}{(x-1)^5 (1+x)^3} H_0 \zeta_5
\Biggr\} \Biggr\} 
+ F_{A,2,1}^{(0)} + F_{A,2,2}^{(0)} \zeta_2 + F_{A,2,3}^{(0)}  \zeta_3 + F_{A,2}^{(0),r},
\end{eqnarray}
with the polynomials
\begin{align}
P^{(12)}_1&=2 x^4-7 x^3-8 x^2-7 x+2,\\P^{(12)}_2&=3 x^4-14 x^3-2 x^2-14 x+3,\\P^{(12)}_3&=5 x^4-30 x^3-4 x^2-26 x+7,\\P^{(12)}_4&=6 x^4-19 
x^3-16 x^2-19 x+6,\\P^{(12)}_5&=14 x^4+35 x^3+54 x^2+35 x+14,\\P^{(12)}_6&=15 x^4-41 x^3-16 x^2-41 x+15,\\P^{(12)}_7&=17 x^4-118 
x^3+58 x^2-118 x+17,\\P^{(12)}_8&=19 x^4-82 x^3-12 x^2-86 x+17,\\P^{(12)}_9&=29 x^4-16 x^3+154 x^2-16 x+29,\\P^{(12)}_{10}&=30 x^4+41 
x^3+32 x^2-83 x-80,\\P^{(12)}_{11}&=49 x^4-2 x^3+56 x^2-6 x+47,\\P^{(12)}_{12}&=65 x^4+5552 x^3-6446 x^2+5552 
x+65,\\P^{(12)}_{13}&=77 x^4-200 x^3+270 x^2-200 x+77,\\P^{(12)}_{14}&=80 x^4-215 x^3+276 x^2-215 x+80,\\P^{(12)}_{15}&=80 x^4-209 
x^3+264 x^2-209 x+80,\\P^{(12)}_{16}&=83 x^4-218 x^3+258 x^2-218 x+83,\\P^{(12)}_{17}&=87 x^4-92 x^3+298 x^2-92 
x+87,\\P^{(12)}_{18}&=105 x^4-182 x^3+334 x^2-182 x+105,\\P^{(12)}_{19}&=135 x^4+110 x^3-98 x^2+110 x+135,\\P^{(12)}_{20}&=1709 
x^4+1708 x^3-690 x^2+716 x+829,\\P^{(12)}_{21}&=2415 x^4-3158 x^3+6274 x^2-3158 x+2415,\\P^{(12)}_{22}&=x^5+25 x^4-8 x^3+16 
x^2-9 x+7,\\P^{(12)}_{23}&=13 x^5-43 x^4+40 x^3-32 x^2+59 x-5,\\P^{(12)}_{24}&=14 x^5-133 x^4+31 x^3-65 x^2+3 
x-10,\\P^{(12)}_{25}&=15 x^5-177 x^4+50 x^3-94 x^2+27 x-21,\\P^{(12)}_{26}&=28 x^5-42 x^4+35 x^3+29 x^2-117 x+55,\\P^{(12)}_{27}&=29 
x^5-33 x^4+50 x^3-34 x^2+65 x-13,\\P^{(12)}_{28}&=391 x^6+502 x^5-1151 x^4-2236 x^3-1151 x^2+502 x+391,\\P^{(12)}_{29}&=10 
x^7-21 x^6-35 x^5+158 x^4+236 x^3+223 x^2+13 x-40,\\P^{(12)}_{30}&=4647 x^7+4667 x^6-12281 x^5-14941 x^4-819 
x^3+4761 x^2-4347 x-3447,\\P^{(12)}_{31}&=27 x^8-92 x^7-404 x^6-356 x^5+66 x^4+28 x^3+76 x^2+36 x-21,\\P^{(12)}_{32}&=31 
x^8-116 x^7-564 x^6-524 x^5+122 x^4-140 x^3-84 x^2+12 x-17,\\P^{(12)}_{33}&=60 x^8-441 x^7-3520 x^6+4719 x^5+4313 
x^4+4773 x^3-3358 x^2-387 x+33,\\P^{(12)}_{34}&=208 x^{10}+10715 x^9-21650 x^8
-31368 x^7+25186 x^6+85658 x^5+25186 x^4
\nonumber\\ &
-31368 x^3-21650 x^2+10715 x+208.
\end{align}

\noindent
The first expansion coefficients of $F_{A,2,i}^{(0)}, i = 1...3$ are given by
\begin{eqnarray}
F_{A,2,1}^{(0)}(x) &=& 
-\frac{82929376}{18225} + \frac{264976}{81 y^2} - \frac{264976}{81 y} +
    \frac{53768196023 y^2}{53581500} 
\nonumber\\ &&
+ \frac{53768196023 y^3}{53581500} + 
  \frac{42082197145871 y^4}{54010152000} + \frac{14983026350279 y^5}{27005076000} 
+ O(y^6)
\nonumber\\
\\
F_{A,2,2}^{(0)}(x) &=& 
\frac{81870064}{18225} + \frac{11132}{27 y^2} - \frac{11132}{27 y} -
    \frac{7224542428 y^2}{7441875} 
\nonumber\\ &&
- \frac{7224542428 y^3}{7441875} -
  \frac{164328135367 y^4}{218791125} - \frac{581273616754 y^5}{1093955625} 
+ O(y^6)
\\
F_{A,2,3}^{(0)}(x) &=& 
\frac{6230776}{6075} - \frac{1024}{27 y^2} + \frac{1024}{27 y} -
    \frac{22054357 y^2}{11907000} 
\nonumber\\ &&
- \frac{22054357 y^3}{11907000} -
  \frac{12191383321 y^4}{266716800} - \frac{19907290871 y^5}{222264000} 
+ O(y^6).
\end{eqnarray}
The $1/y^k$ behaviour of these expressions is cancelled by corresponding terms of other contributions.

The renormalization of the three--loop massive form factors is performed in the same way as in earlier calculations,
cf.~\cite{Ablinger:2018yae,Ablinger:2018zwz}. We use a mixed scheme. The heavy quark mass and wave function have been 
renormalized in the on-shell (OS) renormalization scheme, while the strong coupling 
constant is renormalized
in the $\overline{\rm MS}$ scheme, where we set the universal factor $S_\varepsilon =
\exp(-\varepsilon (\gamma_E - \ln(4\pi))$ for each loop order to one at the end of the calculation. The required 
renormalization constants are available and are denoted by $Z_{m, {\rm OS}}$ \cite{Broadhurst:1991fy, 
Melnikov:2000zc,Marquard:2007uj,Marquard:2015qpa,Marquard:2016dcn}, $Z_{2,{\rm OS}}$ \cite{Broadhurst:1991fy, 
Melnikov:2000zc,Marquard:2007uj,Marquard:2018rwx} and $Z_{a_s}$ 
\cite{Tarasov:1980au,Larin:1993tp,vanRitbergen:1997va,Czakon:2004bu,Baikov:2016tgj,Herzog:2017ohr,
Luthe:2017ttg} for the heavy quark mass, wave function and strong coupling constant, respectively.
The renormalization of the heavy-quark wave function and the strong coupling constant are multiplicative, while the 
renormalization of massive fermion lines has been taken care of by properly considering the 
counter terms. Pseudoscalar and axialvector form factors are related by a Ward identity. We explicitly 
verified our results fulfill this relation.
\subsection{Numerical Results}
\label{sec:56}

\begin{figure}[H]
  \centering
\includegraphics[width=.48\linewidth]{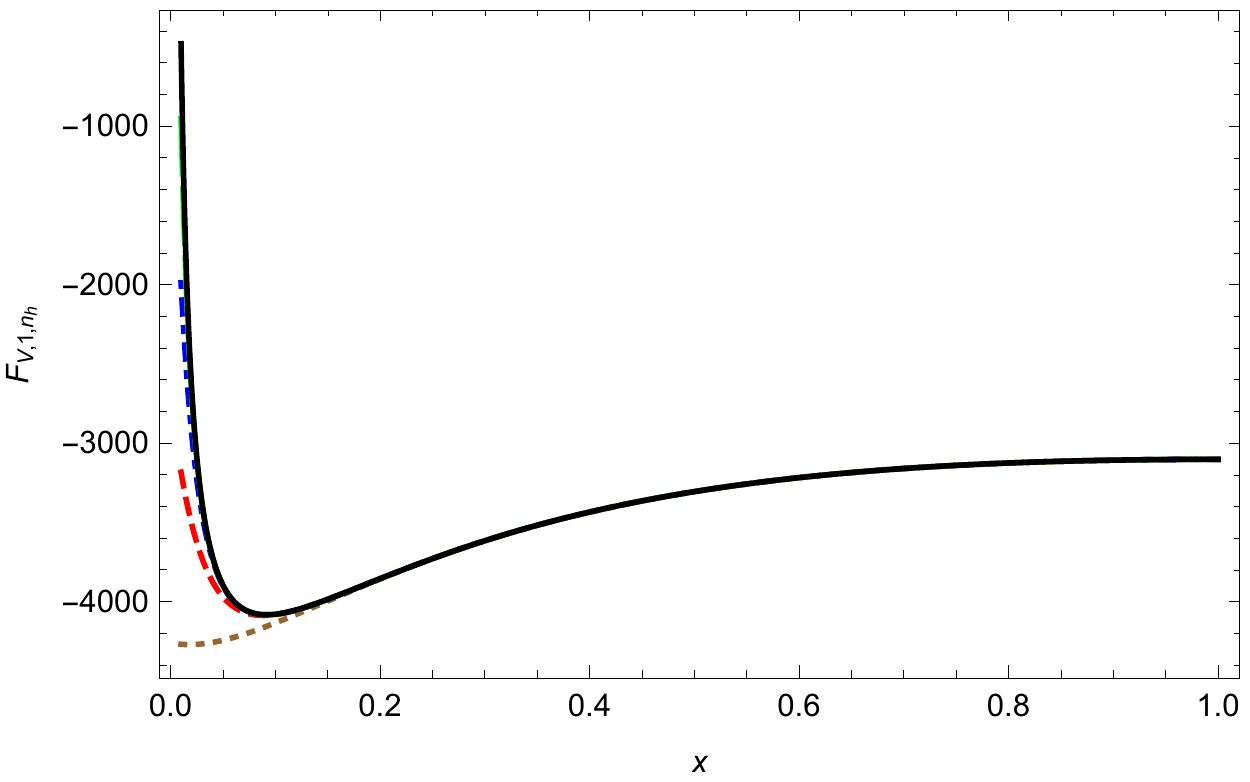}\hfill
\includegraphics[width=.48\linewidth]{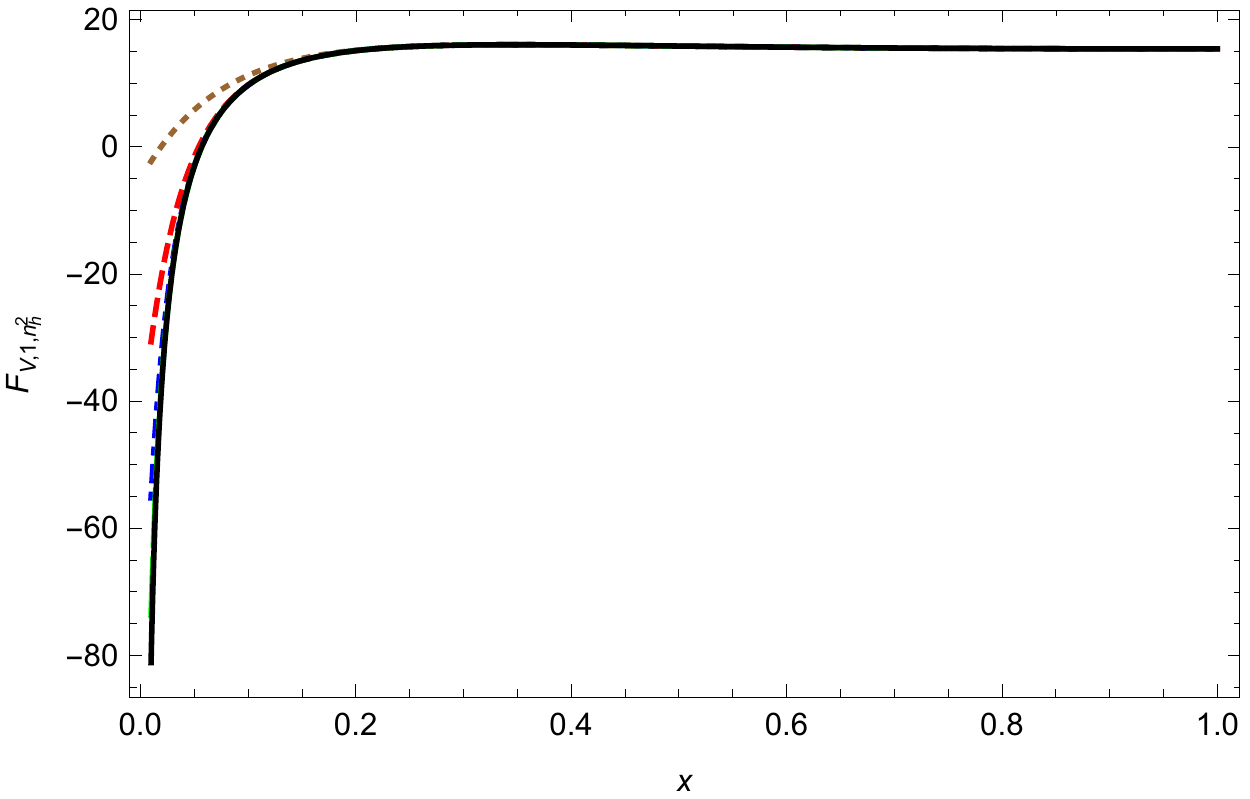}  
  \caption[]{\sf \small Vector form factor $F_{V,1}$: left $\ep^0 n_h^1$, right $\ep^0 n_h^2$, the 
approximation 
with 20, 50, 100, 200, 500 terms is shown in brown, red, blue, green and black, respectively.}
  \label{fig:1}
\end{figure}

\begin{figure}[H]
  \centering
\includegraphics[width=.48\linewidth]{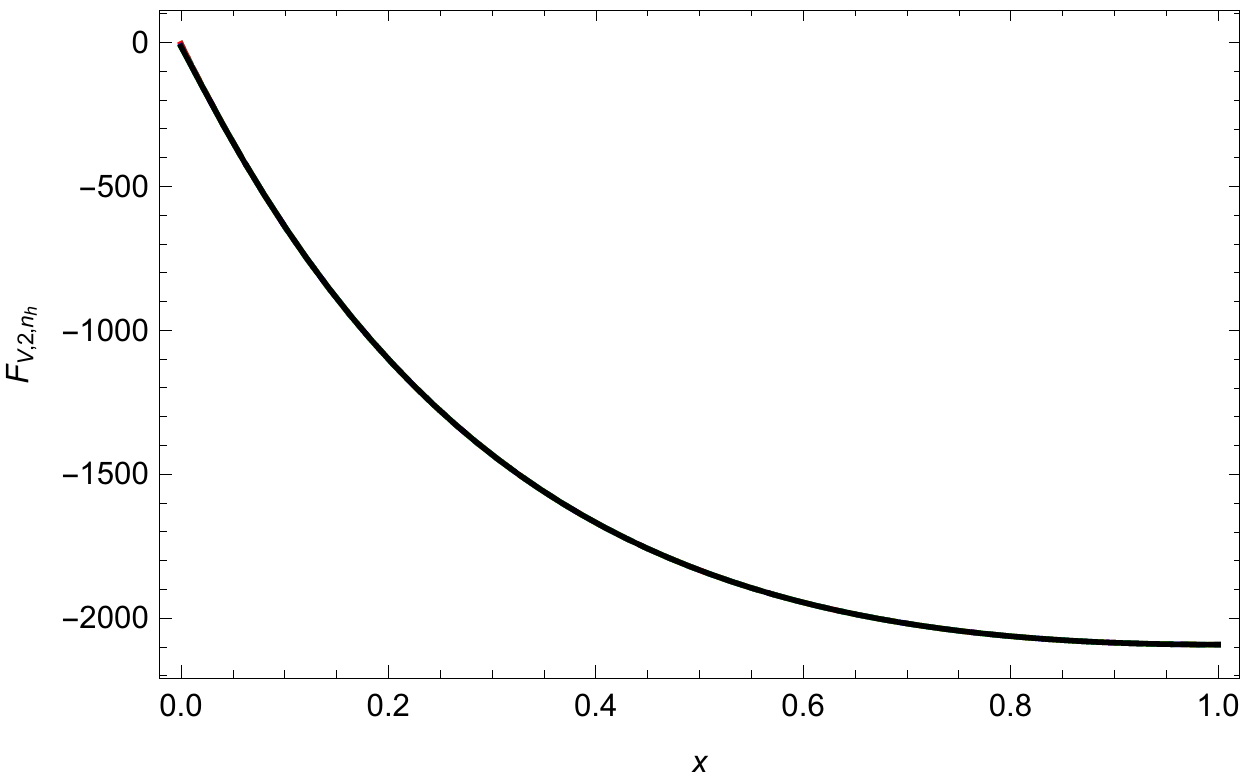}\hfill
\includegraphics[width=.48\linewidth]{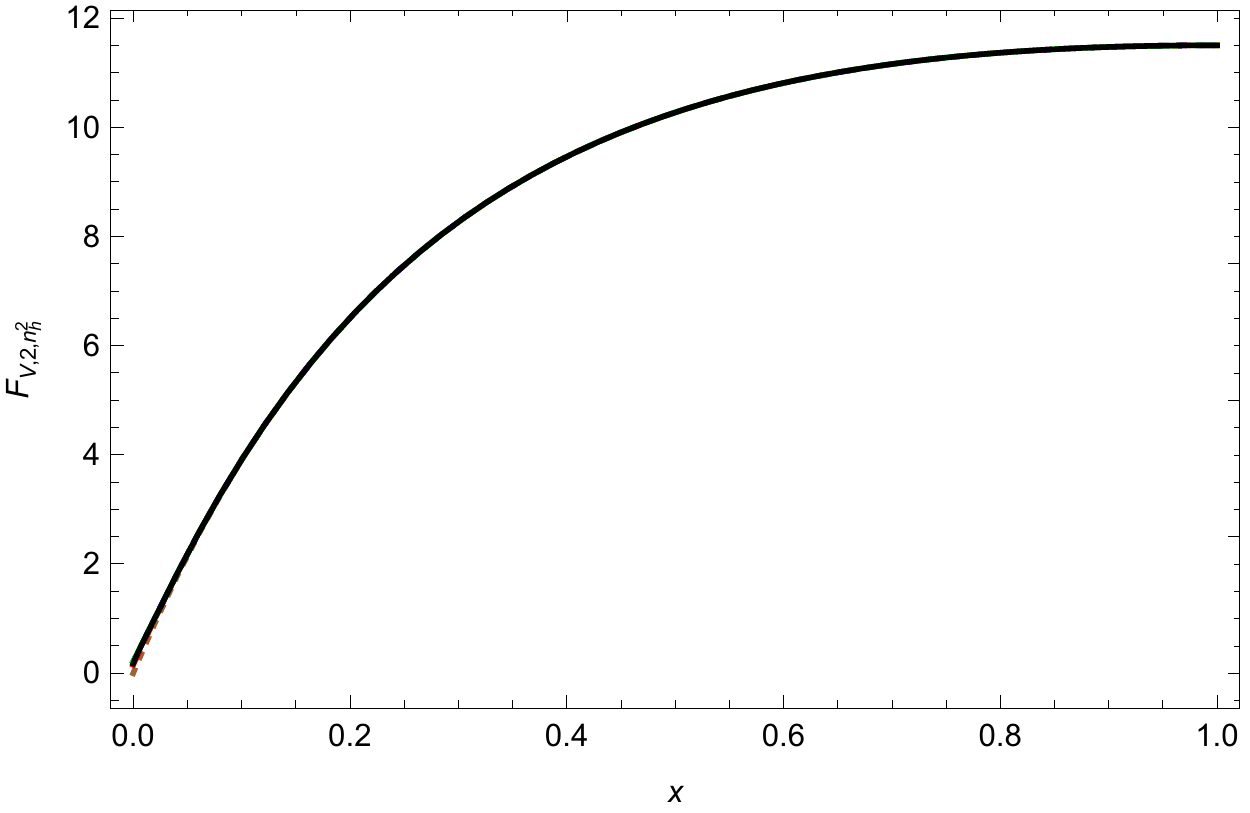}
  \caption[]{\sf \small Vector form factor $F_{V,2}$: left $\ep^0 n_h^1$, right $\ep^0 n_h^2$, the 
approximation with 20, 50, 100, 200, 500 terms is shown in brown, red, blue, green and black, respectively.}
  \label{fig:2}
\end{figure}

\begin{figure}[H]
  \centering
\includegraphics[width=.48\linewidth]{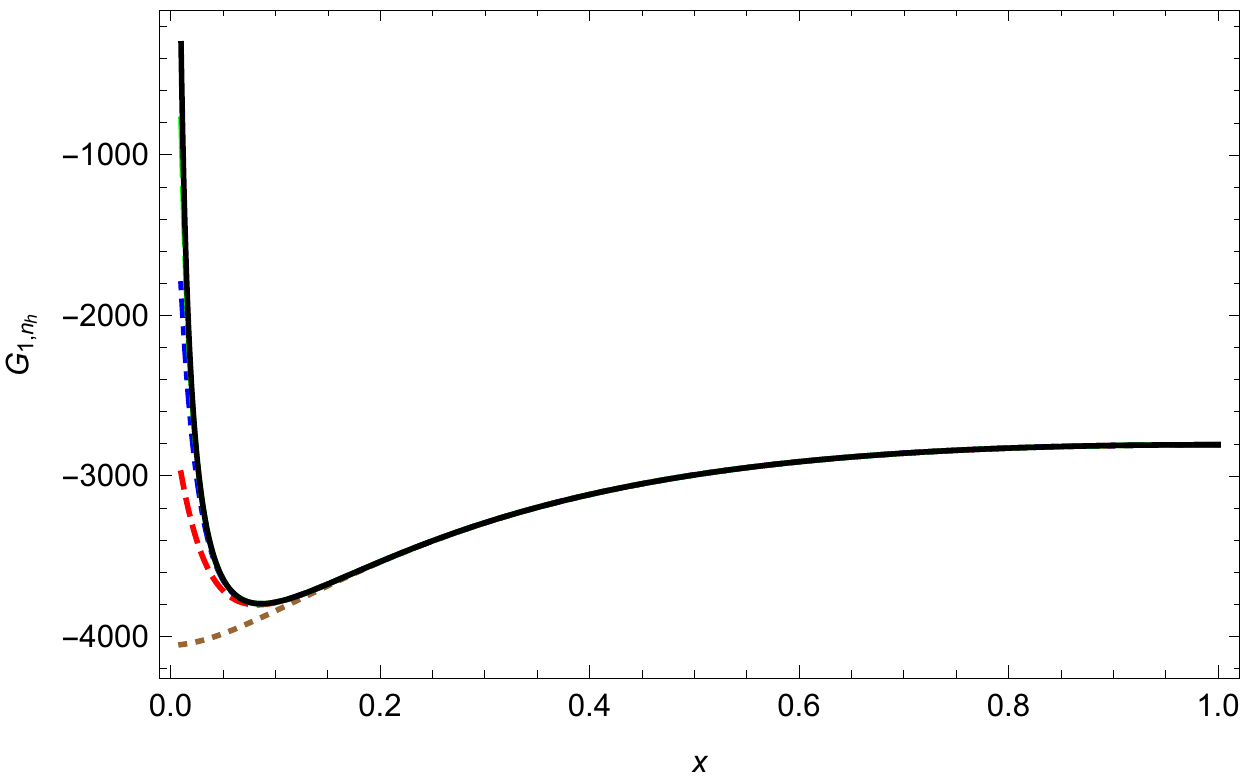}\hfill
\includegraphics[width=.48\linewidth]{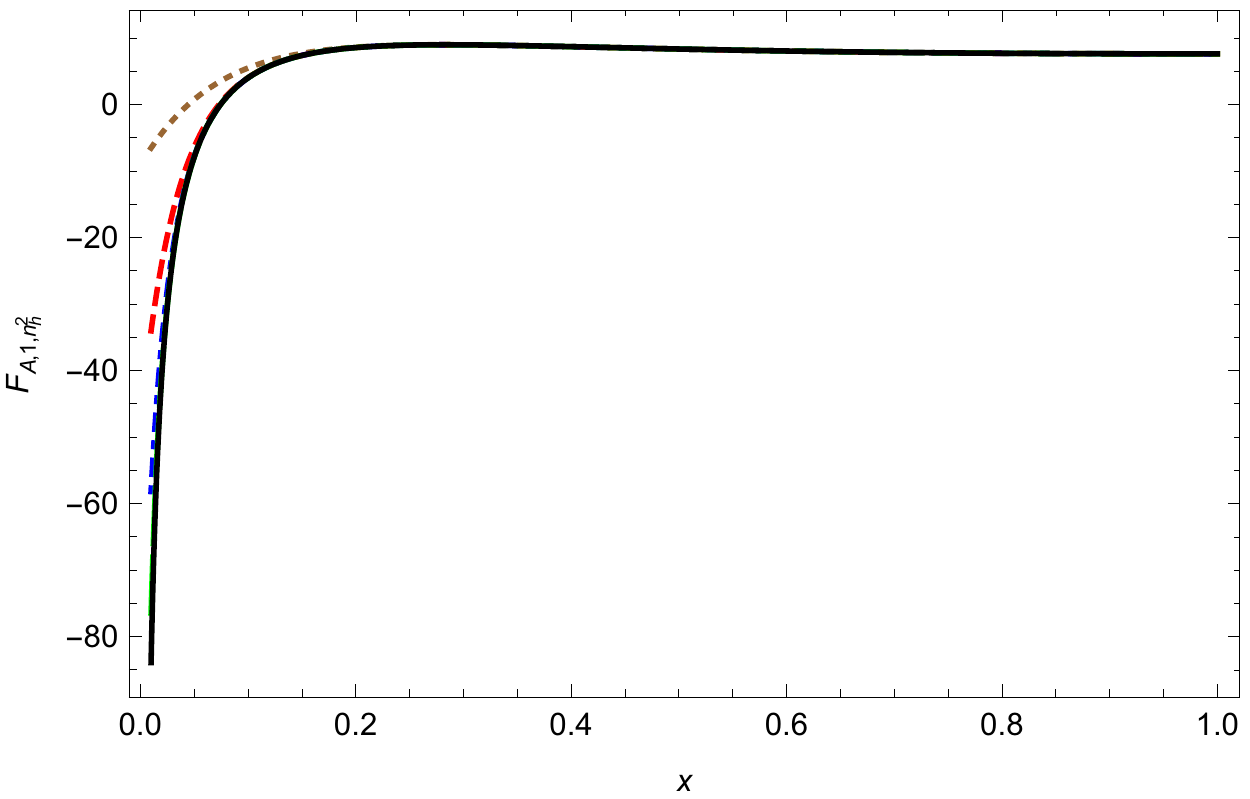}  
  \caption[]{\sf \small Axial vector form factor $F_{A,1}$: left $\ep^0 n_h^1$, right $\ep^0 n_h^2$, the 
approximation with 20, 50, 100, 200, 500 terms is shown in brown, red, blue, green and black, respectively.}
  \label{fig:3}
\end{figure}

\begin{figure}[H]
  \centering
\includegraphics[width=.48\linewidth]{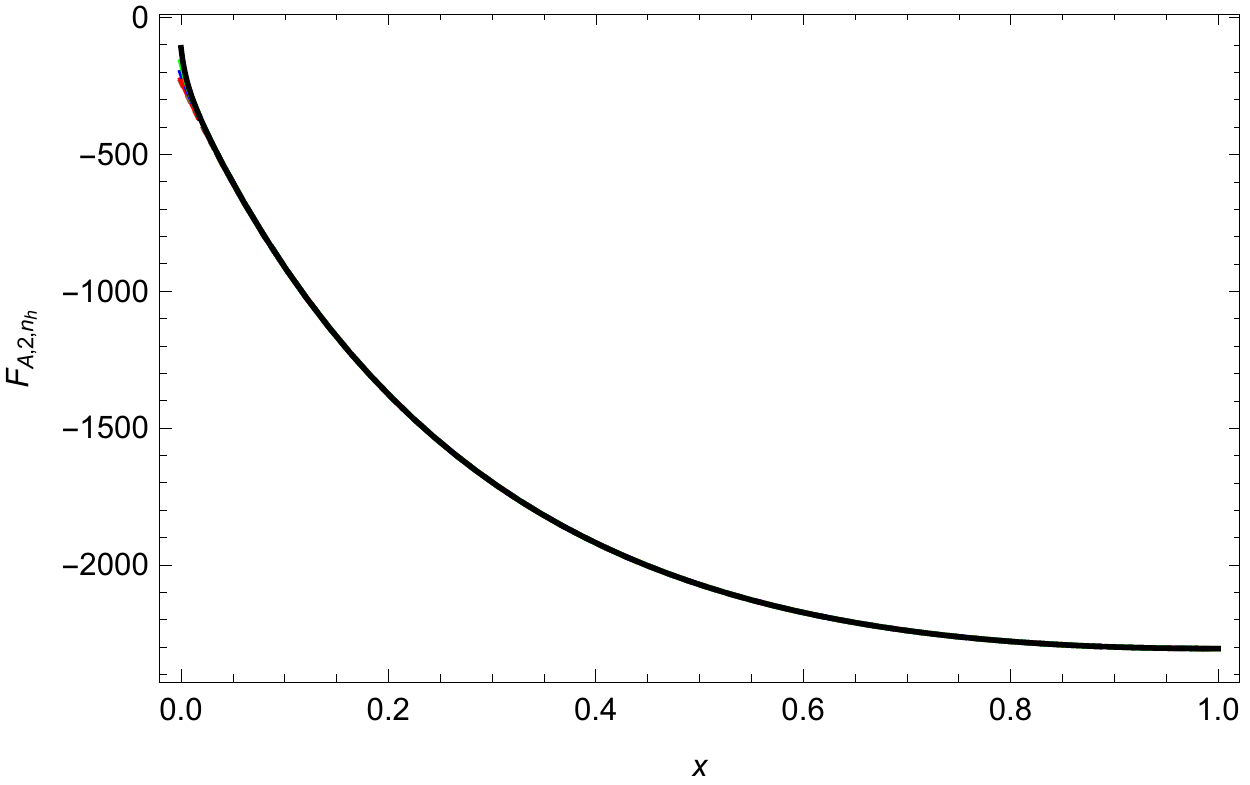}\hfill
\includegraphics[width=.48\linewidth]{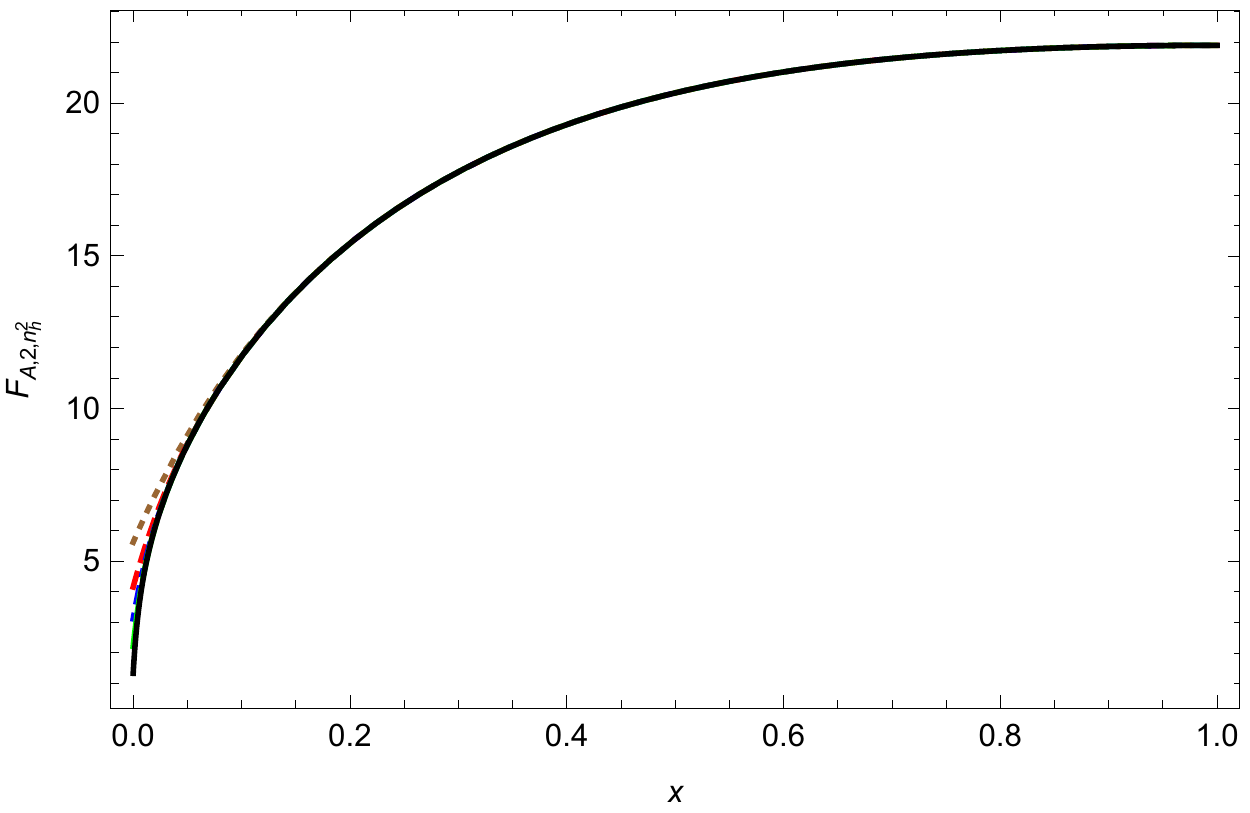}
  \caption[]{\sf \small Axial vector form factor $F_{A,2}$: left $\ep^0 n_h^1$, right $\ep^0 n_h^2$, the 
approximation with 20, 50, 100, 200, 500 terms is shown in brown, red, blue, green and black, respectively.}
  \label{fig:4}
\end{figure}

\begin{figure}[H]
  \centering
\includegraphics[width=.48\linewidth]{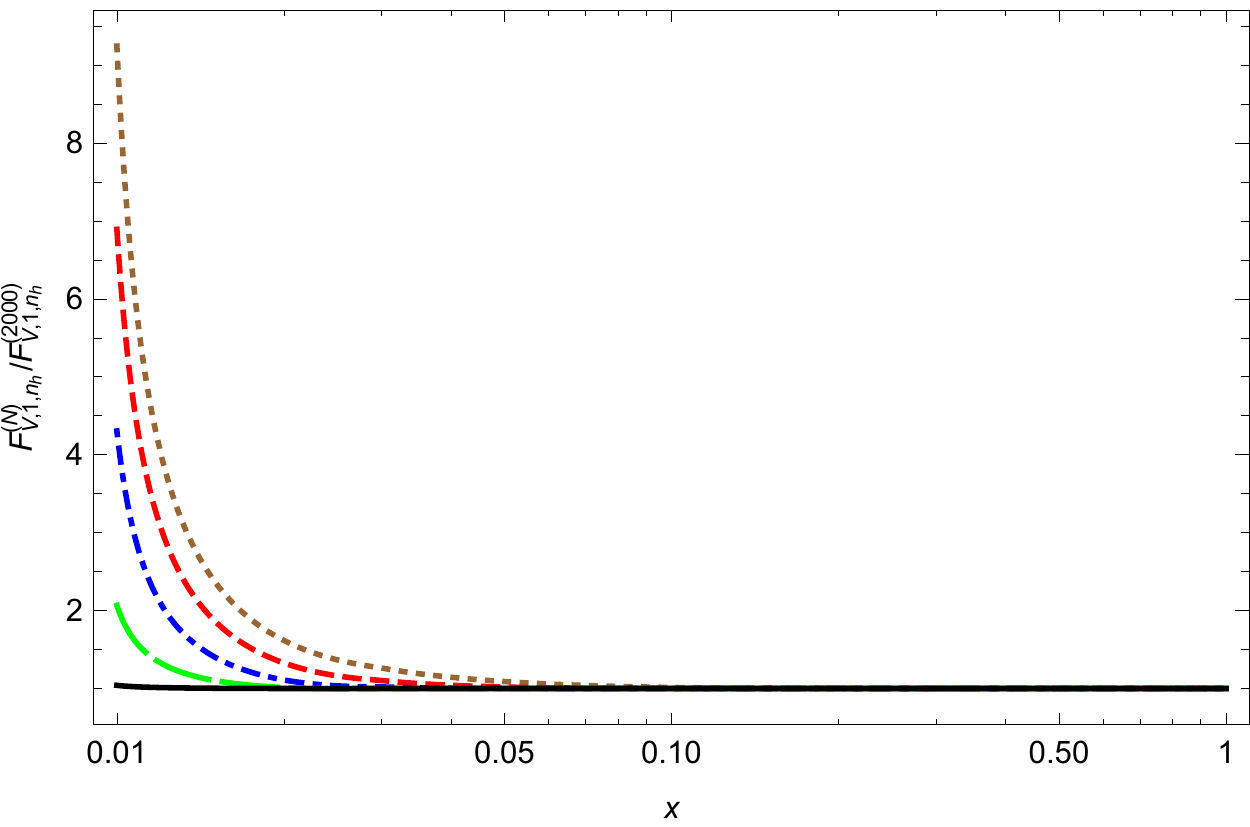}\hfill
\includegraphics[width=.48\linewidth]{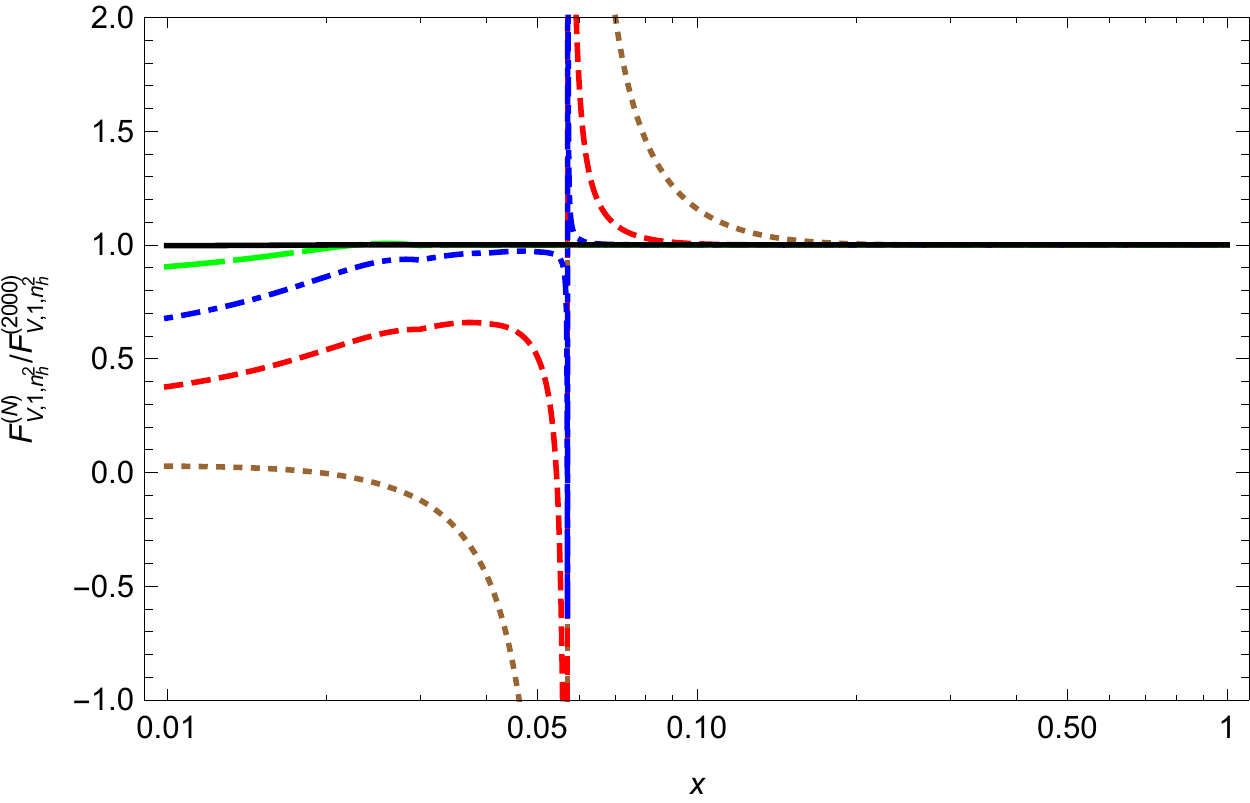}  
  \caption[]{\sf \small Ratios of the approximations with 20, 50, 100, 200, 500 terms and our best approximation 
using 2000 terms 
for the vector form factor $F_{V,1}$. Left $\ep^0 n_h^1$, right $\ep^0 n_h^2$. The ratio using 20, 50, 100, 
200, 500 terms is shown in brown, red, blue, green and black, respectively.}
  \label{fig:5}
\end{figure}

\begin{figure}[H]
  \centering
\includegraphics[width=.48\linewidth]{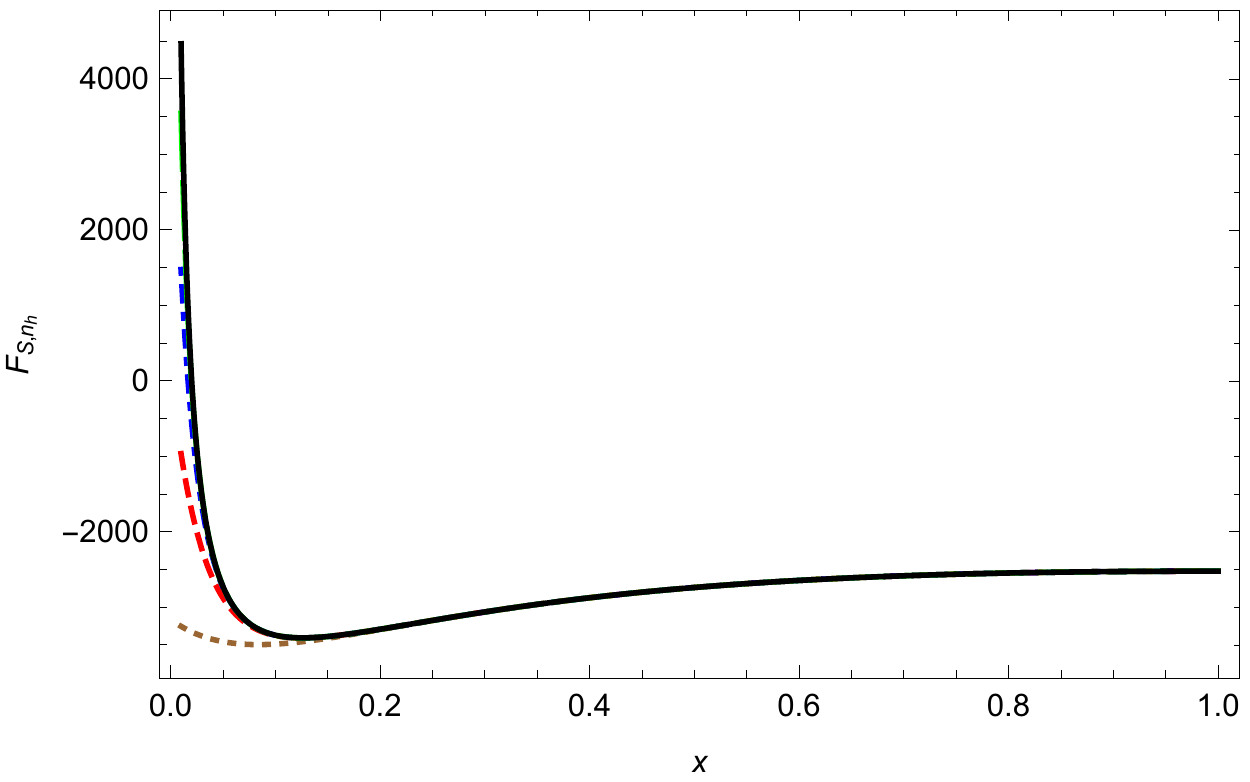}\hfill
\includegraphics[width=.48\linewidth]{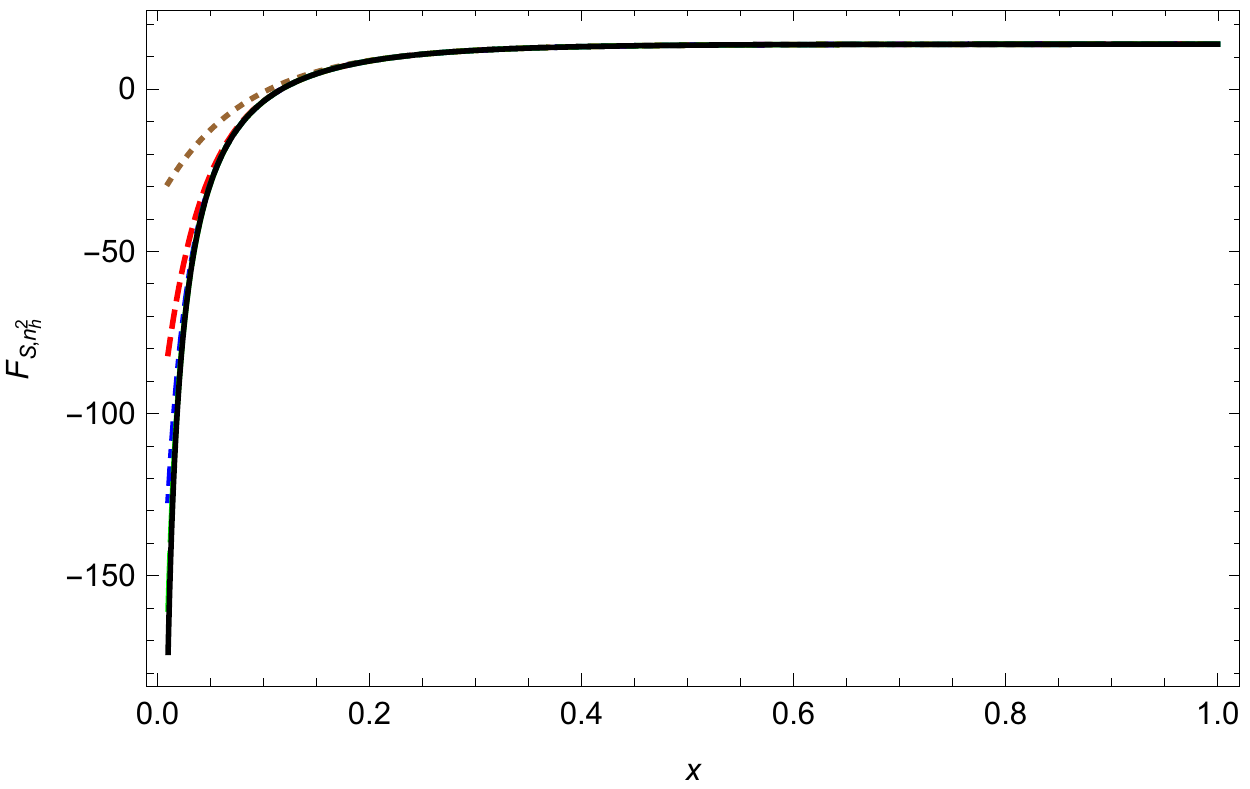}
  \caption[]{\sf small Scalar form factor $F_S$: left $\ep^0 n_h^1$, right $\ep^0 n_h^2$, the approximation 
with 20, 50, 100, 200, 500 terms is shown in brown, red, blue, green and black, respectively.}
  \label{fig:6}
\end{figure}

\begin{figure}[H]
  \centering
\includegraphics[width=.48\linewidth]{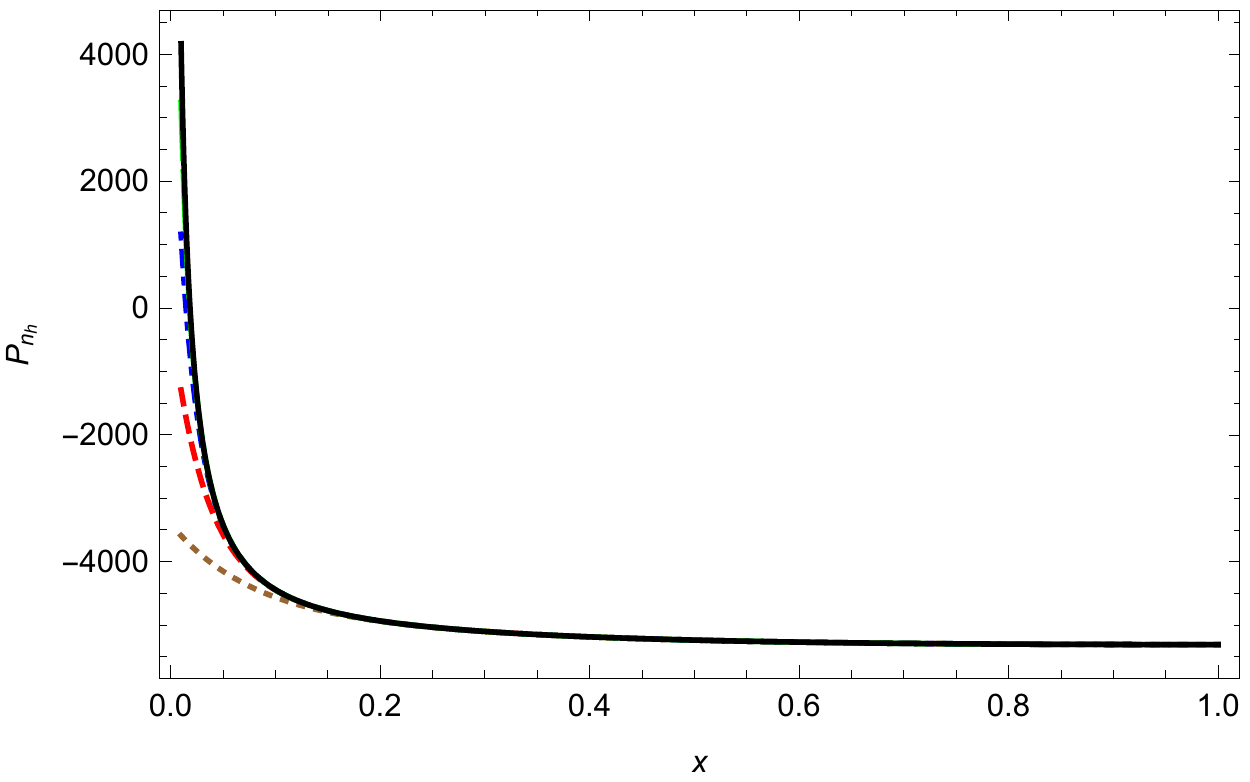}\hfill
\includegraphics[width=.48\linewidth]{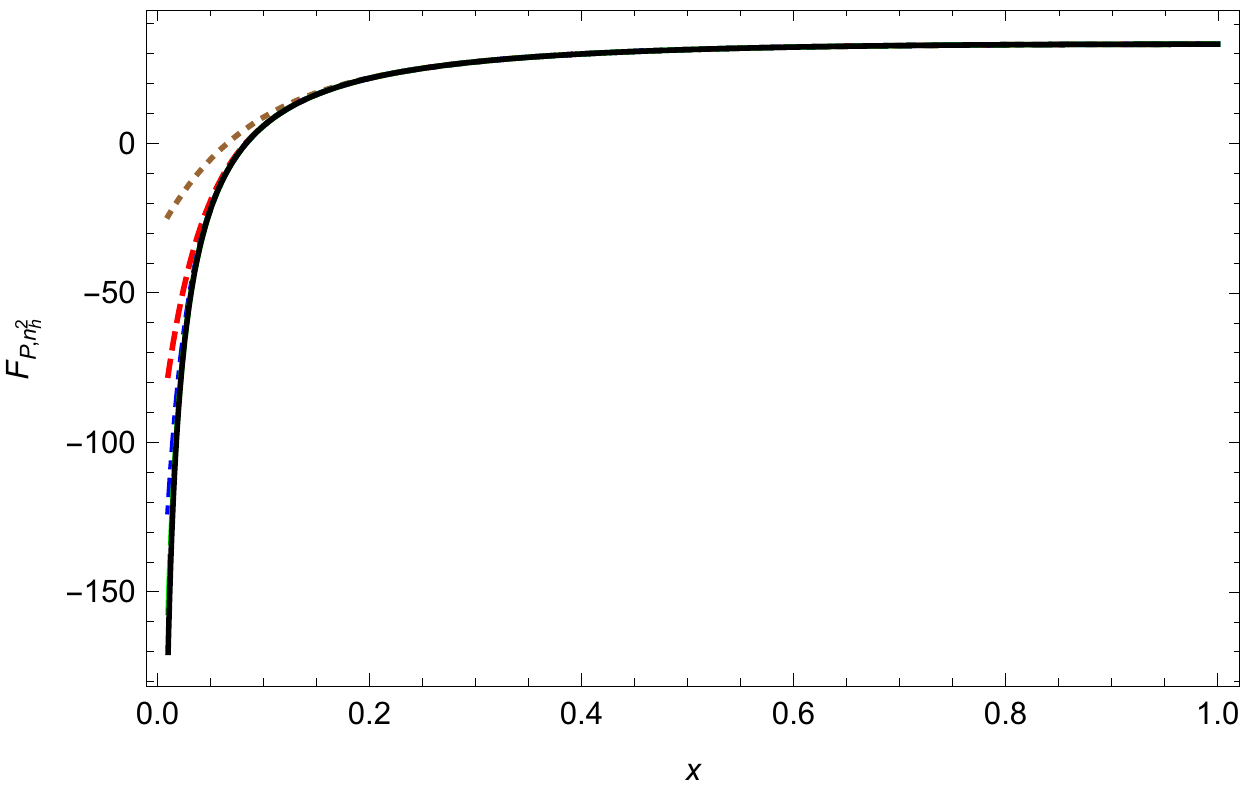}
  \caption[]{\sf \small The pseudoscalar form factor $F_P$: left $\ep^0 n_h^1$, right $\ep^0 n_h^2$, the 
approximation with 20, 50, 100, 200, 500 terms is shown in brown, red, blue, green and black, respectively.}
  \label{fig:7}
\end{figure}

\begin{figure}[H]
  \centering
\includegraphics[width=.48\linewidth]{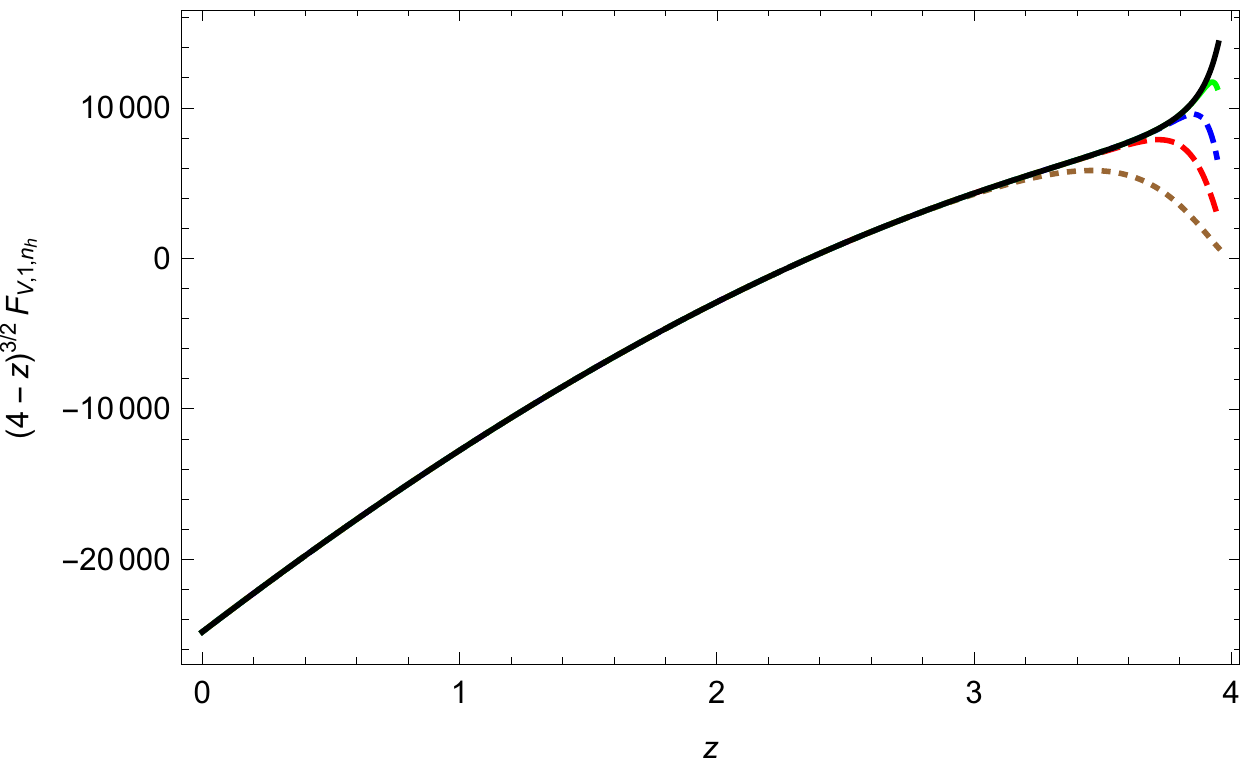}\hfill
\includegraphics[width=.48\linewidth]{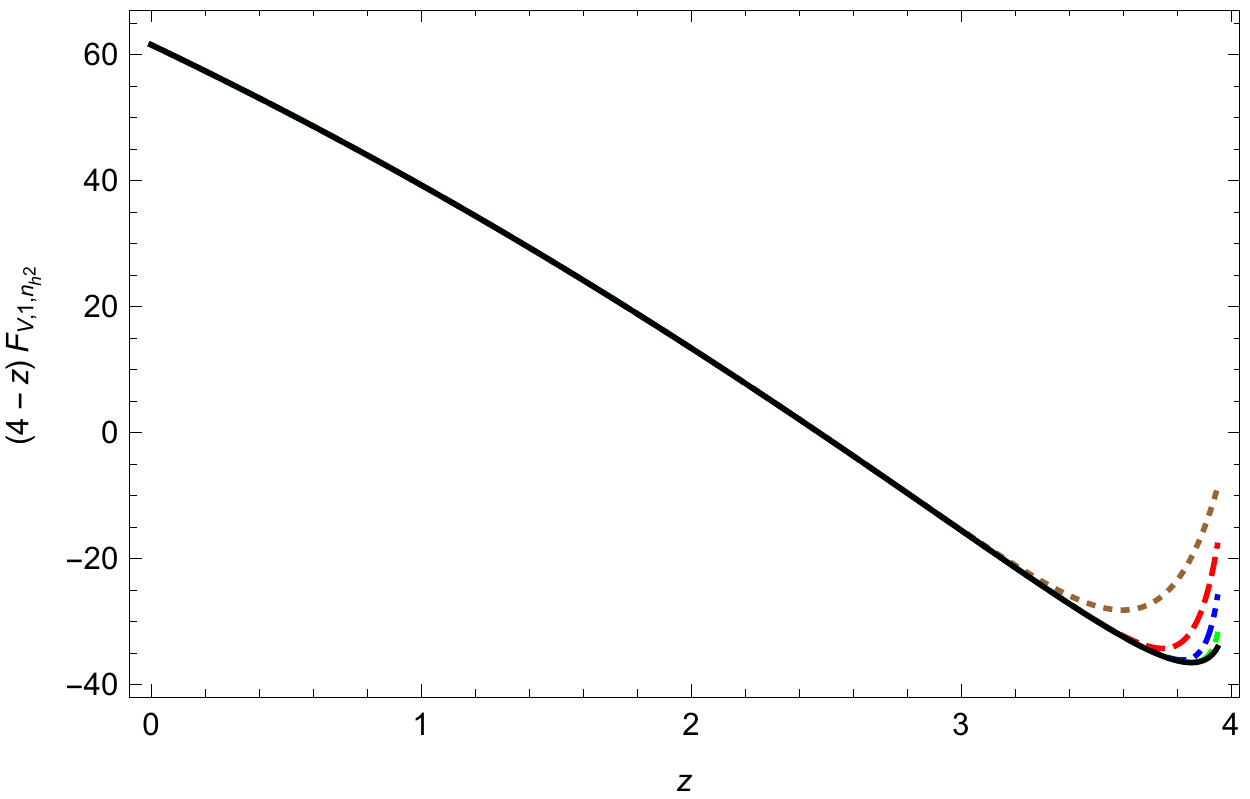}  
  \caption[]{\sf \small The vector form factor $F_{V,1}$ in the threshold region as a function of $z$: left 
$\ep^0 
n_h^1$, right $\ep^0 
n_h^2$, the 
approximation with 20, 50, 100, 200, 500 terms is shown in brown, red, blue, green and black, respectively.}
  \label{fig:8}
\end{figure}

\begin{figure}[H]
  \centering
\includegraphics[width=.48\linewidth]{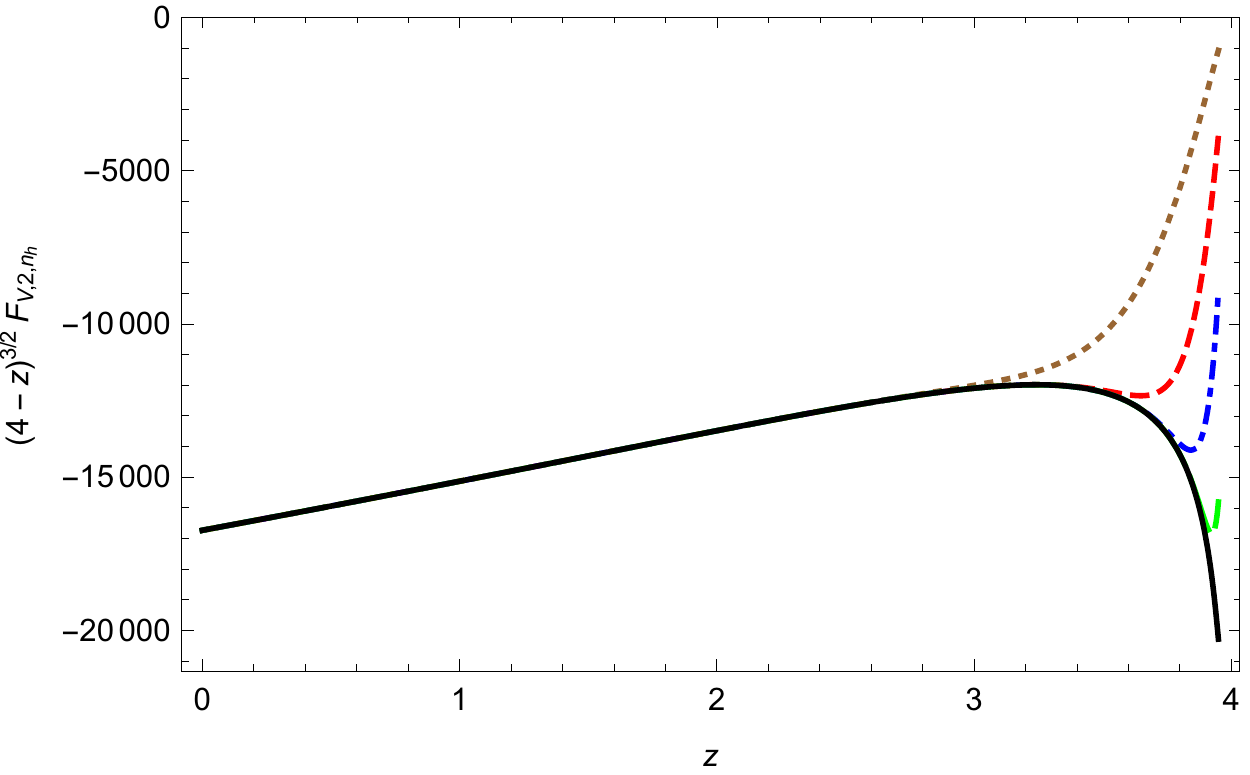}\hfill
\includegraphics[width=.48\linewidth]{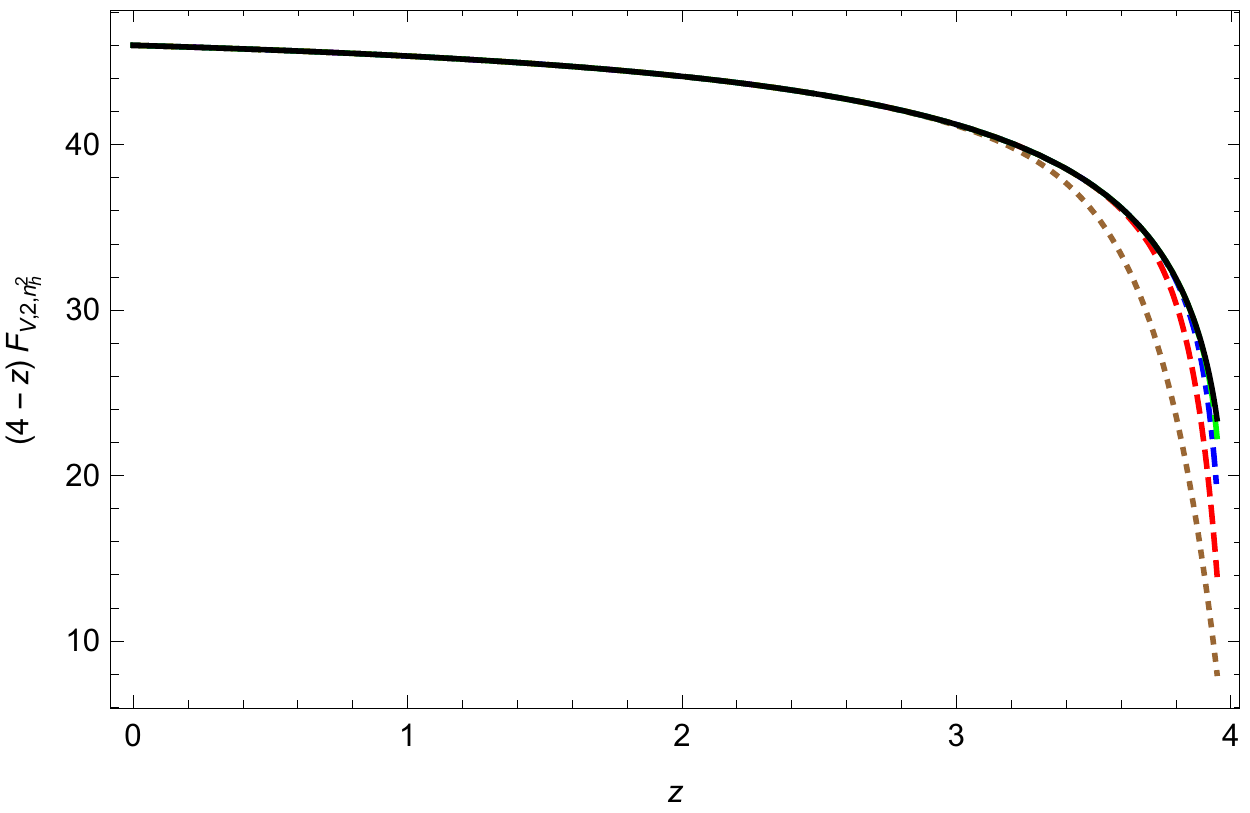}  
  \caption[]{\sf \small The vector form factor $F_{V,2}$ in the threshold region as a function of $z$: left 
$\ep^0 
n_h^1$, right $\ep^0 n_h^2$, the 
approximation with 20, 50, 100, 200, 500 terms is shown in brown, red, blue, green and black, respectively.}
  \label{fig:9}
\end{figure}

\begin{figure}[H]
  \centering
\includegraphics[width=.48\linewidth]{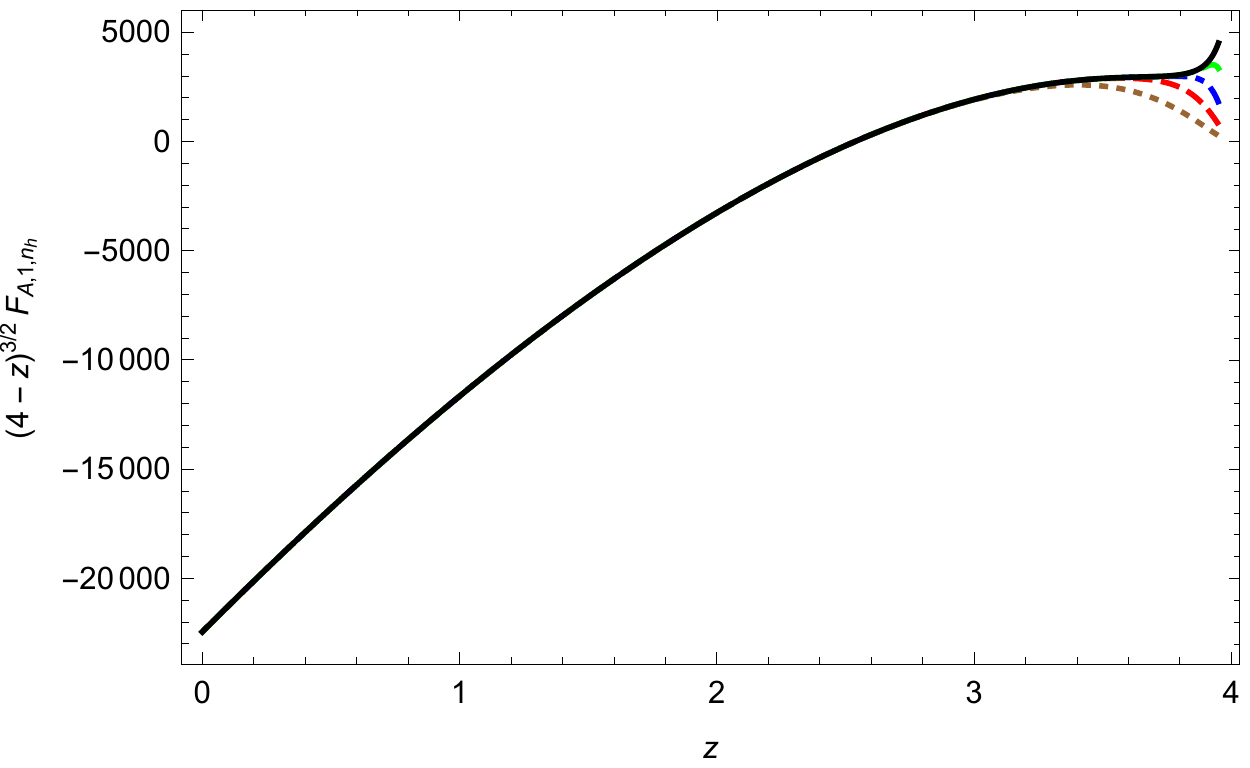}\hfill
\includegraphics[width=.48\linewidth]{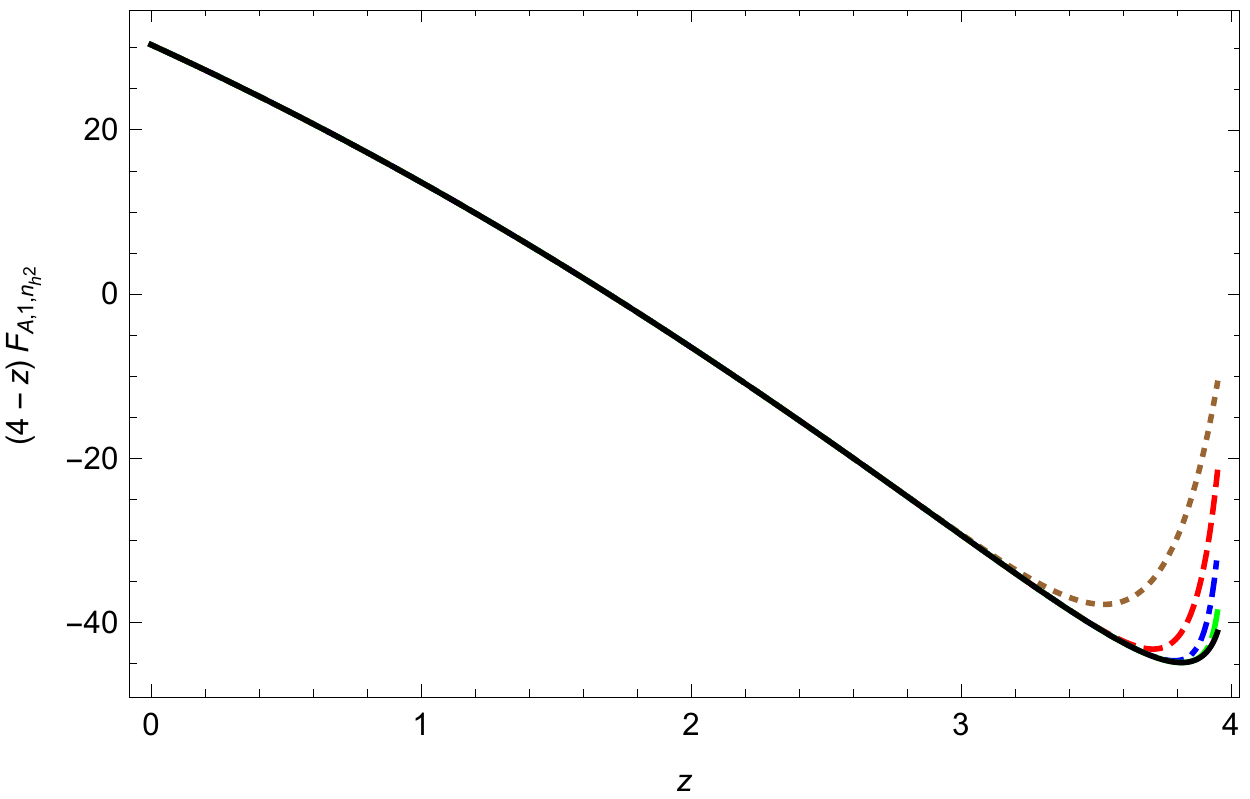}  
  \caption[]{\sf small The axialvector form factor $F_{A,1}$ in the threshold region as a function of $z$: left 
$\ep^0 n_h^1$, right $\ep^0 n_h^2$, the 
approximation with 20, 50, 100, 200, 500 terms is shown in brown, red, blue, green and black, respectively.}
  \label{fig:10}
\end{figure}

\begin{figure}[H]
  \centering
\includegraphics[width=.48\linewidth]{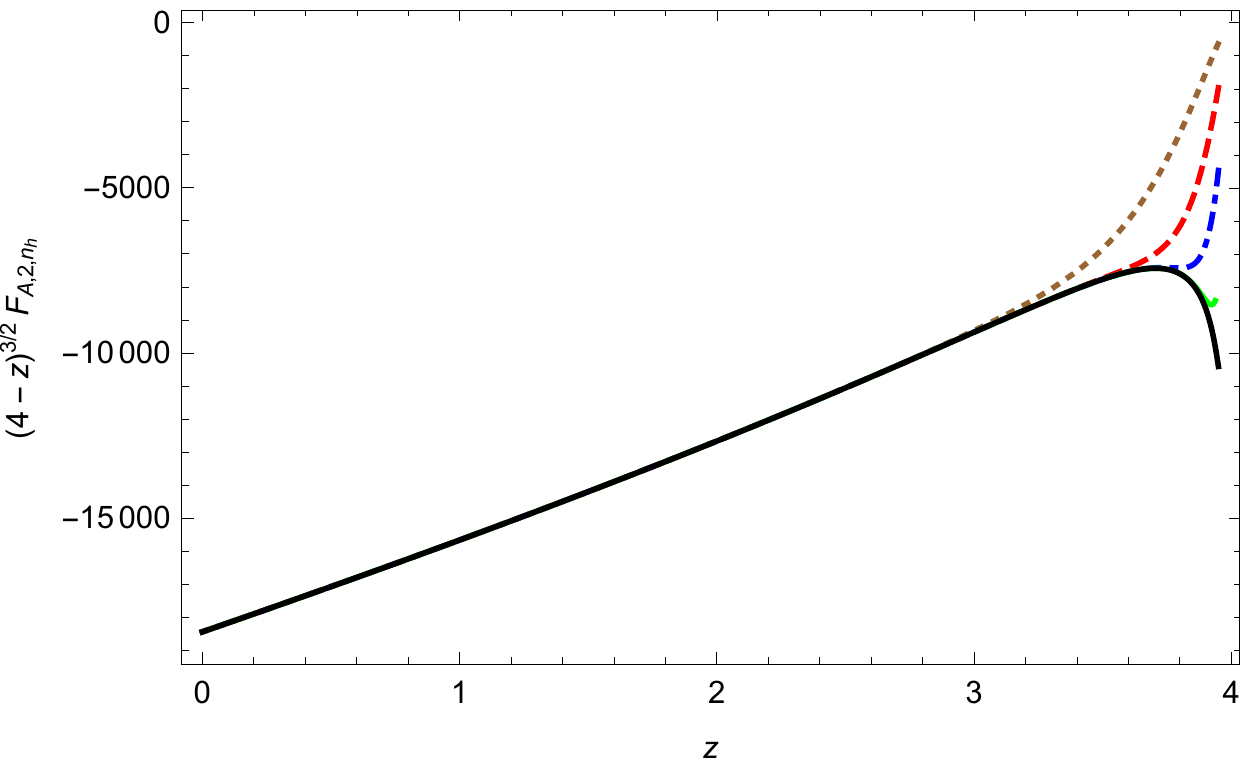}\hfill
\includegraphics[width=.48\linewidth]{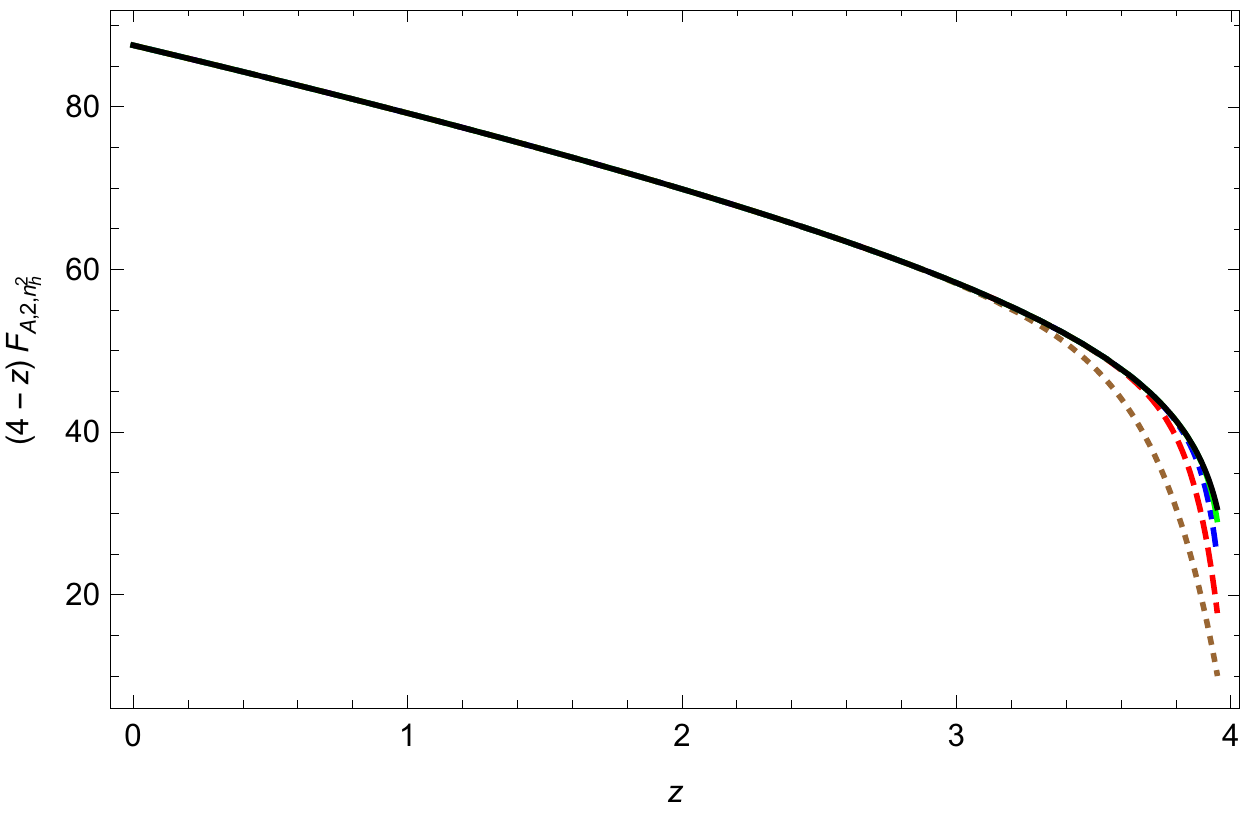}  
  \caption[]{\sf \small The axialvector form factor $F_{A,2}$ in the threshold region as a function of $z$: 
left 
$\ep^0 n_h^1$, right $\ep^0 n_h^2$, the 
approximation with 20, 50, 100, 200, 500 terms is shown in brown, red, blue, green and black, respectively.}
  \label{fig:11}
\end{figure}

\begin{figure}[H]
  \centering
\includegraphics[width=.48\linewidth]{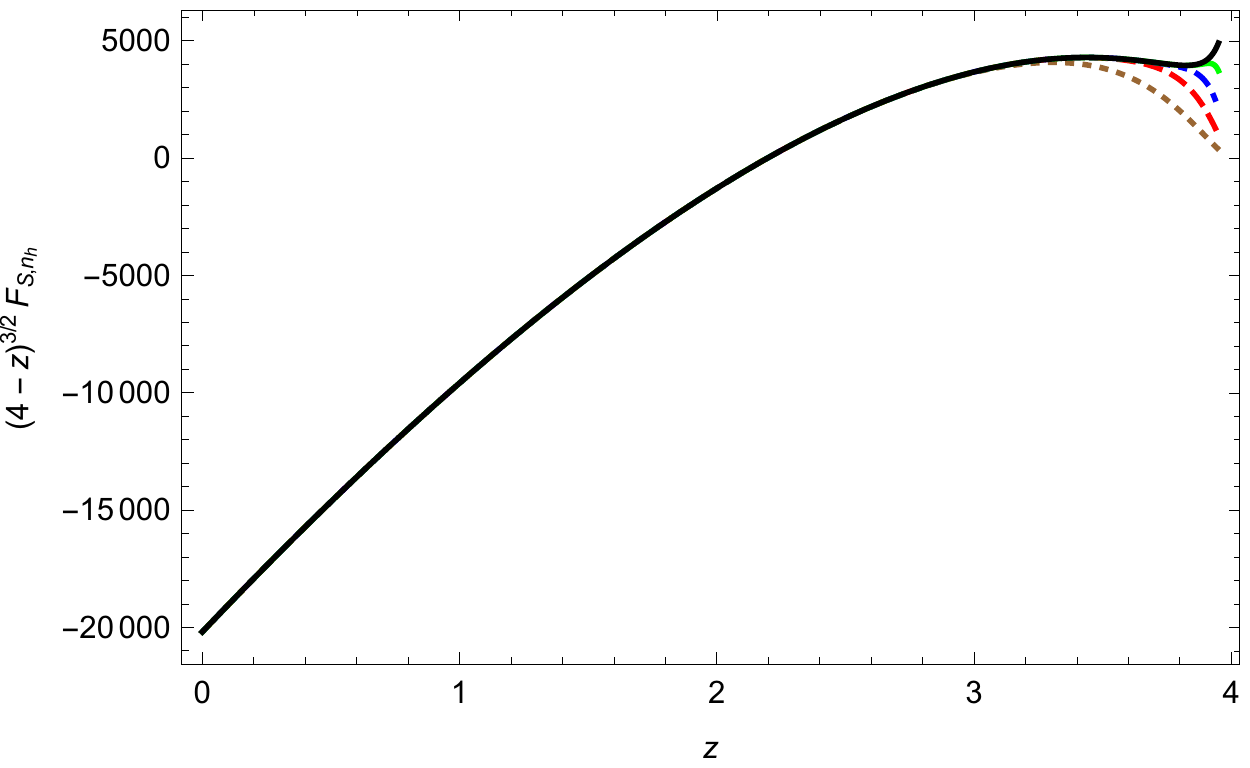}\hfill
\includegraphics[width=.48\linewidth]{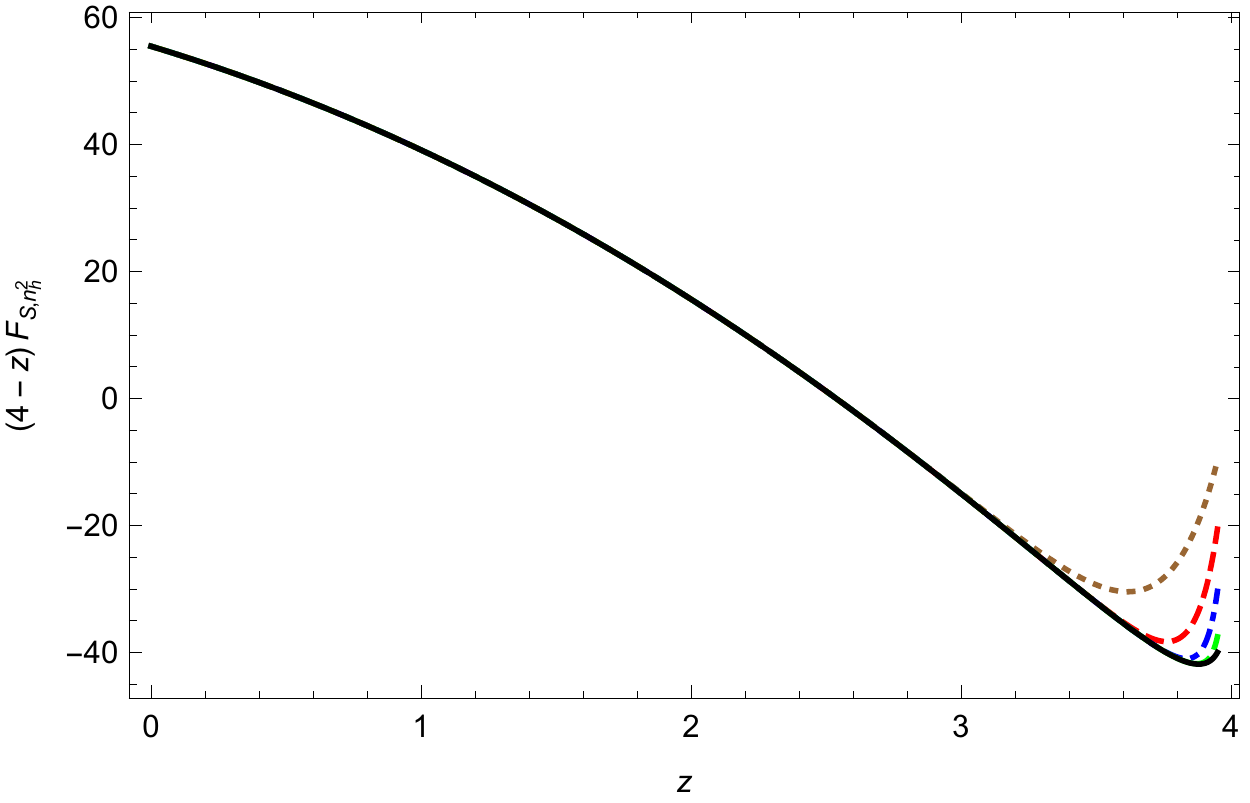}  
  \caption[]{\sf \small The scalar form factor $F_S$ in the threshold region as a function of $z$: left $\ep^0 
n_h^1$, right $\ep^0 n_h^2$, the approximation 
with 20, 50, 100, 200, 500 terms is shown in brown, red, blue, green and black, respectively.}
  \label{fig:12}
\end{figure}

\begin{figure}[H]
  \centering
\includegraphics[width=.48\linewidth]{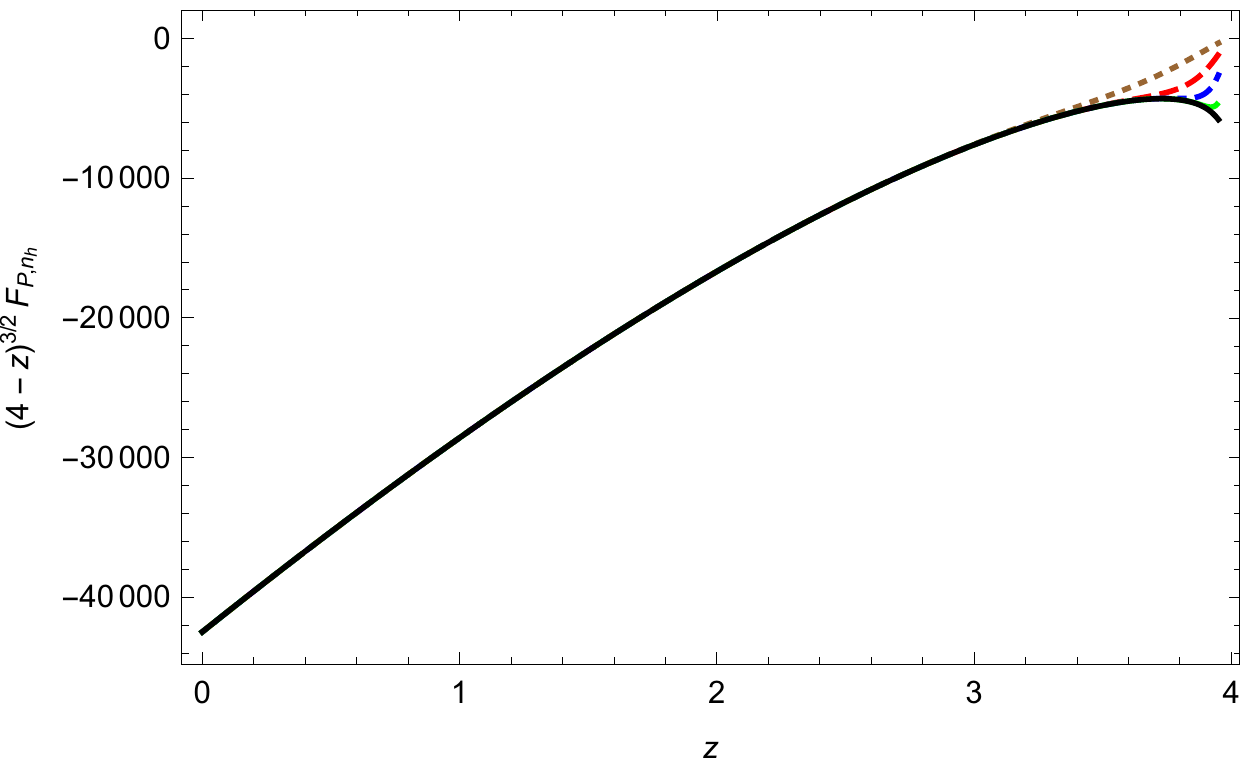}\hfill
\includegraphics[width=.48\linewidth]{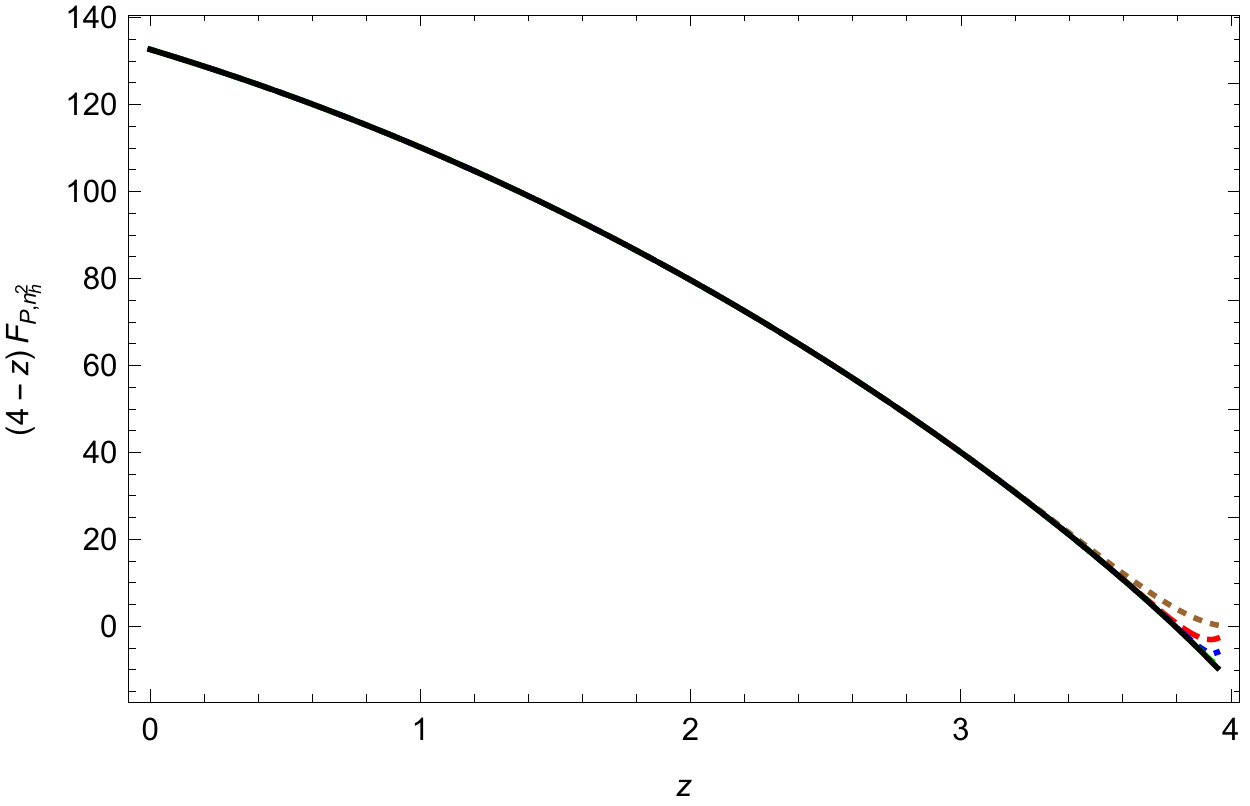}  
  \caption[]{\sf \small The pseudoscalar form factor $F_P$ in the threshold region as a function of $z$: left 
$\ep^0 n_h^1$, right $\ep^0 n_h^2$, the 
approximation with 20, 50, 100, 200, 500 terms is shown in brown, red, blue, green and black, respectively.}
  \label{fig:13}
\end{figure}

\vspace*{1mm}   
\noindent
We present now numerical results for the $\ep^0$ parts of the different unrenormalized form factors.
For comparison the functions $F_{C,i}^{(0)}, i = 1...3$ are shown
using their first 20, 50, 100. 200 and 500 
expansion coefficients. In
Figures~\ref{fig:1}--\ref{fig:7} we show the results in the
Euclidean region $0<x<1$. In Figures~\ref{fig:8}--\ref{fig:13} the results
in the region below threshold ($0 < z < 4$) are shown, expanding around $z=0$.  At small values of $x$ 
the form factor $F_1$ has 
logarithmic singularities both in its $n_h$ (Figure~\ref{fig:1}, left panel) and $n_h^2$ parts 
(Figure~\ref{fig:1}, right panel). Here and in the following we also illustrate taking into account a 
rising 
number of terms $n = 20$ to 500 from the non--first order factorizing contributions to illustrate the degree 
of convergence.  

The vector form factor $F_2$, cf.~Figure~\ref{fig:2} is proportional to $x$, damping out further 
$\ln^k(x)$ contributions. Despite taking only 500 expansion coefficients, one obtains the correct 
representation in the whole $x$ range.
In Figure~\ref{fig:3} we show the behaviour of the axialvector form factor $F_{A,1}$ under the same 
conditions as in Figure~\ref{fig:1}, and for the form factor $F_{A,2}$ in Figure~\ref{fig:4} similar to 
those in Figure~\ref{fig:2}.
In Figure~\ref{fig:5} we illustrate for the ratio of the vector form factor $F_1$ evaluated of $n$ terms 
of the non--first order factorizing contributions for $n = 20, 50, 100, 200$ and $500$ to the case of $n = 
2000$ to see the relative convergence both for the $n_h^2$ and $n_h$ contribution, which approves towards $x 
\rightarrow 0$. However, the complete logarithmic behaviour cannot be resembled by this representation.
\noindent
The contributions to the scalar form Factor $F_S$ are shown in Figure~\ref{fig:6}. Its $x$--dependence 
is 
similar to the one of the vector form factor.
The pseudoscalar form factor is illustrated in Figure~\ref{fig:7}. It has similar behaviour as the 
axialvector form factor.

Now we turn to the illustration of the threshold expansion of the form factors. All form factors are multiplied 
by the factor $(4-z)^{3/2}$ for the $n_h$ terms and by $(4-z)$ for the $n_h^2$ terms for convenience, 
with $z = q^2/m^2$ for the vector- (Figure~\ref{fig:7}, 
\ref{fig:8}), axialvector- (Figures~\ref{fig:9}, \ref{fig:10}), scalar- 
(Figure~\ref{fig:11}) and pseudoscalar form factor (Figure~\ref{fig:12}).
In large $z$  region differences due to the number 
of terms used in the expansion of the non--first order factorizing contributions are seen.
\section{Conclusions}
\label{sec:8}

\vspace*{1mm}
\noindent
We have calculated the $n_h$ contributions to the massive three--loop form factors for vector-, 
axialvector-, scalar- and pseudoscalar currents. The contributing Feynman integrals were reduced to master 
integrals using the package {\tt Crusher} \cite{CRUSHER}. The calculation of the analytic expansion 
coefficients
of the master integrals in $x$ required a new way of decoupling of the differential equations provided 
by the IBP--relations w.r.t. their optimal expansion in the dimensional parameter $\ep$. Otherwise, 
it would have been very demanding to provide the required initial values. Here we used the method of 
arbitrary high moments \cite{Blumlein:2017dxp}, partly requiring to calculate 8000 moments. The recursions were derived using 
the guessing method \cite{GSAGE}. The recursions for the pole terms and a part of the contributions of 
$O(\ep^0)$ are first order factorizing and one obtains representations in terms of harmonic polylogarithms
in the variable $x$. The recursions were solved using the packages {\tt Sigma} \cite{Schneider:2007a,Schneider:2013a} and {\tt EvaluateMultisums} and {\tt SumProduction} 
\cite{Blumlein:2012hg,Schneider:2013zna}. The resulting infinite sums were then converted into harmonic 
polylogarithms using the package {\tt HarmonicSums} \cite{Vermaseren:1998uu,Blumlein:1998if,
Ablinger:2014rba, Ablinger:2010kw, Ablinger:2013hcp, Ablinger:2011te, Ablinger:2013cf, Ablinger:2014bra,
Ablinger:2017Mellin}. For some color-$\zeta$ contributions at $O(\ep^0)$ non--first order factorizing 
parts are contained in the recurrences. Here the largest of the remaining recurrences are of order {\sf o = 
15}, resulting from of original recurrences of order up to {\sf o = 55} and degree {\sf d = 1324}. For those we 
have have obtained analytic polynomial expansions 
in $x$ of degree {\sf d = 2000}, which can be extended in case needed. They already allow for precise
numerical representations in a wide range of $x$. However, in the case of logarithmic divergences for
$x \rightarrow 0$ these representations diverge. The analytic solution of the non--first order 
factorizing recurrences in terms of special higher functions still needs to be performed in the future.
We also considered the leading color contributions for the scalar form factor as an example to see whether
here simplifications can be obtained. This is indeed the case since the most involved of the remaining 
non--first order factorizing recurrences is only of order {\sf o = 5}, stemming from an original recurrence of 
of order {\sf o = 46} and degree {\sf d = 901}. Complete first order factorization can, however, not be 
obtained in this case.

\appendix
\section{Initial Values to High Order in \boldmath $\ep$}
\label{sec:A}

\vspace*{1mm}
\noindent
Boundary values for the master integrals appearing in the present calculation can be obtained using 
three loop propagator integrals as have been dealt with in Ref.~\cite{Melnikov:2000zc} to orders in the
dimensional parameter up to $O(\ep^3)$ and lower. Depending on the method of uncoupling of the associated
differential equations terms up to $O(\ep^9)$, i.e. in total 13 orders in $\ep$, may be necessary. 
For some of the integrals all-order in $\ep$ representations exist, 
\cite{Davydychev:2003mv,
Blumlein:2009cf,
Broadhurst:1991fy,
Broadhurst:1991fi,
Broadhurst:1993mw}. 
For 
the integrals $I_9, I_{11}, I_{12}, I_{15}$ and $I_{16}$ of \cite{Melnikov:2000zc,Broadhurst:1991fi,Broadhurst:1993mw} we have derived the corresponding 
representations. They are
thoroughly given in terms of multiple zeta values, cf.~\cite{Blumlein:2009cf}. The calculation of $\ep$-expansions 
at such depth is very time consuming, despite the fact that the techniques to be used are rather standard ones
available in the packages {\tt Sigma} \cite{Schneider:2007a,Schneider:2013a},
{\tt EvaluateMultisums} and {\tt SumProduction} \cite{Blumlein:2012hg,Schneider:2013zna} together
with the asymptotic expansions of harmonic sums of the package {\tt HarmonicSums}.
In the final results all constants of higher transcendentality will be cancelled. Given the associated calculational 
effort, it is of advantage having decoupling methods, in which only initial values
requiring an expansion in $\ep$ to a far lower order in $\ep$ is needed. This has been possible by the algorithm 
described in Section~\ref{sec:5}.

It is interesting to see that all constants spanning the MZVs \cite{Blumlein:2009cf} are 
contributing to this integral.
\begin{table}[H]\centering
\begin{tabular}{|r|r|}
\hline
{\sf weight} & {\sf \# basis elements} 
\\
\hline
1  &  1 \\
2  &  1 \\
3  &  1 \\
4  &  1 \\
5  &  2 \\
6  &  2 \\
7  &  3 \\
8  &  5 \\
9  &  8 \\
10 & 11 \\
11 & 18 \\
\hline
\end{tabular}
\caption[]{\sf \small Number of basis elements by weight.}
\end{table}

These are (together with their approximate numerical values)
\begin{eqnarray}
\ln(2)                             &\approx& ~~0.69314718055994530942
\\
\zeta_2                            &\approx& ~~1.64493406684822643647
\\
\zeta_3                            &\approx& ~~1.20205690315959428540
\\
\Li_4\left(\frac{1}{2}\right)      &\approx& ~~0.51747906167389938633
\\
\zeta_5                            &\approx& ~~1.03692775514336992633
\\
\Li_5\left(\frac{1}{2}\right)      &\approx& ~~0.50840057924226870746
\\
s_{6} = S_{-5,-1}(\infty)            &\approx& ~~0.98744142640329971377
\\
\Li_6\left(\frac{1}{2}\right)      &\approx& ~~0.50409539780398855069
\\
s_{7a} = S_{-5,1,1}(\infty)        &\approx&  -0.95296007575629860341
\\
s_{7b} = S_{5,-1,-1}(\infty)       &\approx& ~~1.02912126296432453422
\\
\zeta_7                            &\approx& ~~1.00834927738192282684
\\
\Li_7\left(\frac{1}{2}\right)      &\approx& ~~0.50201456332470849457
\\
s_{8a} = S_{5,3}(\infty)           &\approx& ~~1.04178502918279188339
\\
s_{8b} = S_{-7,-1}(\infty)         &\approx& ~~0.99644774839783766598
\\
s_{8c} = S_{-5,-1,-1,-1}(\infty)   &\approx& ~~0.98396667382173367092
\\
s_{8d} = S_{-5,-1,1,1}(\infty)     &\approx& ~~0.99996261346268344770
\\
\Li_8\left(\frac{1}{2}\right)      &\approx& ~~0.50099665909705191056
\\
\zeta_9                            &\approx& ~~1.00200839282608221442
\\
\Li_9\left(\frac{1}{2}\right)      &\approx& ~~0.50049488810595361004
\\
s_{9a} = S_{7,-1,-1}(\infty)       &\approx& ~~1.00640196269235635901
\\
s_{9b} = S_{-7,-1,1}(\infty)       &\approx& ~~0.99842952512288855440
\\
s_{9c}  = S_{-6,-2,-1}(\infty)     &\approx&  -0.98747515763691535588
\\
s_{9d}  = S_{-5,-1,1,1,1}(\infty)  &\approx& ~~1.00219817413397743630
\\
s_{9e}  = S_{-5,-1,-1,-1,1}(\infty) &\approx& ~~0.98591171955244547262
\\
s_{9f}  = S_{-5,-1,-1,1,-1}(\infty) &\approx& ~~0.97848117128116624248
\\
\\
\Li_{10}\left(\frac{1}{2}\right) &\approx& ~~0.50024632060600677501
\\
h_{313111}                       &\approx& ~~0.00000059060152818192
\\
h_{331111}                       &\approx& ~~0.00000040503762995428
\\
h_{3331}                         &\approx& ~~0.00003761590257651408
\\
h_{511111}                       &\approx& ~~0.00000030113797559813
\\
h_{5113}                         &\approx& ~~0.00008812461415599555
\\
h_{5131}                         &\approx& ~~0.00002330610214579640
\\
h_{5311}                         &\approx& ~~0.00001097851878750415
\\
h_{7111}                         &\approx& ~~0.00000654709718938270
\\
h_{73}                           &\approx& ~~0.00735713658369574885
\\
h_{91}                           &\approx& ~~0.00188302419181393899
\\
\zeta_{11}                         &\approx& ~~1.00049418860411946456
\\
\Li_{11}\left(\frac{1}{2}\right)   &\approx& ~~0.50012279152986795519
\\
h_{3113111}                        &\approx& ~~0.00000004259335538990
\\
h_{3131111}                        &\approx& ~~0.00000002903877500717
\\
h_{3311111}                        &\approx& ~~0.00000002146685479915
\\
h_{33131}                          &\approx& ~~0.00000203422625738433
\\
h_{33311}                          &\approx& ~~0.00000095250833508071
\\
h_{5111111}                        &\approx& ~~0.00000001677568330189
\\
h_{51113}                          &\approx& ~~0.00000535991478634246
\\
h_{51131}                          &\approx& ~~0.00000141456111518871
\\
h_{51311}                          &\approx& ~~0.00000066489238255553
\\
h_{53111}                          &\approx& ~~0.00000039564332931785
\\
h_{533}                            &\approx& ~~0.00039368649887395471
\\
h_{551}                            &\approx& ~~0.00010106766377056452
\\
h_{71111}                          &\approx& ~~0.00000026780063665860
\\
h_{731}                            &\approx& ~~0.00004821281485407120
\\
h_{713}                            &\approx& ~~0.00018649660146306540
\\
h_{911}                            &\approx& ~~0.00002219036425770103,
\end{eqnarray}
with
\begin{eqnarray}
h_{abcd} \equiv H_{-a,-b,-c,-d}(1)
\end{eqnarray}
in the (collected) notation as e.g. $H_{-3,-1}(1) \equiv H_{0,0,-1,-1,}(1)$.
Up to weight {\sf w = 9} we follow the representation of {\tt summer.h}. Earlier
a similar representation has been available for the weights {\sf w = 10--12}. These files
do not exist anymore \cite{VERM18}. Therefore we change to the basis which is used in the package 
{\tt HarmonicSums.m}, where also the the respective constant files are now available\footnote{We thank 
J.~Ablinger for making these files available.}
The MZV data mine used another basis for the HPLs at argument one. This basis is of course equivalent.
We used the values $\zeta_{2k+1}$ and $\Li_{k}\left(\frac{1}{2}\right)$ as basis elements, through which one 
of the high weight HPLs is replaced.
\begin{eqnarray}
\Li_{10}\left(\frac{1}{2}\right) &=&
\frac{1}{2} h_{31111111}
-\frac{1}{4} h_{511111}
+\frac{1}{2} h_{7111}
-\frac{17}{8} h_{91}
-\Li_{9}\left(\frac{1}{2}\right) \ln(2)
-\frac{1}{2}\Li_{8}\left(\frac{1}{2}\right) \ln^2(2)
\nonumber\\ &&
-\frac{1}{6}\Li_{7}\left(\frac{1}{2}\right) \ln^3(2)
-\frac{1}{24}\Li_{6}\left(\frac{1}{2}\right) \ln^4(2)
-\frac{1}{120}\Li_{5}\left(\frac{1}{2}\right) \ln^5(2)
-\frac{1}{120}\Li_{4}\left(\frac{1}{2}\right) \ln^6(2)
\nonumber\\ &&
-\frac{1}{43200} \ln^{10}(2)
+\frac{1}{11520} \ln^8(2) \zeta_2
-\frac{789}{8800} \zeta_2^5
-\frac{1}{5760} \ln^7(2) \zeta_3
+\frac{27}{160} \zeta_2^2 \zeta_3^2
+\frac{3}{8} \zeta_2 \zeta_3 \zeta_5
\nonumber\\ &&
+\frac{3}{16} \zeta_5^2
+\frac{3}{8} \zeta_3 \zeta_7
\\
\Li_{11}\left(\frac{1}{2}\right) &=& 
\frac{1}{2} h_{311111111}
-\frac{1}{4} h_{5111111}
+\frac{1}{2} h_{71111}
-\frac{17}{8} h_{911}
-\Li_{1}\left(\frac{1}{2}\right) \ln(2)
-\frac{1}{2} \Li_{9}\left(\frac{1}{2}\right) \ln^2(2)
\nonumber\\ &&
-\frac{1}{6} \Li_{8}\left(\frac{1}{2}\right) \ln^3(2)
-\frac{1}{24} \Li_{7}\left(\frac{1}{2}\right) \ln^4(2)
-\frac{1}{120} \Li_{6}\left(\frac{1}{2}\right) \ln^5(2)
-\frac{1}{720} \Li_{5}\left(\frac{1}{2}\right) \ln^6(2)
\nonumber\\ &&
-\frac{1}{5040} \Li_{4}\left(\frac{1}{2}\right) \ln^7(2)
-\frac{1}{332640} \ln^{11}(2)
+\frac{1}{90720} \ln^9(2) \zeta_2
-\frac{1}{46080} \ln^8(2) \zeta_3
\nonumber\\ &&
-\frac{2533}{5600} \zeta_2^4 \zeta_3
+\frac{5}{48} \zeta_2 \zeta_3^3
-\frac{99}{112} \zeta_2^3 \zeta_5
+\frac{3}{16} \zeta_3^2 \zeta_5
-\frac{57}{40} \zeta_2^2 \zeta_7
-\frac{5}{3} \zeta_2 \zeta_9
+\frac{30945}{2048} \zeta_{11}
\end{eqnarray}

The integrals $I_9$ and  $I_{11}$ are given by 
\begin{eqnarray}
I_9 &=& 
-\frac{2}{3 \textcolor{blue}{\ep^3}}
-\frac{10}{3 \textcolor{blue}{\ep^2}}
+\left(-\frac{26}{3}
-2 \zeta_2\right) \frac{1}{\textcolor{blue}{\ep}}
-2
-\frac{16}{3} \zeta_3
-22 \zeta_2
+\textcolor{blue}{\ep} \Biggl(
        \frac{398}{3}
        -\frac{52}{5} \zeta_2^2
        -\frac{248}{3} \zeta_3
        +96 \ln(2) \zeta_2
\nonumber\\ &&
        -146 \zeta_2
\Biggr)
+\textcolor{blue}{\ep^2} \Biggl(
        1038
        -96 \zeta_5
        -16 \zeta_2 \zeta_3
        -512 \Li_4\left(\frac{1}{2}\right)
        -\frac{64 \ln^4(2)}{3}
        -256 \ln^2(2) \zeta_2
        +\frac{108}{5} \zeta_2^2
\nonumber\\ &&
        -\frac{1888}{3} \zeta_3
        +960 \ln(2) \zeta_2
        -774 \zeta_2
\Biggr)
+\textcolor{blue}{\ep^3} \Biggl(
        \frac{17470}{3}
        -\frac{2944}{35} \zeta_2^3
        -\frac{64}{3} \zeta_3^2
        +2496 \zeta_5
        -4096 \Li_5\left(\frac{1}{2}\right)
\nonumber\\ &&
        +\frac{512 \ln^5(2)}{15}
        +\frac{2048}{3} \ln^3(2) \zeta_2
        -\frac{2944}{5} \ln(2) \zeta_2^2
         +16 \zeta_2 \zeta_3
       -5120 \Li_4\left(\frac{1}{2}\right)
        -2560 \ln^2(2) \zeta_2
\nonumber\\ &&
        -\frac{640 \ln^4(2)}{3}
        +\frac{3004}{5} \zeta_2^2
        -3600 \zeta_3
        +6144 \ln(2) \zeta_2
        -3634 \zeta_2
\Biggr)
+\textcolor{blue}{\ep^4} \Biggl(
        \frac{85562}{3}
        -1328 \zeta_7
\nonumber\\ &&
        -288 \zeta_2 \zeta_5
        -\frac{416}{5} \zeta_2^2 \zeta_3
        -12288 s_{6}
        -32768 \Li_6\left(\frac{1}{2}\right)
        -1536 \Li_4\left(\frac{1}{2}\right) \zeta_2
        -\frac{2048 \ln^6(2)}{45}
\nonumber\\ &&
        -\frac{4288}{3} \ln^4(2) \zeta_2
        +\frac{13696}{5} \ln^2(2) \zeta_2^2
        -2112 \ln(2) \zeta_2 \zeta_3
        +\frac{19728}{5} \zeta_2^3
        +\frac{14680}{3} \zeta_3^2
        -40960 \Li_5\left(\frac{1}{2}\right)
\nonumber\\ &&
        +\frac{1024 \ln^5(2)}{3}
        +\frac{20480}{3} \ln^3(2) \zeta_2
        -5888 \ln(2) \zeta_2^2
        +28512 \zeta_5
        +752 \zeta_2 \zeta_3
        -32768 \Li_4\left(\frac{1}{2}\right)
\nonumber\\ &&
        -\frac{4096 \ln^4(2)}{3}
        -16384 \ln^2(2) \zeta_2
        +\frac{23396}{5} \zeta_2^2
        -\frac{53264}{3} \zeta_3
        +32256 \ln(2) \zeta_2
        -15894 \zeta_2
\Biggr)
\nonumber\\ &&
+\textcolor{blue}{\ep^5} \Biggl(
        \frac{389806}{3}
        -\frac{133968}{175} \zeta_2^4
        -64 \zeta_2 \zeta_3^2
        -768 \zeta_3 \zeta_5
        -\frac{325632 s_{7a}}{7}
        +\frac{362496 s_{7b}}{7}
        -262144 \Li_7\left(\frac{1}{2}\right)
\nonumber\\ &&
        -12288 \Li_5\left(\frac{1}{2}\right) \zeta_2
        -4096 \Li_4\left(\frac{1}{2}\right) \zeta_3
        +\frac{325632 \ln(2) s_{6}}{7}
        +\frac{16384 \ln^7(2)}{315}
        +\frac{34304}{15} \ln^5(2) \zeta_2
\nonumber\\ &&
        -\frac{512}{3} \ln^4(2) \zeta_3
        -\frac{109568}{15} \ln^3(2) \zeta_2^2
        +9472 \ln^2(2) \zeta_2 \zeta_3
        -95232 \ln^2(2) \zeta_5
        -\frac{905344}{35} \ln(2) \zeta_2^3
\nonumber\\ &&
        -\frac{407040}{7} \ln(2) \zeta_3^2
        +\frac{3311468}{7} \zeta_7
        -\frac{421296}{35} \zeta_2^2 \zeta_3
        -\frac{1016280}{7} \zeta_2 \zeta_5
        -122880 s_{6}
        -327680 \Li_6\left(\frac{1}{2}\right)
\nonumber\\ &&
        -15360 \Li_4\left(\frac{1}{2}\right) \zeta_2
        -\frac{4096 \ln^6(2)}{9}
        -\frac{42880}{3} \ln^4(2) \zeta_2
        +27392 \ln^2(2) \zeta_2^2
        -21120 \ln(2) \zeta_2 \zeta_3
\nonumber\\ &&
        +\frac{1489888}{35} \zeta_2^3
        +\frac{149168}{3} \zeta_3^2
        -262144 \Li_5\left(\frac{1}{2}\right)
        +\frac{32768 \ln^5(2)}{15}
        +\frac{131072}{3} \ln^3(2) \zeta_2
\nonumber\\ &&
        -\frac{188416}{5} \ln(2) \zeta_2^2
        +190176 \zeta_5
        +6096 \zeta_2 \zeta_3
        -172032 \Li_4\left(\frac{1}{2}\right)
        -7168 \ln^4(2)
        -86016 \ln^2(2) \zeta_2
\nonumber\\ &&
        +\frac{133996}{5} \zeta_2^2
        -80400 \zeta_3
        +152064 \ln(2) \zeta_2
        -66626 \zeta_2
\Biggr)
+ \textcolor{blue}{\ep^6} \Biggl(
        566958
        -\frac{160960}{9} \zeta_9
        -3984 \zeta_2 \zeta_7
\nonumber\\ &&
        -\frac{512}{9} \zeta_3^3
        -\frac{23552}{35} \zeta_2^3 \zeta_3
        -\frac{7488}{5} \zeta_2^2 \zeta_5
        -\frac{21695376 s_{8a}}{35}
        -\frac{28157952 s_{8b}}{7}
        -\frac{2899968 s_{8c}}{7}
\nonumber\\ &&
        -\frac{2605056 s_{8d}}{7}
        +\frac{1449984 \ln^2(2) s_{6}}{7}
        +\frac{7704576}{7} s_{6} \zeta_2
        -2097152 \Li_8\left(\frac{1}{2}\right)
        -98304 \Li_6\left(\frac{1}{2}\right) \zeta_2
\nonumber\\ && 
       +\frac{4980736}{7} \Li_5\left(\frac{1}{2}\right) \zeta_3
        -\frac{2884608}{35} \Li_4\left(\frac{1}{2}\right) \zeta_2^2
        -\frac{16384 \ln^8(2)}{315}
        -\frac{137216}{45} \ln^6(2) \zeta_2
\nonumber\\ &&
        -\frac{622592}{105} \ln^5(2) \zeta_3
        +\frac{1173376}{105} \ln^4(2) \zeta_2^2
        +\frac{772096}{21} \ln^3(2) \zeta_2 \zeta_3
        +\frac{468736}{35} \ln^2(2) \zeta_2^3
\nonumber\\ && 
       +\frac{407808}{7} \ln^2(2) \zeta_3^2
        +\frac{16493824}{35} \ln(2) \zeta_2^2 \zeta_3
        +\frac{8558784}{7} \ln(2) \zeta_2 \zeta_5
        +\frac{942502872 \zeta_2^4}{6125}
        +\frac{3316480}{7} 
\nonumber\\ &&
\times \zeta_2 \zeta_3^2
        -\frac{3194496}{7} \zeta_3 \zeta_5
        -\frac{3256320 s_{7a}}{7}
        +\frac{3624960 s_{7b}}{7}
        +\frac{3256320 \ln(2) s_{6}}{7}
        -2621440 \Li_7\left(\frac{1}{2}\right)
\nonumber\\ &&
        -122880 \Li_5\left(\frac{1}{2}\right) \zeta_2
        -40960 \Li_4\left(\frac{1}{2}\right) \zeta_3
        +\frac{32768 \ln^7(2)}{63}
        +\frac{68608}{3} \ln^5(2) \zeta_2
        -\frac{5120}{3} \ln^4(2) \zeta_3
\nonumber\\ &&
        -\frac{219136}{3} \ln^3(2) \zeta_2^2
        +94720 \ln^2(2) \zeta_2 \zeta_3
        -952320 \ln^2(2) \zeta_5
        -\frac{1810688}{7} \ln(2) \zeta_2^3
        -\frac{4070400}{7} 
\nonumber\\ &&
\times \ln(2) \zeta_3^2
        +\frac{33458632}{7} \zeta_7
        -\frac{10088208}{7} \zeta_2 \zeta_5
        -\frac{4105216}{35} \zeta_2^2 \zeta_3
        -786432 s_6
        -2097152 \Li_6\left(\frac{1}{2}\right)
\nonumber\\ &&
        -98304 \Li_4\left(\frac{1}{2}\right) \zeta_2
        -\frac{131072 \ln^6(2)}{45}
        -\frac{274432}{3} \ln^4(2) \zeta_2
        +\frac{876544}{5} \ln^2(2) \zeta_2^2
        -135168 
\nonumber\\ &&
\times \ln(2) \zeta_2 \zeta_3
        +\frac{9771392}{35} \zeta_2^3
        +319936 \zeta_3^2
        -1376256 \Li_5\left(\frac{1}{2}\right)
        +\frac{57344 \ln^5(2)}{5}
        +229376 
\nonumber\\ &&
\times \ln^3(2) \zeta_2
        -\frac{989184}{5} \ln(2) \zeta_2^2
        +35440 \zeta_2 \zeta_3
        +1019040 \zeta_5
        -811008 \Li_4\left(\frac{1}{2}\right)
        -33792 \ln^4(2)
\nonumber\\ &&
        -405504 \ln^2(2) \zeta_2
        +\frac{663876}{5} \zeta_2^2
        -\frac{1037584}{3} \zeta_3
        +672768 \ln(2) \zeta_2
        -272326 \zeta_2
\Biggr)
\nonumber\\ &&
+\textcolor{blue}{\ep^7} \Biggl(
        \frac{7221278}{3}
        -6912 \zeta_5^2
        -2304 \zeta_2 \zeta_3 \zeta_5
        -\frac{1956032}{275} \zeta_2^5
        -\frac{1664}{5} \zeta_2^2 \zeta_3^2
        -10624 \zeta_3 \zeta_7
\nonumber\\ &&
        -\frac{114294784 s_{9a}}{7}
        +6750208 s_{9b}
        +\frac{90902528 s_{9c}}{21}
        -\frac{9240576 s_{9d}}{7}
        -\frac{11599872 s_{9e}}{7}
\nonumber\\ &&
        +\frac{143271936}{49} s_{7a} \zeta_2
        -\frac{206985216}{49} s_{7b} \zeta_2
        +\frac{144875904 \ln(2) s_{8a}}{35}
        +\frac{181805056 \ln(2) s_{8b}}{7}
\nonumber\\ &&
        -\frac{1933312 \ln^3(2) s_6}{7}
        -\frac{183871488}{49} \ln(2) s_6 \zeta_2
         + 227696640 s_6 \zeta_3       
         -16777216 \Li_9\left(\frac{1}{2}\right)
\nonumber\\ && 
        -786432 \Li_7\left(\frac{1}{2}\right) \zeta_2
       +\frac{39845888}{7} \Li_6\left(\frac{1}{2}\right) \zeta_3
        -\frac{81076224}{35} \Li_5\left(\frac{1}{2}\right) \zeta_2^2
\nonumber\\ &&
        -\frac{283149312}{7} \Li_4\left(\frac{1}{2}\right) \zeta_5
        +\frac{46116864}{7} \Li_4\left(\frac{1}{2}\right) \zeta_2 \zeta_3
        +\frac{131072 \ln^9(2)}{2835}
        +\frac{1097728}{315} \ln^7(2) \zeta_2
\nonumber\\ &&
        +\frac{2490368}{315} \ln^6(2) \zeta_3
        -\frac{2137088}{525} \ln^5(2) \zeta_2^2
        +\frac{4220416}{21} \ln^4(2) \zeta_2 \zeta_3
        -\frac{11797888}{7} \ln^4(2) \zeta_5
\nonumber\\ &&
        -\frac{3332096}{15} \ln^3(2) \zeta_2^3
        -\frac{543744}{7} \ln^3(2) \zeta_3^2
        -\frac{93715456}{35} \ln^2(2) \zeta_2^2 \zeta_3
        +\frac{102425088}{7} \ln^2(2) \zeta_2 \zeta_5
\nonumber\\ &&
        +606208 \ln^2(2) \zeta_2 \zeta_3
        +\frac{113401856}{7} \ln^2(2) \zeta_7
        -\frac{70934369088 \ln(2) \zeta_2^4}{6125}
\nonumber\\ &&
        +\frac{168809664}{49} \ln(2) \zeta_2 \zeta_3^2
        -\frac{224415616}{7} \ln(2) \zeta_3 \zeta_5
        -\frac{7007214274}{63} \zeta_9
        +\frac{1046207136}{245} \zeta_2^3 \zeta_3
\nonumber\\ &&
        -\frac{48494416}{9} \zeta_3^3
        +\frac{2042564800}{49} \zeta_2^2 \zeta_5
        +\frac{1637594676}{49} \zeta_2 \zeta_7
        -\frac{43390752 s_{8a}}{7}
        -\frac{281579520 s_{8b}}{7}
\nonumber\\ &&
        -\frac{28999680 s_{8c}}{7}
        -\frac{26050560 s_{8d}}{7}
        +\frac{14499840 \ln(2)^2 s_6}{7}
        +\frac{77045760}{7} s_6 \zeta_2
        -20971520 \Li_8\left(\frac{1}{2}\right)
\nonumber\\ &&
        -983040 \Li_6\left(\frac{1}{2}\right) \zeta_2
        +\frac{49807360}{7} \Li_5\left(\frac{1}{2}\right) \zeta_3
        -\frac{5769216}{7} \Li_4\left(\frac{1}{2}\right) \zeta_2^2
        -\frac{32768 \ln^8(2)}{63}
\nonumber\\ &&
        -\frac{274432}{9} \ln^6(2) \zeta_2
        -\frac{1245184}{21} \ln^5(2) \zeta_3
        +\frac{2346752}{21} \ln^4(2) \zeta_2^2
        +\frac{7720960}{21} \ln^3(2) \zeta_2 \zeta_3
\nonumber\\ &&
        +\frac{937472}{7} \ln^2(2) \zeta_2^3
        +\frac{4078080}{7} \ln^2(2) \zeta_3^2
        +\frac{85587840}{7} \ln(2) \zeta_2 \zeta_5
        +\frac{1919703456 \zeta_2^4}{1225}
\nonumber\\ &&
        +\frac{33181376}{7} \zeta_2 \zeta_3^2
        -\frac{31746048}{7} \zeta_3 \zeta_5
        -\frac{20840448 s_{7a}}{7}
        +\frac{23199744 s_{7b}}{7}
\nonumber\\ &&
        +\frac{20840448 \ln(2) s_6}{7}
        -16777216 \Li_7\left(\frac{1}{2}\right)
        -786432 \Li_5\left(\frac{1}{2}\right) \zeta_2
        -262144 \Li_4\left(\frac{1}{2}\right) \zeta_3
\nonumber\\ &&
        +\frac{1048576 \ln^7(2)}{315}
        +\frac{2195456}{15} \ln^5(2) \zeta_2
        -\frac{32768}{3} \ln^4(2) \zeta_3
        -\frac{7012352}{15} \ln^3(2) \zeta_2^2
\nonumber\\ &&
        +\frac{214880784}{7} \zeta_7
        -\frac{5207968}{7} \zeta_2^2 \zeta_3
        -\frac{57942016}{35} \ln(2) \zeta_2^3
        -\frac{64402848}{7} \zeta_2 \zeta_5
\nonumber\\ &&
        -4128768 s_6
        -6094848 \ln^2(2) \zeta_5
        +\frac{32987648}{7} \ln(2) \zeta_2^2 \zeta_3
        -\frac{26050560}{7} \ln(2) \zeta_3^2
\nonumber\\ &&
        -11010048 \Li_6\left(\frac{1}{2}\right) 
        -516096 \Li_4\left(\frac{1}{2}\right) \zeta_2
        -\frac{229376 \ln^6(2)}{15}
        -480256 \ln^4(2) \zeta_2
\nonumber\\ &&
        +\frac{4601856}{5} \ln^2(2) \zeta_2^2
        -709632 \ln(2) \zeta_2 \zeta_3
        +\frac{51932032}{35} \zeta_2^3
        +\frac{5052736}{3} \zeta_3^2
\nonumber\\ &&
        -6488064 \Li_5\left(\frac{1}{2}\right)
        +\frac{270336 \ln^5(2)}{5}
        +1081344 \ln^3(2) \zeta_2
        -\frac{4663296}{5} \ln(2) \zeta_2^2
        +4863456 \zeta_5
\nonumber\\ &&
        +176976 \zeta_2 \zeta_3
        -3588096 \Li_4\left(\frac{1}{2}\right)
        -149504 \ln^4(2)
        -1794048 \ln^2(2) \zeta_2
        +\frac{3033164}{5} \zeta_2^2
\nonumber\\ &&
        -\frac{4325936}{3} \zeta_3
        +2863104 \ln(2) \zeta_2
        -1097042 \zeta_2
\Biggr)
+
\textcolor{blue}{\ep^8}\Biggl(
10037278
+\frac{2261065728}{7} \Li_4\left(\frac{1}{2}\right) 
\nonumber\\ && 
\times \ln(2) \zeta_5
-243216 \zeta_{11}
-\frac{160960}{3} \zeta_2 \zeta_9
-\frac{103584}{5} \zeta_2^2 \zeta_7
-3072 \zeta_3^2 \zeta_5
-\frac{423936}{35} \zeta_2^3 \zeta_5
\nonumber\\ &&
\nonumber\\ &&
-\frac{512}{3} \zeta_2 \zeta_3^3
-\frac{1071744}{175} \zeta_2^4 \zeta_3
+\frac{1202564511072256 h_{91}}{780419}
+\frac{19687710418944 h_{73}}{780419}
\nonumber\\ &&
-\frac{5326405632 h_{7111}}{29}
-\frac{257949696 h_{511111}}{17}
-\frac{24978358272 h_{5113}}{493}
+\frac{15639478272 h_{5131}}{493}
\nonumber\\ &&
+\frac{8004796416 h_{5311}}{493}
-\frac{2606579712 h_{3331}}{493}
+\frac{185597952 h_{331111}}{17}
+\frac{914358272 \ln(2) s_{9a}}{7}
\nonumber\\ &&
-54001664 \ln(2) s_{9b}
-\frac{727220224 \ln(2) s_{9c}}{21}
+\frac{73924608 \ln(2) s_{9d}}{7}
+\frac{92798976 \ln(2) s_{9e}}{7}
\nonumber\\ &&
-13278720 \ln^2(2) s_{8a}
-\frac{553385984 \ln^2(2) s_{8b}}{7}
+\frac{92798976 \ln^2(2) s_{8c}}{7}
+\frac{83361792 \ln^2(2) s_{8d}}{7}
\nonumber\\ &&
-\frac{673097882736}{17255} s_{8a} \zeta_2
-\frac{84473856}{7} s_{8b} \zeta_2
-\frac{8699904}{7} s_{8c} \zeta_2
-\frac{7815168}{7} s_{8d} \zeta_2
\nonumber\\ &&
+\frac{27787264 \ln^3(2) s_{7a}}{7}
-\frac{30932992 \ln^3(2) s_{7b}}{7}
-\frac{1200881664}{49} \ln(2) s_{7a} \zeta_2
+\frac{1716781056}{49} 
\nonumber\\ && 
\times \ln(2) s_{7b} \zeta_2
- \frac{132420673536 s_{7a} \zeta_3}{3451} 
+ \frac{2899968 s_{7b} \zeta_3}{7}
-\frac{1278001152}{7} \ln(2) s_6 \zeta_3 
\nonumber\\ &&
+\frac{164856262656 s_{6} \zeta_2^2}{17255}
-6291456 \ln^4(2) s_{6}
-\frac{227500032}{49} \ln^2(2) s_{6} \zeta_2
-134217728 \Li_{10}\left(\frac{1}{2}\right)
\nonumber\\ &&
-6291456 \Li_8\left(\frac{1}{2}\right) \zeta_2
-2097152 \Li_7\left(\frac{1}{2}\right)\zeta_3
-\frac{333447168}{7} \Li_6\left(\frac{1}{2}\right)\ln(2) \zeta_3
\nonumber\\ && 
-\frac{2555904}{5} \Li_6\left(\frac{1}{2}\right)\zeta_2^2
-\frac{166723584}{7} \Li_5\left(\frac{1}{2}\right)\ln^2(2) \zeta_3
+\frac{630718464}{35} \Li_5\left(\frac{1}{2}\right) \ln(2) \zeta_2^2
\nonumber\\ &&
-589824 \Li_5\left(\frac{1}{2}\right) \zeta_5
+\frac{14942208}{7} \Li_5\left(\frac{1}{2}\right) \zeta_2 \zeta_3
+\frac{83361792}{35} \Li_4\left(\frac{1}{2}\right) \ln(2)^2 \zeta_2^2
\nonumber\\ &&
-\frac{369623040}{7} \Li_4\left(\frac{1}{2}\right)\ln(2) \zeta_2 \zeta_3
-\frac{2015232}{7} \Li_4\left(\frac{1}{2}\right) \zeta_2^3
-16384 \Li_4\left(\frac{1}{2}\right) \zeta_3^2
\nonumber\\ &&
-\frac{524288 \ln^{10}(2)}{14175}
-\frac{1097728}{315} \ln^8(2) \zeta_2
+\frac{41811968}{315} \ln^7(2) \zeta_3
-\frac{32247808}{1575} \ln^6(2) \zeta_2^2
\nonumber\\ &&
+\frac{727211008}{35} \ln^5(2) \zeta_5
-\frac{345571328}{105} \ln^5(2) \zeta_2 \zeta_3
+\frac{297220096}{105} \ln^4(2) \zeta_2^3
\nonumber\\ &&
+\frac{8366080}{3} \ln^4(2) \zeta_3^2
-\frac{1193759744}{7} \ln^3(2) \zeta_7
-145342464 \ln^3(2) \zeta_2 \zeta_5
+\frac{22380544}{3} \ln^3(2) \zeta_2^2 \zeta_3
\nonumber\\ &&
+270697472 \ln^2(2) \zeta_3 \zeta_5
+\frac{21425512704}{245} \ln^2(2) \zeta_2^4
-\frac{2129968640}{49} \ln^2(2) \zeta_2 \zeta_3^2
\nonumber\\ &&
-\frac{12544430784}{49} \ln(2) \zeta_2 \zeta_7
+\frac{56012645392}{63} \ln(2) \zeta_9
-\frac{41723390470912 \ln(2) \zeta_2^2 \zeta_5}{120785}
\nonumber\\ &&
\nonumber\\ &&
+\frac{300811136}{7} \ln(2) \zeta_3^3
-\frac{8089162496}{245} \ln(2) \zeta_2^3 \zeta_3
-\frac{2721028640820472 \zeta_3 \zeta_7}{5462933}
\nonumber\\ &&
-\frac{11685311414928 \zeta_5^2}{26911}
-\frac{1088749068720 \zeta_2 \zeta_3 \zeta_5}{3451}
+\frac{877301874261743216 \zeta_2^5}{4780066375}
\nonumber\\ &&
-\frac{1430424876432 \zeta_2^2 \zeta_3^2}{17255}
-\frac{1142947840 s_{9a}}{7}
+67502080 s_{9b}
+\frac{909025280 s_{9c}}{21}
-\frac{92405760 s_{9d}}{7}
\nonumber\\ &&
-\frac{115998720 s_{9e}}{7}
+\frac{289751808 \ln(2) s_{8a}}{7}
+\frac{1818050560 \ln(2) s_{8b}}{7}
-\frac{19333120 \ln^3(2) s_{6}}{7}
\nonumber\\ &&
-\frac{1838714880}{49} \ln(2) s_{6} \zeta_2
-167772160 \Li_{9}\left(\frac{1}{2}\right)
-7864320 \Li_7\left(\frac{1}{2}\right) \zeta_2
+\frac{398458880}{7} 
\nonumber\\ &&
\times \Li_6\left(\frac{1}{2}\right) \zeta_3
-\frac{162152448}{7} \Li_5\left(\frac{1}{2}\right)\zeta_2^2
-\frac{2831493120}{7} \Li_4\left(\frac{1}{2}\right)\zeta_5
+\frac{461168640}{7} \Li_4\left(\frac{1}{2}\right) \zeta_2 \zeta_3
\nonumber\\ &&
+\frac{262144}{567} \ln^9(2)
+\frac{2195456}{63} \ln^7(2) \zeta_2
+\frac{4980736}{63} \ln^6(2) \zeta_3
-\frac{4274176}{105} \ln^5(2) \zeta_2^2
\nonumber\\ &&
-\frac{117978880}{7} \ln^4(2) \zeta_5
+\frac{42204160}{21} \ln^4(2) \zeta_2 \zeta_3
-\frac{6664192}{3} \ln^3(2) \zeta_2^3
-\frac{5437440}{7} \ln^3(2) \zeta_3^2
\nonumber\\ &&
+\frac{1134018560}{7} \ln^2(2) \zeta_7
+\frac{1024250880}{7} \ln^2(2) \zeta_2 \zeta_5
-\frac{187430912}{7} \ln^2(2) \zeta_2^2 \zeta_3
\nonumber\\ &&
-\frac{2244156160}{7} \ln(2) \zeta_3 \zeta_5
-\frac{141868738176 \ln(2) \zeta_2^4}{1225}
+\frac{1688096640}{49} \ln(2) \zeta_2 \zeta_3^2
\nonumber\\ &&
-\frac{23343484700}{21} \zeta_9
+\frac{16383169752}{49} \zeta_2 \zeta_7
+\frac{102141815744}{245} \zeta_2^2 \zeta_5
-\frac{484925216}{9} \zeta_3^3
\nonumber\\ &&
+\frac{10468171328}{245} \zeta_2^3 \zeta_3
-\frac{1388504064 s_{8a}}{35}
-\frac{1802108928 s_{8b}}{7}
-\frac{185597952 s_{8c}}{7}
-\frac{166723584 s_{8d}}{7}
\nonumber\\ &&
+\frac{1432719360}{49} s_{7a} \zeta_2
-\frac{2069852160}{49} s_{7b} \zeta_2
+\frac{92798976 \ln^2(2) s_{6}}{7}
+\frac{493092864}{7} s_{6} \zeta_2
\nonumber\\ &&
-134217728 \Li_{8}\left(\frac{1}{2}\right)
-6291456 \Li_6\left(\frac{1}{2}\right) \zeta_2
+\frac{318767104}{7} \Li_5\left(\frac{1}{2}\right)\zeta_3
-\frac{184614912}{35} 
\nonumber\\ &&
\times \Li_4\left(\frac{1}{2}\right) \zeta_2^2
-\frac{1048576 \ln^8(2)}{315}
-\frac{8781824}{45} \ln^6(2) \zeta_2
-\frac{39845888}{105} \ln^5(2) \zeta_3
+\frac{75096064}{105} 
\nonumber\\ &&
\times \ln^4(2) \zeta_2^2
+\frac{49414144}{21} \ln^3(2) \zeta_2 \zeta_3
+\frac{29999104}{35} \ln^2(2) \zeta_2^3
+\frac{26099712}{7} \ln^2(2) \zeta_3^2
+\frac{547762176}{7} 
\nonumber\\ &&
\times \ln(2) \zeta_2 \zeta_5
+\frac{1055604736}{35} \ln(2) \zeta_2^2 \zeta_3
-\frac{202743552}{7} \zeta_3 \zeta_5
+\frac{61806558768 \zeta_2^4}{6125}
+\frac{212396736}{7} \zeta_2 \zeta_3^2
\nonumber\\ &&
-15630336 s_{7a}
+17399808 s_{7b}
+15630336 \ln(2) s_{6}
-88080384 \Li_{7}\left(\frac{1}{2}\right)
\nonumber\\ &&
-4128768 \Li_5\left(\frac{1}{2}\right) \zeta_2
-1376256 \Li_4\left(\frac{1}{2}\right)\zeta_3
+\frac{262144 \ln^7(2)}{15}
+\frac{3842048}{5} \ln^5(2) \zeta_2
\nonumber\\ &&
-57344 \ln^4(2) \zeta_3
-\frac{12271616}{5} \ln^3(2) \zeta_2^2
-31997952 \ln^2(2) \zeta_5
+3182592 \ln^2(2) \zeta_2 \zeta_3
\nonumber\\ &&
-\frac{43456512}{5} \ln(2) \zeta_2^3
-19537920 \ln(2)\zeta_3^2
+161445776 \zeta_7
-48240288 \zeta_2 \zeta_5
-\frac{19440544}{5} \zeta_2^2 \zeta_3
\nonumber\\ &&
-19464192 s_{6}
-51904512 \Li_{6}\left(\frac{1}{2}\right)
-2433024 \Li_4\left(\frac{1}{2}\right) \zeta_2
-\frac{360448 \ln^6(2)}{5}
\nonumber\\ &&
-2264064 \ln^4(2) \zeta_2
+\frac{21694464}{5} \ln^2(2) \zeta_2^2
+\frac{22610657280}{7} \Li_4\left(\frac{1}{2}\right) \zeta_5 \ln(2)-3345408 \ln(2) \zeta_2 \zeta_3
\nonumber\\ &&
+\frac{246644352}{35} \zeta_2^3
+7953216 \zeta_3^2
-28704768 \Li_{5}\left(\frac{1}{2}\right)
+\frac{1196032 \ln^5(2)}{5}
+4784128 \ln(2)^3 \zeta_2
\nonumber\\ &&
-\frac{20631552}{5} \ln(2) \zeta_2^2
+21694368 \zeta_5
+812528 \zeta_2 \zeta_3
-15269888 \Li_{4}\left(\frac{1}{2}\right)
-\frac{1908736 \ln^4(2)}{3}
	\nonumber\\ &&
-7634944 \ln^2(2) \zeta_2
+\frac{13199844}{5} \zeta_2^2
-\frac{17697424}{3} \zeta_3
+11894784 \ln(2) \zeta_2
\nonumber\\ &&
-4384758 \zeta_2
+227696640 s_6 \zeta_3
\Biggr)
+ \textcolor{blue}{\ep^9}\Biggl(
        \frac{124045582}{3}
        -\frac{1287680}{9} \zeta_3 \zeta_9
\nonumber\\ &&
        -191232 \zeta_5 \zeta_7
        -31872 \zeta_2 \zeta_3 \zeta_7
        -20736 \zeta_2 \zeta_5^2
        -\frac{59904}{5} \zeta_2^2 \zeta_3 \zeta_5
        -\frac{1024}{9} \zeta_3^4
        -\frac{94208}{35} \zeta_2^3 \zeta_3^2
\nonumber\\&& 
        -\frac{2321418752}{35035} \zeta_2^6
        -\frac{1130454496528082960384 h_{911}}{394189048955}
        +\frac{3788115950202254344192 h_{731}}{8277970028055}
\nonumber\\ &&
        +\frac{1193251487848937005056 h_{713}}{2759323342685}
        -\frac{6057099264 h_{71111}}{731}
        -\frac{226872518626348089344 h_{551}}{1655594005611}
\nonumber\\ &&
        -\frac{3233842372976071696384 h_{533}}{74501730252495}
        +\frac{23750770688 h_{53111}}{731}
        -\frac{2464531349504 h_{51113}}{47515}
\nonumber\\ &&
        +\frac{72970403840 h_{51311}}{731}
        -\frac{2063597568 h_{5111111}}{17}
        +\frac{1484783616 h_{3311111}}{17}
       -\frac{1201267933184 h_{33311}}{47515}
\nonumber\\ &&
        -\frac{9620516088578048 h_{91} \ln(2)}{780419}
        -\frac{157501683351552 h_{73} \ln(2)}{780419}
        +\frac{42611245056 h_{7111} \ln(2)}{29}
\nonumber\\ &&
        +\frac{199826866176 h_{5113} \ln(2)}{493}
        -\frac{125115826176 h_{5131} \ln(2)}{493}
        -\frac{64038371328 h_{5311} \ln(2)}{493}
\nonumber\\ &&
        +\frac{2063597568 h_{511111} \ln(2)}{17}
        +\frac{20852637696 h_{3331} \ln(2)}{493}
        -\frac{1484783616 h_{331111} \ln(2)}{17}
\nonumber\\ &&
        -\frac{3657433088 \ln^2(2) s_{9a}}{7}
        +216006656 \ln^2(2) s_{9b}
        +\frac{2908880896 \ln^2(2) s_{9c}}{21}
\nonumber\\ &&
        -\frac{295698432 \ln^2(2) s_{9d}}{7}
        -\frac{371195904 \ln^2(2) s_{9e}}{7}
        -\frac{20605079846912}{332605} s_{9a} \zeta_2
        +\frac{346075561984}{47515} 
\nonumber\\ &&
\times s_{9b} \zeta_2
        +\frac{90902528}{7} s_{9c} \zeta_2
        -\frac{27721728}{7} s_{9d} \zeta_2
        -\frac{34799616}{7} s_{9e} \zeta_2
        +\frac{933351424 \ln^3(2) s_{8a}}{35}
\nonumber\\ &&
        +\frac{1012137984 \ln^3(2) s_{8b}}{7}
        -\frac{494927872 \ln^3(2) s_{8c}}{7}
        -\frac{444596224 \ln^3(2) s_{8d}}{7}
\nonumber\\ &&
        +\frac{346806855835578375136 s_{8a} \zeta_3}{1970945244775}
        -\frac{4200071353238881144832 s_{8b} \zeta_3}{8277970028055}
        -\frac{23199744}{7} s_{8c} \zeta_3
\nonumber\\ &&
        -\frac{13929023471616}{332605} s_{8d} \zeta_3
        +\frac{5342354835072}{17255} \ln(2) s_{8a} \zeta_2
        +\frac{545415168}{7} \ln(2) s_{8b} \zeta_2
\nonumber\\ &&
        +\frac{34803127545856}{2328235} s_{7a} \zeta_2^2
        -\frac{3021041664}{245} s_{7b} \zeta_2^2
        +\frac{184549376 \ln^5(2) s_6}{5}
\nonumber\\ &&
        +\frac{250412578603008 \ln^2(2) s_6 \zeta_3}{332605}
        -\frac{166723584 \ln^4(2) s_{7a}}{7}
        +\frac{185597952 \ln^4(2) s_{7b}}{7}
\nonumber\\ &&
        +\frac{365283511369728 \ln(2) s_{7a} \zeta_3}{1377935}
        +\frac{4803526656}{49} \ln^2(2) s_{7a} \zeta_2
        -\frac{6867124224}{49} \ln^2(2) s_{7b} \zeta_2
\nonumber\\ && 
       -\frac{7442343481344}{120785} \ln(2) s_6 \zeta_2^2
        -\frac{2356066299904 s_6 \zeta_5}{47515}
        +\frac{3369369600}{49} \ln^3(2) s_6 \zeta_2
\nonumber\\ &&
        +\frac{61328336277504 s_6 \zeta_2 \zeta_3}{332605}
        -1073741824 \Li_{11}\left(\frac{1}{2}\right) 
        -50331648 \Li_9\left(\frac{1}{2}\right) \zeta_2
\nonumber\\ &&
        -16777216 \Li_8\left(\frac{1}{2}\right) \zeta_3
        -\frac{20447232}{5} \Li_7\left(\frac{1}{2}\right) \zeta_2^2
        -4718592 \Li_6\left(\frac{1}{2}\right) \zeta_5
\nonumber\\ &&
        +\frac{119537664}{7} \Li_6\left(\frac{1}{2}\right) \zeta_2 \zeta_3
        +\frac{1333788672}{7} \Li_6\left(\frac{1}{2}\right) \ln^2(2) \zeta_3
        +\frac{27814451740672}{332605} \Li_5\left(\frac{1}{2}\right) \zeta_3^2
\nonumber\\ &&
        -\frac{2522873856}{35} \Li_5\left(\frac{1}{2}\right) \ln^2(2) \zeta_2^2
        -7274496 \Li_5\left(\frac{1}{2}\right) \zeta_2^3
        +\frac{889192448}{7} \Li_5\left(\frac{1}{2}\right) \ln^3(2) \zeta_3
\nonumber\\ &&
\nonumber\\ &&
        -\frac{444596224}{35} \Li_4\left(\frac{1}{2}\right) \ln^3(2) \zeta_2^2
        -\frac{31735658955776}{332605} \Li_4\left(\frac{1}{2}\right) \zeta_2 \zeta_5
\nonumber\\ &&
        +\frac{1478492160}{7} \Li_4\left(\frac{1}{2}\right) \ln^2(2) \zeta_2 \zeta_3
        -1019904 \Li_4\left(\frac{1}{2}\right) \zeta_7
        -\frac{9044262912}{7} \Li_4\left(\frac{1}{2}\right) \ln^2(2) \zeta_5
\nonumber\\ &&
        +\frac{18894660870144}{1663025} \Li_4\left(\frac{1}{2}\right) \zeta_2^2 \zeta_3
        +\frac{4194304 \ln^{11}(2)}{155925}
        +\frac{8781824 \ln^9(2) \zeta_2}{2835}
\nonumber\\ &&
        -\frac{250216448}{315} \ln^8(2) \zeta_3
        +\frac{57638912}{1575} \ln^7(2) \zeta_2^2
        -\frac{10204172288}{105} \ln^6(2) \zeta_5
\nonumber\\ &&
        +\frac{699662336}{45} \ln^6(2) \zeta_2 \zeta_3
         -\frac{85119722061824 \ln^5(2) \zeta_3^2}{4989075}
        -\frac{201367552}{15} \ln^5(2) \zeta_2^3
\nonumber\\ &&
       +\frac{5347831296}{7} \ln^4(2) \zeta_7
        +\frac{602089461569408 \ln^4(2) \zeta_2 \zeta_5}{997815}
        -\frac{53577818647552 \ln^4(2) \zeta_2^2 \zeta_3}{4989075}
\nonumber\\ &&
        -\frac{376156375212032 \ln^3(2) \zeta_3 \zeta_5}{332605}
        +\frac{1403329698221056 \ln^3(2) \zeta_2 \zeta_3^2}{6984705}
        -\frac{224050581568}{63} \ln^2(2) \zeta_9
\nonumber\\ &&
        -\frac{301110728704}{875} \ln^3(2) \zeta_2^4
        +\frac{2527896877478656 \ln^2(2) \zeta_2 \zeta_7}{2328235}
        +\frac{1433322075232256 \ln^2(2) \zeta_2^3 \zeta_3}{11641175}
\nonumber\\ &&
        +2706974720 \ln^2(2) \zeta_3 \zeta_5
        +\frac{100081795637701376 \ln^2(2) \zeta_2^2 \zeta_5}{67518815}
        -\frac{13144557640192 \ln^2(2) \zeta_3^3}{66521}
\nonumber\\ &&
        +\frac{250218587846220419425554176 \ln(2) \zeta_3 \zeta_7}{54288110011131555}
        +\frac{268488970046321792 \ln(2) \zeta_5^2}{75216245}
\nonumber\\ &&
        +\frac{1370557129830144 \ln(2)\zeta_2 \zeta_3 \zeta_5}{567385}
        +\frac{2847832763676416 \ln(2) \zeta_2^2 \zeta_3^2}{3971695}
        -\frac{16178324992}{49} \ln(2) \zeta_2^3 \zeta_3
\nonumber\\ &&
        -\frac{44020721797951110997888 \ln(2) \zeta_2^5}{29392628139875}
        +\frac{276298183385563985523133 \zeta_{11}}{24833910084165}
\nonumber\\ &&
        +\frac{2551544671187332995658 \zeta_2 \zeta_9}{14900346050499}
        -\frac{6411687453232793159672 \zeta_2^2 \zeta_7}{17042879469525}
\nonumber\\ &&
        +\frac{3958383139188142278976 \zeta_3^2 \zeta_5}{2759323342685}
        -\frac{42386525725006839853952 \zeta_2^3 \zeta_5}{51128638408575}
        -\frac{1629519807808 \zeta_2 \zeta_3^3}{66521}
\nonumber\\ &&
        -\frac{6271064327915000770722032 \zeta_2^4 \zeta_3}{7243223774548125}
        +\frac{12025645110722560 h_{91}}{780419}
        +\frac{196877104189440 h_{73}}{780419}
\nonumber\\ &&
        -\frac{53264056320 h_{7111}}{29}
        +\frac{80047964160 h_{5311}}{493}
        -\frac{249783582720 h_{5113}}{493}
        +\frac{156394782720 h_{5131}}{493}
\nonumber\\ &&
        -\frac{2579496960 h_{511111}}{17}
        +\frac{1855979520  h_{331111}}{17}
        -\frac{26065797120 h_{3331}}{493}
        +\frac{9143582720 \ln(2) s_{9a}}{7}
\nonumber\\ &&
\nonumber\\ &&
        -540016640 \ln(2) s_{9b}
        -\frac{7272202240 \ln(2) s_{9c}}{21}
        +\frac{739246080 \ln(2) s_{9d}}{7}
        +\frac{927989760 \ln(2) s_{9e}}{7}
\nonumber\\ &&
        -132787200 \ln^2(2) s_{8a}
        -\frac{5533859840 \ln^2(2) s_{8b}}{7}
        +\frac{927989760 \ln^2(2) s_{8c}}{7}
\nonumber\\ &&
        +\frac{833617920 \ln^2(2) s_{8d}}{7}
        -\frac{1346195765472 s_{8a} \zeta_2}{3451}
        -\frac{844738560}{7} s_{8b} \zeta_2
        -\frac{86999040}{7} s_{8c} \zeta_2
\nonumber\\ &&
        -\frac{78151680}{7} s_{8d} \zeta_2
        +\frac{277872640 \ln^3(2) s_{7a}}{7}
        -\frac{309329920 \ln^3(2) s_{7b}}{7}
        -\frac{1324206735360 s_{7a} \zeta_3}{3451}
\nonumber\\ &&
        +\frac{28999680}{7} s_{7b} \zeta_3
        -\frac{12008816640}{49} \ln(2) s_{7a} \zeta_2
        +\frac{17167810560}{49} \ln(2) s_{7b} \zeta_2
\nonumber\\ &&
        +\frac{329712525312}{3451} s_6 \zeta_2^2
        -62914560 \ln^4(2) s_6
        -\frac{2275000320}{49} \ln^2(2) s_6 \zeta_2
\nonumber\\ &&
        -\frac{12780011520}{7} \ln(2) s_6 \zeta_3
        -1342177280 \Li_{10}\left(\frac{1}{2}\right)
        -62914560 \Li_8\left(\frac{1}{2}\right) \zeta_2
\nonumber\\ &&
        -20971520 \Li_7\left(\frac{1}{2}\right) \zeta_3
        -\frac{3334471680}{7} \Li_6\left(\frac{1}{2}\right) \ln(2)\zeta_3
        -5111808 \Li_6\left(\frac{1}{2}\right) \zeta_2^2
\nonumber\\ && 
\nonumber\\ &&
        +\frac{1261436928}{7} \Li_5\left(\frac{1}{2}\right) \ln(2) \zeta_2^2
        +\frac{149422080}{7} \Li_5\left(\frac{1}{2}\right) \zeta_2 \zeta_3
        -5898240 \Li_5\left(\frac{1}{2}\right) \zeta_5
\nonumber\\ &&
        -\frac{1667235840}{7} \Li_5\left(\frac{1}{2}\right) \ln^2(2) \zeta_3
        +\frac{166723584}{7} \Li_4\left(\frac{1}{2}\right) \ln^2(2) \zeta_2^2
        -163840 \Li_4\left(\frac{1}{2}\right)\zeta_3^2
\nonumber\\ &&
        -\frac{20152320}{7} \Li_4\left(\frac{1}{2}\right) \zeta_2^3
        -\frac{3696230400}{7} \Li_4\left(\frac{1}{2}\right) \ln(2) \zeta_2 \zeta_3
        -\frac{1048576 \ln^{10}(2)}{2835}
\nonumber\\ &&
        -\frac{2195456}{63} \ln^8(2) \zeta_2
        +\frac{83623936}{63} \ln^7(2) \zeta_3
        -\frac{64495616}{315} \ln^6(2) \zeta_2^2
        +\frac{1454422016}{7} \ln^5(2) \zeta_5
\nonumber\\ &&
        -\frac{691142656}{21} \ln^5(2) \zeta_2 \zeta_3
        +\frac{83660800}{3} \ln^4(2) \zeta_3^2
        +\frac{594440192}{21} \ln^4(2) \zeta_2^3
\nonumber\\ &&
        -\frac{11937597440}{7} \ln^3(2) \zeta_7
        -1453424640 \ln^3(2) \zeta_2 \zeta_5
        +\frac{223805440}{3} \ln^3(2) \zeta_2^2 \zeta_3
\nonumber\\ &&
        -\frac{21299686400}{49} \ln^2(2) \zeta_2 \zeta_3^2
        +\frac{42851025408}{49} \ln^2(2) \zeta_2^4
        +\frac{560126453920}{63} \ln(2) \zeta_9
\nonumber\\ &&
        -\frac{125444307840}{49} \ln(2) \zeta_2 \zeta_7
        -\frac{83446780941824 \ln(2) \zeta_2^2 \zeta_5}{24157}
        +\frac{3008111360}{7} \ln(2) \zeta_3^3
\nonumber\\ &&
        +\frac{10803818496}{49} \ln(2) \zeta_2 \zeta_3^2
        -\frac{27208138994797616 \zeta_3 \zeta_7}{5462933}
        -\frac{116846231822496 \zeta_5^2}{26911}
\nonumber\\ &&
        -\frac{10887196496352 \zeta_2 \zeta_3 \zeta_5}{3451}
        -\frac{14304036293152 \zeta_2^2 \zeta_3^2}{17255}
        +\frac{19303408822744671456 \zeta_2^5}{10516146025}
\nonumber\\ &&
        -\frac{7314866176 s_{9a}}{7}
        +432013312 s_{9b}
        +\frac{5817761792 s_{9c}}{21}
        -\frac{591396864 s_{9d}}{7}
        -\frac{742391808 s_{9e}}{7}
\nonumber\\ &&
\nonumber\\ &&
        +\frac{9272057856 \ln(2) s_{8a}}{35}
        +\frac{11635523584 \ln(2) s_{8b}}{7}
        +\frac{9169403904}{49} s_{7a} \zeta_2
        -\frac{13247053824}{49} s_{7b} \zeta_2
\nonumber\\ &&
        -\frac{11767775232}{49} \ln(2) s_6 \zeta_2
        -\frac{123731968 \ln^3(2) s_6}{7}
        +1457258496 s_6 \zeta_3
        -1073741824 \Li_9\left(\frac{1}{2}\right)
\nonumber\\ &&
        -50331648 \Li_7\left(\frac{1}{2}\right) \zeta_2
        +\frac{2550136832}{7} \Li_6\left(\frac{1}{2}\right) \zeta_3
        -\frac{5188878336}{35} \Li_5\left(\frac{1}{2}\right) \zeta_2^2
\nonumber\\ &&
        -\frac{18121555968}{7} \Li_4\left(\frac{1}{2}\right) \zeta_5
        +\frac{2951479296}{7} \Li_4\left(\frac{1}{2}\right) \zeta_2 \zeta_3
        +\frac{8388608 \ln^9(2)}{2835}
\nonumber\\ &&
        +\frac{70254592}{315} \ln^7(2) \zeta_2
        +\frac{159383552}{315} \ln^6(2) \zeta_3
        -\frac{136773632}{525} \ln^5(2) \zeta_2^2
        -\frac{755064832}{7} \ln^4(2) \zeta_5
\nonumber\\ &&
        +\frac{270106624}{21} \ln^4(2) \zeta_2 \zeta_3
        -\frac{34799616}{7} \ln^3(2) \zeta_3^2
        -\frac{213254144}{15} \ln^3(2) \zeta_2^3
        +\frac{7257718784}{7} \ln^2(2) \zeta_7
\nonumber\\ &&
        +\frac{6555205632}{7} \ln^2(2) \zeta_2 \zeta_5
        -\frac{5997789184}{35} \ln^2(2) \zeta_2^2 \zeta_3
        -\frac{4539799621632 \ln(2) \zeta_2^4}{6125}
\nonumber\\ &&
        -\frac{14362599424}{7} \ln(2) \zeta_3 \zeta_5
        -\frac{448104543296}{63} \zeta_9
        +\frac{104867942736}{49} \zeta_2 \zeta_7
        +\frac{653737047104}{245} \zeta_2^2 \zeta_5
\nonumber\\ &&
        -\frac{1034493440}{3} \zeta_3^3
        +\frac{67009518592}{245} \zeta_2^3 \zeta_3
        -\frac{1041378048 s_{8a}}{5}
        -1351581696 s_{8b}
        -139198464 s_{8c}
\nonumber\\ &&
        -125042688 s_{8d}
        +69599232 \ln^2(2) s_6
        +369819648 s_6 \zeta_2
        -704643072 \Li_8\left(\frac{1}{2}\right)
\nonumber\\ &&
        -33030144 \Li_6\left(\frac{1}{2}\right) \zeta_2
        +239075328 \Li_5\left(\frac{1}{2}\right) \zeta_3
        -\frac{138461184}{5} \Li_4\left(\frac{1}{2}\right) \zeta_2^2
        -\frac{262144 \ln^8(2)}{15}
\nonumber\\ &&
        -\frac{15368192}{15} \ln^6(2) \zeta_2
        -\frac{9961472}{5} \ln^5(2) \zeta_3
        +\frac{18774016}{5} \ln^4(2) \zeta_2^2
        +12353536 \ln^3(2) \zeta_2 \zeta_3
\nonumber\\ &&
        +19574784 \ln^2(2) \zeta_3^2
        +\frac{22499328}{5} \ln^2(2) \zeta_2^3
        +410821632 \ln(2) \zeta_2 \zeta_5
        +\frac{791703552}{5} \ln(2) \zeta_2^2 \zeta_3
\nonumber\\ &&
        -151892736 \zeta_3 \zeta_5
        +159311296 \zeta_2 \zeta_3^2
        +\frac{46498767216}{875} \zeta_2^4
        -\frac{515801088 s_{7a}}{7}
        +\frac{574193664 s_{7b}}{7}
\nonumber\\ &&
        +\frac{515801088 \ln(2) s_6}{7}
        -415236096 \Li_7\left(\frac{1}{2}\right)
        -19464192 \Li_5\left(\frac{1}{2}\right) \zeta_2
        -6488064 \Li_4\left(\frac{1}{2}\right) \zeta_3
\nonumber\\ &&
\nonumber\\ &&
        +\frac{2883584 \ln^7(2)}{35}
        +\frac{18112512}{5} \ln^5(2) \zeta_2
\nonumber\\ &&
	-270336 \ln^4(2) \zeta_3
        -\frac{57851904}{5} \ln^3(2) \zeta_2^2
        -150847488 \ln(2)^2 \zeta_5
        +15003648 \ln^2(2) \zeta_2 \zeta_3
\nonumber\\ &&
        -\frac{644751360}{7} \ln(2) \zeta_3^2
        -\frac{1434064896}{35} \ln(2) \zeta_2^3
        +\frac{5333463504}{7} \zeta_7
        -\frac{1590681888}{7} \zeta_2 \zeta_5
\nonumber\\ &&
        -\frac{127947168}{7} \zeta_2^2 \zeta_3
        -86114304 s_6
        -229638144 \Li_6\left(\frac{1}{2}\right)
        -10764288 \Li_4\left(\frac{1}{2}\right)\zeta_2
\nonumber\\ &&
        -\frac{4784128 \ln^6(2)}{15}
        -10016768 \ln^4(2) \zeta_2
        +\frac{61079552}{3} \ln^3(2) \zeta_2
        +\frac{95981568}{5} \ln^2(2) \zeta_2^2
\nonumber\\ &&
        -14800896 \ln(2) \zeta_2 \zeta_3
        +\frac{105679040}{3} \zeta_3^2
        +\frac{1096650368}{35} \zeta_2^3
        -122159104 \Li_5\left(\frac{1}{2}\right)
\nonumber\\ &&
        +\frac{15269888 \ln^5(2)}{15}
	-\frac{87801856}{5} \ln(2) \zeta_2^2
        +92863200 \zeta_5
        +3547600 \zeta_2 \zeta_3
        -63438848 \Li_4\left(\frac{1}{2}\right)
\nonumber\\ &&
        -\frac{7929856 \ln^4(2)}{3}
        -31719424 \ln^2(2) \zeta_2
        +\frac{55731628}{5} \zeta_2^2
        -23863952 \zeta_3
        +48685056 \ln(2) \zeta_2
\nonumber\\ &&
	-17459170 \zeta_2
+\frac{22610657280}{7} \Li_4\left(\frac{1}{2}\right) \zeta_5 \ln(2)
\Biggr)
\end{eqnarray}
\begin{eqnarray}
I_{11} &=&
\frac{1}{\textcolor{blue}{\ep^3}}
+\frac{7}{2 \textcolor{blue}{\ep^2}}
+ \frac{253}{36 \textcolor{blue}{\ep}}
+ \frac{2501}{216}
+ \textcolor{blue}{\ep} \Biggl(
        \frac{59437}{1296}
        -\frac{128 \zeta_2}{3}
\Biggr)
+ \textcolor{blue}{\ep^2} \Biggl(
        \frac{2831381}{7776}
        -\frac{1792}{9} \zeta_3
        +512 \ln(2) \zeta_2
\nonumber\\ && 
\nonumber\\ &&
        -\frac{4544}{9} \zeta_2
\Biggr)
+ \textcolor{blue}{\ep^3} \Biggl(
        \frac{117529021}{46656}
        -\frac{8192}{3} \Li_4\left(\frac{1}{2}\right)
        -\frac{1024}{9} \ln^4(2)
        -\frac{7168}{3} \ln^2(2) \zeta_2
        +\frac{11008}{15} \zeta_2^2
\nonumber\\ &&
        -\frac{63616}{27} \zeta_3
        +\frac{18176}{3} \ln(2) \zeta_2
        -\frac{99680}{27} \zeta_2
\Biggr) 
+ \textcolor{blue}{\ep^4} \Biggl(
        \frac{4081770917}{279936}
        -32768 \Li_5\left(\frac{1}{2}\right)
        +\frac{87296}{3} \zeta_5
\nonumber\\ &&
        +\frac{4096}{15} \ln^5(2)
        +\frac{28672}{3} \ln^3(2) \zeta_2
        -\frac{44032}{5} \ln(2) \zeta_2^2
        +\frac{3584}{3} \zeta_2 \zeta_3
        -\frac{290816}{9} \Li_4\left(\frac{1}{2}\right)
\nonumber\\ &&
        -\frac{36352}{27} \ln^4(2)
        -\frac{254464}{9} \ln^2(2) \zeta_2
        +\frac{390784}{45} \zeta_2^2
        -\frac{1395520}{81} \zeta_3
        +\frac{398720}{9} \ln(2) \zeta_2
\nonumber\\ &&
        -\frac{1750448}{81} \zeta_2
\Biggr)
+ \textcolor{blue}{\ep^5} \Biggl(
        \frac{125873914573}{1679616}
        -180224 s_6
        -393216 \Li_6\left(\frac{1}{2}\right)
        -\frac{8192}{15} \ln^6(2) 
\nonumber\\ &&
        +\frac{633344}{9} \zeta_3^2
        -28672 \ln^4(2) \zeta_2
        +\frac{264192}{5} \ln^2(2) \zeta_2^2
        -14336 \ln(2) \zeta_2 \zeta_3
        +\frac{745472}{15} \zeta_2^3
\nonumber\\ &&
        -\frac{1163264}{3} \Li_5\left(\frac{1}{2}\right)
        +\frac{3099008}{9} \zeta_5
        +\frac{145408}{45} \ln^5(2)
        +\frac{1017856}{9} \ln^3(2) \zeta_2
        -\frac{1563136}{15} \ln(2) \zeta_2^2
\nonumber\\ &&
        +\frac{127232}{9} \zeta_2 \zeta_3
        -\frac{6379520}{27} \Li_4\left(\frac{1}{2}\right)
        -\frac{797440}{81} \ln^4(2)
        -\frac{5582080}{27} \ln^2(2) \zeta_2
        +\frac{1714496}{27} \zeta_2^2
\nonumber\\ &&
        -\frac{24506272}{243} \zeta_3
        +\frac{7001792}{27} \ln(2) \zeta_2
        -\frac{27091736}{243} \zeta_2
\Biggr)
+ \textcolor{blue}{\ep^6} \Biggl(
        \frac{3593750577461}{10077696}
        -\frac{19922944}{21} s_{7a}
\nonumber\\ &&
        +\frac{25493504}{21} s_{7b}
        -4718592 \Li_7\left(\frac{1}{2}\right)
        +\frac{19922944}{21} \ln(2) s_6
        +\frac{72259840}{7} \zeta_7
        +\frac{32768}{35} \ln^7(2)
\nonumber\\ &&
        +\frac{344064}{5} \ln^5(2) \zeta_2
        -\frac{1056768}{5} \ln^3(2) \zeta_2^2
        + 86016 \ln^2(2) \zeta_2 \zeta_3
        -2095104 \ln^2(2) \zeta_5
\nonumber\\ &&
        -\frac{5758976}{15} \ln(2) \zeta_2^3
        -\frac{24903680}{21} \ln(2) \zeta_3^2
        -\frac{7158784}{21} \zeta_2^2 \zeta_3
        -\frac{76080640}{21} \zeta_2 \zeta_5
        -\frac{6397952}{3} s_6
\nonumber\\ &&
        -4653056 \Li_6\left(\frac{1}{2}\right)
        -\frac{290816}{45} \ln^6(2)
        -\frac{1017856}{3} \ln^4(2) \zeta_2
        +\frac{3126272}{5} \ln^2(2) \zeta_2^2
\nonumber\\ &&
        -\frac{508928}{3}  \ln(2) \zeta_2 \zeta_3
        +\frac{26464256}{45} \zeta_2^3
        +\frac{22483712}{27} \zeta_3^2
        -\frac{25518080}{9} \Li_5\left(\frac{1}{2}\right)
        +\frac{67981760}{27} \zeta_5
\nonumber\\ &&
        +\frac{637952}{27} \ln^5(2)
        +\frac{22328320}{27} \ln^3(2) \zeta_2
        -\frac{6857984}{9} \ln(2) \zeta_2^2
        +\frac{2791040}{27} \zeta_2 \zeta_3
\nonumber\\ &&
        -\frac{112028672}{81} \Li_4\left(\frac{1}{2}\right) 
        -\frac{14003584}{243} \ln^4(2)
        -\frac{98025088}{81} \ln^2(2) \zeta_2
        +\frac{150538528}{405} \zeta_2^2
\nonumber\\ &&
        +\frac{108366944}{81} \ln(2) \zeta_2
        -\frac{379284304}{729} \zeta_3
        -\frac{387541868}{729} \zeta_2
\Biggr)
+ \textcolor{blue}{\ep^7} \Biggr(
\frac{97480790072029}{60466176}
\nonumber\\ &&
-\frac{2183312384}{105} s_{8a}
-\frac{897384448}{7} s_{8b}
-\frac{101974016}{7} s_{8c}
-\frac{79691776}{7} s_{8d}
+\frac{50987008}{7} \ln^2(2) s_6
\nonumber\\ &&
+\frac{250216448}{7} \zeta_2 s_6
-56623104 \Li_8\left(\frac{1}{2}\right)
+\frac{159383552}{7} \zeta_3 \Li_5\left(\frac{1}{2}\right)
-\frac{79691776}{35} \zeta_2^2 \Li_4\left(\frac{1}{2}\right)
\nonumber\\ &&
-\frac{49152}{35}      \ln^8(2)
-\frac{688128}{5} \ln^6(2) \zeta_2
-\frac{19922944}{105} \ln^5(2) \zeta_3
+\frac{56614912}{105} \ln^4(2) \zeta_2^2
\nonumber\\ &&
+\frac{32620544}{21} \ln^3(2) \zeta_2 \zeta_3
-\frac{4751360}{7} \ln^2(2) \zeta_2^3
+\frac{14340096}{7} \ln^2(2) \zeta_3^2
+\frac{572674048}{35} \ln(2) \zeta_2^2 \zeta_3
\nonumber\\ &&
+\frac{305018880}{7} \ln(2) \zeta_2 \zeta_5
-\frac{378638336}{21} \zeta_3 \zeta_5
+\frac{22032280576}{6125} \zeta_2^4
+\frac{331879424}{21} \zeta_2 \zeta_3^2
\nonumber\\ &&
-\frac{707264512}{63}  s_{7a}
+\frac{905019392}{63} s_{7b}
+\frac{707264512}{63} \ln(2) s_6
-55836672 \Li_7\left(\frac{1}{2}\right)
\nonumber\\ &&
+\frac{2565224320}{21} \zeta_7
+\frac{1163264}{105}   \ln^7(2)
+\frac{4071424}{5} \ln^5(2) \zeta_2
-\frac{12505088}{5} \ln^3(2) \zeta_2^2
\nonumber\\ &&
-24792064 \ln^2(2) \zeta_5
+1017856 \ln^2(2) \zeta_2 \zeta_3
-\frac{204443648}{45} \ln(2) \zeta_2^3
-\frac{884080640}{63} \ln(2) \zeta_3^2
\nonumber\\ &&
-\frac{2700862720}{63} \zeta_2 \zeta_5
-\frac{254136832}{63} \zeta_2^2 \zeta_3
-\frac{140349440}{9} s_6
-\frac{102072320}{3} \Li_6\left(\frac{1}{2}\right)
\nonumber\\ &&
-\frac{1275904}{27}    \ln^6(2)
-\frac{22328320}{9} \ln^4(2) \zeta_2
+\frac{13715968}{3} \ln^2(2) \zeta_2^2
-\frac{11164160}{9} \ln(2) \zeta_2 \zeta_3
\nonumber\\ &&
+\frac{493216640}{81} \zeta_3^2
+\frac{116107264}{27} \zeta_2^3
-\frac{448114688}{27} \Li_5\left(\frac{1}{2}\right)
+\frac{56014336}{405}  \ln^5(2)
\nonumber\\ &&
+\frac{392100352}{81} \ln^3(2) \zeta_2
-\frac{602154112}{135} \ln(2) \zeta_2^2
+\frac{1193805536}{81} \zeta_5
+\frac{49012544}{81} \zeta_2 \zeta_3
\nonumber\\ &&
-\frac{1733871104}{243} \Li_4\left(\frac{1}{2}\right)
-\frac{216733888}{729} \ln^4(2)
-\frac{1517137216}{243} \ln^2(2) \zeta_2
+\frac{2329889296}{1215} \zeta_2^2
\nonumber\\ &&
-\frac{5425586152}{2187} \zeta_3
+\frac{1550167472}{243} \ln(2) \zeta_2
-\frac{5263826150}{2187} \zeta_2
\Biggr) 
+ \textcolor{blue}{\ep^8} \Biggl(
\frac{2553476823634373}{362797056}
\nonumber\\ &&
-\frac{16011624448}{21} s_{9a}
+\frac{1157496832}{3} s_{9b}
+\frac{13187219456}{63} s_{9c}
-\frac{344457216}{7}s_{9d}
\nonumber\\ &&
-\frac{611844096}{7} s_{9e}
+\frac{7005710336}{35} \ln(2) s_{8a}
+\frac{26374438912}{21} \ln(2) s_{8b}
+\frac{5416157184}{49} s_{7a} \zeta_2
\nonumber\\ &&
-\frac{11319115776}{49} s_{7b} \zeta_2
+1117716480 s_6 \zeta_3
-\frac{101974016}{7} s_6 \ln^3(2)
-\frac{7557611520}{49} s_6 \ln(2) \zeta_2
\nonumber\\ &&
-679477248 \Li_9\left(\frac{1}{2}\right)
+\frac{1912602624}{7}  \Li_6\left(\frac{1}{2}\right) \zeta_3
-\frac{4015521792}{35} \Li_5\left(\frac{1}{2}\right) \zeta_2^2
\nonumber\\ &&
+\frac{1722286080}{7} \Li_4\left(\frac{1}{2}\right) \zeta_2 \zeta_3
-\frac{13673512960}{7} \Li_4\left(\frac{1}{2}\right)  \zeta_5
-\frac{380773366144}{63} \zeta_9
+\frac{65536}{35} \ln^9(2)
\nonumber\\ &&
+\frac{1179648}{5} \ln^7(2) \zeta_2
+\frac{39845888}{105} \ln^6(2) \zeta_3
-\frac{98992128}{175} \ln^5(2) \zeta_2^2
+\frac{39141376}{7} \ln^4(2) \zeta_2 \zeta_3
\nonumber\\ &&
-\frac{1709189120}{21} \ln^4(2) \zeta_5
-\frac{249135104}{35} \ln^3(2) \zeta_2^3
-\frac{28680192}{7} \ln^3(2) \zeta_3^2
\nonumber\\ &&
-\frac{4011491328}{35} \ln^2(2) \zeta_2^2 \zeta_3
+\frac{4822073344}{7} \ln^2(2) \zeta_2 \zeta_5
+\frac{15886533632}{21} \ln^2(2) \zeta_7
\nonumber\\ &&
-\frac{3435955044352}{6125} \ln(2) \zeta_2^4
+\frac{6177746944}{49} \ln(2) \zeta_2 \zeta_3^2
-\frac{32555948032}{21} \ln(2) \zeta_3 \zeta_5
\nonumber\\ &&
+\frac{61344280576}{245} \zeta_2^3 \zeta_3
-\frac{7109786624}{27} \zeta_3^3
+\frac{1581235817984}{735} \zeta_2^2 \zeta_5
+\frac{78334057728}{49} \zeta_2 \zeta_7
\nonumber\\ &&
-\frac{77507589632}{315} s_{8a}
-\frac{31857147904}{21} s_{8b}
-\frac{3620077568}{21} s_{8c}
-\frac{2829058048}{21} s_{8d}
\nonumber\\ &&
-670040064 \Li_8\left(\frac{1}{2}\right)
+\frac{1810038784}{21} s_6 \ln^2(2) 
+\frac{8882683904}{21} s_6 \zeta_2
+\frac{5658116096}{21} \Li_5\left(\frac{1}{2}\right) \zeta_3
\nonumber\\ &&
-\frac{2829058048}{105} \Li_4\left(\frac{1}{2}\right) \zeta_2^2
-\frac{581632}{35} \ln^8(2)
-\frac{392100352}{27} \ln^4(2) \zeta_2
-\frac{8142848}{5} \ln^6(2) \zeta_2
\nonumber\\ &&
-\frac{707264512}{315} \ln^5(2) \zeta_3
+\frac{2009829376}{315} \ln^4(2) \zeta_2^2
+\frac{1158029312}{63} \ln^3(2) \zeta_2 \zeta_3
\nonumber\\ &&
-\frac{168673280}{21} \ln^2(2) \zeta_2^3
+\frac{169691136}{7} \ln^2(2) \zeta_3^2
+\frac{20329928704}{105} \ln(2) \zeta_2^2 \zeta_3
\nonumber\\ &&
+\frac{3609390080}{7} \ln(2) \zeta_2 \zeta_5
+\frac{782145960448}{18375} \zeta_2^4
+\frac{11781719552}{63} \zeta_2 \zeta_3^2
-\frac{13441660928}{63} \zeta_3 \zeta_5
\nonumber\\ &&
-\frac{2216427520}{27}  s_{7a}
+\frac{2836152320}{27}  s_{7b}
-408289280 \Li_7\left(\frac{1}{2}\right)
+\frac{2216427520}{27} \ln(2) s_6
\nonumber\\ &&
+\frac{8038907200}{9} \zeta_7
+\frac{729088}{9} \ln^7(2)
+\frac{17862656}{3} \ln^5(2) \zeta_2
-\frac{54863872}{3} \ln^3(2) \zeta_2^2
\nonumber\\ &&
+\frac{22328320}{3} \ln^2(2) \zeta_2 \zeta_3
-\frac{543854080}{3} \ln^2(2) \zeta_5
-\frac{896960512}{27} \ln(2) \zeta_2^3
-\frac{2770534400}{27} \ln(2) \zeta_3^2
\nonumber\\ &&
-\frac{796414720}{27} \zeta_2^2 \zeta_3
-\frac{8463971200}{27} \zeta_2 \zeta_5
-\frac{1792458752}{9} \Li_6\left(\frac{1}{2}\right)
-\frac{2464630784}{27}  s_6
\nonumber\\ &&
-\frac{112028672}{405} \ln^6(2)
+\frac{1204308224}{45} \ln^2(2) \zeta_2^2
-\frac{196050176}{27} \ln(2) \zeta_2 \zeta_3
+\frac{10194609152}{405} \zeta_2^3
\nonumber\\ &&
+\frac{8661216704}{243} \zeta_3^2
-\frac{6935484416}{81} \Li_5\left(\frac{1}{2}\right)
+\frac{866935552}{1215} \ln^5(2)
+\frac{6068548864}{243} \ln^3(2) \zeta_2
\nonumber\\ &&
-\frac{9319557184}{405} \ln(2) \zeta_2^2
+\frac{758568608}{243} \zeta_2 \zeta_3
+\frac{18476563952}{243} \zeta_5
-\frac{24802679552}{729} \Li_4\left(\frac{1}{2}\right)
\nonumber\\ &&
-\frac{3100334944}{2187} \ln^4(2)
-\frac{21702344608}{729} \ln^2(2) \zeta_2
+\frac{33328600648}{3645} \zeta_2^2
-\frac{73693566100}{6561} \zeta_3
\nonumber\\ &&
+\frac{21055304600}{729} \ln(2) \zeta_2
-\frac{69020223371}{6561} \zeta_2
\Biggr) 
+ \textcolor{blue}{\ep^9} \Biggl(
\frac{65282718863433709}{2176782336}
\nonumber\\ &&
+\frac{14684258304}{17} h_{331111}
-\frac{19013284397056}{54723} h_{3331}
-\frac{24574427136}{17} h_{511111}
\nonumber\\ &&
-\frac{407096785895424}{127687} h_{5113}
+\frac{38026568794112}{18241} h_{5131}
+\frac{128886518054912}{127687} h_{5311}
\nonumber\\ &&
-\frac{77744179249152}{7511} h_{7111}
+\frac{982846081352335360}{606385563} h_{73}
+\frac{498389163900928000}{5462933} h_{91}
\nonumber\\ &&
-8153726976 \Li_{10}\left(\frac{1}{2}\right)
-8040480768               \Li_9\left(\frac{1}{2}\right)
-4899471360                \Li_8\left(\frac{1}{2}\right)
\nonumber\\ &&
-\frac{7169835008}{3}      \Li_7\left(\frac{1}{2}\right)
-\frac{27741937664}{27}    \Li_6\left(\frac{1}{2}\right)
-\frac{99210718208}{243}   \Li_5\left(\frac{1}{2}\right)
\nonumber\\ &&
-\frac{336884873600}{2187} \Li_4\left(\frac{1}{2}\right)
-\frac{393216}{175} \ln^{10}(2)
+\frac{2326528}{105} \ln^9(2)
-\frac{364544}{3}     \ln^8(2)
\nonumber\\ &&
+\frac{64016384}{135} \ln^7(2)
-\frac{1733871104}{1215} \ln^6(2)
+\frac{12401339776}{3645} \ln^5(2) 
-\frac{42110609200}{6561} \ln^4(2)
\nonumber\\ &&
-467140608 s_6 \ln^4(2)
+\frac{1912602624}{7} s_{7a} \ln^3(2)
-\frac{38921961472}{81} s_{7a}
-\frac{2447376384}{7}  s_{7b} \ln^3(2)
\nonumber\\ &&
+\frac{49804746752}{81} s_{7b}
-\frac{3620077568}{21} s_6 \ln^3(2)
+\frac{5672304640}{9} s_6 \ln^2(2)
\nonumber\\ &&
+\frac{38921961472}{81} s_6 \ln(2)
-\frac{38145164288}{81} s_6
-\frac{4524146688}{5} s_{8a} \ln^2(2)
\nonumber\\ &&
+\frac{248702716928}{105} s_{8a} \ln(2)
-\frac{48578700544}{27} s_{8a}
-\frac{99834019840}{9} s_{8b}
\nonumber\\ &&
+\frac{936292581376}{63} s_{8b} \ln(2)
-\frac{40886075392}{7} s_{8b} \ln^2(2)
-\frac{11344609280}{9}  s_{8c}
\nonumber\\ &&
+\frac{7342129152}{7} s_{8c} \ln^2(2)
-\frac{8865710080}{9} s_{8d}
+\frac{5737807872}{7} s_{8d} \ln^2(2)
-\frac{568412667904}{63} s_{9a}
\nonumber\\ &&
+\frac{64046497792}{7} s_{9a} \ln(2)
+\frac{41091137536}{9} s_{9b}
-4629987328 s_{9b} \ln(2)
\nonumber\\ &&
+\frac{468146290688}{189} s_{9c}
-\frac{52748877824}{21} s_{9c} \ln(2)
-\frac{4076077056}{7} s_{9d}
+\frac{4133486592}{7} s_{9d} \ln(2)
\nonumber\\ &&
-\frac{7240155136}{7} s_{9e}
+\frac{7342129152}{7} s_{9e} \ln(2)
-\frac{1766128887523}{39366} \zeta_2
\nonumber\\ &&
+\frac{276080893484}{2187} \ln(2) \zeta_2
-\frac{294774264400}{2187} \ln^2(2) \zeta_2
+\frac{86809378432}{729} \ln^3(2) \zeta_2
\nonumber\\ &&
-\frac{6068548864}{81} \ln^4(2) \zeta_2
+\frac{1568401408}{45} \ln^5(2) \zeta_2
-\frac{35725312}{3}    \ln^6(2) \zeta_2
+\frac{13959168}{5} \ln^7(2) \zeta_2
\nonumber\\ &&
-\frac{1769472}{5} \ln^8(2) \zeta_2
+\frac{27836579840}{9} s_6 \zeta_2
-\frac{89431736320}{49} s_6 \ln(2) \zeta_2
-\frac{35417751552}{49} s_6 \ln^2(2) \zeta_2
\nonumber\\ &&
+\frac{64091193344}{49} s_{7a} \zeta_2
-\frac{64993886208}{49} s_{7a} \ln(2) \zeta_2
-\frac{133942870016}{49} s_{7b} \zeta_2
\nonumber\\ &&
+\frac{135829389312}{49} \ln(2) s_{7b} \zeta_2
-\frac{42561416368128}{18241} s_{8a} \zeta_2
+\frac{90537809780}{2187} \zeta_2^2
\nonumber\\ &&
-\frac{47517007872}{35} \Li_5\left(\frac{1}{2}\right) \zeta_2^2
-\frac{1773142016}{9} \Li_4\left(\frac{1}{2}\right) \zeta_2^2
-\frac{133314402592}{1215} \ln(2) \zeta_2^2
\nonumber\\ &&
+\frac{48186261504}{35} \Li_5\left(\frac{1}{2}\right) \ln(2) \zeta_2^2
+\frac{18639114368}{135} \ln^2(2) \zeta_2^2
+\frac{5737807872}{35} \Li_4\left(\frac{1}{2}\right) \ln^2(2) \zeta_2^2
\nonumber\\ &&
-\frac{4817232896}{45} \ln^3(2) \zeta_2^2
+\frac{1259681792}{27} \ln^4(2) \zeta_2^2
-\frac{1171406848}{175} \ln^5(2) \zeta_2^2
\nonumber\\ &&
-\frac{279773184}{175} \ln^6(2) \zeta_2^2
+\frac{50887098236928}{127687} s_6 \zeta_2^2
+\frac{157782270464}{1215} \zeta_2^3
-\frac{78756156416}{405} \ln(2) \zeta_2^3
\nonumber\\ &&
-\frac{528588800}{9} \ln^2(2) \zeta_2^3
-\frac{8844296192}{105} \ln^3(2) \zeta_2^3
+\frac{7067271168}{35} \ln^4(2) \zeta_2^3
+\frac{490218242816}{1575} \zeta_2^4
\nonumber\\ &&
-\frac{121976404074496}{18375} \ln(2) \zeta_2^4
+\frac{39645136330752}{6125} \ln^2(2) \zeta_2^4
+\frac{262144173134771720192}{23822289975} \zeta_2^5
\nonumber\\ &&
-\frac{966283127194}{19683} \zeta_3
+\frac{17731420160}{9} \Li_5\left(\frac{1}{2}\right) \zeta_3
+\frac{22632464384}{7} \Li_6\left(\frac{1}{2}\right)   \zeta_3
\nonumber\\ &&
-\frac{22951231488}{7} \Li_6\left(\frac{1}{2}\right) \ln(2) \zeta_3
-\frac{11475615744}{7} \Li_5\left(\frac{1}{2}\right)  \ln^2(2) \zeta_3
-\frac{443285504}{27} \ln^5(2) \zeta_3
\nonumber\\ &&
+\frac{1414529024}{315} \ln^6(2) \zeta_3
+ \frac{318767104}{35} \ln^7(2) \zeta_3
+13226311680 s_6 \zeta_3
\nonumber\\ &&
-13412597760 s_6 \ln(2) \zeta_3
-\frac{305322589421568}{127687} s_{7a} \zeta_3
+\frac{10851172304}{729} \zeta_2 \zeta_3
\nonumber\\ &&
+\frac{20380385280}{7} \Li_4\left(\frac{1}{2}\right)  \zeta_2 \zeta_3
-\frac{3034274432}{81} \ln(2) \zeta_2 \zeta_3
-\frac{20667432960}{7} \Li_4\left(\frac{1}{2}\right) \ln(2) \zeta_2 \zeta_3
\nonumber\\ &&
+\frac{392100352}{9} \ln^2(2) \zeta_2 \zeta_3
+\frac{3629035520}{27} \ln^3(2) \zeta_2 \zeta_3
+\frac{1389518848}{21} \ln^4(2) \zeta_2 \zeta_3
\nonumber\\ &&
-\frac{6783172608}{35} \ln^5(2) \zeta_2 \zeta_3
-\frac{13985579392}{81} \zeta_2^2 \zeta_3
+\frac{12741997568}{9} \ln(2) \zeta_2^2 \zeta_3
\nonumber\\ &&
-\frac{47469314048}{35} \ln^2(2) \zeta_2^2 \zeta_3
+\frac{2068316160}{7} \ln^3(2) \zeta_2^2 \zeta_3
+\frac{2177721960448}{735} \zeta_2^3 \zeta_3
\nonumber\\ &&
-\frac{736131366912}{245} \ln(2) \zeta_2^3 \zeta_3
+\frac{134049909728}{729} \zeta_3^2
-\frac{48652451840}{81} \ln(2) \zeta_3^2
\nonumber\\ &&
+177259520 \ln^2(2) \zeta_3^2
-\frac{339382272}{7} \ln^3(2) \zeta_3^2
+182403072 \ln^4(2) \zeta_3^2
+\frac{36921585920}{27} \zeta_2 \zeta_3^2
\nonumber\\ &&
+\frac{219310016512}{147} \ln(2) \zeta_2 \zeta_3^2
-\frac{129888706560}{49} \ln^2(2) \zeta_2 \zeta_3^2
-\frac{9358852028348416}{1915305} \zeta_2^2 \zeta_3^2
\nonumber\\ &&
-\frac{252397425152}{81} \zeta_3^3
+\frac{28439146496}{9} \ln(2) \zeta_3^3
+\frac{264303553976}{729} \zeta_5
\nonumber\\ &&
-\frac{485409710080}{21} \Li_4\left(\frac{1}{2}\right) \zeta_5
+\frac{164082155520}{7} \Li_4\left(\frac{1}{2}\right)  \ln(2) \zeta_5
-\frac{9550444288}{9} \ln^2(2) \zeta_5
\nonumber\\ &&
-\frac{60676213760}{63} \ln^4(2) \zeta_5
+\frac{53190565888}{35} \ln^5(2) \zeta_5
-\frac{148633040320}{81} \zeta_2 \zeta_5
\nonumber\\ &&
+\frac{11311116800}{3} \ln(2) \zeta_2 \zeta_5
+\frac{171183603712}{21} \ln^2(2) \zeta_2 \zeta_5
-10360356864 \ln^3(2) \zeta_2 \zeta_5
\nonumber\\ &&
+\frac{56133871538432}{2205} \zeta_2^2 \zeta_5
-\frac{118824071572576256}{4469045} \ln(2) \zeta_2^2 \zeta_5
-\frac{42123514880}{27} \zeta_3 \zeta_5
\nonumber\\ &&
-\frac{1155736155136}{63} \ln(2) \zeta_3 \zeta_5
+19901587456 \ln^2(2) \zeta_3 \zeta_5
-\frac{2399840466176000}{127687} \zeta_2 \zeta_3 \zeta_5
\nonumber\\ &&
-\frac{186051331138705408}{6969949} \zeta_5^2
+\frac{141168629920}{27} \zeta_7
+\frac{563971943936}{63} \ln^2(2) \zeta_7
\nonumber\\ &&
-\frac{84356968448}{7} \ln^3(2) \zeta_7
+\frac{926953016448}{49} \zeta_2 \zeta_7
-\frac{940008692736}{49} \ln(2) \zeta_2 \zeta_7
\nonumber\\ &&
-\frac{18265279696831990784}{606385563} \zeta_3 \zeta_7
-\frac{13517454498112}{189} \zeta_9
+\frac{1523093464576}{21} \ln(2) \zeta_9
\Biggr)
\end{eqnarray}
\begin{align}
I_{12} &= \frac{2}{\textcolor{blue}{\ep^3}}
+\frac{23}{3 \textcolor{blue}{\ep^2}}
+\frac{35}{2 \textcolor{blue}{\ep}}
+\frac{275}{12}
+\frac{7}{24} \textcolor{blue}{\ep} \big(
        -81+128 \zeta_3\big)
+\textcolor{blue}{\ep^2} \Biggl(
        -\frac{14917}{48}
        -\frac{136 \pi ^4}{45}
        -\frac{32}{3} \pi ^2 \ln^2(2)
\nonumber\\ &  
      +\frac{32 \ln^4(2)}{3}
        +256 \Li_4\left(\frac{1}{2}\right)
        +280 \zeta_3
\Biggr)
+\textcolor{blue}{\ep^3} \Biggl(
        -\frac{48005}{32}
        -\frac{68 \pi ^4}{3}
        +\frac{272 \pi ^4 \ln(2)}{15}
        -80 \pi ^2 \ln^2(2)
\nonumber\\ & 
        +\frac{64}{3} \pi ^2 \ln^3(2)
        +80 \ln^4(2)
        -\frac{64 \ln^5(2)}{5}
       +1920 \Li_4\left(\frac{1}{2}\right)
        +1536 \AXF
        +\frac{4060 \zeta_3}{3}
        -1240 \zeta_5
\big)
\nonumber\\ & 
+ \textcolor{blue}{\ep^4} \Biggl(
        -\frac{1108525}{192}
        -\frac{986 \pi ^4}{9}
        -\frac{32 \pi ^6}{5}
        +136 \pi ^4 \ln(2)
        -\frac{1160}{3} \pi ^2 \ln^2(2)
        -\frac{272}{5} \pi ^4 \ln^2(2)
\nonumber\\ & 
       +160 \pi ^2 \ln^3(2)
        +\frac{1160 \ln^4(2)}{3}
        -32 \pi ^2 \ln^4(2)
        -96 \ln^5(2)
        +\frac{64 \ln^6(2)}{5}
        +9280 \Li_4\left(\frac{1}{2}\right)
\nonumber\\ & 
        +11520 \AXF
+9216 \Li_6\left(\frac{1}{2}\right) 
+3840 s_6
        +5390 \zeta_3
        -\frac{4880 \zeta_3^2}{3}
        -9300 \zeta_5
\Biggr)
\nonumber\\ &
+\textcolor{blue}{\ep^5}
\Biggl(
        -\frac{2570029}
        {128}
        -
        \frac{1309 \pi ^4}{3}
        -48 \pi ^6
        +\frac{1972 \pi ^4 \ln(2)}{3}
        +\frac{3824 \pi ^6 \ln(2)}{135}
        -1540 \pi ^2 \ln^2(2)
\nonumber\\ & 
        -408 \pi ^4 \ln^2(2)
        +\frac{2320}{3} \pi ^2 \ln^3(2)
        +\frac{544}{5} \pi ^4 \ln^3(2)
        +1540 \ln^4(2)
        -240 \pi ^2 \ln^4(2)
        -464 \ln^5(2)
\nonumber\\ & 
        +\frac{192}{5} \pi ^2 \ln^5(2)
      +96 \ln^6(2)
        -\frac{384 \ln^7(2)}{35}
        +36960 \Li_4\left(\frac{1}{2}\right)
        +55680 \AXF
        +69120 \Li_6\left(\frac{1}{2}\right)
\nonumber\\ & 
        +55296 \AXS
        +28800 s_6
       -\frac{74240}{7} \ln(2) s_6
        +\frac{74240 s_{\text{7a}}}{7}
        -\frac{87040 s_{\text{7b}}}{7}
        +\frac{57967 \zeta_3}{3}
        +\frac{720 \pi ^4 \zeta_3}{7}
\nonumber\\ & 
        -12200 \zeta_3^2
       +\frac{92800}{7} \ln(2) \zeta_3^2
        -44950 \zeta_5
        +\frac{130360 \pi ^2 \zeta_5}{21}
        +22320 \ln^2(2) \zeta_5
        -\frac{772868 \zeta_7}{7}
\Biggr)
\nonumber\\ &
+\textcolor{blue}{\ep^6} \Biggl(
        -\frac{50743957}{768}
        -\frac{140777 \pi ^4}{90}
        -232 \pi ^6
        -\frac{593716 \pi ^8}{33075}
        +2618 \pi ^4 \ln(2)
        +\frac{1912 \pi ^6 \ln(2)}{9}
\nonumber\\ & 
 -\frac{16562}{3} \pi ^2 \ln^2(2)
        -1972 \pi ^4 \ln^2(2)
        -\frac{3328}{315} \pi ^6 \ln^2(2)
        +3080 \pi ^2 \ln^3(2)
        +816 \pi ^4 \ln^3(2)
\nonumber\\ & 
        +\frac{16562 \ln^4(2)}
        {3}
      -1160 \pi ^2 \ln^4(2)
        -
        \frac{46768}{315} \pi ^4 \ln^4(2)
        -1848 \ln^5(2)
        +288 \pi ^2 \ln^5(2)
        +464 \ln^6(2)
\nonumber\\ &      
  -\frac{192}{5} \pi ^2 \ln^6(2)
        -\frac{576 \ln^7(2)}{7}
     +\frac{288 \ln^8(2)}{35}
        +132496 \Li_4\left(\frac{1}{2}\right)
        +\frac{7424 \pi ^4 \Li_4\left(\frac{1}{2}\right)}{21}
\nonumber\\ & 
        +221760 \AXF
        +334080 \Li_6\left(\frac{1}{2}\right)
        +414720 \AXS
    +331776 \Li_8\left(\frac{1}{2}\right)
        +139200 s_6
\nonumber\\ & 
        -\frac{229120}{7} \pi ^2 s_6
        -\frac{556800}{7} \ln(2) s_6
        -\frac{261120}{7} \ln^2(2) s_6
        +\frac{556800 s_{\text{7a}}}{7}       -\frac{652800 s_{\text{7b}}}{7}
\nonumber\\ &     
    +\frac{767504 s_{\text{8a}}}{7}
        +\frac{4862976 s_{\text{8b}}}{7}
        +\frac{522240 s_{\text{8c}}}{7}
        +\frac{445440 s_{\text{8d}}}{7}
        +\frac{130095 \zeta_3}{2}
       +\frac{5400 \pi ^4 \zeta_3}{7}
\nonumber\\ &        
 -\frac{51952}{21} \pi ^4 \ln(2) \zeta_3
        -\frac{37120}{21} \pi ^2 \ln^3(2) \zeta_3
        +\frac{7424}{7} \ln^5(2) \zeta_3
        -\frac{890880}{7} \AXF \zeta_3
        -\frac{176900 \zeta_3^2}{3}
\nonumber\\ & 
       -\frac{96080}{7} \pi ^2 \zeta_3^2
        +\frac{696000}{7} \ln(2) \zeta_3^2
        -\frac{73440}{7} \ln^2(2) \zeta_3^2
        -179025 \zeta_5
        +\frac{325900 \pi ^2 \zeta_5}{7}
\nonumber\\ & 
       -\frac{261120}{7} \pi ^2 \ln(2) \zeta_5
        +167400 \ln^2(2) \zeta_5
        +\frac{636208 \zeta_3 \zeta_5}{7}
        -\frac{5796510 \zeta_7}{7}
\Biggr)
\nonumber\\ & + \textcolor{blue}{\ep^7}
\Biggl(
        -\frac{107716245}{512}
        -\frac{21063 \pi ^4}{4}
        -924 \pi ^6
        -\frac{296858 \pi ^8}{2205}
        +\frac{140777 \pi ^4 \ln(2)}{15}
        +\frac{27724 \pi ^6 \ln(2)}{27}
\nonumber\\ & 
      +\frac{38646953 \pi ^8 \ln(2)}{33075}
        -18585 \pi ^2 \ln^2(2)
        -7854 \pi ^4 \ln^2(2)
        -\frac{1664}{21} \pi ^6 \ln^2(2)
        +\frac{33124}{3} \pi ^2 \ln^3(2)
\nonumber\\ &
        +3944 \pi ^4 \ln^3(2)
      +\frac{43376}{315} \pi ^6 \ln^3(2)
        +18585 \ln^4(2)
        -4620 \pi ^2 \ln^4(2)
        -\frac{23384}
        {21} \pi ^4 \ln^4(2)
\nonumber\\ &        
 - \frac{33124 \ln^5(2)}{5}
        +1392 \pi ^2 \ln^5(2)
      +\frac{22112}{175} \pi ^4 \ln^5(2)
        +1848 \ln^6(2)
        -288 \pi ^2 \ln^6(2)
\nonumber\\ &        
 -\frac{2784 \ln^7(2)}{7}
        +\frac{1152}{35} \pi ^2 \ln^7(2)
        +\frac{432 \ln^8(2)}{7}
        -\frac{192 \ln^9(2)}{35}
        +446040 \Li_4\left(\frac{1}{2}\right)
\nonumber\\ &         
+\frac{18560 \pi ^4 \Li_4\left(\frac{1}{2}\right)}{7}
        +794976 \AXF
        +\frac{58368 \pi ^4}{7} \AXF
        +1330560 \Li_6\left(\frac{1}{2}\right)
\nonumber\\ &       
  +2004480 \AXS
      +2488320 \Li_8\left(\frac{1}{2}\right)
        +1990656 \AXN
        +554400 s_6
        -\frac{1718400}{7} \pi ^2 s_6
\nonumber\\ &         
-\frac{2691200}{7} \ln(2) s_6
        +\frac{3878400}{49} \pi ^2 \ln(2) s_6
        -\frac{1958400}{7} \ln^2(2) s_6
        +\frac{261120}{7} \ln^3(2) s_6
\nonumber\\ &       
  +\frac{2691200 s_{\text{7a}}}{7}
        -\frac{2964480}{49} \pi ^2 s_{\text{7a}}
        -\frac{3155200 s_{\text{7b}}}{7}
        +\frac{4830720 \pi ^2 s_{\text{7b}}}{49}
        +\frac{5756280 s_{\text{8a}}}{7}
\nonumber\\ &        
 -\frac{3787464}{7} \ln(2) s_{\text{8a}}
        +\frac{36472320 s_{\text{8b}}}{7}
        -\frac{23764480}{7} \ln(2) s_{\text{8b}}
        +\frac{3916800 s_{\text{8c}}}{7}
        +\frac{3340800 s_{\text{8d}}}{7}
\nonumber\\ &     
    +\frac{14821888 s_{\text{9a}}}{7}
        -951808 s_{\text{9b}}
       -\frac{11882240 s_{\text{9c}}}{21}
        +\frac{1105920 s_{\text{9d}}}{7}
        +\frac{1566720 s_{\text{9e}}}{7}
        +\frac{2526055 \zeta_3}
        {12}
\nonumber\\ &       
  +
        \frac{26100 \pi ^4 \zeta_3}{7}
        -\frac{18482398 \pi ^6 \zeta_3}{6615}
        -\frac{129880}{7} \pi ^4 \ln(2) \zeta_3
        +\frac{67912}{7} \pi ^4 \ln^2(2) \zeta_3
        -\frac{92800}{7} \pi ^2 \ln^3(2) \zeta_3
\nonumber\\ &     
    -\frac{19840}{7} \pi ^2 \ln^4(2) \zeta_3
        +\frac{55680}{7} \ln^5(2) \zeta_3
        -\frac{7424}{7} \ln^6(2) \zeta_3
        -\frac{921600}{7} \pi ^2 \Li_4\left(\frac{1}{2}\right) \zeta_3
\nonumber\\ &        
 -\frac{6681600}{7} \AXF \zeta_3
        -\frac{5345280}{7} \Li_6\left(\frac{1}{2}\right) \zeta_3
       -3011328 s_6 \zeta_3
        -234850 \zeta_3^2
        -\frac{720600}{7} \pi ^2 \zeta_3^2
\nonumber\\ &        
 +\frac{3364000}{7} \ln(2) \zeta_3^2
        -\frac{3448560}{49} \pi ^2 \ln(2) \zeta_3^2
        -\frac{550800}{7} \ln^2(2) \zeta_3^2
        +\frac{73440}{7} \ln^3(2) \zeta_3^2
\nonumber\\ &      
   +\frac{6408524 \zeta_3^3}{9}
        -\frac{1283555 \zeta_5}{2}
        +\frac{4725550 \pi ^2 \zeta_5}{21}
       -\frac{690281971 \pi ^4 \zeta_5}{4410}
        -\frac{1958400}{7} \pi ^2 \ln(2) \zeta_5
\nonumber\\ &       
  +809100 \ln^2(2) \zeta_5
        -\frac{2232728}{7} \pi ^2 \ln^2(2) \zeta_5
        +\frac{1545608}{7} \ln^4(2) \zeta_5
        +\frac{37094592 \Li_4\left(\frac{1}{2}\right) \zeta_5}{7}
\nonumber\\ &        
 +\frac{4771560 \zeta_3 \zeta_5}{7}
        +\frac{29334280}{7} \ln(2) \zeta_3 \zeta_5
        -\frac{28016465 \zeta_7}{7}
        -\frac{70452569}{98}
         \pi ^2 \zeta_7
\nonumber\\ &        
-\frac{14706092}{7} \ln^2(2) \zeta_7
        +\frac{216126121 \zeta_9}{14}
\Biggr)
\end{align}
\begin{eqnarray}
I_{15} &=& 
\frac{1}{2 \textcolor{blue}{\ep^3}}
+\frac{7}{4 \textcolor{blue}{\ep^2}}
+\frac{1}{24 \textcolor{blue}{\ep}} \Biggl(75+8 \pi ^2\Biggr)
-\frac{5}{16}
+\frac{7 \pi ^2}{6}
+4 \zeta_3
+\textcolor{blue}{\ep} \Biggl(
        -\frac{959}{32}
        +\frac{25 \pi ^2}{12}
        +\frac{16 \pi ^4}{45}
        +14 \zeta_3
\Biggr)
\nonumber\\ && 
+\textcolor{blue}{\ep^2} \Biggl(
        -\frac{10493}{64}
        +\frac{56 \pi ^4}{45}
        +\pi ^2 \Biggl(
                -\frac{5}{24}
                +\frac{8 \zeta_3}{3}
        \Biggr)
        +25 \zeta_3
        +72 \zeta_5
\Biggr)
+\textcolor{blue}{\ep^3} \Biggl(
        -\frac{85175}{128}
        +\frac{20 \pi ^4}{9}
\nonumber\\ && 
        +\frac{458 \pi ^6}{945}
        +\pi ^2 \big(
                -\frac{959}{48}
                +\frac{28 \zeta_3}{3}
        \Biggr)
        -\frac{5}{2} \zeta_3
        +16 \zeta_3^2
        +252 \zeta_5
\Biggr)
+\textcolor{blue}{\ep^4} \Biggl(
        -\frac{610085}{256}
        +\frac{229 \pi ^6}{135}
\nonumber\\ && 
        +\pi ^2 \Biggl(
                -\frac{10493}{96}
                +\frac{50 \zeta_3}{3}
                +48 \zeta_5
        \Biggr)
        +\pi ^4 \Biggl(
                -\frac{2}{9}
                +\frac{128 \zeta_3}{45}
        \Biggr)
        -\frac{959}{4} \zeta_3
        +56 \zeta_3^2
        +450 \zeta_5
\nonumber\\ && 
        +996 \zeta_7
\Biggr)
+\textcolor{blue}{\ep^5} \Biggl(
        -\frac{4087919}{512}
        +\frac{1145 \pi ^6}{378}
        +\frac{3337 \pi ^8}{4725}
        +\pi ^2 \Biggl(
                -\frac{85175}{192}
                -\frac{5}{3} \zeta_3
                +\frac{32}{3} \zeta_3^2
\nonumber\\ && 
                +168 \zeta_5
        \Biggr)
        +\pi ^4 \Biggl(
                -\frac{959}
                {45}
                +
                \frac{448 \zeta_3}{45}
        \Biggr)
        +\Biggl(
                -\frac{10493}{8}
                +576 \zeta_5
        \Biggr) \zeta_3
        +100 \zeta_3^2
        -45 \zeta_5
        +3486 \zeta_7
\Biggr)
\nonumber\\ && 
+\textcolor{blue}{\ep^6} \Biggl(
        -\frac{26332493}{1024}
        +\frac{3337 \pi ^8}{1350}
        +\pi ^4 \Biggl(
                -\frac{10493}{90}
                +\frac{160 \zeta_3}{9}
                +\frac{256 \zeta_5}{5}
        \Biggr)
        +\pi ^2 \Biggl(
                -\frac{610085}{384}
\nonumber\\ && 
                -\frac{959}{6} \zeta_3
                +\frac{112}{3} \zeta_3^2
                +300 \zeta_5
                +664 \zeta_7
        \Biggr)
        +\pi ^6 \Biggl(
                -\frac{229}{756}
                +\frac{3664 \zeta_3}{945}
        \Biggr)
        -\Biggl(
                \frac{85175}{16}
                -2016 \zeta_5
        \Biggr) \zeta_3
\nonumber\\ && 
        -10 \zeta_3^2
        +\frac{128}{3} \zeta_3^3
        -\frac{8631}{2} \zeta_5
        +6225 \zeta_7
        +\frac{40240}{3} \zeta_9
\Biggr)
+\textcolor{blue}{\ep^7} \Biggl(
        -\frac{165503975}{2048}
        +\frac{3337 \pi ^8}{756}
\nonumber\\ && 
        +\frac{70336 \pi ^{10}}{66825}
        +\pi ^4 \Biggl(
                -\frac{17035}{36}
                -\frac{16}{9} \zeta_3
                +\frac{512}{45} \zeta_3^2
                +\frac{896}{5} \zeta_5
        \Biggr)
        +\pi ^2 \Biggl(
                -\frac{4087919}{768}
+\Biggl(
                        -\frac{10493}{12}
\nonumber\\ &&                 
       +384 \zeta_5
                \Biggr) \zeta_3
                +\frac{200}{3} \zeta_3^2
                -30 \zeta_5
                +2324 \zeta_7
        \Biggr)
        +\pi ^6 \Biggl(
                -\frac{31373}{1080}
                +\frac{1832 \zeta_3}
                {135}
        \Biggr)
        +\Biggl(
                -
                \frac{610085}{32}
\nonumber\\ && 
                +3600 \zeta_5
                +7968 \zeta_7
        \Biggr) \zeta_3
        -959 \zeta_3^2
        +\frac{448}{3} \zeta_3^3
        -\frac{94437}{4} \zeta_5
        +5184 \zeta_5^2
        -\frac{1245}{2} \zeta_7
        +\frac{140840}{3} \zeta_9
\Biggr)
\nonumber\\ && 
+\textcolor{blue}{\ep^8} \Biggl(
        -\frac{1023933365}{4096}
        +\frac{246176 \pi ^{10}}{66825}
        +\pi ^6 \Biggl(
                -\frac{343271}{2160}
                +\frac{4580 \zeta_3}{189}
                +\frac{7328 \zeta_5}{105}
        \Biggr)
\nonumber\\ && 
        +\pi ^4 \Biggl(
                -\frac{122017}{72}
                -\frac{7672}{45} \zeta_3
                +\frac{1792}{45} \zeta_3^2
                +320 \zeta_5
                +\frac{10624}{15} \zeta_7
        \Biggr)
        +\pi ^2 \Biggl(
                -\frac{26332493}{1536}
\nonumber\\ &&                
 +\Biggl(
                        -\frac{85175}{24}
                        +1344 \zeta_5
                \Biggr) \zeta_3
                -\frac{20}{3} \zeta_3^2
                +\frac{256}{9} \zeta_3^3
                -2877 \zeta_5
                +4150 \zeta_7
                +\frac{80480}{9} \zeta_9
        \Biggr)
\nonumber\\ && 
        +\pi ^8 \Biggl(
                -\frac{3337}{7560}
                +\frac{26696 \zeta_3}{4725}
        \Biggr)
        +\Biggl(
                -\frac{4087919}{64}
                -360 \zeta_5
                +27888 \zeta_7
        \Biggr) \zeta_3
        +\Biggl(
                -\frac{10493}{2}
\nonumber\\ && 
                +2304 \zeta_5
        \Biggr) \zeta_3^2
        +\frac{800}{3} \zeta_3^3
        -\frac{766575}{8} \zeta_5
        +18144 \zeta_5^2
        -\frac{238791}{4} \zeta_7
        +\frac{251500}{3} \zeta_9
        +182412 \zeta_{11}
\Biggr)
\nonumber\\ 
\\
I_{16} &=&
-\frac{1}{6 \textcolor{blue}{\ep^2}}
-\frac{35}{36 \textcolor{blue}{\ep}}
-\frac{559}{216}
-\frac{\pi ^2}{3}
+\textcolor{blue}{\ep} \Biggl(
        \frac{2737}{1296}
        -\frac{35 \pi ^2}{18}
        -\frac{16}{3} \zeta_3
\Biggr)
+\textcolor{blue}{\ep^2} \Biggl(
        \frac{552041}{7776}
        -\frac{559 \pi ^2}{108}
        -\frac{37 \pi ^4}{45}
\nonumber\\ && 
        -\frac{280}{9} \zeta_3
\Biggr)
+\textcolor{blue}{\ep^3} \Biggl(
        \frac{25027345}{46656}
        -\frac{259 \pi ^4}{54}
        +\pi ^2 \Biggl(
                \frac{2737}{648}
                -\frac{32 \zeta_3}{3}
        \Biggr)
        -\frac{2236}{27} \zeta_3
        -208 \zeta_5
\Biggr)
\nonumber\\ && 
+\textcolor{blue}{\ep^4} \Biggl(
        \frac{855963737}{279936}
        -\frac{20683 \pi ^4}{1620}
        -\frac{2318 \pi ^6}{945}
        +\pi ^2 \Biggl(
                \frac{552041}{3888}
                -\frac{560 \zeta_3}{9}
        \Biggr)
        +\frac{5474}{81} \zeta_3
        -\frac{256}{3} \zeta_3^2
\nonumber\\ && 
        -\frac{3640}{3} \zeta_5
\Biggr)
+\textcolor{blue}{\ep^5} \Biggl(
        \frac{25728647329}{1679616}
        -\frac{1159 \pi ^6}{81}
        +\pi ^2 \Biggl(
                \frac{25027345}{23328}
                -\frac{4472 \zeta_3}{27}
                -416 \zeta_5
        \Biggr)
\nonumber\\ && 
        +\pi ^4 \Biggl(
                \frac{101269}{9720}
                -\frac{1184 \zeta_3}{45}
        \Biggr)
        +\frac{552041}{243} \zeta_3
        -\frac{4480}{9} \zeta_3^2
        -\frac{29068}{9} \zeta_5
        -6168 \zeta_7
\Biggr)
\nonumber\\ && 
+\textcolor{blue}{\ep^6} \Biggl(
        \frac{718599153833}{10077696}
        -\frac{647881 \pi ^6}{17010}
        -\frac{37189 \pi ^8}{4725}
        +\pi ^2 \Biggl(
                \frac{855963737}{139968}
                +\frac{10948}{81} \zeta_3
                -\frac{512}{3} \zeta_3^2
\nonumber\\ && 
                -\frac{7280}{3} \zeta_5
        \Biggr)
        +\pi ^4 \Biggl(
                \frac{20425517}{58320}
                -\frac{4144 \zeta_3}{27}
        \Biggr)
        +\Biggl(
                \frac{25027345}{1458}
                -6656 \zeta_5
        \Biggr) \zeta_3
        -\frac{35776}
        {27} \zeta_3^2
\nonumber\\ && 
        +
        \frac{71162}{27} \zeta_5
        -35980 \zeta_7
\Biggr)
+\textcolor{blue}{\ep^7} \Biggl(
        \frac{19166358676465}{60466176}
        -\frac{37189 \pi ^8}{810}
        +\pi ^4 \Biggl(
                \frac{185202353}{69984}
\nonumber\\ && 
            -\frac{165464 \zeta_3}{405}
                -\frac{15392 \zeta_5}{15}
        \Biggr)
        +\pi ^2 \Biggl(
                \frac{25728647329}{839808}
                +\frac{1104082}{243} \zeta_3
                -\frac{8960}{9} \zeta_3^2
                -\frac{58136}{9} \zeta_5
\nonumber\\ &&               
  -12336 \zeta_7
        \Biggr)
        +\pi ^6 \Biggl(
                \frac{453169}{14580}
                -\frac{74176 \zeta_3}{945}
        \Biggr)
        +\Biggl(
                \frac{855963737}{8748}
                -\frac{116480 \zeta_5}{3}
        \Biggr) \zeta_3
        +\frac{87584}{81} \zeta_3^2
\nonumber\\ && 
        -\frac{8192}{9} \zeta_3^3
        +\frac{7176533}{81} \zeta_5
        -\frac{287326}{3} \zeta_7
        -\frac{1629280}{9} \zeta_9
\Biggr)
+\textcolor{blue}{\ep^8} \Biggl(
        \frac{495815946702713}{362797056}
\nonumber\\ && 
        -\frac{20788651 \pi ^8}{170100}
        -\frac{1738426 \pi ^{10}}{66825}
        +\pi ^4 \Biggl(
                \frac{31670658269}{2099520}
                +\frac{405076 \zeta_3}{1215}
                -\frac{18944}{45} \zeta_3^2
                -\frac{53872}{9} \zeta_5
        \Biggr)
\nonumber\\ && 
        +\pi ^2 \Biggl(
                \frac{718599153833}{5038848}
                +\Biggl(
                        \frac{25027345}{729}
                        -13312 \zeta_5
                \Biggr) \zeta_3
                -\frac{71552}{27} \zeta_3^2
                +\frac{142324}{27} \zeta_5
                -71960 \zeta_7
        \Biggr)
\nonumber\\ && 
        +\pi ^6 \Biggl(
                \frac{91402217}{87480}
                -\frac{37088 \zeta_3}{81}
        \Biggr)
        +\Biggl(
                \frac{25728647329}{52488}
                -\frac{930176 \zeta_5}{9}
                -197376 \zeta_7
        \Biggr) \zeta_3
\nonumber\\ &&         
        +\frac{8832656}{243} \zeta_3^2
-\frac{143360}{27} \zeta_3^3
        +\frac{325355485}
        {486} \zeta_5
        -129792 \zeta_5^2
        +\frac{703409}{9} \zeta_7
        -\frac{28512400}{27} \zeta_9
\Biggr)
\nonumber\\
\end{eqnarray}

\vspace{2ex}
\noindent
{\bf Acknowledgment.}~We would like to thank J.~Ablinger, A.~Behring, A. De Freitas, K.~Sch\"onwald and J.~Vermaseren 
for discussions. This work was supported in part by the Austrian Science Fund (FWF) grant SFB F50 (F5009-N15), by the 
EU TMR network SAGEX Marie Sk\l{}odowska-Curie grant agreement No. 764850 and COST action CA16201: Unraveling new 
physics at the LHC through the precision frontier.


\end{document}